\documentclass[12pt]{article}
\usepackage{eqsection,subeqnarray,indent,amsfonts,amssymb}
\usepackage{bm}    
\usepackage{cite}  
\usepackage{epic,eepic}
\usepackage{graphicx}

\begin{document}

\bibliographystyle{plain}

\date{\vspace*{-4mm} December 30, 2003 \\
      revised October 19, 2004 \\[1.5mm]
      final revision February 26, 2005 \vspace*{-7mm}}

\title{\vspace*{-2.6cm}
       Spanning forests and the \\ $q$-state Potts model in the limit $q\to 0$
       \vspace*{-3mm}
      }

\author{
  {\small Jesper Lykke Jacobsen}               \\[-0.2cm]
  {\small\it Laboratoire de Physique Th\'eorique et Mod\`eles Statistiques}
                                               \\[-0.2cm]
  {\small\it Universit\'e Paris-Sud}           \\[-0.2cm]
  {\small\it B\^atiment 100}                   \\[-0.2cm]
  {\small\it F-91405 Orsay, FRANCE }           \\[-0.2cm]
  {\small\tt JACOBSEN@IPNO.IN2P3.FR}           \\[1mm]
  {\small Jes\'us Salas}                       \\[-0.2cm]
  {\small\it Grupo de Modelizaci\'on y Simulaci\'on Num\'erica}    \\[-0.2cm]
  {\small\it Universidad Carlos III de Madrid} \\[-0.2cm]
  {\small\it Avda.\  de la Universidad, 30}    \\[-0.2cm]
  {\small\it 28911 Legan\'es, SPAIN}           \\[-0.2cm]
  {\small\tt JSALAS@MATH.UC3M.ES}              \\[1mm]
  {\small Alan D.~Sokal}                       \\[-0.2cm]
  {\small\it Department of Physics}            \\[-0.2cm]
  {\small\it New York University}              \\[-0.2cm]
  {\small\it 4 Washington Place}               \\[-0.2cm]
  {\small\it New York, NY 10003 USA}           \\[-0.2cm]
  {\small\tt SOKAL@NYU.EDU}                    \\[-0.2cm]
  {\protect\makebox[5in]{\quad}}  
  \\
}

\maketitle
\thispagestyle{empty}   

\begin{abstract}
We study the $q$-state Potts model
with nearest-neighbor coupling $v=e^{\beta J}-1$
in the limit $q,v \to 0$ with the ratio $w = v/q$ held fixed.
Combinatorially, this limit gives rise to the generating polynomial
of spanning forests;
physically, it provides information about the Potts-model phase diagram
in the neighborhood of $(q,v) = (0,0)$.
We have studied this model on the square and triangular lattices,
using a transfer-matrix approach at both real and complex values of $w$.
For both lattices, we have computed the symbolic transfer matrices
for cylindrical strips of widths $2\leq L \leq 10$,
as well as the limiting curves ${\cal B}$
of partition-function zeros in the complex $w$-plane.
For real $w$, we find two distinct phases separated by
a transition point $w=w_0$,
where $w_0 =-1/4$ (resp.\ $w_0=-0.1753 \pm 0.0002$)
for the square (resp.\ triangular) lattice. 
For $w>w_0$ we find a non-critical disordered phase
that is compatible with the predicted asymptotic freedom as $w \to +\infty$.
For $w<w_0$ our results are compatible with
a massless Berker--Kadanoff phase with central charge $c=-2$
and leading thermal scaling dimension $x_{T,1} = 2$
(marginally irrelevant operator).
At $w=w_0$ we find a ``first-order critical point'':
the first derivative of the free energy is discontinuous at $w_0$,
while the correlation length diverges as $w\downarrow w_0$
(and is infinite at $w=w_0$).
The critical behavior at $w=w_0$ seems to be the same for both lattices
and it differs from that of the Berker--Kadanoff phase:
our results suggest that the central charge is $c=-1$,
the leading thermal scaling dimension is $x_{T,1}=0$,
and the critical exponents are $\nu=1/d=1/2$ and $\alpha=1$.  
\end{abstract}


\clearpage

\newcommand{\be}{\begin{equation}}
\newcommand{\ee}{\end{equation}}
\newcommand{\<}{\langle}
\renewcommand{\>}{\rangle}
\newcommand{\widebar}{\overline}
\def\reff#1{(\protect\ref{#1})}
\def\spose#1{\hbox to 0pt{#1\hss}}
\def\ltapprox{\mathrel{\spose{\lower 3pt\hbox{$\mathchar"218$}}
 \raise 2.0pt\hbox{$\mathchar"13C$}}}
\def\gtapprox{\mathrel{\spose{\lower 3pt\hbox{$\mathchar"218$}}
 \raise 2.0pt\hbox{$\mathchar"13E$}}}
\def\textprime{${}^\prime$}
\def\proof{\par\medskip\noindent{\sc Proof.\ }}
\def\qed{\hbox{\hskip 6pt\vrule width6pt height7pt depth1pt \hskip1pt}\bigskip}
\def\proofof#1{\bigskip\noindent{\sc Proof of #1.\ }}
\def\half{ {1 \over 2} }
\def\third{ {1 \over 3} }
\def\twothird{ {2 \over 3} }
\def\smfrac#1#2{\textstyle{#1\over #2}}
\def\smhalf{ \smfrac{1}{2} }
\newcommand{\real}{\mathop{\rm Re}\nolimits}
\renewcommand{\Re}{\mathop{\rm Re}\nolimits}
\newcommand{\imag}{\mathop{\rm Im}\nolimits}
\renewcommand{\Im}{\mathop{\rm Im}\nolimits}
\newcommand{\sgn}{\mathop{\rm sgn}\nolimits}
\newcommand{\tr}{\mathop{\rm tr}\nolimits}
\newcommand{\diag}{\mathop{\rm diag}\nolimits}
\newcommand{\Gal}{\mathop{\rm Gal}\nolimits}
\newcommand{\mycup}{\mathop{\cup}}
\newcommand{\Arg}{\mathop{\rm Arg}\nolimits}
\def\hboxscript#1{ {\hbox{\scriptsize\em #1}} }
\def\zhat{ {\widehat{Z}} }
\def\phat{ {\widehat{P}} }
\def\qtilde{ {\widetilde{q}} }
\newcommand{\mod}{\mathop{\rm mod}\nolimits}
\renewcommand{\emptyset}{\varnothing}

\def\scra{\mathcal{A}}
\def\scrb{\mathcal{B}}
\def\scrc{\mathcal{C}}
\def\scrd{\mathcal{D}}
\def\scrf{\mathcal{F}}
\def\scrg{\mathcal{G}}
\def\scrl{\mathcal{L}}
\def\scro{\mathcal{O}}
\def\scrp{\mathcal{P}}
\def\scrq{\mathcal{Q}}
\def\scrr{\mathcal{R}}
\def\scrs{\mathcal{S}}
\def\scrt{\mathcal{T}}
\def\scrv{\mathcal{V}}
\def\scrz{\mathcal{Z}}

\def\q{{\sf q}}

\def\Z{{\mathbb Z}}
\def\R{{\mathbb R}}
\def\C{{\mathbb C}}
\def\Q{{\mathbb Q}}

\def\T{{\mathsf T}}
\def\H{{\mathsf H}}
\def\V{{\mathsf V}}
\def\D{{\mathsf D}}
\def\J{{\mathsf J}}
\def\P{{\mathsf P}}
\def\QQ{{\mathsf Q}}
\def\RR{{\mathsf R}}

\def\bsigma{{\boldsymbol{\sigma}}}
\def\bone{{\mathbf 1}}
\def\vv{{\bf v}}
\def\uu{{\bf u}}
\def\w{{\bf w}}

\newtheorem{theorem}{Theorem}[section]
\newtheorem{definition}[theorem]{Definition}
\newtheorem{proposition}[theorem]{Proposition}
\newtheorem{lemma}[theorem]{Lemma}
\newtheorem{corollary}[theorem]{Corollary}
\newtheorem{conjecture}[theorem]{Conjecture}


\newenvironment{sarray}{
          \textfont0=\scriptfont0
          \scriptfont0=\scriptscriptfont0
          \textfont1=\scriptfont1
          \scriptfont1=\scriptscriptfont1
          \textfont2=\scriptfont2
          \scriptfont2=\scriptscriptfont2
          \textfont3=\scriptfont3
          \scriptfont3=\scriptscriptfont3
        \renewcommand{\arraystretch}{0.7}
        \begin{array}{l}}{\end{array}}

\newenvironment{scarray}{
          \textfont0=\scriptfont0
          \scriptfont0=\scriptscriptfont0
          \textfont1=\scriptfont1
          \scriptfont1=\scriptscriptfont1
          \textfont2=\scriptfont2
          \scriptfont2=\scriptscriptfont2
          \textfont3=\scriptfont3
          \scriptfont3=\scriptscriptfont3
        \renewcommand{\arraystretch}{0.7}
        \begin{array}{c}}{\end{array}}

\tableofcontents
\clearpage

%
%
\section{Introduction} \label{sec.intro}

\subsection{Phase diagram of the Potts model}  \label{sec1.1}

The Potts model \cite{Potts_52,Wu_82,Wu_84}
on a regular lattice ${\cal L}$ is characterized by two parameters:
the number $q$ of Potts spin states,
and the nearest-neighbor coupling $v = e^{\beta J}-1$.\footnote{
   Here we are considering only the {\em isotropic}\/ model,
   in which each nearest-neighbor edge is assigned the same coupling $v$.
   In a more refined analysis, one could put (for example)
   different couplings $v_1,v_2$ on the horizontal and vertical edges
   of the square lattice, different couplings $v_1,v_2,v_3$ on the
   three orientations of edges of the triangular or hexagonal lattice, etc.
}
Initially $q$ is a positive integer
and $v$ is a real number in the interval $[-1,+\infty)$,
but the Fortuin--Kasteleyn representation
(reviewed in Section~\ref{sec.setup.1})
shows that the partition function $Z_G(q,v)$ of the $q$-state Potts model
on any finite graph $G$ is in fact a {\em polynomial}\/ in $q$ and $v$.
This allows us to interpret $q$ and $v$ as taking arbitrary
real or even complex values,
and to study the phase diagram of the Potts model in
the real $(q,v)$-plane or in complex $(q,v)$-space.

According to the Yang--Lee picture of phase transitions \cite{Yang-Lee_52},
information about the possible loci of phase transitions can be obtained
by investigating the zeros of the partition function
for finite subsets of the lattice ${\cal L}$
when one or more physical parameters (e.g.\ temperature or magnetic field)
are allowed to take {\em complex}\/ values;
the accumulation points of these zeros in the infinite-volume limit
constitute the phase boundaries.
For the Potts model, therefore,
by studying the zeros of $Z_G(q,v)$ in complex $(q,v)$-space
for larger and larger pieces of the lattice ${\cal L}$,
we can learn about the phase diagram of the Potts model in
the real $(q,v)$-plane and more generally in complex $(q,v)$-space.

Since the problem of computing the phase diagram in
complex $(q,v)$-space is difficult, it has proven convenient
to study first certain ``slices'' through $(q,v)$-space,
in which one parameter is fixed (usually at a real value)
while the remaining parameter is allowed to vary in the complex plane.
Thus, the authors and others (notably Shrock and collaborators)
have in previous work studied the chromatic polynomial ($v=-1$),
which corresponds to the zero-temperature limit
of the Potts antiferromagnet\footnote{
   See \cite{transfer1} for an extensive list of references
   through December 2000;
   and see \cite{transfer2,transfer3} for more recent work.
};
the flow polynomial ($v=-q$), which is dual to the chromatic polynomial
\cite{Shrock_flow};
the $q$-plane behavior for fixed real $v$
in both the ferromagnetic and antiferromagnetic regions
\cite{Shrock_00_SQ,Chang_00_TRI,Chang_01_SQNNN,Chang_01_HEX,Chang_01_SQ,%
Tutte_sq,Tutte_tri};
and the $v$-plane behavior for fixed real $q$,
notably either $q=2,3,4,\ldots$
\cite{Matveev_95a,Matveev_96,Matveev_95b,%
Martin_86,Feldmann_97,Feldmann_98a,Feldmann_98b,Shrock_00_SQ,%
Chang_00_TRI,Chang_01_SQNNN,Chang_01_HEX,Chang_01_SQ,Tutte_sq,Tutte_tri},
$q \to 1$ \cite{Tutte_sq,Tutte_tri}
or $q=0$ \cite{Tutte_sq,Tutte_tri,Shrock_reliability}.

In this paper we will study yet another slice,
namely the limit $q,v \to 0$ with $w = v/q$ finite.\footnote{
   We stress that, despite the fact that $v \to 0$,
   this is {\em not}\/ an ``infinite-temperature'' limit
   in any relevant physical sense, because $v$ and $q$ are
   {\em simultaneously}\/ varying, and one must take account
   of the {\em joint}\/ effect of the two parameters.
   Indeed, it turns out that for $q$ small it is $w=v/q$, and not $v$ itself,
   that plays the role of an ``inverse temperature''.
   Thus, the radius of convergence of the small-$v$ expansion
   is asymptotically proportional to $|q|$ when $q \to 0$;
   that is why the small-$w$ expansion in the spanning-forest model
   is convergent for small $|w|$ but {\em not}\/ in general for all $w$
   (see Section~\ref{sec.smallw} and Appendix~\ref{sec_series_analysis}).
   In particular, phase transitions can and do occur at finite values
   of real or complex $w$, as we shall see in this paper.

   For the two-dimensional lattices considered here,
   it will turn out that there is no phase transition
   for {\em positive real}\/ $w$.
   But this expresses a deep fact about the critical behavior
   for these lattices, namely the asymptotic freedom \cite{Caracciolo_PRL}.
   The situation is likely to be quite different for
   three-dimensional lattices,
   for which there may well exist a ``ferromagnetic'' phase transition
   at some finite positive real $w_{\rm crit}$.
}
{}From a combinatorial point of view, this limit corresponds to
the generating polynomial of {\em spanning forests}\/
(see Section~\ref{sec.setup.2}).
{}From a physical point of view, this limit corresponds to
investigating the behavior of the phase diagram in a small neighborhood
of the point $(q,v)=(0,0)$ ---
more precisely, to investigating those phase-transition curves
that pass through $(q,v)=(0,0)$ with finite slope $w$.\footnote{
  By a standard duality transformation 
  [see \protect\reff{eq.duality}--\protect\reff{eq.duality3} below], 
  the spanning-forest model on the lattice $\mathcal{L}$ at parameter $w$
  is equivalent to the $q=0$ model on the dual lattice $\mathcal{L}^*$
  at parameter $v=1/w$. In particular, our results concerning the
  spanning-forest model on the square and triangular lattices can be 
  immediately translated into results for the $q=0$ Potts model at fixed $v$
  on the square and hexagonal lattices. We leave these straightforward 
  translations to the reader.
}
This limit takes on additional interest in light of the
recent discoveries \cite{Caracciolo_PRL}
that (a) it can be mapped onto a fermionic theory
containing a Gaussian term and a special four-fermion coupling,
and (b) this latter theory is equivalent,
to all orders in perturbation theory in $1/w$,
to the $N$-vector model at $N=-1$ with $\beta=-w$,
and in particular is perturbatively asymptotically free in two dimensions,
analogously to two-dimensional $\sigma$-models
and four-dimensional nonabelian gauge theories.

Further motivation for this study
comes from our ongoing work \cite{JSS_RGflow}
on the phase diagram and renormalization-group flows of
the Potts model on the square and triangular lattices.
These phase diagrams have been actively studied
(see e.g.\ \cite{Salas_97} for an extensive set of references);
but certain aspects of the phase diagram in the antiferromagnetic regime
remain unclear, notably on the triangular lattice.
Let us begin, therefore, by giving a brief summary
of what is known and what is mysterious.
We shall parametrize the interval $0 \le q \le 4$ by
\be
  q \;=\; 4 \cos^2 (\pi/\delta)
  \qquad\hbox{with } 2 \le \delta \le \infty  \;.
\label{def_delta}
\ee

\bigskip

\begin{figure}
\centering
\includegraphics[width=400pt]{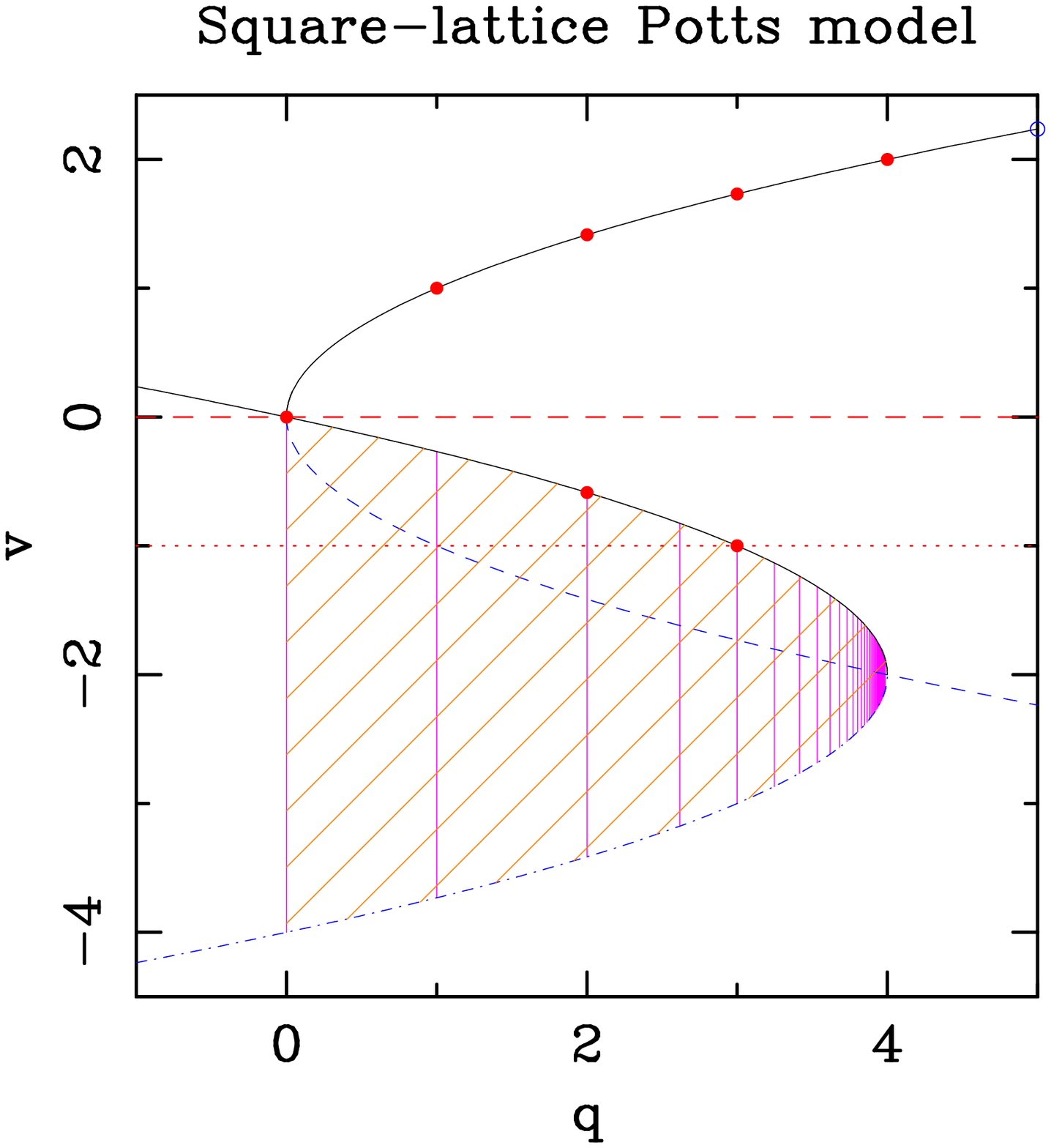}
\caption{\label{figure_sq}
Phase diagram for the square-lattice Potts model in the $(q,v)$-plane.
The solid curves show the ferromagnetic ($v>0$) and antiferromagnetic
($v < 0$) phase-transition curves.
The dots $\bullet$ (resp.\ circles $\circ$)
indicate the known second-order (resp.\  first-order) transition values
$v_c(q)$ for integer $q$.
The dashed and dot-dashed curves represent
(\ref{curve_sq_1}$_-$) and (\ref{curve_sq_2}$_-$), respectively.
The hatched region corresponds to the conjectured Berker--Kadanoff phase.
The horizontal dashed line corresponds to infinite temperature ($v=0$),
and the horizontal dotted line corresponds to the zero-temperature
antiferromagnet ($v=-1$). The pink vertical lines show the Beraha numbers
$q=4\cos^2(\pi/n)$ ($n=2,3,\ldots$); at these values the Berker--Kadanoff
phase is {\em not} defined.
}
\end{figure}

{\em Square lattice}\/:
Baxter \cite{Baxter_book,Baxter_82} has determined the exact free energy
for the square-lattice Potts model on two special curves in the $(q,v)$-plane:
\begin{eqnarray}
v &=&    \pm \sqrt{q}   \label{curve_sq_1} \\
v &=& -2 \pm \sqrt{4-q} \label{curve_sq_2}
\end{eqnarray}
These curves are plotted in Figure~\ref{figure_sq}.
Curve (\ref{curve_sq_1}$_+$) is known \cite{Baxter_book}
to correspond to the ferromagnetic phase-transition point,
whose critical behavior is by now well understood
\cite{Nienhuis_84,Itzykson_collection,DiFrancesco_97,Henkel_99}:
for $0 \le q \le 4$ the critical ferromagnet
is described by a conformal field theory (CFT)
\cite{Itzykson_collection,DiFrancesco_97,Henkel_99}
with central charge
\be
   c  \;=\;  1 - {6 \over \delta(\delta-1)}
 \label{def_c_ferro}
\ee
and thermal scaling dimensions\footnote{
   For a scaling operator ${\cal O}_i$,
   we denote by $y_i$ its renormalization-group eigenvalue,
   so that $y_i > 0$ (resp.\ $=0$, $<0$) corresponds
   to a relevant (resp.\ marginal, irrelevant) operator.
   Then $x_i = d-y_i$ is the corresponding scaling dimension,
   where $d$ denotes the system's spatial dimension.
   When $d=2$ (as in this paper),
   we can use the language of conformal field theory (CFT):
   if a scaling operator ${\cal O}_i$
   has conformal weight $h_i$ (resp.\ $\bar{h}_i$) with respect to
   the holomorphic (resp.\ antiholomorphic) variable,
   then $x_i = h_i + \bar{h}_i$ is the scaling dimension
   and $s_i = h_i - \bar{h}_i$ is the spin.
}${}^,$\footnote{
   See \protect\cite[Appendix A.1]{Salas-Sokal_potts4}
   for a convenient summary of the critical properties
   of the two-dimensional ferromagnetic Potts model.
   The parameter $x$ used there is related to $\delta$ by $x=-2/\delta$.
   Note also that there is a typographical error in equation (A.10)
   of \protect\cite{Salas-Sokal_potts4}, which should read
   $\Delta_{r,s}=([2(s-r)+s x]^2 -x^2)/(8[(2+x)]$.
      \par\quad
   In CFT language, the $n$-th thermal scaling dimension $x_{T,n}$
   is twice the conformal weight of the operator $\phi_{n+1,1}$
   obtained by repeated fusion of the fundamental thermal operator
   $\phi_{2,1}$.
   Please note that for {\em integer}\/ $\delta \ge 4$,
   the central charge \reff{def_c_ferro} coincides with that
   of a unitary minimal model,
   for which some zero coefficients appear in the usual Coulomb-gas
   fusion rules.  This means that the operator $\phi_{n+1,1}$ will
   be present in the Kac table only for $n=1,2,\ldots,\delta-3$.
   In particular, for $q=2$ (resp.\ $q=3$),
   the operator with exponent \reff{def_xT_ferro}
   should be a {\em local}\/ observable in terms of the Potts
   spin variables only for $n=1$ (resp.\ $n=1,2,3$).
   However, nothing would seem to prevent it from being observable
   in terms of the Fortuin--Kasteleyn clusters
   (see Section~\ref{sec.setup.1} below),
   whose definition is nonlocal in terms of the spins.
   Indeed, it is conceivable that the $n=2$ operator
   is indeed present in the two-dimensional Ising model
   and causes $L^{-4/3}$ corrections to scaling
   in the Fortuin--Kasteleyn bond observables
   \cite{Nienhuis_82,Blote_88,Salas-Sokal_Ising_v1}.
}
\be
   x_{T,n}  \;=\;  {n(n\delta+2) \over 2(\delta-1)}
   \qquad [n=1,2,3,\ldots] \;.
 \label{def_xT_ferro}
\ee

Baxter \cite{Baxter_82} conjectured that curve (\ref{curve_sq_2}$_+$)
with $0 \le q \le 4$
corresponds to the antiferromagnetic critical point.
For $q=2$ this gives the known exact value \cite{Onsager_44};
for $q=3$ it predicts a zero-temperature critical point ($v_c = -1$),
in accordance with strong analytical and numerical evidence
\cite{Lenard_67,Baxter_70,Nijs_82,Burton_Henley_97,%
Salas_98,deQueiroz_99,Ferreira_99,Cardy_01};
and for $q>3$ it predicts that the putative critical point lies in the
unphysical region $v < -1$,
so that the entire physical region $-1 \le v \le 0$
lies in the disordered phase,
in agreement with numerical evidence for $q=4$ \cite{Ferreira_99}.

Saleur \cite{Saleur_90,Saleur_91} pursued the investigation
of the phase diagram and critical behavior in the
antiferromagnetic ($-1 \le v \le 0$) and unphysical ($v < -1$) regimes.
Firstly, he extended Baxter's conjecture by suggesting \cite{Saleur_91}
that the critical antiferromagnetic Potts model (\ref{curve_sq_2}$_+$)
with $0 \le q \le 4$ is described by a CFT with central charge
\be
  c \;=\; 2 - 6/\delta \;.
\label{def_c_antiferro}
\ee
Furthermore, Saleur conjectured that the
leading thermal scaling dimension $x_T$ in this CFT is given by
\be
x_T \;=\; 4/\delta  \;,
     \label{def_xT_antiferro}
\ee
so that the associated critical exponent is
\be
\nu \;\equiv\; {1 \over 2 - x_T} \;=\; {\delta \over 2(\delta-2)}  \;.
   \label{def_nu_antiferro}
\ee
This conjecture agrees with the known value at $q=2$ (namely, $\nu=1$);
on the other hand, it predicts the value $\nu=3/4$ at $q=3$,
which is incorrect \cite{Ferreira_99,Cardy_01}.
However, after discussing the representation of the 3-state model as a
free bosonic field with $g=1/3$, Saleur finds another operator with
$x_T = 3/2$ and $\nu=2$, which is the correct answer
\cite{Ferreira_99,Cardy_01}.
Clearly, this latter operator (and not the initially predicted one)
is the leading thermal operator in the square-lattice
3-state Potts antiferromagnet \cite{Cardy_01}.\footnote{
  In the Coulomb gas picture \protect\cite{Nienhuis_84}
  employed by Saleur \protect\cite{Saleur_91},
  there are two basic types of operators:
  electric (or vertex) operators of electric charge $e$,
  and magnetic (or vortex) operators of magnetic charge $m$.
  The thermal scaling field associated to a general operator with electric 
  charge $e \in \Z/3$ and magnetic charge $m \in 3 \Z$ is given by 
  $$
  x_T \; = \; {e^2 \over 2g} \,+\, {g m^2 \over 2}
  $$
  where $g$ is the coupling constant of the free bosonic field
  onto which the original model renormalizes.
  In the square-lattice 3-state Potts antiferromagnet,
  $g=1/3$ \protect\cite{Saleur_91}.
  Saleur initially conjectured the thermal operator
  to be an electric operator of charge $e=2/3$,
  thus leading to $x_T = 2/3$ and $\nu=3/4$.
  The alternate conjecture ($x_T=3/2$ and $\nu=2$) comes from identifying the
  thermal operator with a magnetic one with charge $m=3$.
  This result agrees with the identification (made in
  Ref.~\protect\cite{Cardy_01}) of the thermal operator as a {\em vortex}\/
  operator with the smallest possible topological charge. (The normalization
  conventions in Refs.~\protect\cite{Saleur_91} and \protect\cite{Cardy_01}
  differ: in Ref.~\protect\cite{Cardy_01}, one has
  $x_T = {1 \over 4\pi} ( \alpha^2/K + K \beta^2 )$
  with the correspondence $\alpha = \pi e$, $\beta = 2m$ and $K = \pi g/2$.)
}

It is not clear whether \reff{def_xT_antiferro}/\reff{def_nu_antiferro}
should be expected to be correct for $q \neq 3$.
We defer testing the validity of these expressions, as a function of $q$,
to a subsequent paper \cite{JSS_RGflow}. However, in the limit $q\to 0$
along the curve (\ref{curve_sq_2}$_+$), we shall find that there are
{\em at least two}\/ thermal-type operators that are more relevant than
\reff{def_xT_antiferro},
namely one with $x_T = 0$ and another with $x_T \approx 0.3$--0.4
(see Sections~\ref{sec.res.sq3} and \ref{sec.res.tri3} for more details).

Saleur \cite{Saleur_91} also investigated the meaning of
the other two special curves,
(\ref{curve_sq_1}$_-$) and (\ref{curve_sq_2}$_-$).
He suggested that there exists a Berker--Kadanoff phase \cite{Berker_80}
--- i.e.\ a massless low-temperature phase with algebraically decaying
correlation functions ---
extending between the curves (\ref{curve_sq_2}$_\pm$)
in the range $0 \le q \le 4$
[i.e.\ throughout the hatched region in Figure~\ref{figure_sq}]
except when $q$ is a Beraha number $B_n = 4 \cos^2 (\pi/n)$
[$n=2,3,4,\ldots$, corresponding to the pink vertical lines in
Figure~\ref{figure_sq}],
and that the critical behavior of this phase
is determined by an attractive fixed point
lying on the unphysical self-dual line (\ref{curve_sq_1}$_-$).
He further conjectured that the model on the line (\ref{curve_sq_1}$_-$)
with $0 \le q \le 4$
--- and hence throughout the Berker--Kadanoff phase ---
is described by a CFT with central charge
\be
  c \;=\; 1 - {6 (\delta-1)^2 \over \delta}   \;,
\label{def_c_BK}
\ee
provided that $\delta$ is not an integer.
Finally, he conjectured that the
leading thermal scaling dimension in this CFT is
\be
x_T \;=\; {3\delta-2\over 2}  \;.
      \label{def_xT_BK} \\
\ee
Since $x_T \geq 2$, the energy is an {\em irrelevant}\/ operator in
this phase (except at $q=0$, where it is marginal),
in accordance with the fact that there is
an entire {\em interval}\/ of critical points,
all governed by a single renormalization-group fixed point.\footnote{
   It is worth noticing \cite{Saleur_91} that there is a unified
   way of looking at the ferromagnetic and Berker--Kadanoff phases as
   continuations of one another.  Let us parametrize
   (\ref{curve_sq_1}$_\pm$) by
   $q = 4 \cos^2(\pi u/2)$ and $v  = 2 \cos(\pi u/2)$,
   with $0 \le u \le 1$ for the ferromagnetic critical curve
   and $1 \le u \le 2$ for the Berker--Kadanoff curve.
   (We thus have $u=2/\delta$ for the ferromagnetic phase,
   and $u = 2 - 2/\delta$ for the Berker--Kadanoff phase.)
   Then we can write the central charge as
   $c = 1 - 3u^2/(2-u)$, and the thermal scaling dimensions read
   $x_{T,n} = n(n+u)/(2-u)$.
   Moreover, continuing these formulae to $-1 \le u \le 0$
   gives the central charge and thermal scaling dimensions
   for the \emph{tricritical} Potts model.
   (This variable $u$ corresponds to the negative of the variable $x$
    employed in \protect\cite[Appendix A.1]{Salas-Sokal_potts4}.)
}

Finally, (\ref{curve_sq_2}$_-$) is the dual of
the antiferromagnetic critical curve (\ref{curve_sq_2}$_+$).
Therefore, the transfer matrices for (\ref{curve_sq_2}$_\pm$)
with cylindrical boundary conditions
are identical up to multiplication by a constant.
If we assume (as seems likely) that the different endgraphs
needed in the two cases do not lead to any zero amplitudes, the theories
(\ref{curve_sq_2}$_\pm$) should therefore be completely equivalent;
in particular, they should have the same central charge
and the same thermal scaling dimensions.
(However, a local operator in one theory could correspond to
 a nonlocal operator in the dual theory.)
This equivalence is corroborated by the fact that in CFT,
the complete operator content is linked to the modular-invariant
partition function on the {\em torus}\/ \cite{DiFrancesco_97};
obviously, in this geometry the lattice coincides with its dual.

Let us tentatively accept these conjectures
and determine their implications for the limit $q,v\to 0$
with $w = v/q$ fixed.
The self-dual curves (\ref{curve_sq_1}$_\pm$) pass through $(q,v)=(0,0)$
with slope $w=\pm\infty$,
while the antiferromagnetic critical curve (\ref{curve_sq_2}$_+$)
passes through $(q,v)=(0,0)$ with slope $w=-1/4$.
Thus, all values $-1/4 < w < +\infty$ are predicted to lie in the
high-temperature phase and hence be noncritical,
while all values $-\infty < w < -1/4$ are predicted to lie in the
Berker--Kadanoff phase and hence be critical with
central charge $c=-2$ and leading thermal scaling dimension $x_T=2$
(corresponding to a marginally irrelevant operator)
as given by \reff{def_c_BK}/\reff{def_xT_BK} with $\delta=2$.
The transition between these two behaviors occurs at
\be
  w \;=\; w_0({\rm sq}) \;=\; -1/4  \;,
\label{def_w0_sq}
\ee
where we expect a critical theory with
central charge $c=-1$ as given by \reff{def_c_antiferro} with $\delta=2$.
If \reff{def_xT_antiferro}/\reff{def_nu_antiferro} is correct,
we should expect the thermal operator at $w = -1/4$ to be marginal:
$x_T = 2$ and hence $\nu = \infty$.
But as we shall see (Sections~\ref{sec.res.sq3} and \ref{sec.res.tri3}),
the prediction \reff{def_xT_antiferro}/\reff{def_nu_antiferro}
is {\em wrong}\/,
and a more likely scenario is $x_T=0$, so that $\nu = 1/2$.
Finally, $w = +\infty$ is a ferromagnetic critical point
with central charge $c=-2$ and leading thermal scaling dimension $x_T=2$
(corresponding to a marginally relevant operator),
as given by \reff{def_c_ferro}/\reff{def_xT_ferro} with $\delta=2$.
In a separate paper \cite{Caracciolo_PRL}
we have shown that the $w = +\infty$ theory
can be represented in terms of a pair of free scalar fermions,
while the theory at finite $w$
can be mapped onto a fermionic theory that contains a Gaussian term
and a special four-fermion coupling;
furthermore, this latter theory is perturbatively asymptotically free
and is in fact equivalent, to all orders in perturbation theory in $1/w$,
to the $N$-vector model at $N=-1$ with $\beta=-w$.

\bigskip

\begin{figure}
\centering
\includegraphics[width=400pt]{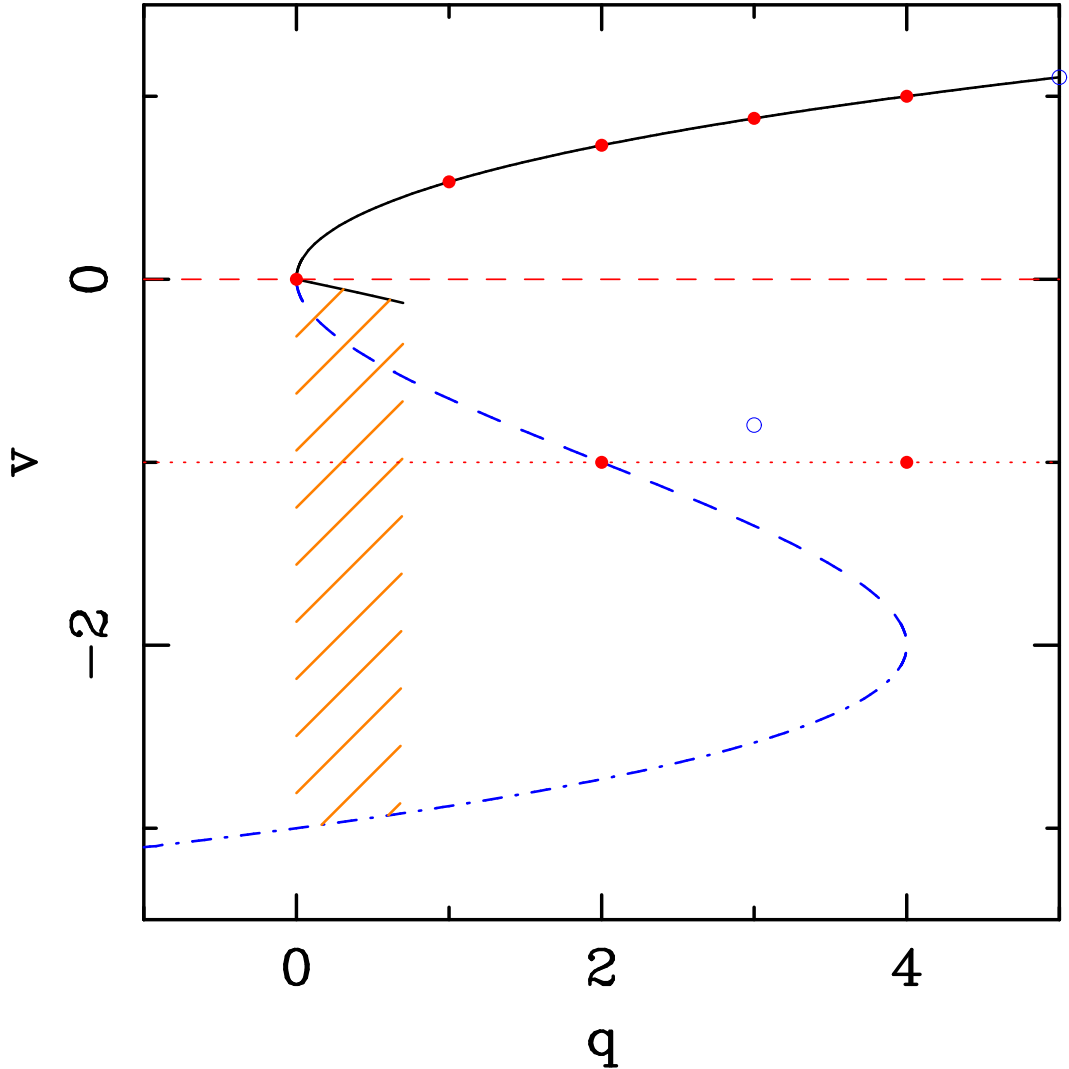}
\caption{\label{figure_tri}
Phase diagram for the triangular-lattice Potts model in the $(q,v)$-plane.
The upper solid curve shows the ferromagnetic ($v>0$) phase-transition curve.
The dots $\bullet$ (resp.\ circles $\circ$)
indicate the known second-order (resp.\  first-order) transition values
$v_c(q)$ for integer $q$.
The lower solid curve shows the hypothesized new phase-transition curve
for small positive $q$, as estimated numerically in this paper
(see Section~\ref{sec.res.tri1}).
The dashed and dot-dashed curves represent
the middle and lower branches of (\protect\ref{curve_tri_1}), respectively.
The hatched region corresponds to (a portion of)
the conjectured Berker--Kadanoff phase.
The horizontal dashed line corresponds to infinite temperature ($v=0$),
and the horizontal dotted line corresponds to the zero-temperature
antiferromagnet ($v=-1$).
}
\end{figure}

{\em Triangular lattice}\/:
Baxter and collaborators \cite{Baxter_78,Baxter_86,Baxter_87} have determined
the exact free energy for the triangular-lattice Potts model on two special
curves in the $(q,v)$-plane:
\begin{eqnarray}
v^3+3v^2 -q  &=&  0   \label{curve_tri_1} \\
v &=& -1              \label{curve_tri_2}
\end{eqnarray}
These curves are plotted in Figure~\ref{figure_tri}.
The uppermost branch ($v\geq 0$) of curve \reff{curve_tri_1} is known
to correspond to the ferromagnetic phase-transition point,
whose critical behavior is identical (thanks to universality)
to that of the square-lattice ferromagnet
\cite{Itzykson_collection,DiFrancesco_97,Henkel_99,Nienhuis_84}.
The significance of the other two branches of \reff{curve_tri_1}
is not clear.
The curve \reff{curve_tri_2} is not critical in general,
but it does contain the zero-temperature critical points
at $q=2$ \cite{Stephenson} and $q=4$ \cite{Henley}
(see \cite{transfer3} for further discussion and references).
The existence of an antiferromagnetic critical curve
for the triangular-lattice Potts model is at present not established;
{\em a fortiori} its location, if it exists, is unknown.

Consideration of the renormalization-group flow for the
triangular-lattice Potts model has led the present authors to
hypothesize \cite{JSS_RGflow}
that there exists an additional curve of RG fixed points
--- repulsive in the temperature direction --- emanating from the
point $(q,v)=(0,0)$ and extending into the antiferromagnetic region
$v<0$ as $q$ grows.
In a subsequent paper \cite{JSS_RGflow}
we shall present numerical estimates of the location and properties
(e.g.\ critical vs.\ first-order)
of this new phase-transition curve and
discuss how it combines with the known antiferromagnetic
critical points at $q=2,4$ with $v=-1$, and with the first-order phase
transition at $q=3$, $v \approx -0.79692 \pm 0.00003$
\cite{Adler_95,Tutte_tri}, to form a consistent phase diagram.
One interesting issue is whether this curve (or part of it)
might constitute a locus of critical points,
in analogy with the case of the square lattice.
For the time being we limit ourselves to the conjecture, in analogy with the
square-lattice phase diagram, that the region lying in the range $0
\le q \le 4$ between the new phase-transition curve and the lower branch of
curve \reff{curve_tri_1} will constitute a Berker--Kadanoff phase
whose critical behavior is determined by an attractive fixed point
lying on the middle branch of curve \reff{curve_tri_1}.

Note that Saleur \cite[p.~669]{Saleur_90} expects universality both for
the Berker--Kadanoff phase and for the critical theories forming its
upper and lower boundaries. This would suggest, in particular, that
the central charge in the Berker--Kadanoff phase of the triangular-lattice
model might also be given by \reff{def_c_BK},
and that the thermal scaling dimension might be given by \reff{def_xT_BK}.
We have numerical evidence of the former result, which will be published
elsewhere \cite{JSS_RGflow}.
Moreover, if Saleur's conjecture is true, one would expect the lower
branch of curve \reff{curve_tri_1} to be governed by the same critical
continuum theory as the curve (\ref{curve_sq_2}${}_\pm$) of the square-lattice
Potts model.

Let us denote by $w_0({\rm tri})$ the slope at $q=0$
of the new phase-transition curve.
Then, if we consider the limit $q,v\to 0$ with $w=v/q$ fixed,
we find two different regimes:
all values $w_0({\rm tri}) < w < +\infty$ are predicted to lie in the
high-temperature phase and hence be noncritical,
while all values $-\infty < w < w_0({\rm tri})$ are predicted to lie in the
Berker--Kadanoff phase and hence be critical
with central charge $c=-2$ as given by \reff{def_c_BK} with $\delta=2$.
The transition between these behaviors occurs at $w = w_0({\rm tri})$,
which may possibly constitute a critical theory of unknown type.

Let us observe, finally, that the analytical results of \cite{Caracciolo_PRL}
show that the conjectured universality of the Berker--Kadanoff phase
{\em does}\/ hold at least in the limit $q,v\to 0$ with $w=v/q$ fixed.
Indeed, for all two-dimensional lattices, the Berker--Kadanoff phase
at $-\infty < w < w_0$ is simply the $c=-2$ theory
of a pair of free scalar fermions, perturbed by a four-fermion operator
that is (in this phase) marginally irrelevant.

\bigskip

{\bf Remark.}
It should be stressed that the Potts spin model has a
probabilistic interpretation (i.e., has nonnegative weights)
only when $q$ is a positive integer and $v \ge -1$.
Likewise, the Fortuin--Kasteleyn random-cluster model
[cf.\ \reff{eq1.2} below],
which reformulates the Potts model and extends it to noninteger $q$,
has a probabilistic interpretation only when $q \ge 0$ and $v \ge 0$
(or in the limit considered here, $w \ge 0$).
In all other cases, the model belongs to the ``unphysical'' regime
(i.e., the weights can be negative or complex),
and the ordinary statistical-mechanical properties need not hold.
For instance, the free energy need not possess the usual convexity properties;
the leading eigenvalue of the transfer matrix need not be simple;
and phase transitions can occur even in one-dimensional systems
with short-range interactions.\footnote{
   For a recent pedagogical discussion of the conditions under which
   one-dimensional systems with short-range interactions
   can or cannot have a phase transition, see \cite{Cuesta_04}.
}
Nevertheless, there is a long history of studying statistical-mechanical
models in ``unphysical'' regimes:
examples include
the hard-core lattice gas at its negative-fugacity critical point
\cite{Kurtze_79,Shapir_82,Dhar_83,Poland_84,Baram_87,Guttmann_87,%
Lai_95,Park_99,Todo_99,Brydges_03a,Brydges_03b};
the closely related \cite{Kurtze_79,Shapir_82,Lai_95,Park_99}
problem of the Yang--Lee edge singularity
\cite{Yang-Lee_52,Lee_Yang,Kortman_71,Fisher_78,Alcantara_81,Parisi_81,%
Cardy_85,Itzykson_86};
and the low-temperature $N$-vector model at $N<1$,
with application to dense polymers
\cite{Wheeler_81,Gujrati_81,Gujrati_82,Obukhov_82,%
Nienhuis_82b,Griffiths_83,Gujrati_85,Gaspari_86,%
Duplantier_86,Saleur_87,Duplantier_87,Duplantier_88}.
Indeed, the previously cited papers of
Baxter \cite{Baxter_82,Baxter_86,Baxter_87}
and Saleur \cite{Saleur_90,Saleur_91}
deal in part with ``unphysical'' regimes in the Potts model.
And though one must be especially careful in such studies,
it generally turns out that the ``unphysical'' regime
can be understood using the standard tools of statistical mechanics,
appropriately modified.
In particular, conformal field theory (CFT) seems to apply
also in the ``unphysical'' regime,
although there is (as yet) no rigorous understanding
of why this should be the case:
well-studied examples include
the Yang--Lee edge singularity \cite{Cardy_85,Itzykson_86}
and dense polymers \cite{Duplantier_86,Saleur_87,Duplantier_87}.
Note that the ``unphysical'' nature of these
models means that the corresponding CFT is non-unitary.

Some aspects of the studies made in the present paper
of the regime $w < 0$ in the spanning-forest model
--- notably, Sections~\ref{sec.res.sq2b}, \ref{sec.res.sq3},
   \ref{sec.res.tri2b} and \ref{sec.res.tri3} ---
must therefore be understood as relying implicitly on
such a {\em conjectured extension}\/ of conformal field theory,
analogously to the just-cited studies.
On the other hand, the internal consistency of our results
provides additional evidence for the validity of such an extension.

\subsection{Outline of this paper}   \label{sec1.2}

The purpose of this paper is to shed light on these phase diagrams
in the neighborhood of $(q,v)=(0,0)$
by studying the $q=0$ Potts-model partition function in the
complex $w$-plane for lattice strips of width $m$ and length $n$,
using a transfer-matrix method.
For fixed width $m$ and arbitrary length $n$,
this partition function can be expressed via a transfer matrix of
fixed size $M\times M$ (which unfortunately grows rapidly with
the strip width $m$):
\begin{subeqnarray}
   Z_{m \times n}(w)  & = &   \tr[ A(w) \, T(w)^{n-1} ]  \\[2mm]
               & = &   \sum\limits_{k=1}^{M}
                           \alpha_k(w) \, \lambda_k(w)^{n-1}  \;.
\label{general_form_P}
\end{subeqnarray}
Here the transfer matrix $T(w)$ and the boundary-condition matrix $A(w)$
are polynomials in $w$, so that the eigenvalues $\{ \lambda_k \}$ of $T$
and the amplitudes $\{ \alpha_k \}$ are algebraic functions of $w$.
We can of course use $T(w)$ and $A(w)$ to compute the zeros of the
partition function for any {\em finite}\/ strip $m\times n$;
but more importantly, we can compute
the accumulation points of these zeros in the limit $n\rightarrow\infty$,
i.e.~for the case of an {\em semi-infinite}\/ strip
\cite{Beraha_79,Beraha_80,Shrock_97a,Shrock_98a,transfer1,transfer2,transfer3}.
According to the Beraha--Kahane--Weiss theorem
\cite{BKW_75,BKW_78,Sokal_chromatic_roots},
the accumulation points of zeros when $n\rightarrow\infty$ can either
be isolated limiting points (when the amplitude associated to the
dominant eigenvalue vanishes, or when all eigenvalues vanish simultaneously)
or belong to a limiting curve ${\cal B}$
(when two dominant eigenvalues cross in modulus).
As the strip width $m$ tends to infinity, the curve $\scrb = \scrb_m$
is expected to tend to a thermodynamic-limit curve $\scrb_\infty$,
which we interpret as a phase boundary in the complex $w$-plane.
In particular, $\scrb_\infty$ is expected to cross the real axis
precisely at the physical phase-transition point $w_0$.

Our approach is, therefore, to compute the curves $\scrb_m$
up to as large a value of $m$ as our computer is able to handle,
and then extrapolate these curves to $m = \infty$.
One output of our study is a numerical estimate of $w_0$.
For the square lattice, we find
\be
w_0({\rm sq}) \;=\; -0.2501 \pm 0.0002   \;,
\ee
in excellent agreement with the prediction $w_0({\rm sq}) = -1/4$.
We are therefore justified in assuming,
in our subsequent analysis, that $w_0({\rm sq}) = -1/4$ exactly.
For the triangular lattice, we find
\be
w_0({\rm tri}) \;=\; -0.1753 \pm 0.0002 \;.
\ee
We do not yet know whether the exact value of $w_0({\rm tri})$
is given by any simple closed-form expression.
Nor do we know whether there exists a simple exact formula
for the location of the critical curve $\scrb_\infty$
in the complex $w$-plane,
for either the square or triangular lattice.

In order to shed further light on the phase diagram
of the square- and triangular-lattice systems,
we have studied the free energy (and its derivatives with respect to $w$),
the central charge $c$ and the thermal scaling dimension $x_T$
as a function of $w$ on the {\em real}\/ $w$-axis.
We find that at $w=w_0$ there is a {\em first-order}\/ phase transition:
the free energy has a discontinuous first derivative with respect to $w$
at $w=w_0$. From our numerical work it is not clear whether the discontinuity
in this first derivative is finite or infinite:
the analysis seems to favor a finite limit, but a weak divergence such as
$f'(w) \sim \log (w-w_0)$ or $f'(w) \sim \log \log (w-w_0)$
is also possible. The two phases are characterized as follows:

\begin{itemize}
\item $w<w_0$: In this case the system is critical and it can be described by
    a conformal field theory with central charge $c=-2$.
    The leading thermal scaling exponent is $x_T=2$, so that the energy
    is a marginal operator.
    Both results agree with Saleur's prediction
    \reff{def_c_BK}/\reff{def_xT_BK} for the Berker--Kadanoff phase.
\item $w>w_0$: In this case the system is non-critical,
    i.e., the correlation length is finite.
\end{itemize}

The behavior at $w=w_0$ is rather special.
On the one hand, it is a coexistence point of two different phases;
on the other hand, it is itself a critical point,
which belongs to a {\em different}\/ universality class from that
of the Berker--Kadanoff phase,
namely the one corresponding to the $q \to 0$ limit
along the antiferromagnetic critical curve (\ref{curve_sq_2}$_+$).
Thus, at least for the square lattice, we expect that this point
will be described by a conformal field theory of central charge $c=-1$
in accordance with Saleur's prediction \reff{def_c_antiferro}.

Our numerical conclusions concerning the behavior at $w=w_0$
are drawn principally from the results on the square lattice
(as in this case we know the exact location of $w_0$),
but we expect them to hold also for the triangular lattice.
On the square lattice we have clear evidence that $w_0({\rm sq})=-1/4$
is indeed critical, and we can give rough estimates of the
central charge and thermal scaling dimensions:
\begin{itemize}
   \item[1)] Extrapolation of small-$w$ expansions using differential
      approximants (Appendix~\ref{sec_series_analysis})
      shows that the second derivative of the free energy
      diverges for $w\downarrow w_0$ as ${(w-w_0)}^{\approx -0.91(2)}$.
      This exponent is close to the theoretical prediction $\alpha=1$
      for a first-order critical point (see below),
      i.e.\ $f''(w) \sim (w-w_0)^{-1}$,
      and is consistent with it if one does not take the alleged
      error bar too seriously
      (the error estimates in series extrapolation have no strong
       theoretical basis).
      We have checked (Section~\ref{sec.res.sq2a})
      that our data from finite-width strips
      are consistent with this latter behavior,
      possibly modified by a multiplicative logarithmic correction.
   \item[2)] The central charge at $w=w_0$ is different from both
      the Berker--Kadanoff value ($c=-2$) and the noncritical value ($c=0$).
      Indeed, we get estimates around $c \approx -1.3$,
      which seem to be tending roughly towards $c=-1$
      as the strip width grows (Section~\ref{sec.res.sq2b}).
   \item[3)] The correlation length at $w=w_0$ is infinite for all even
      strip widths, reflecting the fact that $w=w_0$ is
      an exact endpoint of the limiting curve ${\cal B}$
      for even widths.
      For odd strip widths, the correlation length at $w=w_0$
      is finite and seems to be tending to infinity
      as the strip width tends to infinity (Section~\ref{sec.res.sq3}).
   \item[4)] Because the correlation length at $w=w_0$ is infinite
      for all even strip widths,
      we conclude from standard CFT arguments \cite{Cardy_84}
      that the leading thermal scaling dimension $x_{T,1}$ equals 0,
      and hence that $\nu=1/2$ (Section~\ref{sec.res.sq3}).
      The scaling law $d\nu=2-\alpha$ then yields the
      specific-heat exponent $\alpha=1$, in agreement with the
      results from small-$w$ expansions.  
   \item[5)] The second thermal scaling dimension $x_{T,2}$ can be obtained 
      from the second gap for even widths (Section~\ref{sec.res.sq3}).
      We obtain a value  $x_{T,2}\ltapprox 0.47$ that is clearly different
      from the Berker--Kadanoff value $x_{T,1}=2$.
      This result is close to the one obtained from the first gap
      for odd widths: $x_{T,2}\ltapprox 0.37$.
\end{itemize}

In conclusion, the phase transition at $w=w_0$ is rather unusual.
The point $w=w_0$ is what Fisher and Berker \cite[pp.~2510--2511]{Fisher_82}
have called a {\em first-order critical point}\/:
namely, it is both a first-order transition point
(the first derivative of the free energy is discontinuous at $w=w_0$)
and critical (the correlation length is infinite for $w\leq w_0$
and diverges as $w\downarrow w_0$).
The critical exponents take the values $\alpha=1$ and $\nu=1/d$,
in agreement both with finite-size-scaling theory for
first-order transitions ($\nu=1/d$ and $\alpha/\nu=d$)
\cite{Fisher_82,Borgs-Kotecky}
and with the hyperscaling law $d\nu=2-\alpha$ for critical points.\footnote{
   Please note that any two of the four equations
   $\alpha=1$, $\nu=1/d$, $\alpha/\nu=d$ and $d\nu=2-\alpha$
   imply the other two.
}

This sort of phase transition seems to be rare;  indeed, we are aware of
only two other equilibrium-model examples:

1) {\em One-dimensional $q$-state clock model with a $\theta$-term}\/
\cite{asorey_91,asorey_93a,asorey_93b,asorey_94}.
In this model, the first derivative of the free energy has a
finite discontinuity at the transition points,
and the correlation length diverges there with exponent $\nu = 1/d=1$.
However, the specific heat does not diverge at the transition points:
it is a discontinuous but bounded function of the temperature
(see \cite{asorey_94} for a computation of the specific heat
 in the limit $q\to\infty$). 
This behavior does not contradict the Fisher--Berker scaling theory for 
first-order critical points \cite[p.~2511]{Fisher_82}:
``first-order critical points are no more than ordinary critical points in 
which either one or both of the relevant eigenvalue exponents attains its 
thermodynamically allowed limiting value $\lambda=d$''.
Unfortunately the magnetic exponent 
was not considered in ref.~\cite{asorey_94}.  

2) {\em Six-vertex model}\/
(which contains the KDP model of a ferroelectric \cite{Wu_67,Lieb_67}
 as a special case).\footnote{
   We are grateful to an anonymous referee for bringing this model
   to our attention.
}
We follow the notations of Baxter's book
\cite[Sections 8.10 and 8.11]{Baxter_book}.
The transition between any of the two ferroelectrically ordered phases
(regimes I and II in Baxter's notation) and the critical phase (regime III)
is characterized by the following properties:
\begin{itemize}
  \item[a)] The first derivative of the free energy with respect to
      the temperature, $f'_{6V}$, has a finite jump discontinuity
      on the transition line.
  \item[b)] The second derivative of the free energy with respect to
      the temperature, $f''_{6V}$, is identically zero in the
      ferroelectrically ordered phases and diverges like $(T-T_c)^{-1/2}$
      as we approach the transition line from the critical phase.
  \item[c)] The correlation length is infinite in the critical phase
      (i.e., the correlations decay algebraically to zero)
      and identically zero in the ferroelectrically ordered phases.
  \item[d)] The electric polarization is identically zero
      in the critical phase and identically 1 in the
      ferroelectrically ordered phases.
\end{itemize}
Because of (a) and (b), it is unclear whether we should say
$\alpha=1$ or $\alpha=1/2$.
And because of (c), the critical exponent $\nu$ cannot be sensibly defined.
Taken together, the properties (a)--(d) are rather unusual.

It would be interesting to know of other examples of
first-order critical points.\footnote{
   The one-dimensional Ising ferromagnet with
   $1/r^2$ long-range interaction,
   cited by Fisher and Berker \cite[p.~2510]{Fisher_82}
   as an example of this phenomenon,
   does not, strictly speaking, qualify,
   as the (exponential) correlation length is $+\infty$ for all $\beta > 0$
   (by Griffiths' second inequality,
    $\< \sigma_x \sigma_y \> \ge \tanh J_{xy} \approx \beta/|x-y|^2$),
   so that a critical exponent $\nu$ cannot be defined.
   M\"uller \cite{Muller_93} studied a one-dimensional model of
   $U(N)$- or $SU(N)$-valued spins
   with a complex nearest-neighbor interaction,
   generalizing the work on the $q$-state clock model with a $\theta$-term.
   He showed that for $U(N)$, the model undergoes a sequence of
   first-order phase transitions;  however, it is not clear to us whether
   the correlation length diverges at those transition points.
   Finally, Oliveira and coworkers \cite{Oliveira_93,Oliveira_03}
   have found that some {\em non-equilibrium}\/ models exhibit a
   first-order critical point with $\nu=1/d$.
}

The plan of this paper is as follows:
In Section~\ref{sec.setup} we review the needed background
concerning the Potts-model partition function
and the combinatorial polynomials that can be obtained from it.
In Section~\ref{sec.theor} we review the small-$w$ and large-$w$ expansions
for the $q \to 0$ Potts model.
In Section~\ref{sec.transfer} we summarize how we computed
the transfer matrices.
In Sections~\ref{sec.sq} and \ref{sec.tri}
we report our results for square-lattice and triangular-lattice
strips, respectively.
In Section~\ref{sec.res} we analyze the data to extract
estimates of the critical point $w_0$,
the free energy $f(w)$,
the central charge $c(w)$,
and the thermal scaling dimensions $x_{T,i}(w)$
for each of the two lattices.
Finally, in Section~\ref{sec.discussion}
we discuss some open questions.
In the Appendix
we discuss how we performed  the analysis of the small-$w$ series expansions
obtained in Section~\ref{sec.theor}.

%
%
\section{Basic set-up} \label{sec.setup}

We begin by reviewing the Fortuin--Kasteleyn representation
of the $q$-state Potts model (Section~\ref{sec.setup.1}).
Then we discuss the various polynomials that can be obtained
from the $q$-state Potts model
by taking the limit $q \to 0$ (Section~\ref{sec.setup.2}).
After this, we define the specific quantities that we will be
studying in our work on spanning forests on the
square and triangular lattices,
and review the basic principles of finite-size scaling
in conformal field theory (Section~\ref{sec.setup.3}).
Finally, we review briefly the Beraha--Kahane--Weiss theorem
(Section~\ref{sec.setup.4}).

\subsection{Fortuin--Kasteleyn representation}  \label{sec.setup.1}

Let $G=(V,E)$ be a finite undirected graph
with vertex set $V$ and edge set $E$;
let $\{ J_e \} _{e \in E}$ be a set of couplings;
and let $q$ be a positive integer.
At each site $i \in V$ we place a spin $\sigma_i \in \{ 1,2,\ldots,q \}$,
and we write $\bsigma = \{\sigma_i\}_{i \in V}$ to denote the
spin configuration.
The Hamiltonian of the $q$-state Potts model on $G$ is
\be
   H(\bsigma)  \;=\;  - \sum_{e=ij \in E} J_e \delta(\sigma_i,\sigma_j)
   \;,
\ee
where $\delta$ denotes the Kronecker delta.
The partition function $Z = \sum_{\bsigma} e^{-\beta H(\bsigma)}$
can be written in the form
\be
   Z_G(q,{\bf v})  \;=\;
   \sum\limits_{\bsigma}  \,  \prod_{e=ij \in E}  \,
      \biggl[ 1 + v_e \delta(\sigma_i, \sigma_j) \biggr]
   \;,
 \label{eq1.1}
\ee
where
\be
   v_e \;=\; e^{\beta J_e} - 1
\ee
and ${\bf v} = \{v_e\}_{e \in E}$.
A coupling $J_e$ (or $v_e$)
is called {\em ferromagnetic}\/ if $J_e \ge 0$ ($v_e \ge 0$),
{\em antiferromagnetic}\/ if $-\infty \le J_e \le 0$ ($-1 \le v_e \le 0$),
and {\em unphysical}\/ if $v_e \notin [-1,\infty)$.

At this point $q$ is still a positive integer.
However, we now assert that $Z_G(q,{\bf v})$
is in fact the evaluation at $q \in \Z_+$ of
a {\em polynomial}\/ in $q$ and $\{v_e\}$
(with coefficients that are indeed either 0 or 1).
To see this, we proceed as follows:
In \reff{eq1.1}, expand out the product over $e \in E$,
and let $A \subseteq E$ be the set of edges for which the term
$v_e \delta(\sigma_i, \sigma_j)$ is taken.
Now perform the sum over configurations $\bsigma$:
in each connected component of the subgraph $(V,A)$
the spin value $\sigma_i$ must be constant,
and there are no other constraints.
Therefore
\be
   Z_G(q,{\bf v}) \;=\;
   \sum_{ A \subseteq E }  q^{k(A)}  \prod_{e \in A}  v_e
   \;,
 \label{eq1.2}
\ee
where $k(A)$ is the number of connected components
(including isolated vertices) in the subgraph $(V,A)$.
The subgraph expansion \reff{eq1.2} was discovered by
Birkhoff \cite{Birkhoff_12} and Whitney \cite{Whitney_32a}
for the special case $v_e = -1$ (see also Tutte \cite{Tutte_47,Tutte_54});
in its general form it is due to
Fortuin and Kasteleyn \cite{Kasteleyn_69,Fortuin_72}
(see also \cite{Edwards-Sokal}).
Henceforth we take \reff{eq1.2}
as the {\em definition}\/ of $Z_G(q,{\bf v})$
for arbitrary complex $q$ and ${\bf v}$.
When $v_e$ takes the same value $v$ for all edges $e$, we write $Z_G(q,v)$
for the corresponding two-variable polynomial.

Let us observe, for future reference, that \reff{eq1.2} can alternatively
be rewritten as
\be
   Z_G(q,{\bf v}) \;=\;
   q^{|V|} \sum_{ A \subseteq E }  q^{c(A)}  \prod_{e \in A} {v_e \over q}
   \;,
 \label{eq1.2cycles}
\ee
where we use the notation $|S|$ to denote the number of elements
of a finite set $S$, and
\be
   c(A)  \;=\;  |A| + k(A) - |V|
 \label{eq.cyclomatic}
\ee
is the cyclomatic number of the subgraph $(V,A)$,
i.e.\ the number of linearly independent circuits in $(V,A)$.

\bigskip

{\bf Remark.}  In the mathematical literature, these formulae
are usually written in terms of the Tutte polynomial $T_G$
defined by \cite{Welsh_93}
\be
   T_G(x,y)  \;=\;
   \sum_{ A \subseteq E }  (x-1)^{k(A)-k(G)}  (y-1)^{c(A)}
 \label{eq.Tutte.1}
\ee
where $k(G) = k(E)$ is the number of connected components in $G$,
and $c(A)$ is the cyclomatic number defined in \reff{eq.cyclomatic}.
Comparison with \reff{eq1.2} shows that
\be
   T_G(x,y)   \;=\;   (x-1)^{-k(G)} \, (y-1)^{-|V|} \,
                        Z_G \bigl( (x-1)(y-1), \, y-1 \bigr)   \;.
 \label{eq.Tutte.2}
\ee
In other words, the Tutte polynomial $T_G(x,y)$ and the
Potts-model partition function $Z_G(q,v)$ are essentially equivalent
under the change of variables
\begin{subeqnarray}
   x  & = &  1 + q/v   \\[2mm]
   y  & = &  1 + v     \\[5mm]
   q  & = &  (x-1)(y-1) \\[2mm]
   v  & = &  y-1
 \label{eq.Tutte.3}
\end{subeqnarray}
The advantage of the Tutte notation is that it allows a slightly smoother
treatment of the $q \to 0$ limit
(see Remark 1 at the end of the next subsection).
The disadvantage is that the use of the variables $x$ and $y$
conceals the fact that the particular combinations $q$ and $v$
play very different roles:
$q$ is a global variable,
while $v$ can be given separate values $v_e$ on each edge.
This latter freedom is extremely important in many contexts
(e.g.\ in using the series and parallel reduction formulae).
We therefore strongly advocate the multivariable approach in which
$Z_G(q,{\bf v})$ is considered the fundamental quantity,
even if one is ultimately interested in a particular two-variable
or one-variable specialization.

\subsection[$q\to0$ limits]{\mbox{\protect\boldmath $q \to 0$} limits}
  \label{sec.setup.2}

Let us now consider the different ways in which a meaningful
$q \to 0$ limit can be taken in the $q$-state Potts model.

The simplest limit is to take $q \to 0$ with fixed couplings ${\bf v}$.
{}From \reff{eq1.2} we see that this selects out the subgraphs $A \subseteq E$
having the smallest possible number of connected components;
the minimum achievable value is of course $k(G)$ itself
(= 1 in case $G$ is connected).
We therefore have
\be
   \lim_{q \to 0} q^{-k(G)} Z_G(q,{\bf v}) \;=\;  C_G({\bf v})
   \;,
\ee
where
\be
   C_G({\bf v}) \;=\;  \sum\limits_{\begin{scarray}
                                       A \subseteq E \\
                                       k(A) = k(G)
                                    \end{scarray}}
                       \prod_{e \in A}  v_e
\ee
is the generating polynomial of ``maximally connected spanning subgraphs''
(= connected spanning subgraphs in case $G$ is connected).\footnote{
   A subgraph is called {\em spanning}\/ if its vertex set
   is the entire set $V$ of vertices of $G$
   (as opposed to some proper subset of $V$).
   All of the subgraphs arising in this paper are spanning subgraphs,
   since we consider subsets $A$ of {\em edges}\/ only,
   and retain all the vertices.
}

A different limit can be obtained by taking $q \to 0$
with fixed values of ${\bf w} = {\bf v}/q$.
{}From \reff{eq1.2cycles} we see that this selects out
the subgraphs $A \subseteq E$ having the smallest possible cyclomatic number;
the minimum achievable value is of course 0.
We therefore have \cite{Stephen_76,Wu_77}
\be
   \lim_{q \to 0} q^{-|V|} Z_G(q,q{\bf w}) \;=\;  F_G({\bf w})
   \;,
 \label{eq.limit.FG}
\ee
where
\be
   F_G({\bf w}) \;=\;  \sum\limits_{\begin{scarray}
                                       A \subseteq E \\
                                       c(A) = 0
                                    \end{scarray}}
                       \prod_{e \in A}  w_e
   \label{def_F}
\ee
is the generating polynomial of {\em spanning forests}\/,
i.e.\ spanning subgraphs not containing any circuits.

Finally, suppose that in $C_G({\bf v})$
we replace each edge weight $v_e$ by $\lambda v_e$
and then take $\lambda \to 0$.
This obviously selects out,
from among the maximally connected spanning subgraphs,
those having the fewest edges:
these are precisely the maximal spanning forests
(= spanning trees in case $G$ is connected),
and they all have exactly $|V| - k(G)$ edges.
Hence
\be
   \lim_{\lambda \to 0} \lambda^{k(G)-|V|} C_G(\lambda {\bf v}) \;=\;
   T_G({\bf v})
   \;,
\ee
where
\be
   T_G({\bf v}) \;=\;  \sum\limits_{\begin{scarray}
                                       A \subseteq E \\
                                       k(A) = k(G) \\
                                       c(A) = 0
                                    \end{scarray}}
                       \prod_{e \in A}  v_e
 \label{def.tree}
\ee
is the generating polynomial of maximal spanning forests
(= spanning trees in case $G$ is connected).\footnote{
   We trust that there will be no confusion between the
   generating polynomial $T_G({\bf v})$
   and the Tutte polynomial $T_G(x,y)$.
   We have used here the letter $T$ because in the most important applications
   the graph $G$ is connected, so that $T_G({\bf v})$ is the generating
   polynomial of spanning {\em trees}\/.
}
Alternatively, suppose that in $F_G({\bf w})$
we replace each edge weight $w_e$ by $\lambda w_e$
and then take $\lambda \to \infty$.
This obviously selects out, from among the spanning forests,
those having the greatest number of edges:
these are once again the maximal spanning forests.
Hence
\be
   \lim_{\lambda \to \infty} \lambda^{k(G)-|V|} F_G(\lambda {\bf w}) \;=\;
   T_G({\bf w})
   \;.
\ee

In summary, we have the following scheme for the $q \to 0$ limits
of the Potts model:
\be
  \begin{array}{ccccc}
        &   &  C_G({\bf v})   &     &    \\  
        &   \setlength{\unitlength}{1pt}
            \begin{picture}(70,50)(0,0)
            \drawline(0,0)(25.2,18)
            \put(0,23){{\small $q \to 0,\, {\bf v} \, {\rm fixed}$}}
            \drawline(49,35)(70,50)
            \drawline(70,50)(64.34,41.76)
            \drawline(70,50)(60.37,47.32)
            \end{picture} &
        &   \setlength{\unitlength}{1pt}
            \begin{picture}(70,50)(0,0)
            \drawline(0,50)(21,35)
            \put(20,23){{\small ${\bf v} \, {\rm infinitesimal}$}}
            \drawline(44.8,18)(70,0)
            \drawline(70,0)(60.37,2.68)
            \drawline(70,0)(64.34,8.24)
            \end{picture} &              \\
   Z_G(q,{\bf v})  &    &      &    &   T_G({\bf v} \: {\rm or} \: {\bf w}) \\
        &   \setlength{\unitlength}{1pt}
            \begin{picture}(70,50)(0,0)
            \drawline(0,50)(21,35)
            \put(0,23){{\small $q \to 0,\, {\bf w} = {\bf v}/q \, {\rm fixed}$}}
            \drawline(44.8,18)(70,0)
            \drawline(70,0)(60.37,2.68)
            \drawline(70,0)(64.34,8.24)
            \end{picture} &
        &   \setlength{\unitlength}{1pt}
            \begin{picture}(70,50)(0,0)
            \drawline(0,0)(25.2,18)
            \put(20,23){{\small ${\bf w} \, {\rm infinite}$}}
            \drawline(49,35)(70,50)
            \drawline(70,50)(64.34,41.76)
            \drawline(70,50)(60.37,47.32)
            \end{picture} & \\
        &   &  F_G({\bf w})   &     &
  \end{array}
\ee

Finally, maximal spanning forests (= spanning trees in case $G$ is connected)
can also be obtained directly from $Z_G(q,{\bf v})$ by a one-step process
in which the limit $q \to 0$ is taken at fixed ${\bf x} = {\bf v}/q^\alpha$,
where $0 < \alpha < 1$ \cite{Fortuin_72,Stephen_76,Wu_77,Haggstrom_95}.
Indeed, simple manipulation of \reff{eq1.2} and \reff{eq.cyclomatic} yields
\be
   Z_G(q, q^\alpha {\bf x}) \;=\;
   q^{\alpha|V|} \sum_{ A \subseteq E }  q^{\alpha c(A) + (1-\alpha)k(A)}
       \prod_{e \in A} x_e
   \;.
 \label{eq1.2alpha}
\ee
The quantity $\alpha c(A) + (1-\alpha)k(A)$ is minimized on (and only on)
maximal spanning forests, where it takes the value $(1-\alpha)k(G)$.
Hence
\be
   \lim_{q \to 0} q^{-\alpha|V| - (1-\alpha)k(G)} Z_G(q, q^\alpha {\bf x})
   \;=\;  T_G({\bf x})
   \;.
\ee

\bigskip

{\bf Remarks.}
1.  Let us rewrite the formulae of this subsection
in terms of the Tutte polynomial $T_G(x,y)$
[cf.\ \reff{eq.Tutte.1}--\reff{eq.Tutte.3}].
The limit $q,v \to 0$ with $w = v/q$ fixed corresponds
to $x = 1 + 1/w$, $y=1$;
simple algebra using \reff{eq.Tutte.2} and \reff{eq.limit.FG} gives
\be
   T_G(1+ 1/w, 1)  \;=\;  w^{k(G) - |V|} F_G(w)   \;.
\ee
In particular, several of the evaluations of $T_G(1+ 1/w, 1)$
have interesting combinatorial interpretations:
\begin{itemize}
\item $T_G(1,1)$ (i.e., $w=\infty$) counts the number of
      maximal spanning forests in $G$ (= spanning trees if $G$ is connected).

\item $T_G(2,1)$ (i.e., $w=1$) counts the number of spanning forests in $G$.

\item $T_G(1+1/k,1)$ (i.e., $w=k$) for $k=1,2,3,\ldots\,$ is,
      up to a prefactor, the number of possible ``score vectors''
      in a tournament of constant-sum games
      with scores lying in the set $\{0,1,\ldots,k\}$
      \cite[Propositions 6.3.19 and 6.3.25]{Brylawski_92}.

\item $T_G(0,1)$ (i.e., $w=-1$):  If $G$ is a directed graph having a
      fixed ordering on its edges, $|T_G(0,1)|$ counts the number of totally
      cyclic reorientations $\tau$ of $G$ such that in each cycle of $\tau$
      the lowest edge is not reoriented.
      For a planar graph with no isthmuses,
      $|T_G(0,1)|$ counts the number of totally cyclic orientations
      in which there is no clockwise cycle.
      See \cite[Examples 6.3.30 and 6.3.31]{Brylawski_92}.
\end{itemize}
We emphasize, however, that from a physical point of view
there is nothing special about the particular values
$w=\infty,1,2,3,\ldots,-1$.
Rather, it is important to study $F_G(w)$ {\em as a function of}\/
the real or complex variable $w$.
The ``special'' values of $w$ are those lying
on the phase boundary ${\cal B}$;
they are determined only {\em a posteriori}\/.

\medskip

2. The polynomial $F_G(w)$ also equals the Ehrhart polynomial
of a particular unimodular zonotope determined by the graph $G$:
see \cite[Section XI.A]{Welsh_00} for details.

\medskip

3. Suppose that $G=(V,E)$ is a connected {\em planar}\/ graph;
then we can define a dual graph $G^* = (V^*,E^*)$
by the usual geometric construction.\footnote{
   More precisely, consider a particular plane representation of $G$,
   and define $G^*$ by placing one vertex in each face of $G$
   and then drawing an edge of $G^*$ through each edge of $G$.
   The dual graph $G^*$ is not necessarily unique;
   nonisomorphic plane representations of $G$
   (which can arise if $G$ is not 3-connected)
   can give rise to nonisomorphic duals
   (see e.g.\ \cite[p.~114]{Harary_69} for an example).
   In any case, {\em each}\/ of the dual graphs $G^*$
   satisfies the relations \reff{eq.duality}--\reff{eq.duality3} below.
}
Moreover, there is a one-to-one correspondence between the edges of $G$
and their corresponding dual edges in $G^*$;
we can therefore identify $E^*$ with $E$,
and assign the same weights ${\bf v} = \{v_e\}_{e \in E}$
to the edges of $G$ and $G^*$.
We then have the {\em fundamental duality relation}\/ \cite{Wu_82}
\be
   Z_{G^*}(q, {\bf v})  \;=\;
   q^{1-|V|} \left( \prod\limits_{e \in E} v_e \right)  Z_G(q, q/{\bf v})
 \label{eq.duality}
\ee
[where $q/{\bf v}$ of course denotes the vector $\{q/v_e\}_{e \in E}$].
In particular, we have
\begin{eqnarray}
   C_{G^*}({\bf v})   & = &   \left( \prod\limits_{e \in E} v_e \right)
                                  F_G(1/{\bf v})
     \label{eq.duality1} \\[2mm]
   F_{G^*}({\bf w})   & = &   \left( \prod\limits_{e \in E} w_e \right)
                                  C_G(1/{\bf w})
     \label{eq.duality2} \\[2mm]
   T_{G^*}({\bf v})   & = &   \left( \prod\limits_{e \in E} v_e \right)
                                  T_G(1/{\bf v})
     \label{eq.duality3}
\end{eqnarray}

\medskip

4.  Suppose that $G$ is connected;
and let us consider $G$ as a communications network
with unreliable communication channels,
in which edge $e$ is operational with probability $p_e$
and failed with probability $1-p_e$,
independently for each edge.
Let $R_G({\bf p})$ be the probability that
every node is capable of communicating with every other node
(this is the so-called {\em all-terminal reliability}\/).
Clearly we have
\be
   R_G({\bf p})  \;=\;
   \sum_{\begin{scarray}
           A \subseteq E \\
           (V,A) \, {\rm connected}
         \end{scarray}}
   \prod_{e \in A} p_e  \prod_{e \in E \setminus A} (1-p_e)
   \;,
\ee
where the sum runs over all connected spanning subgraphs of $G$.
The polynomial $R_G({\bf p})$ is called
the (multivariate) {\em reliability polynomial}\/
\cite{Colbourn_87} for the graph $G$.
Modulo trivial prefactors, it is equivalent to $C_G({\bf v})$
under the change of variables
\begin{subeqnarray}
   v_e  & = &  {p_e \over 1-p_e}  \\[2mm]
   p_e  & = &  {v_e \over 1+v_e}
\end{subeqnarray}
The reliability polynomial is therefore one of the objects
obtainable as a $q \to 0$ limit of the Potts model.

\medskip

5.  Brown and Colbourn \cite{Brown_92},
followed by Wagner \cite{Wagner_00} and Sokal \cite{Sokal_chromatic_bounds},
have studied the possibility that the complex roots of the
reliability polynomial might satisfy a theorem of Lee--Yang type.
To state what is at issue, let us say that a graph $G$ has
\begin{itemize}
 \item  the {\em univariate Brown--Colbourn property}\/
    if $C_G(v) \neq 0$ whenever $|1 + v| < 1$;
 \item  the {\em multivariate Brown--Colbourn property}\/
    if $C_G({\bf v}) \neq 0$ whenever $|1 + v_e| < 1$ for all edges $e$;
 \item  the {\em univariate dual Brown--Colbourn property}\/
    if $F_G(w) \neq 0$ whenever $\real w < -1/2$;
 \item  the {\em multivariate dual Brown--Colbourn property}\/
    if $F_G({\bf w}) \neq 0$ whenever $\real w_e < -1/2$ for all edges $e$.
\end{itemize}
Here the word ``dual'' refers to the fact
that a {\em planar}\/ graph $G$ has the (univariate or multivariate)
Brown--Colbourn property if and only if its dual graph $G^*$
(which is also planar)
has the (univariate or multivariate) dual Brown--Colbourn property:
this is an immediate consequence of the identities
\reff{eq.duality1}/\reff{eq.duality2}.

Brown and Colbourn \cite{Brown_92},
having studied the univariate reliability polynomial $R_G(p)$
in a number of examples, conjectured that every loopless graph
has the univariate Brown--Colbourn property.\footnote{
   A {\em loop}\/ is an edge connecting a vertex $i$ to itself.
   Loops must be excluded because a loop $e$ multiples $C_G$
   by a factor $1+v_e$ and therefore places a root at $v_e = -1$,
   violating the Brown--Colbourn property.
}
(Of course, they didn't call it that!)
Subsequently, Sokal \cite{Sokal_chromatic_bounds}
made the stronger conjecture that every loopless graph
has the multivariate Brown--Colbourn property.
Moreover, Sokal \cite{Sokal_chromatic_bounds} proved this latter conjecture
for the special case of series-parallel graphs
(which are a subset of planar graphs).\footnote{
   A graph is called {\em series-parallel}\/ if it can be obtained
   from a forest by a sequence of series and parallel extensions
   (i.e.\ replacing an edge by two edges in series or two edges in parallel).
   The proof that every loopless series-parallel graph has the
   multivariate Brown--Colbourn property
   is an almost trivial two-line induction;
   it can be found in \cite[Remark 3 in Section 4.1]{Sokal_chromatic_bounds}.
   Earlier, Wagner \cite{Wagner_00} had proven,
   using an ingenious and complicated construction,
   that every loopless series-parallel graph
   has the {\em univariate}\/ Brown--Colbourn property.
}
And since the class of series-parallel graphs is self-dual,
it follows immediately that every bridgeless series-parallel graph
has the multivariate dual Brown--Colbourn property.\footnote{
   A {\em bridge}\/ is an edge whose removal increases (by 1)
   the number of connected components of the graph.
   Bridges must be excluded because a bridge $e$ multiplies $F_G$
   by a factor $1+w_e$ and therefore places a root at $w_e = -1$,
   violating the dual Brown--Colbourn property.
   Please note that a planar graph $G$ is loopless (resp.\ bridgeless)
   if and only if its dual graph $G^*$ is bridgeless (resp.\ loopless).
}
Finally, Sokal's conjecture would imply that every bridgeless
{\em planar}\/ graph (series-parallel or not)
has the multivariate dual Brown--Colbourn property.
These conjectures, if true, would constitute a powerful result
of Lee--Yang type, constraining the complex zeros of the
corresponding partition functions.

Recently, however, Royle and Sokal \cite{Royle-Sokal}
have discovered --- to their amazement ---
that there exist planar graphs for which all these properties fail!
Indeed, the multivariate Brown--Colbourn and dual Brown--Colbourn properties
fail already for the simplest non-series-parallel graph,
namely the complete graph on four vertices ($K_4$).
The univariate properties fail for certain graphs that can be obtained
from $K_4$ by series and/or parallel extensions.
In fact, Royle and Sokal \cite{Royle-Sokal} show that a graph $G$
has the multivariate Brown--Colbourn property {\em if and only if}\/
it is series-parallel.

In addition, Chang and Shrock \cite{Shrock_reliability},
building on one of the Royle--Sokal examples,
have devised families of strip graphs
in which the limiting curve ${\cal B}$ of zeros of $C_G(v)$
penetrates into the ``forbidden region'' $|1+v| < 1$.

Nonetheless, the strip graphs studied in this paper do seem to possess
at least the univariate dual Brown--Colbourn property:
all the roots we find lie in the region $\real w \ge -1/2$.

\subsection{Quantities to be studied}   \label{sec.setup.3}

The graphs $G$ to be considered in this paper are $m \times n$
strips of the square or triangular lattice,
with periodic boundary conditions in the first (transverse) direction
and free boundary conditions in the second (longitudinal) direction.
We therefore denote these strips as $m_{\rm P} \times n_{\rm F}$,
and call this {\em cylindrical boundary conditions}\/.\footnote{
   This accords with the terminology of Shrock and collaborators
   \cite{Shrock_01_review} for the various boundary conditions:
   free ($m_{\rm F} \times n_{\rm F}$),
   cylindrical ($m_{\rm P} \times n_{\rm F}$),
   cyclic ($m_{\rm F} \times n_{\rm P}$),
   toroidal ($m_{\rm P} \times n_{\rm P}$),
   M\"obius ($m_{\rm F} \times n_{\rm TP}$) and
   Klein bottle ($m_{\rm P} \times n_{\rm TP}$).
   Here F denotes ``free'', P denotes ``periodic'',
   and TP denotes ``twisted periodic''
   (i.e.\ the longitudinal ends are identified with a
   reversal of orientation).
}
We shall also use the letter $L$
as an alternative name for the strip width $m$.
All these graphs are planar.

In this paper we will be focussing on $F_G({\bf w})$,
the generating polynomial of spanning forests.
Since we will be assigning the same weight $w$ to all nearest-neighbor edges,
we have a univariate polynomial $F_G(w)$.
We will refer to $F_G(w)$ as the ``partition function''
(since that is basically what it is).
Our principal goal is to study the behavior of $F_G(w)$
in the thermodynamic limit $m,n \to\infty$.

Let us now define the free energy (or ``entropy'') per site
for a finite lattice\footnote{
   Note that our ``free energy'' is the {\em negative}\/ of the usual free
   energy.
}
\be
   f_{m,n}(w)  \;=\;
       {1 \over mn} \log F_{m_{\rm P} \times n_{\rm F}}(w)
 \label{def.fmn}
\ee
and its limiting values for a semi-infinite strip
\be
   f_{m}(w)   \;=\;
       \lim\limits_{n \to \infty}
       {1 \over mn} \log F_{m_{\rm P} \times n_{\rm F}}(w)
 \label{def.fm}
\ee
and for the infinite lattice
\be
   f(w)  \;=\;
       \lim\limits_{m,n \to \infty}
       {1 \over mn} \log F_{m_{\rm P} \times n_{\rm F}}(w)
   \;.
 \label{def.f}
\ee
Here we are assuming, of course, that the indicated limits exist
and that in \reff{def.f} the limit is independent of the way that
$m$ and $n$ tend to infinity.\footnote{
   For $w \ge 0$, it is not hard to prove rigorously that the
   limit \reff{def.f} exists, at least if we insist that
   $m$ and $n$ tend to infinity in such a way that
   the ratio $m/n$ stays bounded away from zero and infinity.
   The proof is based on the submultiplicativity of $F_G(w)$
   for disjoint regions of the lattice,
   together with a standard result on subadditive functions
   \cite[Proposition A.4]{vanEnter_93}.

   This proof, by itself, says nothing about $w$ negative or complex.
   But the convergence can in some cases be extended to part
   of the complex $w$-plane, by using a normal-families argument
   \cite{Montel_27,Schiff_93}.
   Indeed, suppose that $D \subset \C$ is a connected open set
   having a nonempty intersection with the positive real axis,
   on which the partition function $F_{m \times n}(w)$
   is nonvanishing for all (or all sufficiently large) $m,n$.
   Then the trivial bound $|F_{m \times n}(w)| \le (1+|w|)^{2mn}$
   guarantees that $\real f_{m,n}(w)$ is uniformly bounded above
   on compact subsets of $D$,
   from which it follows \cite[Example~2.3.9]{Schiff_93}
   that the analytic functions $\{ f_{m,n} \}$ form a normal family on $D$.
   Then a Vitali-like argument
   \cite[Lemma 3.5]{Sokal_chromatic_roots}
   shows that the convergence for $w \ge 0$ extends to all of $D$.
   Simon \cite[p.~343]{Simon_74} calls this reasoning ``log exp Vitali''.
}
In particular, in \reff{def.f} we can allow $n$ to tend to infinity first
and then take $m \to \infty$, so that
\be
   f(w)  \;=\; \lim\limits_{m \to \infty} f_{m}(w)  \;.
\ee
Let us also note that for $w$ negative or complex these formulae may contain
some ambiguities about the branch of the logarithm.
We nevertheless expect $f_{m}(w)$ and $f(w)$ to be well-defined
analytic functions in the complex $w$-plane minus certain branch cuts;
but for simplicity we shall mostly focus on the {\em real part}\/
of the free energy, which does not suffer from any ambiguities.

For each fixed width $m$,
the partition function $F_{m_{\rm P} \times n_{\rm F}}(w)$
for strips of arbitrary length $n$
can be expressed in terms of a transfer matrix:
\begin{subeqnarray}
F_{m_{\rm P} \times n_{\rm F}}(w)
     &=& \uu(w)^{T} \T(w)^{n-1} \vv_{\rm id}(w)
  \slabel{def_transfer_matrix} \\
     &=& \sum\limits_{k=1}^M \alpha_k(w) \lambda_k(w)^{n-1}
  \slabel{def_transfer_matrix.b}
\end{subeqnarray}
where $M = \dim \T$ is the dimension of the transfer matrix.
This is explained in detail in Ref.~\cite{transfer1}
and is summarized very briefly in Section~\ref{sec.transfer} below.
The elements of the transfer matrix $\T(w)$
and of the left and right vectors $\uu(w)$ and $\vv_{\rm id}(w)$
are polynomials in the complex variable $w$.
Therefore, the eigenvalues $\lambda_k(w)$ and the amplitudes $\alpha_k(w)$
are algebraic functions of $w$. We have numerically checked for $m\leq 9$
that none of the amplitudes $\alpha_k(w)$ vanishes identically.
This is important in order to compute the limiting curves ${\cal B}$,
as one would get the {\em wrong}\/ curve $\scrb$
if there were an identically vanishing amplitude
for an eigenvalue that happened to be dominant in some region
of the $w$-plane.

Let us denote by $\lambda_\star(w)$ the eigenvalue of $T(w)$
having largest modulus, whenever it is unique.
(Typically there is a unique dominant eigenvalue at all points
 in the complex $w$-plane with the exception of
 a finite union of real algebraic curves ${\cal B}$.)
It follows from \reff{def_transfer_matrix.b} that
the strip free energy \reff{def.fm} exists at all such points $w$
--- except possibly at isolated points where the amplitude $\alpha_k$
corresponding to the dominant eigenvalue vanishes ---
and equals
\be
   f_{L}(w)   \;=\;  {1 \over L} \log \lambda_\star(w)   \;.
\label{def_fm}
\ee
[In particular, $\real f_L(w) = (1/L) \log |\lambda_\star(w)|$.]
We then expect $f_L(w)$ to converge as $L \to\infty$ to the
bulk free energy $f(w)$;
but the {\em rate}\/ at which it converges depends on whether
the model at $w$ is critical or not.
If $w$ is a noncritical point, we expect an exponentially rapid convergence:
\be
f_L(w) \;=\; f(w) \,+\,  O(e^{-L/\xi}) \;,
\label{def_fm_non_critical}
\ee
where $\xi < \infty$ is the correlation length of the system.
If $w$ is a critical point, then we expect that
its long-distance behavior can be described by a conformal field theory (CFT)
\cite{Itzykson_collection,DiFrancesco_97,Henkel_99}
with some central charge $c(w)$;
the general principles of CFT then predict \cite{Bloete_86,Affleck_86}
that\footnote{
  Note that the $1/L^2$ correction in \protect\reff{def_fm_critical}
  has the opposite sign from~\protect\cite[equation 1]{Bloete_86}.
  This change compensates the global change of sign introduced in our
  definition of the free energy \protect\reff{def.fmn}, so that the
  central charge $c(w)$ in \protect\reff{def_fm_critical} has the
  conventional sign.
}
\be
f_L(w) \;=\; f(w) \,+\, {\pi G \over 6} {c(w) \over L^2}  \,+\, \ldots  \;,
\label{def_fm_critical}
\ee
where $G$ is a geometrical factor depending on the lattice structure,
\be
G = \left\{ \begin{array}{ll}
             1                & \quad \hbox{\rm square lattice} \\
             \sqrt{3}/2 \quad & \quad \hbox{\rm triangular lattice}
            \end{array} \right.
\label{def_G}
\ee
and the dots stand for higher-order corrections.
These higher-order corrections always include a $1/L^4$ term
(with a nonuniversal amplitude) coming from the operator $T \bar{T}$,
where $T$ is the stress-energy tensor;
sometimes irrelevant operators may give additional corrections
in-between $1/L^2$ and $1/L^4$.

Let the eigenvalues of the transfer matrix
(at some particular value of $w$ and some particular strip width $L$)
be ordered in modulus as
$|\lambda_\star| \ge |\lambda_1| \ge |\lambda_2| \ge \ldots$,
and let us define correlation lengths $\xi_1 \ge \xi_2 \ge \ldots$ by
\be
   \xi_i^{-1}  \;=\;  \log \left| {\lambda_\star \over \lambda_i} \right|  \;.
\label{def_inverse_xi}
\ee
Then CFT predicts \cite{Cardy_84} that,
at a critical point, the correlation lengths behave for large $L$ as
\be
   \xi_i^{-1}  \;=\;  2\pi G {x_i \over L} \,+\, \ldots  \;,
\label{def_xi_critical}
\ee
where $x_i$ is the scaling dimension of an appropriate scaling operator,
$G$ is the geometrical factor \reff{def_G} \cite{Indekeu_86},
and the dots stand again for higher-order corrections.

\bigskip

{\bf Remarks.}
1. In order to compare our results more directly with
those of mathematicians (e.g., \cite{Calkin_03}),
it is convenient to introduce also the quantities
\begin{subeqnarray}
n_m(w) &=& e^{\real f_m(w)} = |\lambda_\star(w)|^{1/m}  \slabel{def_nm_w} \\
n(w)   &=& \lim\limits_{m\to\infty} n_m(w) = e^{\real f(w)} \slabel{def_n_w}
\end{subeqnarray}

2.  For the square lattice at $w=1$,
Calkin {\em et al.}\/ \cite{Calkin_03} have proven the bounds
\be
  1.29335 \,\approx\, \log 3.64497  \;\le\;  f({\rm sq},w=1)  \;\le\;
  \log 3.74101 \,\approx\, 1.31936   \;.
\label{Noy_bounds}
\ee
Their proof uses an $m \times m$ lattice with free boundary conditions
in both directions, but the same bound for cylindrical boundary conditions
is an easy corollary.
Weaker bounds of the same type were proven earlier by
Merino and Welsh \cite{Merino_99}.
The upper bound in \reff{Noy_bounds} comes
from the inequality \cite[Theorem 5.3]{Calkin_03}
\be
f({\rm sq},w)  \;\le\;
   {1 \over m} \log |(1+w) \lambda^{({\rm free},m)}_\star(w)|
   \qquad\hbox{for all $m > 0$ and $w \ge 0$,}
\label{upper_bound_n_sq}
\ee
where $\lambda^{({\rm free},m)}_\star(w)$ is the dominant eigenvalue of the
transfer matrix for a square-lattice strip of width $m$
with free boundary conditions.
(The proof given in \cite{Calkin_03} for $w=1$
 generalizes immediately to any $w \ge 0$.)
The upper bound in \reff{Noy_bounds} was obtained by evaluating
the right-hand side of \reff{upper_bound_n_sq} for $m=8$.
We have slightly improved this upper bound by computing
the dominant eigenvalue at $w=1$ for $m=10$:\footnote{
  We have obtained the symbolic form of the transfer matrices for
  square-lattice strips with free boundary conditions up to widths
  $L\leq 9_{\rm F}$. The dimensions of such matrices were obtained in closed
  form in \cite[Theorem~5]{Tutte_sq}, and they are listed on column
  SqFree($L$) in Table~\ref{table_dimensions}. For $L= 10_{\rm F}$ we have
  obtained numerically the transfer matrix at $w=1$. Details of these
  transfer matrices will be given elsewhere.
}
\be
  1.29335 \,\approx\, \log 3.64497  \;\le\;  f({\rm sq},w=1)  \;\le\;
  \log 3.73264 \,\approx\, 1.31711   \;.
\label{Noy_bounds_improved}
\ee

3. For the triangular lattice one can compute a similar upper bound for
$f({\rm tri},w)$ by mimicking the derivation of \reff{upper_bound_n_sq}.
One obtains the following rigorous upper bound:
\be
f({\rm tri},w)  \;\le\;
             {1 \over m} \log |(1+w)^2 \lambda^{({\rm free},m)}_\star(w)|
   \qquad\hbox{for all $m > 0$ and $w \ge 0$.}
\label{upper_bound_n_tri}
\ee
We have obtained a numerical bound by computing the leading eigenvalue
of the transfer matrix at $w=1$ for $m=9$.\footnote{
  We have obtained the symbolic form of the transfer matrices for
  triangular-lattice strips with free boundary conditions up to widths
  $L\leq 8_{\rm F}$. The transfer-matrix dimension is the Catalan number $C_L$;
  they are listed on column TriFree($L$) in Table~\ref{table_dimensions}.
  For $L=9_{\rm F}$ we have obtained numerically the transfer matrix
  at $w=1$. Details of these transfer matrices will be given elsewhere.
}
A trivial lower bound can be obtained in terms of the entropy per site
for spanning trees on the triangular lattice, $f_0({\rm tri})$,
given in \reff{def_f0_tri} below
\cite{Merino_99,Wu_77,Shrock_Wu_00,Glasser_Wu_03}.
Putting these bounds together, we have
\be
1.61533 \approx f_0({\rm tri}) \;\le\;  f({\rm tri},w=1)
      \;\le\;\log 5.77546 \,\approx\, 1.75362 \;.
\label{Noy_bounds_tri}
\ee

\subsection{Beraha--Kahane--Weiss theorem}   \label{sec.setup.4}

A central role in our work is played by a theorem
on analytic functions due to
Beraha, Kahane and Weiss \cite{BKW_75,BKW_78,Beraha_79,Beraha_80}
and generalized slightly by one of us \cite{Sokal_chromatic_roots}.
The situation is as follows:
Let $D$ be a domain (connected open set) in the complex plane,
and let $\alpha_1,\ldots,\alpha_M,\lambda_1,\ldots,\lambda_M$ ($M \ge 2$)
be analytic functions on $D$, none of which is identically zero.
For each integer $n \ge 0$, define
\be
   f_n(z)   \;=\;   \sum\limits_{k=1}^M \alpha_k(z) \, \lambda_k(z)^n
   \;.
   \label{def_fn}
\ee
We are interested in the zero sets
\be
   \scrz(f_n)   \;=\;   \{z \in D \colon\;  f_n(z) = 0 \}
\ee
and in particular in their limit sets as $n\to\infty$:
\begin{eqnarray}
   \liminf \scrz(f_n)   & = &  \{z \in D \colon\;
   \hbox{every neighborhood $U \ni z$ has a nonempty intersection} \nonumber\\
      & & \qquad \hbox{with all but finitely many of the sets } \scrz(f_n) \}
   \\[4mm]
   \limsup \scrz(f_n)   & = &  \{z \in D \colon\;
   \hbox{every neighborhood $U \ni z$ has a nonempty intersection} \nonumber\\
      & & \qquad \hbox{with infinitely many of the sets } \scrz(f_n) \}
\end{eqnarray}
Let us call an index $k$ {\em dominant at $z$}\/ if
$|\lambda_k(z)| \ge |\lambda_l(z)|$ for all $l$ ($1 \le l \le M$);
and let us write
\be
   D_k  \;=\;  \{ z \in D \colon\;  k \hbox{ is dominant at } z  \}
   \;.
\ee
Then the limiting zero sets can be completely characterized as follows:

\begin{theorem}
  {\bf \protect\cite{BKW_75,BKW_78,Beraha_79,Beraha_80,Sokal_chromatic_roots}}
   \label{BKW_thm}
Let $D$ be a domain in $\C$,
and let $\alpha_1,\ldots,\alpha_M$, $\lambda_1,\ldots,\lambda_M$ ($M \ge 2$)
be analytic functions on $D$, none of which is identically zero.
Let us further assume a ``no-degenerate-dominance'' condition:
there do not exist indices $k \neq k'$
such that $\lambda_k \equiv \omega \lambda_{k'}$ for some constant $\omega$
with $|\omega| = 1$ and such that $D_k$ ($= D_{k'}$)
has nonempty interior.
For each integer $n \ge 0$, define $f_n$ by
$$
   f_n(z)   \;=\;   \sum\limits_{k=1}^M \alpha_k(z) \, \lambda_k(z)^n
   \;.
$$
Then $\liminf \scrz(f_n) = \limsup \scrz(f_n)$,
and a point $z$ lies in this set if and only if either
\begin{itemize}
   \item[(a)]  There is a unique dominant index $k$ at $z$,
       and $\alpha_k(z) =0$;  or
   \item[(b)]  There are two or more dominant indices at $z$.
\end{itemize}
\end{theorem}
Note that case (a) consists of isolated points in $D$,
while case (b) consists of curves
(plus possibly isolated points where
all the $\lambda_k$ vanish simultaneously).
Henceforth we shall denote by $\scrb$ the locus of points
satisfying condition (b).

\bigskip

\begin{figure}
\centering
\begin{tabular}{cc}
   \includegraphics[width=200pt]{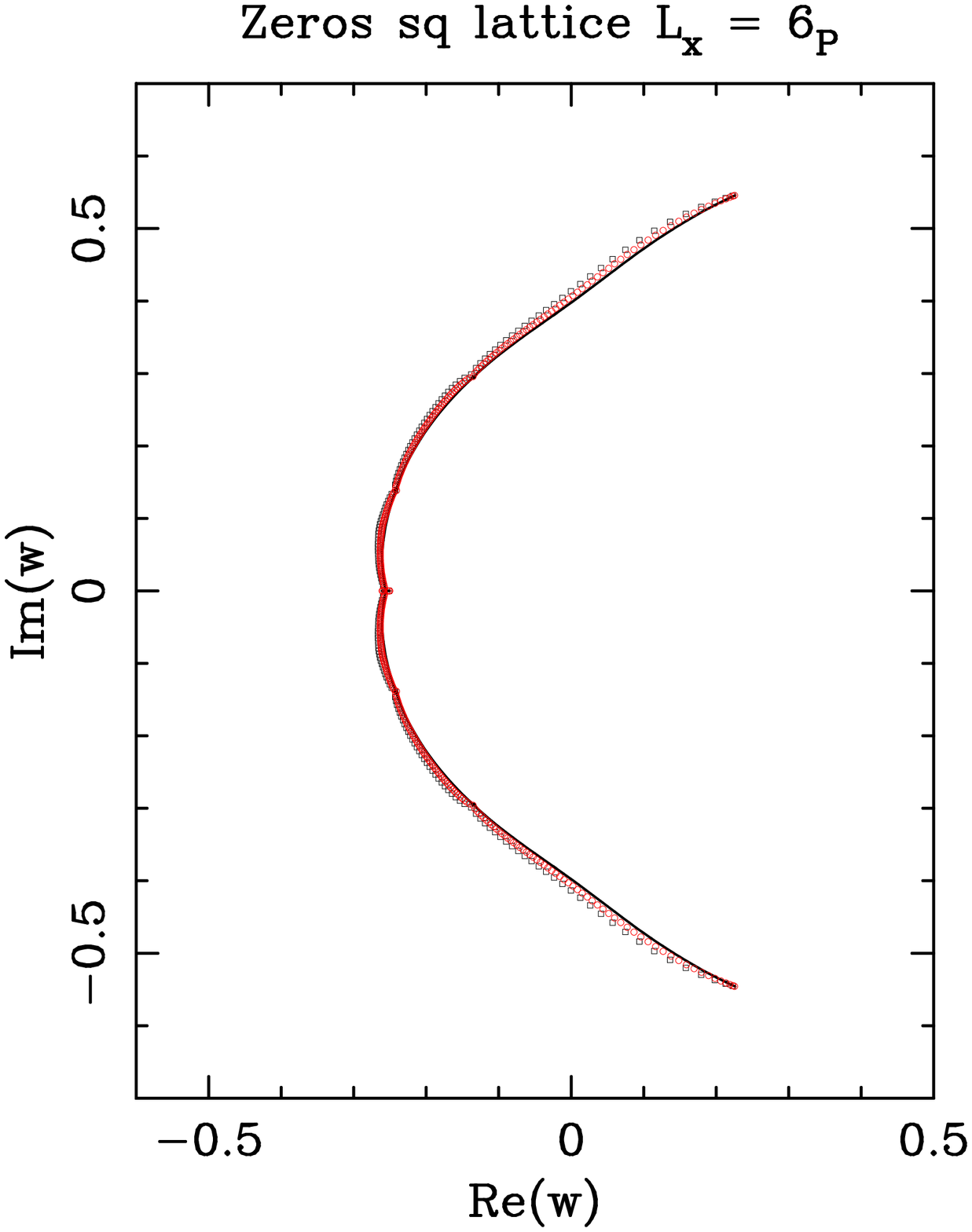} &
   \includegraphics[width=200pt]{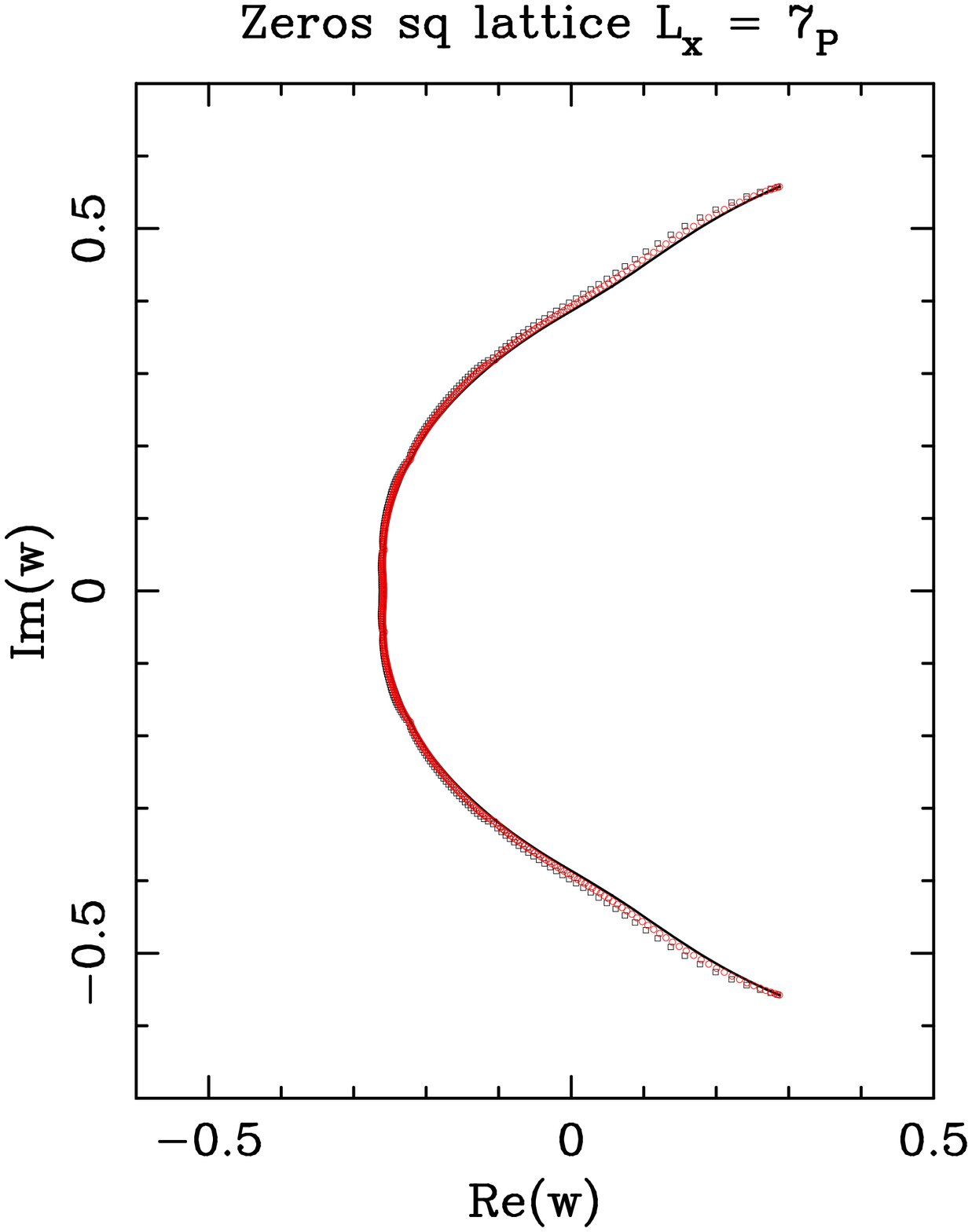} \\[1mm]
   \phantom{(((a)}(a)    & \phantom{(((a)}(b) \\[5mm]
   \includegraphics[width=200pt]{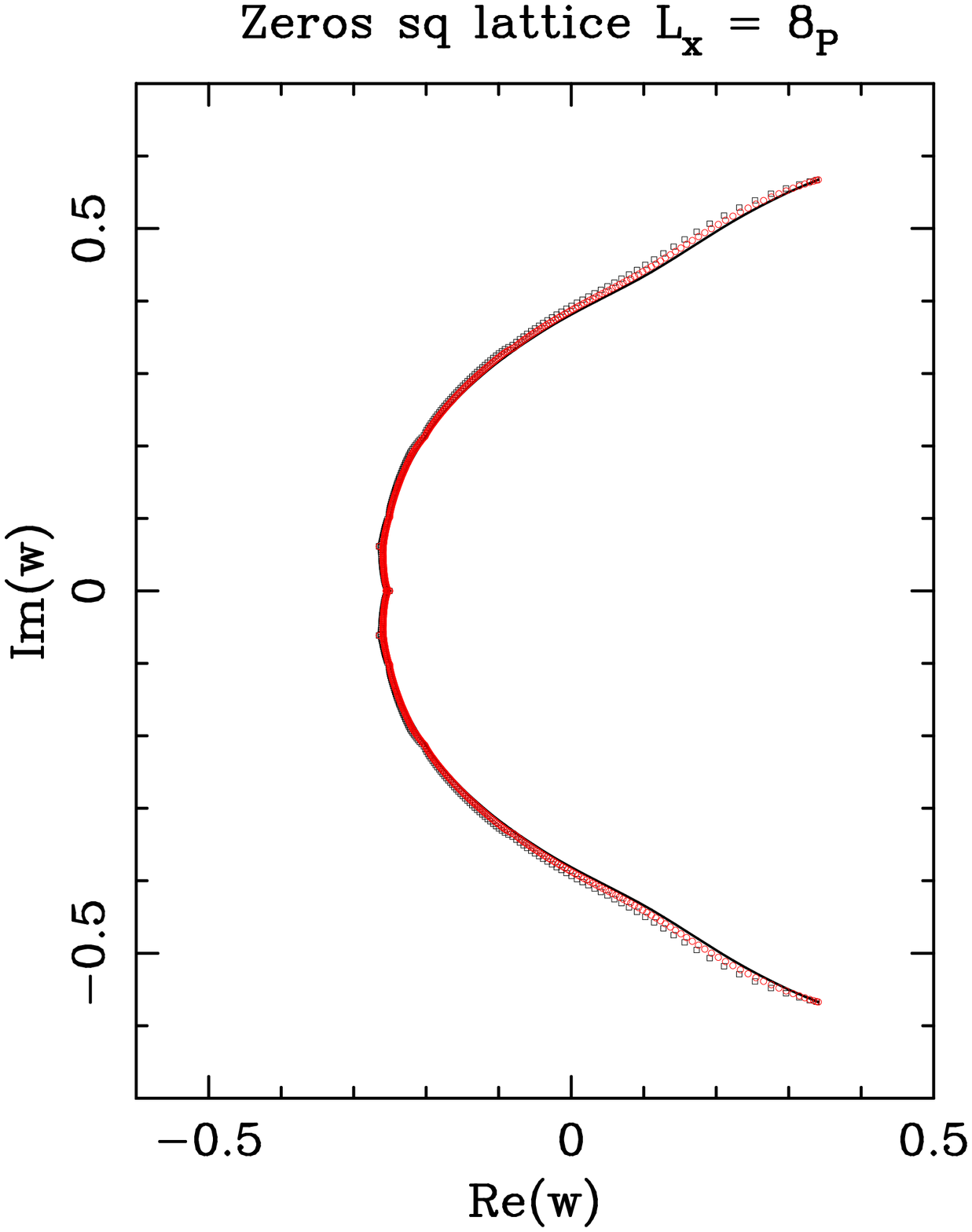} &
   \includegraphics[width=200pt]{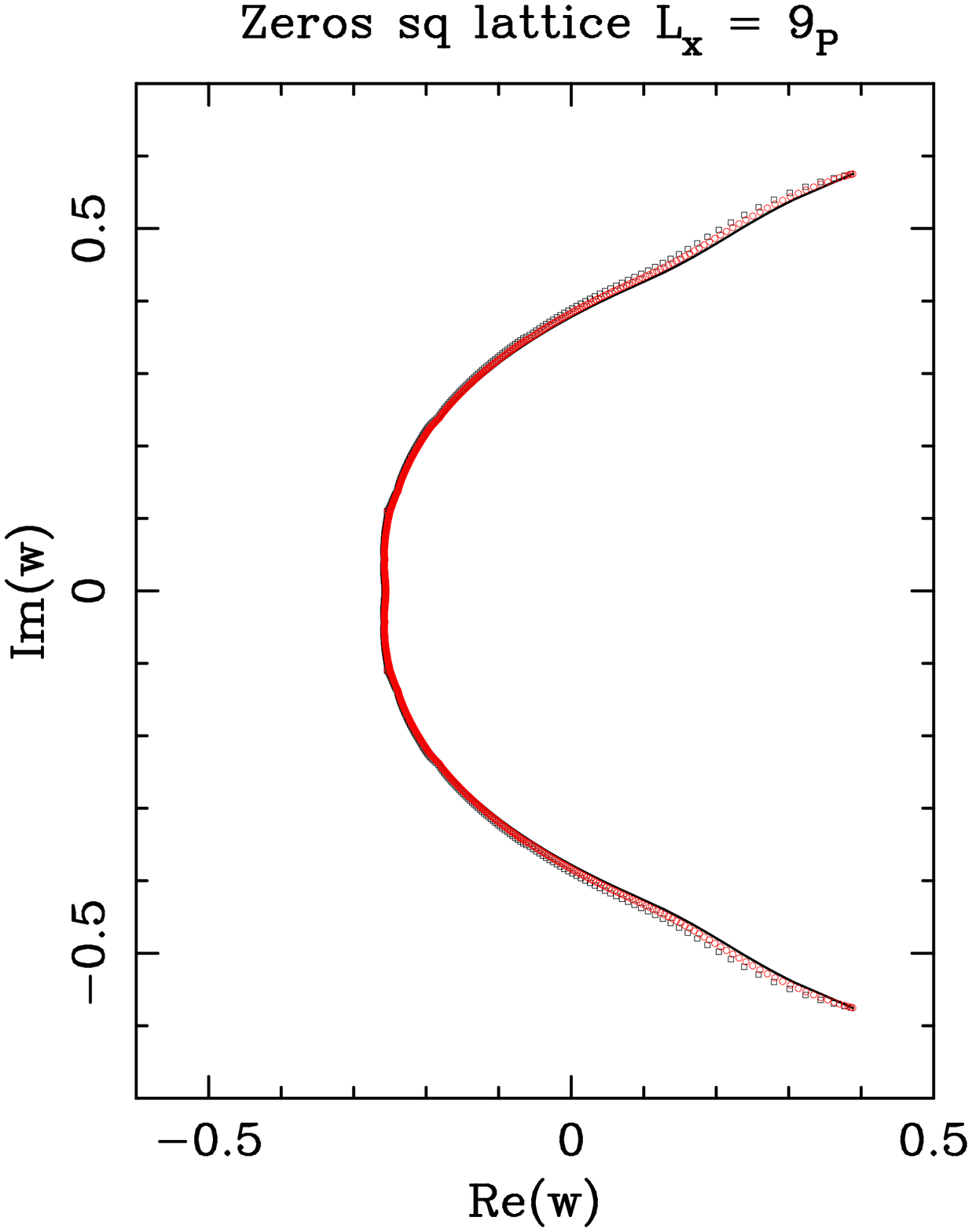} \\[1mm]
   \phantom{(((a)}(c)   & \phantom{(((a)}(d) \\
\end{tabular}
\caption{\label{figure_sq_2}
Limiting curves for square-lattice strips of width (a) $L=6$, (b) $L=7$,
(c) $L=8$, and (d) $L=9$ with cylindrical boundary conditions. We also show the
zeros for the strips $L_{\rm P} \times (5L)_{\rm F}$ (black $\Box$) and
$L_{\rm P} \times (10L)_{\rm F}$ (red $\circ$) for the same values of $L$.
}
\end{figure}

\noindent
{\bf Remark.}
For the strip lattices considered in this paper,
we have looked for isolated limiting points
using the determinant criterion \cite{transfer1}.
For $L\leq 5$, we have found that there are no such limiting points: the
dominant amplitude does not vanish anywhere in the complex $w$-plane.
For $L\geq 6$, we were unable to compute the needed determinants
due to memory limitations.
But our computations of zeros for finite-length strips
(Figures~\ref{figure_sq_2} and \ref{figure_tri_2})
give no indication of any isolated limiting points.
We therefore conjecture that there are no isolated limiting points for
any of the strip lattices studied in this paper.

%
%
\section{Small-$w$ and large-$w$ expansions} \label{sec.theor}

In this section we discuss the small-$w$ and large-$w$ expansions
for the spanning-forest model.

\subsection{Small-$w$ (high-temperature) expansion}  \label{sec.smallw}

Let us begin by considering spanning forests on an $L \times L$ square lattice
with periodic boundary conditions in both directions.
For small $k$, it is easy to count the number of ways of making a
$k$-edge spanning forest;
the result is a polynomial in the volume $V=L^2$
provided that one restricts attention to $L \ge k+1$
so as to avoid clusters that ``wind around the lattice''.
For example, doing this for $k \le 8$ one finds
\begin{eqnarray}
   Z_{L_{\rm P} \times L_{\rm P}}^{\rm sq}(w) &=&
          1 \,+\, 2V w \,+\,
                 {2V \choose 2} w^2 \,+\,
                 {2V \choose 3} w^3
    \,+\, \left[ {2V \choose 4} - V \right] w^4  \nonumber \\
& & \,+\, \left[ {2V \choose 5} - V(2V-4) \right] w^5 \,+\,
          \left[ {2V \choose 6} - V(2V^2 - 9V + 12) \right] w^6 \nonumber \\
& & \,+\, \left[ {2V \choose 7} - {2V\over 3}
                 (2V^3 - 15V^2 + 43V - 54) \right] w^7 \nonumber \\
& & \,+\, \left[ {2V \choose 8} - {V\over 6}
                 (4V^4 - 44 V^3 + 203 V^2 - 526 V + 687)\right] w^8 \nonumber\\
& & \,+\, O(w^9) \;,
\label{series_Z_sq0}
\end{eqnarray}
valid for $L \ge 9$.
Taking the logarithm, one finds that each coefficient is proportional to $V$:
all higher powers of $V$ have cancelled.
Dividing by $V$,
one obtains the small-$w$ expansion of the bulk free energy:
\be
   f({\rm sq},w)  \;=\;
   2w - w^2 + {2 \over 3} w^3 - {3 \over 2} w^4 + {22\over 5} w^5
   - {37\over 3} w^6 + {254\over 7} w^7 -{459\over 4} w^8
   +  O(w^9)  \;.
\label{series_f_sq0}
\ee

Of course, this derivation assumes that the limit $L \to\infty$
can validly be interchanged with expansion in $w$.
This can presumably be proven with additional work,
by methods of the cluster expansion \cite{Borgs}.
It can presumably also be proven, by the same methods,
that this small-$w$ expansion is convergent in some disc $|w| < \epsilon$.
Note, finally, that we have for simplicity used here
periodic boundary conditions in both directions,
in contrast to the cylindrical boundary conditions
used elsewhere in this paper;
but this change should make no difference,
because the bulk free energy is expected
to be independent of boundary conditions.
This can be proven rigorously for $w \ge 0$, by ``soft'' methods;
and by cluster-expansion methods it can presumably be proven
also in some disc $|w| < \epsilon$.

For the triangular lattice one can perform a similar small-$w$ expansion
of the spanning-forest partition function. On an $L \times L$ triangular
lattice with periodic boundary conditions in both directions, the result
for $k\leq 5$ is
\begin{eqnarray}
   Z_{L_{\rm P} \times L_{\rm P}}^{\rm tri}(w) &=&
   1 \,+\, 3Vw \,+\,
                  {3V \choose 2}             w^2 \,+\,
           \left[ {3V \choose 3} - 2V\right] w^3 \nonumber \\
 & & \,+\, \left[ {3V \choose 4} - 3V(2V-1) \right] w^4  \nonumber \\
 & & \,+\, \left[ {3V \choose 5} - 3V^2(3V-4) \right] w^5 \,+\,
      O(w^6) \;,
\label{series_Z_tri0}
\end{eqnarray}
valid for $L \ge 6$.
Taking the logarithm, one finds again that each coefficient is proportional
to $V$. Dividing by $V$, one obtains the small-$w$ expansion of the bulk
free energy for the triangular lattice:
\be
   f({\rm tri},w)  \;=\;
   3w - {3\over 2} w^2 - w^3 + {9 \over 4} w^4 + {3\over 5} w^5
   +  O(w^6)  \;.
\label{series_f_tri0}
\ee

If one wants to extend these expansions to higher order,
direct enumeration becomes increasingly tedious and cumbersome.
A vastly more efficient approach is the finite-lattice method \cite{Neef_77}.
For the square lattice, one can write the small-$w$ expansion of
the infinite-volume free energy as \cite{Neef_77,Guttmann_94}
\be
   f(w)  \;\equiv\;
   \lim_{N\to\infty} {1 \over N^2} \log Z^{\rm free}_{N \times N}(w)
   \;=\;
   \sum_{(q,r) \in B(k)} \alpha_k(q,r) \log Z^{\rm free}_{q \times r}(w)
     \,+\, O(w^{2k-2})
   \;,
\label{flm_sq}
\ee
where $Z^{\rm free}_{q \times r}(w)$ is the partition function for a
square-lattice grid of size $q \times r$ with {\em free}\/ boundary conditions,
and the sum is taken over all rectangles belonging to the set
\be
   B(k)  \;=\;  \{(q,r) \colon\;  q \le r \hbox{ and } q+r \le k \} \;.
\ee
The weights $\alpha_k(q,r)$ are defined as
\be
\alpha_k(q,r)  \;=\; \left\{ \begin{array}{ll}
                 2W_k(q,r) & \qquad \mbox{\rm for $q<r$} \\
                  W_k(q,q) & \qquad \mbox{\rm for $q=r$} \\
                 0       & \qquad \mbox{\rm for $q>r$}
                 \end{array}
         \right.
\label{def_V}
\ee
where
\be
W_k(q,r) \;=\; \left\{ \begin{array}{ll}
                 1 & \qquad \mbox{\rm for $q+r=k$}   \\
                 -3& \qquad \mbox{\rm for $q+r=k-1$} \\
                  3& \qquad \mbox{\rm for $q+r=k-2$} \\
                 -1& \qquad \mbox{\rm for $q+r=k-3$} \\
                  0& \qquad \mbox{\rm otherwise} \\
                 \end{array}
         \right.
\label{def_W}
\ee
The error term in \reff{flm_sq} is given by the smallest connected graph
with no vertices of order 0 or 1 that does not fit into any of the
rectangles in $B(k)$ \cite{Neef_77}. For the square lattice, this graph
is a rectangle of perimeter $2k-2$, hence the error term $w^{2k-2}$.
The main limiting factor on this method is the maximum strip width
$L_{\rm max}$ we are able to handle (due essentially to memory constraints).
In particular, if we set the cut-off $k$ to $2L_{\rm max}+1$,
then formula \reff{flm_sq} gives the free-energy series correct through order
$w^{4L_{\rm max}-1}$.\footnote{
  We have confirmed this fact empirically by doing runs for different
  values of $L_{\rm max}$ and checking to what order they agree.
}

The extension of this method to the triangular lattice is straightforward:
the free energy can be approximated using \reff{flm_sq}, and the
weights are given by \reff{def_V}/\reff{def_W} \cite{Guttmann_97}.
However, the method is less efficient than for the square lattice,
as the error term is now larger than that of \reff{flm_sq}.
If $L_{\rm max}$ is the maximum width we can compute and if we set
the cut-off $k=2L_{\rm max}+1$ as before, then the smallest graphs
not fitting into any of the rectangles in $B(k)$ are trapezoids that
have inclined sides of lengths $L_{\rm max}$ and $L_{\rm max}-1$, respectively,
together with one vertical side of length 1
and one horizontal side of length 1.
The perimeter of such graphs is $2L_{\rm max}+1$, which implies
that \reff{flm_sq} gives the free energy for the triangular-lattice
model correct through order $w^{2L_{\rm max}}$.\footnote{
  Once again, we have confirmed this fact empirically by doing runs
  for different values of $L_{\rm max}$ and checking to what order they agree.
}
This series can be improved slightly by noticing that
the correct term of order $w^{2L_{\rm max}+1}$
can be obtained by adding the quantity $-2 w^{2L_{\rm max}+1}$ to the
finite-lattice-method result.\footnote{
  We discovered this fact empirically by comparing the series obtained for
  two consecutive values of $L_{\rm max}$.
  In all cases, we found that the difference was precisely
  $-2 w^{2L_{\rm max}+1} + O(w^{2L_{\rm max}+2})$.
  It would be interesting to find a combinatorial proof of this fact
  (if indeed it is true in general).
} 

We have computed the partition functions $Z^{\rm free}_{q \times r}(w)$
for both the square and triangular lattices,
by using two complementary methods based on the transfer-matrix approach
(see Section~\ref{sec.transfer}).
We obtained the explicit symbolic form of the transfer matrix for
square-lattice strips of widths $L\leq 9_{\rm F}$ and for
triangular-lattice strips of widths $L\leq 8_{\rm F}$;
we then computed the partition functions $Z^{\rm free}_{q \times r}(w)$
using the standard formulae \cite{transfer1}.
For larger widths (up to $L_{\rm max}=12_{\rm F}$),
we were unable to compute the transfer matrix explicitly;
rather, we used two independent programs (written in C and Perl, respectively)
to build the partition functions layer-by-layer. To handle the
large integers that occur in the computation of the partition function for
large lattices, we used modular arithmetic and the Chinese remainder theorem.
For the square (resp.\ triangular) lattice, we obtained the first 47
(resp.\  25) terms of the free-energy series
\be
f(w) = \sum\limits_{k=1}^\infty f_k w^k  \;.
\ee
The results are displayed in Table~\ref{table_series}.
It is interesting to note that for all the free-energy coefficients $f_k$
we have computed, the quantity $k f_k$ is an integer;
it would be interesting to understand combinatorially why this is so.
{}From these series expansions one can easily obtain
the corresponding small-$w$ expansions
for the derivatives of the free energy (by differentiating the series
term-by-term with respect to $w$).
The analysis of these series is deferred to
Appendix~\ref{sec_series_analysis}.

\subsection{Large-$w$ (perturbative) expansion}  \label{sec.largew}

On any graph $G=(V,E)$,
the generating polynomial of spanning forests can be written as
\be
   F_G(w)  \;=\;  w^{|V|-1} \sum_{k=1}^\infty {c_k(G) \over w^{k-1}}
   \;,
\ee
where $c_k(G)$ is the number of $k$-component spanning forests on $G$
(in particular, $c_1(G)$ is the number of spanning trees of $G$).
Taking logarithms, we find
\be
   {1 \over |V|} \log F_G(w)  \;=\;
   {|V|-1 \over |V|} \log w  \,+\, {1 \over |V|} \log c_1(G) \,+\,
   {1 \over |V|} \log\left[ 1 \,+\, \sum_{k=2}^\infty {c_k(G) \over c_1(G)}
                                                      \, w^{-(k-1)} \right]
   \;.
 \label{eq.largew.2}
\ee
In particular, if we take $G$ to be a large piece of a regular lattice
(with a not-too-eccentric shape), then in the infinite-volume limit
we expect the coefficients of this series to tend term-by-term to limits.
If this indeed occurs, we obtain a large-$w$ expansion of the form
\be
f(w) \;=\; \log w \,+\, f_0 \,+\, {f_1 \over w} \,+\, {f_2 \over w^2} \,+\,
   \ldots  \;.
\ee

Let us now demonstrate,
in the case of the first nontrivial term,
that this convergence indeed occurs;
we shall explicitly compute the limiting value $f_0$,
which is the entropy per site for spanning trees on $G$.
By the matrix-tree theorem \cite[Corollary 6.5]{Biggs_93},
$c_1(G)$ equals $1/|V|$ times the product of the nonzero eigenvalues
of the Laplacian matrix of $G$.
For an $L \times \cdots \times L$ simple-hypercubic lattice
in $d$ dimensions with periodic boundary conditions in all directions,
the eigenvectors of the Laplacian are Fourier modes,
and the eigenvalues can be written explicitly \cite{Shrock_Wu_00};
we have
\be
   c_1(L \times \cdots \times L)  \;=\;
   {1 \over L^d} \!
   \prod_{\begin{scarray}
             p \in \{0, 2\pi/L, \ldots, 2(L-1)\pi/L \}^d  \\
             p \neq 0
          \end{scarray}}
   \left[ 2 \sum\limits_{i=1}^d (1- \cos p_i) \right]
   \;.
\ee
Therefore, in the infinite-volume limit we obtain\footnote{
   The same formula holds for the infinite-volume limit of
   free boundary conditions, but the proof is more complicated:
   see e.g.\ \cite{Lyons_02} and the references cited in
   \cite[footnote 1]{Lyons_02}.
}
\be
   \lim_{L \to \infty} {1 \over L^d} \log c_1(L \times \cdots \times L)
   \;=
   f_0
   \;=\;
   \int\limits_{[-\pi,\pi]^d} \!\! {d^d p \over (2\pi)^d} \,
     \log\!\left[ 2 \sum\limits_{i=1}^d (1- \cos p_i) \right]
   \;.
\ee
(This integral is infrared-convergent in any dimension $d>0$.)
Analogous formulae hold for other regular lattices
\cite{Wu_77,Shrock_Wu_00,Felker-Lyons,Glasser_Wu_03}.
For the standard two-dimensional lattices one can carry out the integrals
explicitly, yielding \cite{Wu_77,Shrock_Wu_00,Glasser_Wu_03}
\begin{eqnarray}
   f_0({\rm sq})  & = &
   {4 \over \pi} \sum_{k=0}^\infty {(-1)^k \over (2k+1)^2}
   \;=\;  {4{\bf G} \over \pi} \;\approx\; 1.166\,243\,616\,123\ldots
   \slabel{def_f0_sq}
      \\[4mm]
   f_0({\rm tri})  & = &
   {3\sqrt{3} \over \pi}
   \sum_{k=0}^\infty \left( {1 \over (6k+1)^2} - {1 \over (6k+5)^2} \right)
   \;=\;
   {\sqrt{3} \over 12\pi} \, [\psi'(1/6) - \psi'(5/6)]
      \nonumber \\
   & & \hphantom{{3\sqrt{3} \over \pi}
   \sum_{k=0}^\infty \left( {1 \over (6k+1)^2} - {1 \over (6k+5)^2} \right)}
   \;\approx\; 1.615\,329\,736\,097\ldots
   \slabel{def_f0_tri}
\end{eqnarray}
where ${\bf G}$ is Catalan's constant
and $\psi'(z) = (d^2/dz^2) \log\Gamma(z)$ is the
first derivative of the digamma function.
See also \cite{Felker-Lyons} for high-precision computations of $f_0$
on the simple hypercubic lattice in dimensions $3 \le d \le 20$,
as well as on the body-centered hypercubic lattice in dimensions 3 and 4.

We defer to a later paper \cite{Polin-Sokal} the computation of
$f_1, f_2, \ldots$,
which can be obtained by perturbative expansion
in a fermionic theory that contains a Gaussian term
and a special four-fermion coupling (see \cite{Caracciolo_PRL} for details).
It is worth mentioning that this perturbation expansion can sometimes be
infrared-divergent, reflecting the fact that the coefficients
of \reff{eq.largew.2} may in fact {\em diverge}\/ as $L \to\infty$.
For example, in dimension $d=1$ with periodic boundary conditions,
we have the exact formula
\be
   F_{L_{\rm P}}(w)  \;=\;  (1+w)^L - w^L  \;=\;
      L w^{L-1} \left[ 1 \,+\, {L-1 \over 2} \, w^{-1} \,+\,
                               {(L-1)(L-2) \over 6} \, w^{-2} \,+\, \ldots
                \right]  \;,
\ee
so that
\be
   {1 \over L} \log F_{L_{\rm P}}(w)  \;=\;
   {L-1 \over L} \log w  \,+\, {\log L \over L} \,+\, {L-1 \over 2L} \, w^{-1}
      \,+\, {(L-1)(L-5) \over 24L} \, w^{-2} \,+\, \ldots  \;.
 \label{eq.F.d=1}
\ee
We see that $f_1 = 1/2$, but $f_2$ is infrared-divergent
(as are $f_3$ and subsequent terms).
We do not yet know whether $f_2,f_3,\ldots$ are infrared-finite
in dimension $d=2$.

We can estimate the first terms $f_i$ by computing
$f(w)$ numerically at a set of (large) values of $w$,
and trying to fit it to a polynomial Ansatz in $1/w$:
\be
\real f(w) - \log |w| - f_0  \;=\;
 \sum\limits_{k=1}^{k_{\rm max}} {f_k \over w^k}   \;.
\label{Ansatz_large_w}
\ee
For this calculation, we began by computing the finite-width free energies
$f_L(w)$ at $w= \pm 2, \pm 3,\ldots, \pm 10$
on strips up to $L=14_{\rm P}$ (square) and $L=13_{\rm P}$ (triangular).
We then estimated the infinite-volume free energy $f(w)$
by extrapolating the finite-width data using the
Ansatz~\reff{def_ansatz_f_bis} below
(see Sections~\ref{sec.res.sq2a} and~\ref{sec.res.tri2a} for details).
Finally, we used a polynomial fit \reff{Ansatz_large_w} with $k_{\rm max}=5$
to obtain the subleading coefficients $f_k$ for $k=1,\ldots,5$,
fitting separately the data with $w>0$ and $w<0$.
More precisely, each sign of $w$ and each value of $w_{\rm min}=2,\ldots,6$,
we fit the points with $|w| = w_{\rm min}, w_{\rm min}+1,\ldots,w_{\rm min}+4$
to the Ansatz \reff{Ansatz_large_w} with $k_{\rm max}=5$.
The observed variations in the parameter estimates are due to
the neglected higher-order terms $k>k_{\rm max}$ in \reff{Ansatz_large_w}.
These fits are displayed in the rows labelled ``Non-Biased''
in Tables~\ref{table_fit_sq_large_w} (square lattice)
and~\ref{table_fit_tri_large_w} (triangular lattice).

It is clear that the coefficient $f_1$ is close to
$0.25=1/8$ for the square lattice and to
$0.08333 \approx 1/12$ for the triangular lattice.
This motivates the following conjecture:

\begin{conjecture} \label{conjecture_f1}
For any regular lattice in dimension $d \ge 2$
with coordination number $r$,
the large-$w$ series expansion of the spanning--forest free energy
takes the form
\be
f(w)  \;=\; \log |w| \,+\, f_0 \,+\, {1\over 2r} {1\over w} \,+\, O(w^{-2})
\label{def_large_w_series}
\ee
where $f_0$ gives the entropy per site for spanning trees on the lattice
in question.
\end{conjecture}

\noindent
Note that the formula \reff{def_large_w_series} is {\em false}\/
in dimension $d=1$:  as seen from \reff{eq.F.d=1},
we have $f_1 = 1/2$, not $1/4$.

We have redone the above fits by fixing $f_1$ to its conjectured value
$1/(2r)$.
Then, for each sign of $w$ and each value of $w_{\rm min}=2,\ldots,7$,
we fit the points with $|w|=w_{\rm min},\ldots,w_{\rm min}+3$ to this
biased Ansatz. The results are shown in rows labelled ``Biased'' in
Tables~\ref{table_fit_sq_large_w} and~\ref{table_fit_tri_large_w}. The next
coefficient $f_2$ appears to take the same value for the fits with positive and
negative values of $w$: $f_2({\rm sq})\approx 0.0098$ and
$f_2({\rm tri})\approx 0.0025$.

\bigskip

{\bf Remark.}  We have very recently proven Conjecture~\ref{conjecture_f1},
and have also understood why $d=1$ is an exception.
These results will be reported elsewhere \cite{Polin-Sokal}.

%
%
\section{Transfer matrices}   \label{sec.transfer}

The first step in our analysis is to compute
the transfer matrix $\T$ and
the vectors $\uu$ and $\vv_{\rm id}$ that appear in
\reff{def_transfer_matrix}.
We can proceed in three alternative ways:
\begin{itemize}
   \item[(a)]  Compute the transfer matrix for the
       general $q$-state Potts model,
       symbolically as a polynomial in $q$ and $v$.
       (See \cite{transfer1} for the theory,
        and \cite{Tutte_sq,Tutte_tri} for the computations.)
       Then perform the limit $q,v\to0$ with $w=v/q$ fixed.
   \item[(b)]  Take the limit $q,v\to0$ with $w=v/q$ fixed
       right at the beginning, and compute the transfer matrix
       for the spanning-forest problem,
       symbolically as a polynomial in $w$.
   \item[(c)]  Same as (b), but compute the transfer matrix
       numerically for a specified (but arbitrary) value of $w$.
\end{itemize}
For $L\leq 9$ we used methods (a) and (b), and verified that they
give the same answer (this is an important check on the correctness
of our programs). For $L=10$ we could use only method (b).
For $L\geq 11$ we were unable to perform the computation symbolically;
we therefore performed the computations numerically (using machine
double-precision) for $11 \le L \le 16$ at selected real values of $w$.

The dimension of the transfer matrix for a square-lattice strip
of width $L$ and cylindrical boundary conditions is given by SqCyl($L$) in
Table~\ref{table_dimensions}.
It coincides with the dimension of the transfer matrix
for the full Tutte polynomial on square-lattice strips
with the same boundary conditions \cite{Tutte_sq},
and equals the number of equivalence classes
modulo translation and reflection
of non-crossing partitions of $\{1,2,\ldots,L\}$.
This number is denoted by
$N_{Z,{\rm sq},{\rm PF},L}$ in Refs.~\cite{Tutte_sq,Tutte_tri};
see \reff{eq.SqCyl} below for an exact formula.

The dimension of the transfer matrix for a triangular-lattice strip
of width $L$ and cylindrical boundary conditions is given by TriCyl$'(L)$ in
Table~\ref{table_dimensions}.
It coincides with the dimension of the transfer matrix
for the full Tutte polynomial on triangular-lattice strips
with the same boundary conditions \cite{Tutte_tri},
and equals the number of equivalence classes modulo translation
of non-crossing partitions of $\{1,2,\ldots,L\}$.
In Ref.~\cite{Tutte_tri} TriCyl$'(L)$ is denoted $N_{Z,{\rm tri},{\rm PF},L}$,
and an exact formula is provided:
\be
\hbox{\rm TriCyl}'(L) = {1\over L} \left[ {1 \over L+1} {2L \choose L}
    \,+\, \sum\limits_{d|L,\, 1 \le d < L} \phi(L/d) {2d \choose d} \right]
\ee
where $d|L$ means that $d$ divides $L$,
and $\phi(n)$ is the Euler totient function
(i.e.\ the number of positive integers $k \le n$
that are relatively prime to $n$).
The sequence TriCyl$'(L)$ is given as sequence A054357 in \cite{Sloane};
it equals the number of bi-colored unlabelled plane trees having $L$ edges
\cite{Bona_00}.

As in previous work on other cases of the Potts model
\cite{transfer3,Tutte_tri},
we can reduce the dimension of this transfer matrix
down to TriCyl($L$) $=$ SqCyl($L$) by noting that in
the translation-invariant subspace of connectivities, the transfer matrix
{\em does}\/ commute with reflections
(see \cite[beginning of Section 4]{transfer3} for a detailed explanation).
Thus, in a new basis with connectivities
that are either odd or even under reflection, the transfer matrix takes a
block-diagonal form: the block that is even under reflection has dimension
TriCyl($L$) = SqCyl($L$) and a non-vanishing amplitude; the block that is odd
under reflection has a zero amplitude and hence
does not contribute to the partition function.

The dimensions SqCyl $=$ TriCyl and TriCyl$'$ are related by the following
equation (proved in \cite{Tutte_tri}):
\be
{\rm SqCyl} (L) = {\rm TriCyl}(L) = {1\over2} \left[ {\rm TriCyl}'(L)
  + { L \choose\lfloor L/2 \rfloor } \right]
 \label{eq.SqCyl}
\ee
The numerical values of SqCyl($L$) and TriCyl$'$($L$) up to $L=20$
are shown in Table~\ref{table_dimensions}.
Both grow asymptotically like $4^L L^{-5/2}$ as $L \to\infty$.

%
%
\section{Square-lattice strips with cylindrical boundary conditions}
\label{sec.sq}

For each lattice width $L$ up to $L=10$,
we have computed symbolically the transfer matrix $\T$ and
the vectors $\uu$ and $\vv_{\rm id}$.
Then we computed the zeros of the partition function
$F_{m_{\rm P} \times n_{\rm F}}(w)$
for strips of aspect ratio $\rho=5$ and $\rho=10$;
and for $L \le 8$ we also computed
the accumulation set ${\cal B}$ of partition-function zeros
in the limit $\rho\to\infty$.
For $L=9,10$ we were unable to compute the full limiting curve ${\cal B}$,
but we did compute some selected points along it.

\begin{figure}
\centering
\begin{tabular}{cc}
   \includegraphics[width=200pt]{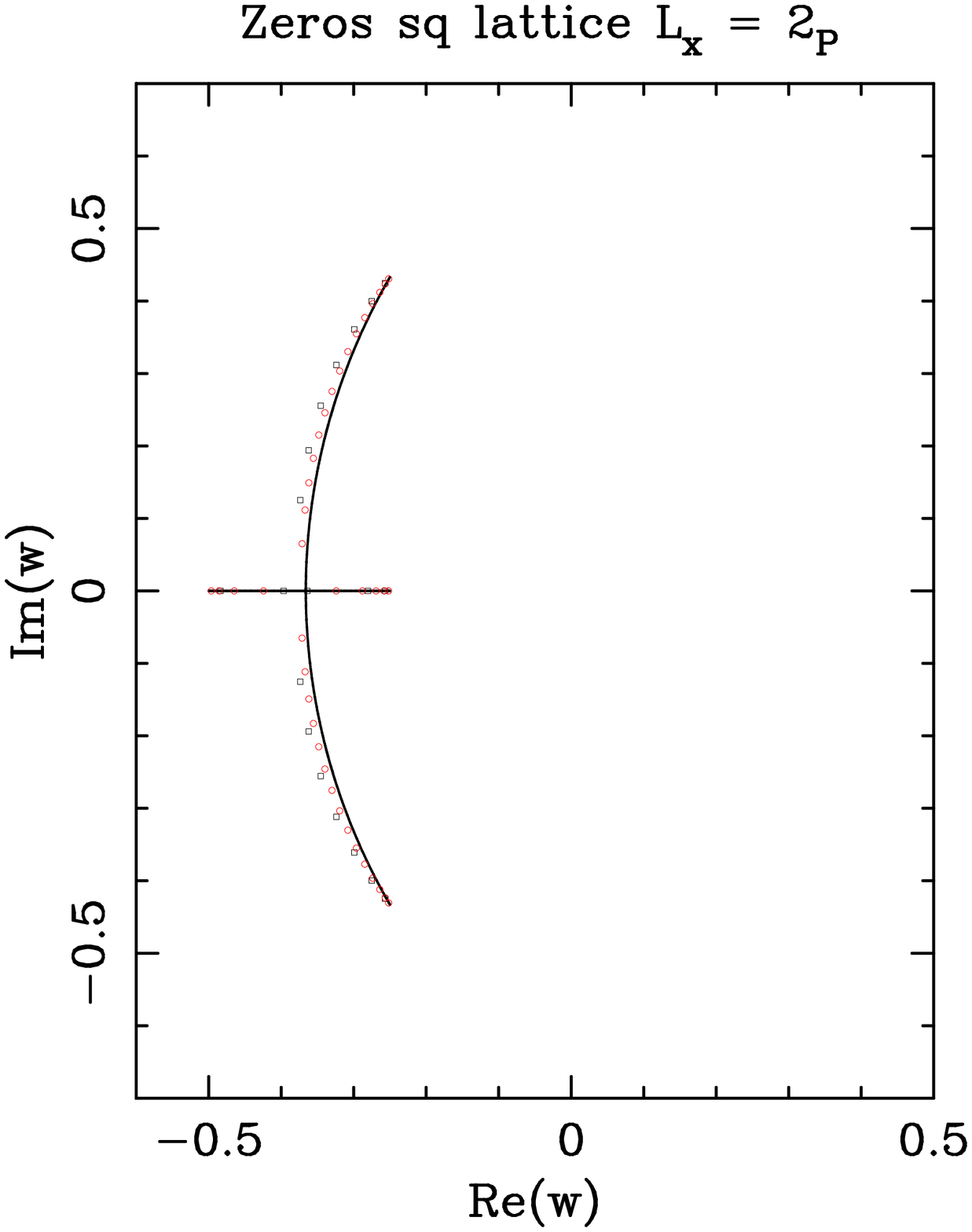} &
   \includegraphics[width=200pt]{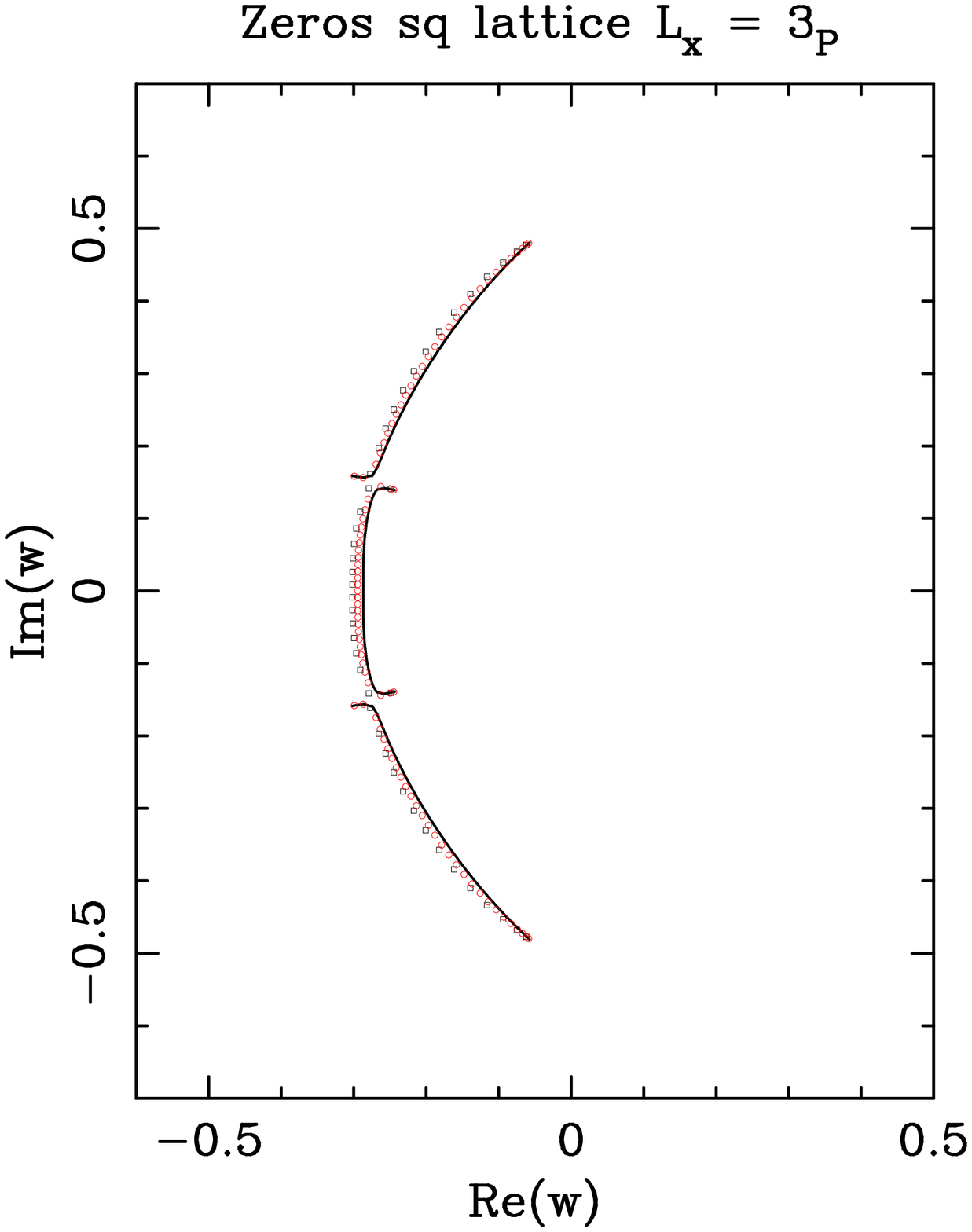} \\[1mm]
   \phantom{(((a)}(a)    & \phantom{(((a)}(b) \\[5mm]
   \includegraphics[width=200pt]{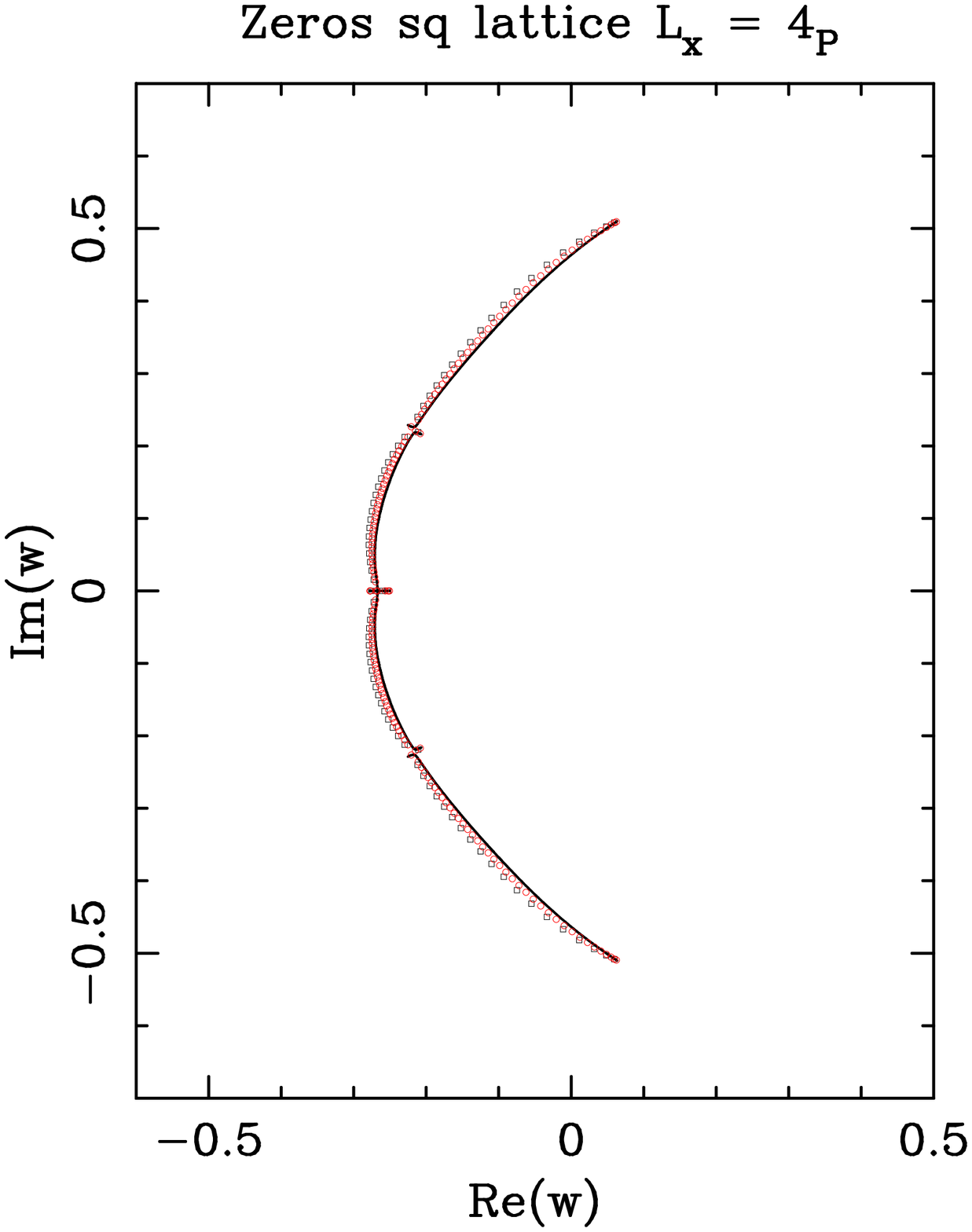} &
   \includegraphics[width=200pt]{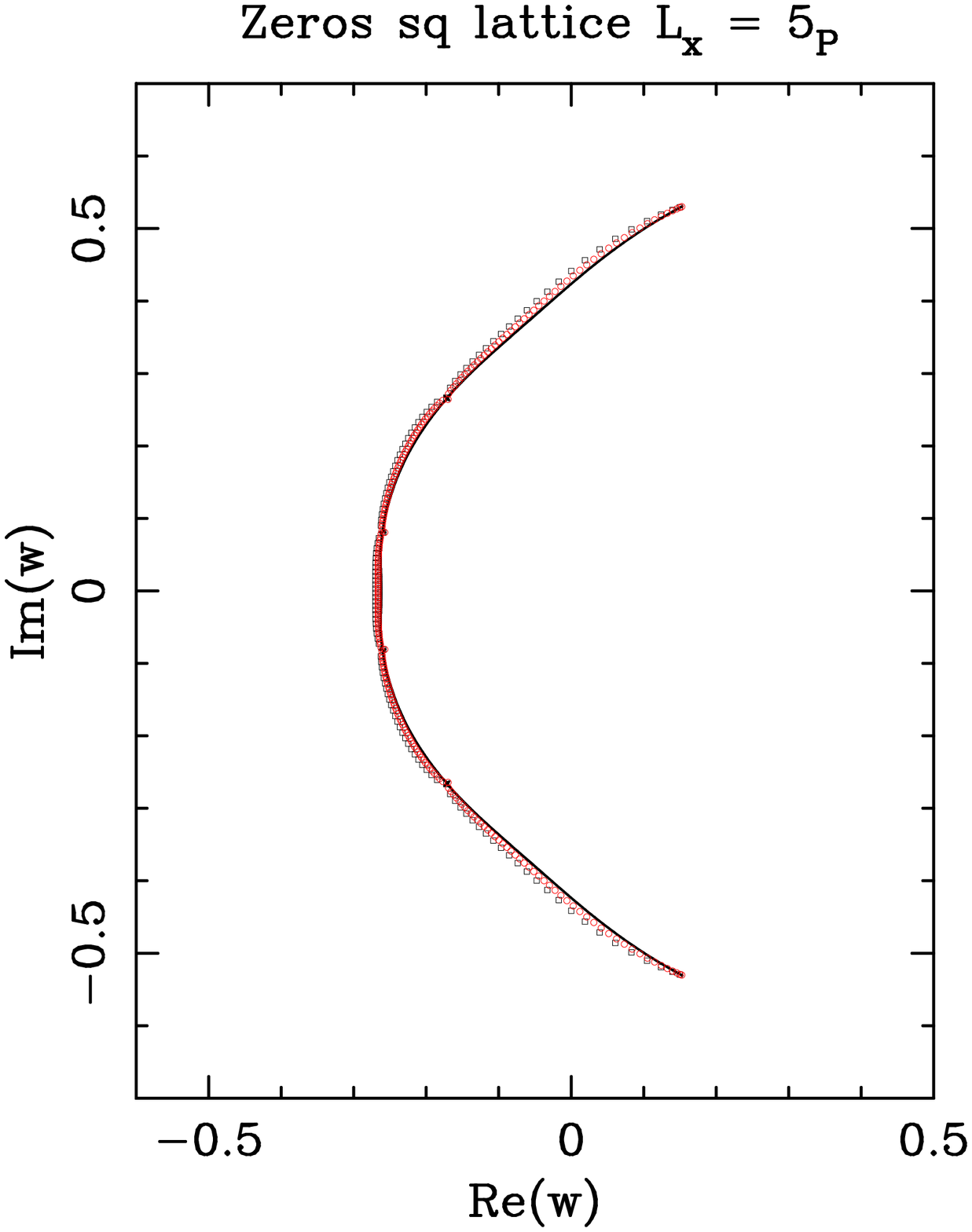} \\[1mm]
   \phantom{(((a)}(c)    & \phantom{(((a)}(d) \\
\end{tabular}
\caption{\label{figure_sq_1}
Limiting curves for square-lattice strips of width (a) $L=2$, (b) $L=3$,
(c) $L=4$, and (d) $L=5$ with cylindrical boundary conditions. We also show the
zeros for the strips $L_{\rm P} \times (5L)_{\rm F}$ (black $\Box$) and
$L_{\rm P} \times (10L)_{\rm F}$ (red $\circ$) for the same values of $L$.
}
\end{figure}

\begin{figure}
\centering
\includegraphics[width=400pt]{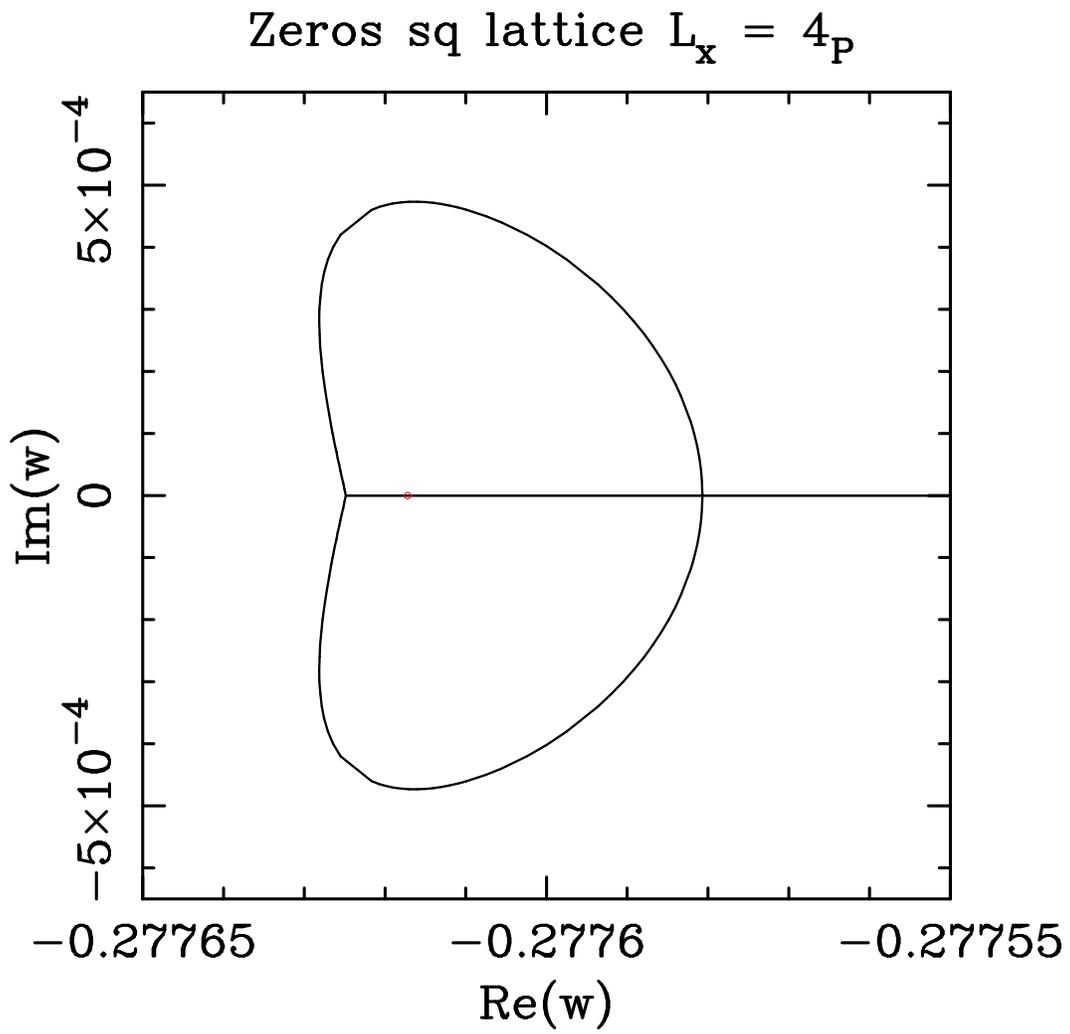}
\caption{\label{figure_sq_4P_zoom}
Blow-up of Figure~\protect\ref{figure_sq_1}(c) around the
T point $w_{0-} \approx -0.2776248333$.
The multiple point is located at $w\approx -0.2775806860$.
}
\end{figure}

\begin{figure}
\centering
\includegraphics[width=400pt]{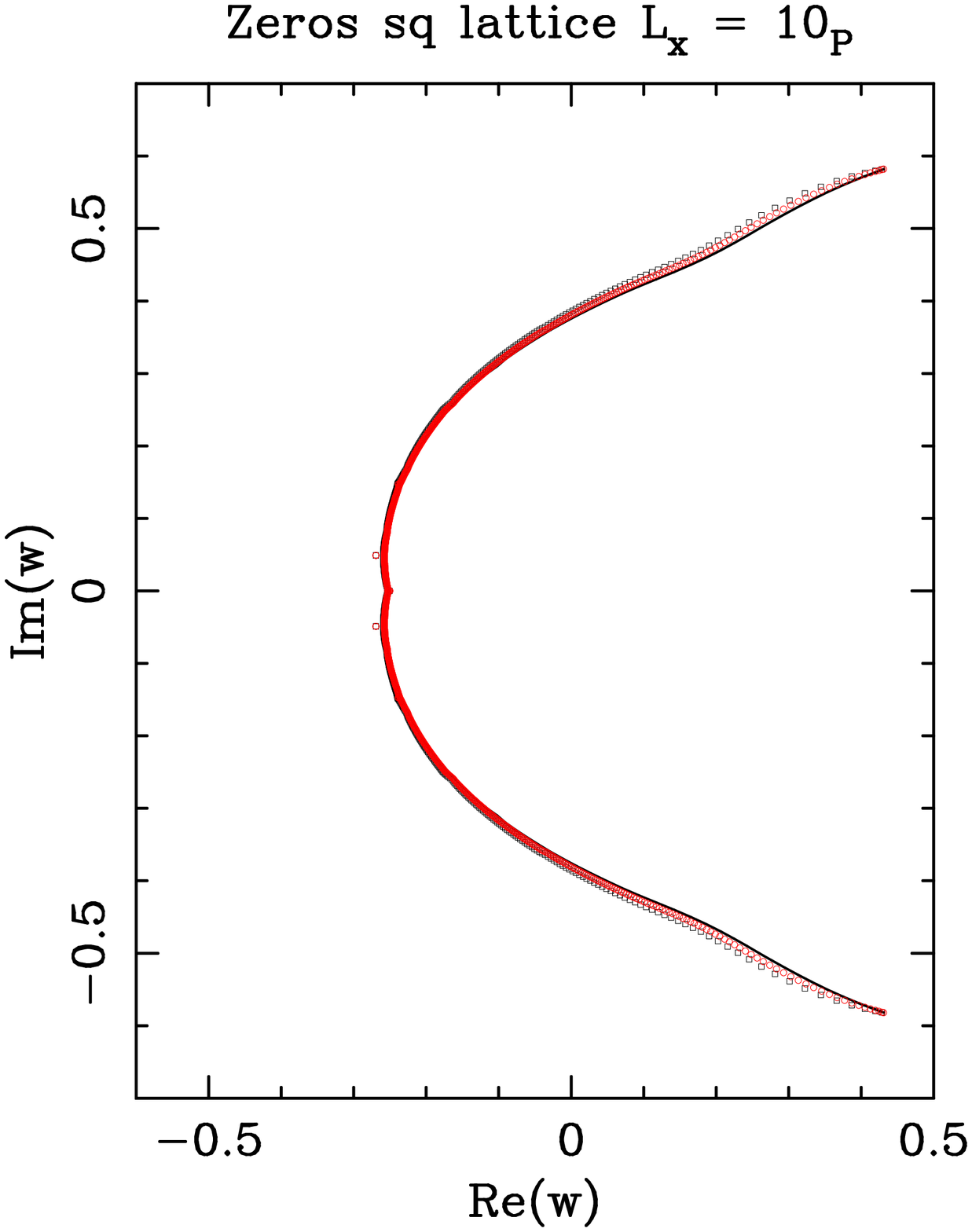}
\caption{\label{figure_sq_3}
Limiting curve for the square-lattice strip of width $L=10$ with cylindrical
boundary conditions. We also show the
zeros for the strips $10_{\rm P} \times 50_{\rm F}$ (black $\Box$) and
$10_{\rm P} \times 100_{\rm F}$ (red $\circ$).
}
\end{figure}

The limiting curves $\scrb$ resulting from these computations
are shown in Figures~\ref{figure_sq_1}--\ref{figure_sq_3}
(superposed in Figure~\ref{figure_sq_all}),
and their principal characteristics
are summarized in Table~\ref{table_sq}.
One interesting feature of ${\cal B}$ is the point(s)
where it crosses the real $w$-axis.
For odd width $L$, it turns out that there is only one such point,
which we shall denote $w_{0Q}(L)$.
For even width $L$, by contrast,
${\cal B}$ contains a real segment $[w_{0-}(L), \, w_{0+}(L)]$.
This segment contains a multiple point of ${\cal B}$,
where an arc of ${\cal B}$ lying at $\imag w \neq 0$ crosses the real axis;
we shall denote this point $w_{0Q}(L)$.
It is a curious fact that for the
even-width square lattices considered here ($L=2,4,6,8,10$),
we find $w_{0+}(L) = -1/4$ {\em exactly}\/.
Finally, we define $w_B(L)$ to be the complex-conjugate pair of endpoints
of ${\cal B}$ with the largest real part.

For $L\leq 5$ we computed the limiting curve ${\cal B}$
using the resultant method explained in \cite[Section 4.1.1]{transfer1}.
For $L=6$, we computed the endpoints using this method, and the rest of
$\scrb$
using the direct-search method explained in \cite[Section 4.1.2]{transfer1}.
In all these cases we can be sure that we did not miss any endpoints
or connected components of ${\cal B}$.
For $L\geq 7$, by contrast, we were unable to compute the resultant;
we therefore located ${\cal B}$ using the direct-search method.
Here we could easily have missed some small components of ${\cal B}$.
Our reports in Table~\ref{table_sq}
for the number of connected components (\#C), endpoints (\#E)
and multiple points (\#Q) must therefore be viewed for $L \ge 7$
as lower bounds.
For $L\geq9$ the computation of the limiting curve ${\cal B}$ is very
time-consuming;  we therefore computed only a few important points.

%
%
\subsection{$L=2$} \label{sec.sq2P}

In this case the connectivity basis is two-dimensional.
The transfer matrix $\T$ and the vectors $\vv_{\rm id}$ and $\uu$ are given by
\begin{subeqnarray}
\T &=&
   \left( \begin{array}{cc}
         w^2   & 2 w^3 \\
       2 w +1  & 5 w^2 + 4 w + 1
          \end{array} \right) \\
\uu^{\rm T}           &=&  \left( 1, 2w+1 \right) \\
\vv_{\rm id}^{\rm T}  &=&  \left( 0, 1    \right)
\end{subeqnarray}
The zeros of the polynomials $F_{2_{\rm P} \times n_{\rm F}}(w)$
with $n=10,20$ are displayed in Figure~\ref{figure_sq_1}(a).
In the same figure we also show the limiting curve ${\cal B}$.
The curve ${\cal B}$ is connected:
it is the union of a horizontal segment
running from $w_{0-} = -1/2$ to $w_{0+} = -1/4$
and an arc running between the complex-conjugate endpoints
$w_B = -1/4 \pm (\sqrt{3}/4)i$.
The segment and the arc cross
at the multiple point at $w_{0Q} \approx -0.3660254038$.

It is worth noting that, even though the curve $\scrb$
contains the point $w=-1/2$ as an endpoint,
the value $w=-1/2$ is {\em not}\/ a zero of the partition function
for any finite strip $2_{\rm P}\times n_{\rm F}$
other than the trivial case $n=1$.
As a matter of fact, the partition function takes the value
\be
F_{2_{\rm P} \times n_{\rm F}}(w=-1/2) \;=\; - (n-1)/4^{n-1}
   \;.
\ee
This observation supports the conjecture that
there are no roots in the half-plane $\real w < -1/2$,
i.e.\ that the family $2_{\rm P}\times n_{\rm F}$ of strip graphs
possesses the univariate dual Brown-Colbourn property
for all finite lengths $n$.

%
%
\subsection{$L=3$} \label{sec.sq3P}

The connectivity basis is three-dimensional; the transfer matrix
$\T$ and the vectors $\vv_{\rm id}$ and $\uu$ are given by
\begin{subeqnarray}
\T &=&
   \left( \begin{array}{ccc}
          w^3     & 6 w^4              & 3 w^5  \\
          w^2     & w^2(7 w + 1)       & w^3(4 w + 1) \\
          3 w + 1 & 3(8 w^2 + 5 w + 1) & 16 w^3 + 15 w^2 + 6 w + 1
\end{array} \right) \\
\uu^{\rm T}          &=&  \left( 1, 3(2w + 1), 3w^2+3w+1 \right) \\
\vv_{\rm id}^{\rm T} &=&  \left( 0,         0,         1 \right)
\end{subeqnarray}
The zeros of the polynomials $F_{3_{\rm P} \times m_{\rm F}}(w)$
with $n=15,30$ are displayed in Figure~\ref{figure_sq_1}(b),
along with the limiting curve ${\cal B}$.
The curve ${\cal B}$ has three connected components.
One of them runs between the complex-conjugate endpoints
$-0.2432796623 \pm 0.1389560739\, i$
and intersects the real $w$-axis at $w_{0Q} \approx -0.2868497019$.
The other two run from
$w\approx -0.3020539886 \pm 0.1587843374\, i$
to $w_B \approx -0.0574443351 \pm 0.4806253161\, i$.

%
%
\subsection{$L=4$} \label{sec.sq4P}

The connectivity basis is six-dimensional; the transfer matrix
$\T$ and the vectors $\vv_{\rm id}$ and $\uu$ are given by
\begin{subeqnarray}
\T &=&
   \left( \begin{array}{cccccc}
w^4 & 8w^5      & 4w^5      & 12w^6        & 8w^6         & 4w^7\\
w^3 & w^3 C_9   & 4w^4      & 2w^4 C_7     & 2w^4 C_5     & w^5 C_5\\
0   & 0         & w^4       & 2w^5         & 0            & w^6\\
w^2 & 2w^2 C_5  & w^2 C_6   & w^2 D_{21,8} & 4w^3 C_3     & w^3 D_{9,5}\\
w^2 & 2w^2 C_5  & 4w^3      & 4w^3 C_4     & w^2 D_{13,6} & 2w^4 C_3 \\
4w+1& 4 D_{11,6}& 2 D_{12,6}& 4 C_3 D_{8,4}&2 C_3 D_{10,5}& T_{6,6} \\

\end{array} \right) \\
\uu^{\rm T}          &=&  \left(1, 4C_2, 2C_2, 4D_{3,3}, 2C_2^2, C_2 D_{2,2}
                         \right) \\
\vv_{\rm id}^{\rm T} &=&  \left(0,0,0,0,0,1 \right)
\end{subeqnarray}
where we have used the shorthand notations
\begin{subeqnarray}
C_k(w)     &=& k w + 1         \slabel{def_Ck} \\
D_{m,n}(w) &=& m w^2 + n w + 1 \slabel{def_Dk} \\
T_{6,6}(w) &=& 45w^4+52w^3+28w^2+8w+1
\end{subeqnarray}

The zeros of the polynomials $F_{4_{\rm P} \times n_{\rm F}}(w)$
with $n=20,40$ are displayed in Figure~\ref{figure_sq_1}(c), along with
the limiting curve ${\cal B}$.
The curve ${\cal B}$ has three connected components.
One of them is the union of a horizontal segment
running from $w_{0-} \approx -0.2776248333$ to $w_{0+} = -1/4$
and an arc running between the complex-conjugate endpoints
$w \approx -0.2252016334 \pm 0.2287233280\, i$.
The segment and the arc cross
at the multiple point at $w_{0Q} \approx -0.2670015604$.
The other two components are complex-conjugate arcs running from
$w\approx -0.2068057238 \pm 0.2160687519\, i$
to $w_B \approx 0.0632994010  \pm 0.5099839130 \, i$.
Finally, there is a pair of very small complex-conjugate bulb-like regions
emerging from the endpoint $w_{0-} \approx -0.2776248333$ ---
which is therefore a T point --- and going back
to the real $w$-axis at the multiple point $w \approx -0.2775806860$:
see the blow-up picture in Figure~\ref{figure_sq_4P_zoom}.

%
%
\subsection{$L=5,6,7,8,9,10$} \label{sec.sq5P}

The transfer matrices for $L\geq 5$ are too lengthy to be quoted here.
Those for $L \le 9$ can be found in the {\sc Mathematica} file
{\tt forests\_sq\_2-9P.m} that is available with the electronic version
of this paper in the cond-mat archive at {\tt arXiv.org}.
The file for $L=10$, which is 13.6 MB long, can be obtained on request
from the authors.

We have plotted for each $L$ ($=5,6,7,8,9,10$) the zeros
of $F_{L_{\rm P}\times (\rho L)_{\rm F}}(w)$ for aspect ratios
$\rho=5,10$ as well as the limiting curves ${\cal B}$ ($\rho=\infty)$.
See Figure~\ref{figure_sq_1}(d) for $L=5$,
Figures~\ref{figure_sq_2}(a,b,c,d) for $L=6,7,8,9$, and
Figure~\ref{figure_sq_3} for $L=10$.

The principal features of the limiting curves ${\cal B}$ are summarized in
Table~\ref{table_sq}. For $L=5$, there are ten endpoints located at
$w \approx -0.2645566722 \pm  0.0812460198\, i$,
$ -0.2562049781 \pm 0.0804259250\, i$,
$ -0.1745330113 \pm 0.2687656693\, i$,
$ -0.1684195011 \pm 0.2629234881\, i$, and  \linebreak
$  0.1533657968 \pm 0.5306112949\, i$
(the latter is $w_B$).
The limiting curve crosses the real axis at $w_{0Q} \approx -0.2620754678$.

For $L=6$ there is a multiple point at $w_{0Q} \approx -0.2563792782$.
This point belongs to a horizontal segment running from
$w_{0-} \approx -0.2609570768$ to $w_{0+} =-1/4$.
There are ten more endpoints at
$w\approx -0.2441302665 \pm 0.1392115396\, i$,
$-0.2401194057 \pm  0.1380752982\, i$,
$-0.1359169474 \pm  0.2964285650\, i$,
$-0.1338793838 \pm  0.2939739183\, i$, and  \linebreak
$ 0.2262944917 \pm  0.5460127254\, i$
(the latter is $w_B$).

For $L=7$ there are fourteen endpoints located at
$w\approx -0.258797 \pm  0.056312\, i$,
$-0.256989 \pm 0.056335\, i$,
$-0.222450 \pm 0.180872\, i$,
$-0.220899 \pm 0.180261\, i$,
$-0.103677 \pm 0.316136\, i$,
$-0.104346 \pm 0.317122\, i$, and
$ 0.2879810252 \pm 0.5579895735\, i$
(the latter is $w_B$).
The limiting curve crosses the real $w$-axis at
$w_{0Q} \approx -0.2559077691$.\footnote{
   As explained at the beginning of this section,
   we located the endpoints using the resultant method for $L\leq 6$
   and the direct-search method for $L \ge 7$.
   The direct-search method is quite efficient for locating
   a {\em real}\/ endpoint, but is extremely tedious
   for locating a complex endpoint
   (since we have to search a two-dimensional space).
   This explains why the precision obtained for the {\em complex}\/
   endpoints with $L \ge 7$ (error $\approx 10^{-6}$)
   is inferior to that obtained both for the endpoints with $L\leq 6$
   and for the real endpoints with $L \ge 7$ (error $\approx 10^{-10}$).
   However, we have put an extra effort in computing the rightmost complex
   endpoints $w_B$ for $L\geq 7$ with error $\approx 10^{-10}$
   (see Section~\protect\ref{sec.res}).
}

For $L=8$, we find a multiple point at $w_{0Q} \approx -0.2532620078$,
which belongs to a horizontal segment running from
$w_{0-} \approx -0.2556827597$ to $w_{0+} =-1/4$.
There are fourteen additional endpoints at
$w\approx -0.250381 \pm 0.101684\, i$,
         $-0.249500 \pm 0.101496\, i$,
         $-0.201140 \pm 0.212546\, i$,
         $-0.200485 \pm 0.212232\, i$,
         $-0.077462 \pm 0.333326\, i$,
         $-0.077248 \pm 0.332940\, i$, and
     $ 0.3415981731 \pm 0.5675966263\, i$
(the latter is $w_B$).

For $L=9$ we were unable to compute the full limiting curve ${\cal B}$,
but we were able to compute the points corresponding to fixed values of
$|\imag w|=0,0.01,0.02$, etc.
These points give us a rough idea of the shape of the limiting curve.
In carrying out this computation,
it is important to keep track of the quantity
$\theta = |\imag \log(\lambda_\star/\lambda'_\star)|$
as we move along the limiting curve ${\cal B}$
(here $\lambda_\star$ and $\lambda'_\star$
 denote the dominant equimodular eigenvalues),
since $\theta$ vanishes at endpoints \cite{transfer1}.
By careful monitoring of the angle $\theta$,
we can obtain a lower bound on
the number of endpoints and connected components of the curve ${\cal B}$.
In particular, for $L=9$ we have
found that there are at least 18 endpoints and 9 connected components
(see Table~\ref{table_sq}).
We have also computed the point where the the limiting curve crosses
the real $w$-axis $w_{0Q} \approx -0.2534832041$, and the rightmost endpoints
$w_B \approx 0.3890914478 \pm 0.5754959494\, i$.

For $L=10$ the computation of the limiting curve ${\cal B}$
using arbitrary-precision {\sc Mathematica} scripts
is beyond our computer facilities. However,
we have been able to compute some points along this curve
(those corresponding to $|\imag w|=0,0.01,0.02$, etc.)\
by using the double-precision Fortran subroutines
of the {\sc arpack} package \cite{arpack}.\footnote{
   The arithmetic precision of the {\sc arpack} package is thus
   less than that of {\sc Mathematica}.
   However, we have checked in a difficult but manageable case ($L=9$)
   that the results obtained by the two methods agree to
   at least ten decimal digits. We have also checked
   the performance of the {\sc arpack} routines at some specific points
   ($w_{0+}$, $w_{0-}$ and $w_{0Q}$)
   for $L=10$, and the disagreement with {\sc Mathematica}
   is again less than $10^{-10}$.
}
We find a multiple point at $w_{0Q} \approx -0.2519570283$. This point belongs
to a horizontal segment running from $w_{0-}\approx -0.2534268314$ to
$w_{0+}=-1/4$. Using the same method as in $L=9$ we have concluded that
${\cal B}$ is formed by at least 9 connected components and contains at least
20 endpoints. Finally, we have computed the rightmost endpoints
$w_B \approx 0.4317571213 \pm 0.5821234989\, i$.

%
%
\section{Triangular-lattice strips with cylindrical boundary conditions}
\label{sec.tri}

\begin{figure}
\centering
\begin{tabular}{cc}
   \includegraphics[width=200pt]{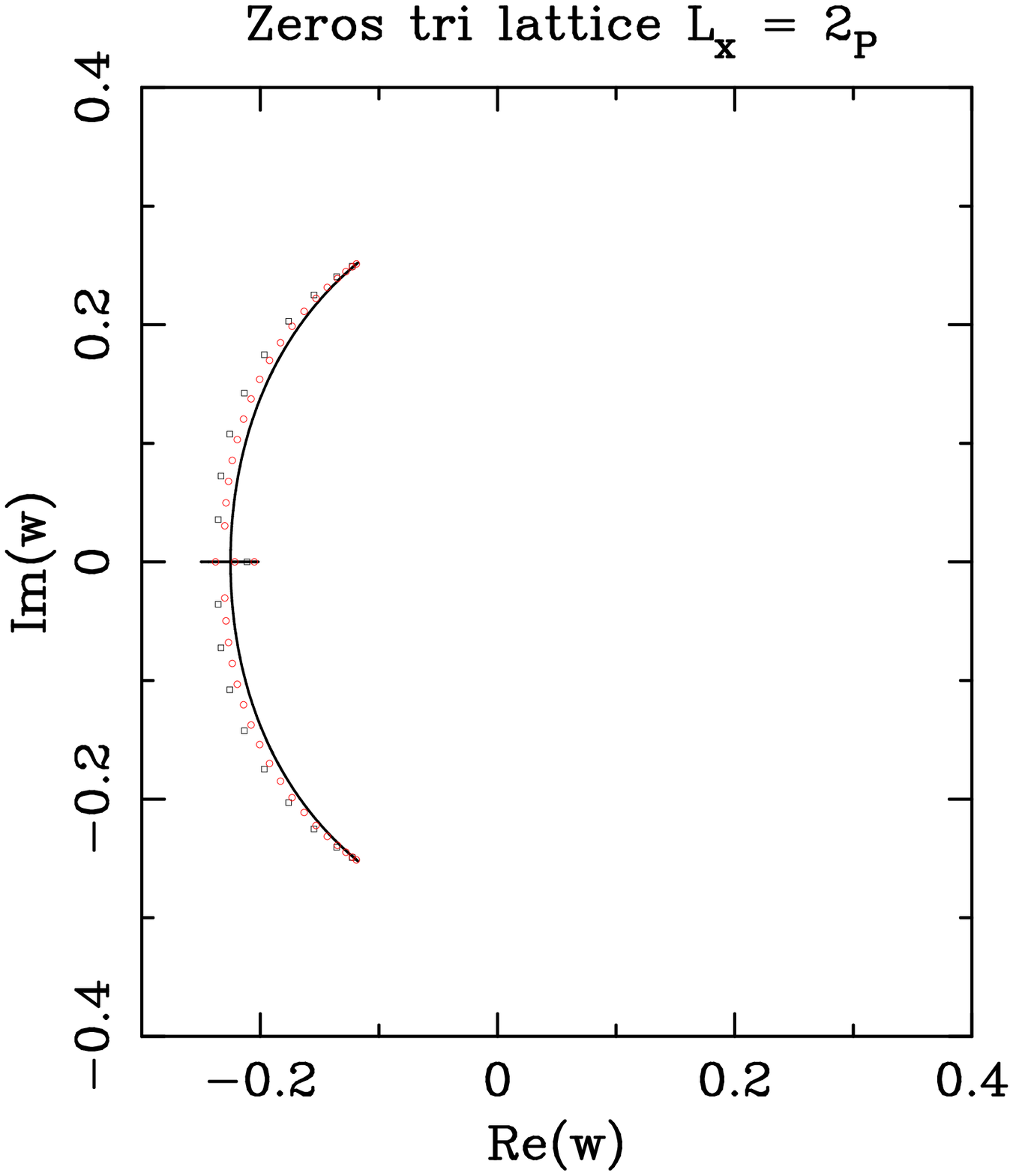} &
   \includegraphics[width=200pt]{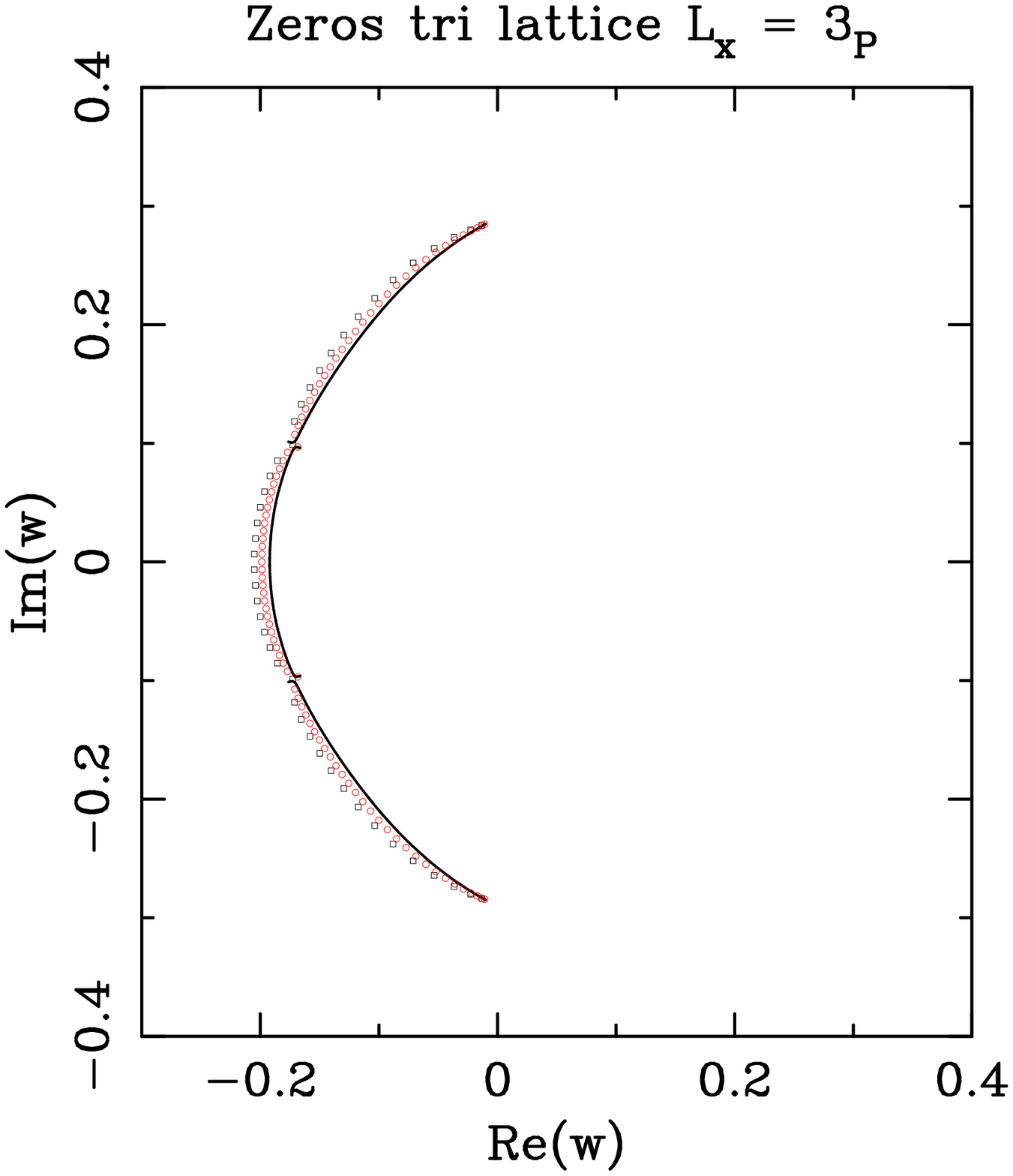} \\[1mm]
   \phantom{(((a)}(a)    & \phantom{(((a)}(b) \\[5mm]
   \includegraphics[width=200pt]{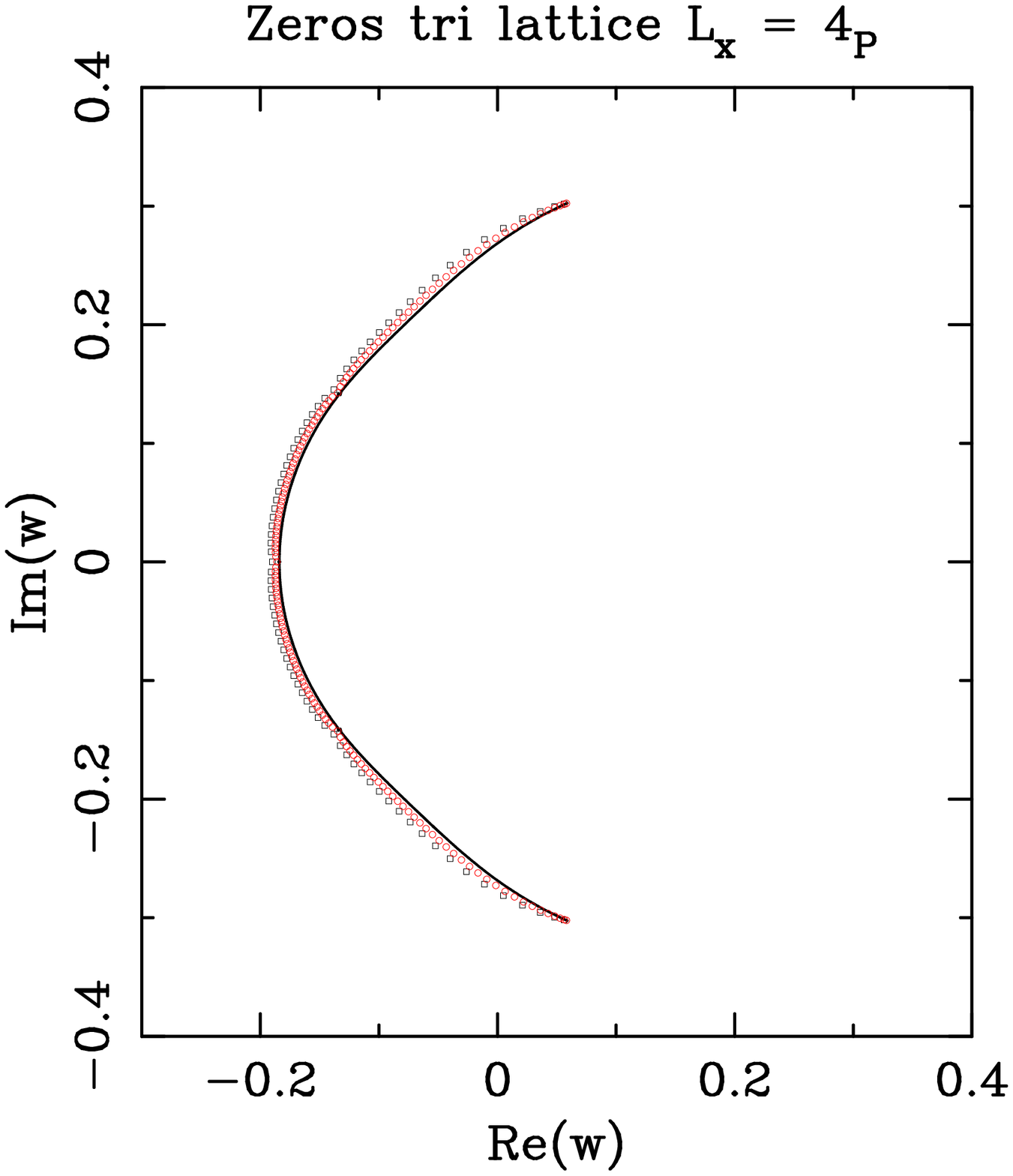} &
   \includegraphics[width=200pt]{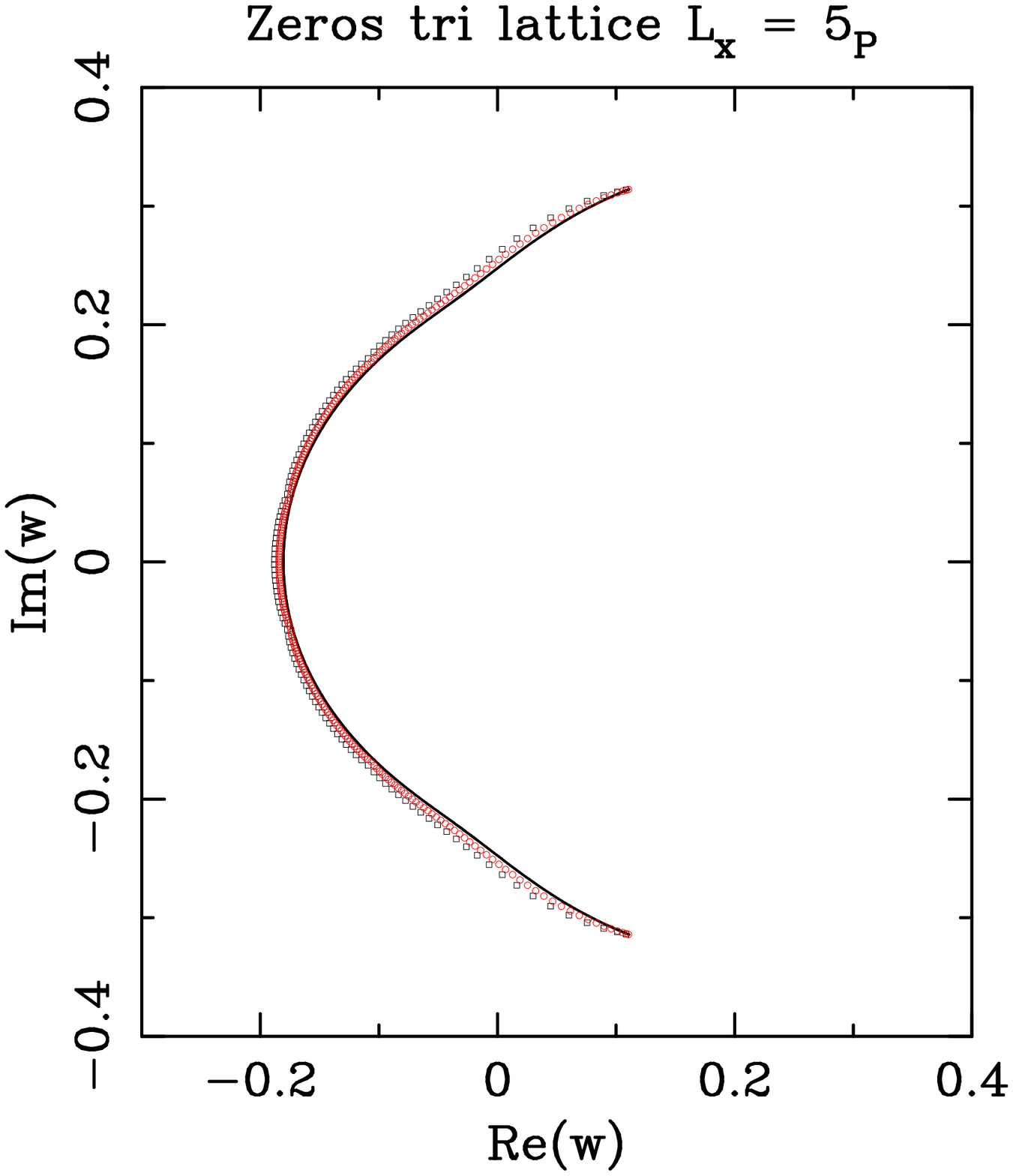}
   \\[1mm]
   \phantom{(((a)}(c)    & \phantom{(((a)}(d) \\
\end{tabular}
\caption{\label{figure_tri_1}
Limiting curves for triangular-lattice strips of width (a) $L=2$, (b) $L=3$,
(c) $L=4$, and (d) $L=5$ with cylindrical boundary conditions. We also show the
zeros for the strips $L_{\rm P} \times (5L)_{\rm F}$ (black $\Box$) and
$L_{\rm P} \times (10L)_{\rm F}$ (red $\circ$) for the same values of $L$.
}
\end{figure}

\begin{figure}
\centering
\includegraphics[width=400pt]{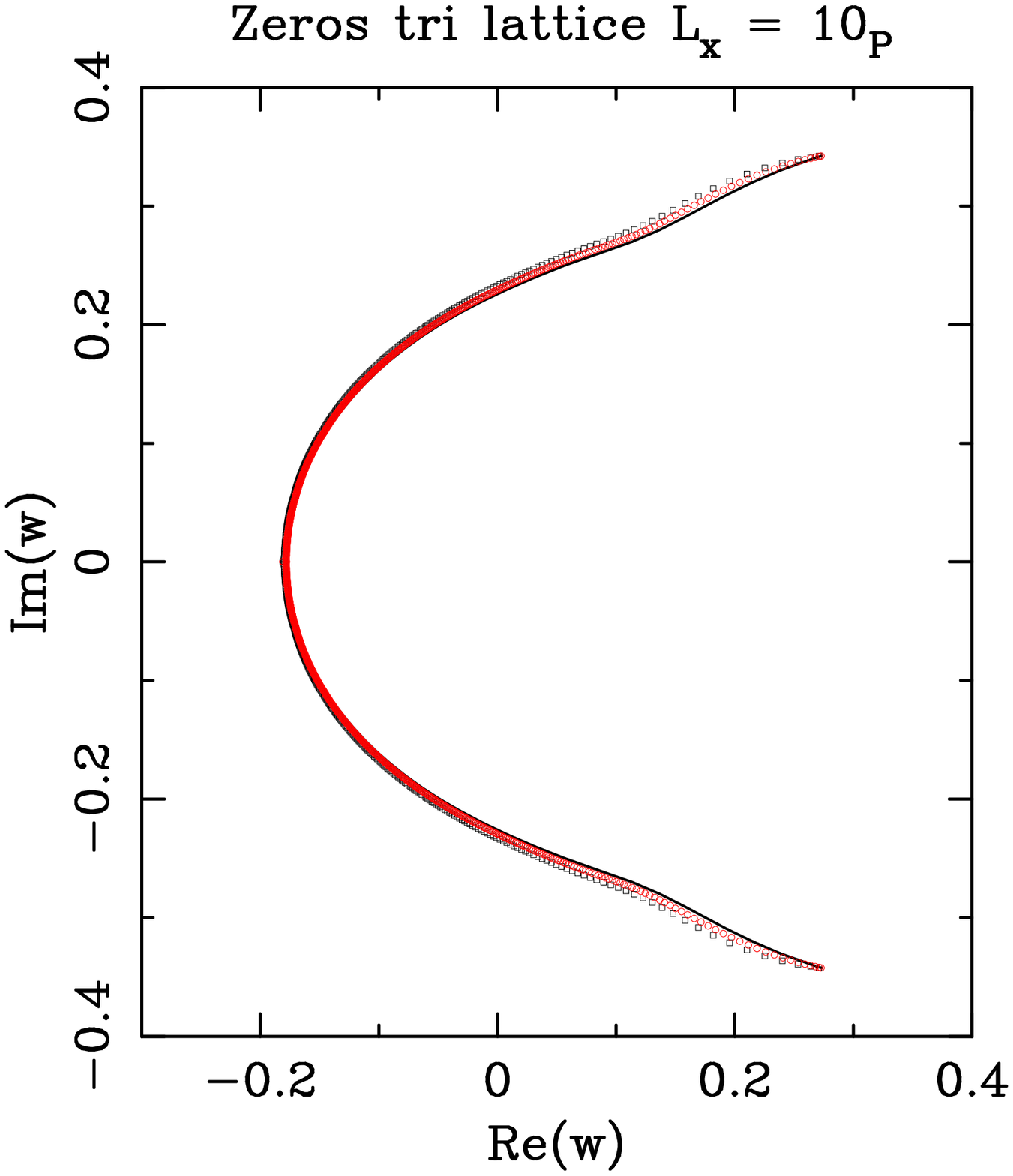}
\caption{\label{figure_tri_3}
Limiting curve for the triangular-lattice strip of width $L=10$ with
cylindrical boundary conditions. We also show the
zeros for the strips $10_{\rm P} \times 50_{\rm F}$ (black $\Box$) and
$10_{\rm P} \times 100_{\rm F}$ (red $\circ$).
}
\end{figure}

We performed for the triangular lattice the same calculations
as for the square lattice.
Thus, for each strip width $L \leq 10$ we computed
the transfer matrix and the associated left and right vectors;
from these we obtained the partition-function zeros
for strips with aspect ratio $\rho=5,10$
as well as the limiting curves ${\cal B}$
corresponding to the limit $\rho\to\infty$.
The limiting curves $\scrb$ resulting from these computations
are shown in Figures~\ref{figure_tri_1}--\ref{figure_tri_3}
(superposed in Figure~\ref{figure_tri_all}),
and their principal features are summarized in Table~\ref{table_tri}.
Once again, the full curves ${\cal B}$ are computed by the
resultant method for $L \le 5$,
by a combination of the resultant method (endpoints only)
and direct search for $L=6$,
and by the direct-search method for $L \ge 7$.

%
%
\subsection{$L=2$} \label{sec.tri2P}

The connectivity basis is two-dimensional, and the transfer matrix
$\T$, and the vectors $\vv_{\rm id}$ and $\uu$ are given by
\begin{subeqnarray}
\T &=&
   \left( \begin{array}{cc}
       4 w^2   & 2 w^2 (6w+1)\\
       4 w +1  & 12 w^2 + 6 w + 1
          \end{array} \right) \\
\uu^{\rm T}            &=&  \left( 1, 2w+1 \right) \\
\vv_{\rm id}^{\rm T}   &=&  \left( 0, 1    \right)
\end{subeqnarray}
We display in Figure~\ref{figure_tri_1}(a) the zeros of the polynomials
$F_{2_{\rm P} \times n_{\rm F}}(w)$ with $n=10,20$, along with
the corresponding limiting curve ${\cal B}$.
The curve ${\cal B}$ is connected:
it is the union of a horizontal segment
running from $w_{0-} = -1/4$ to $w_{0+} \approx -0.2017782928$
and an arc running between the complex-conjugate endpoints
$w_B \approx -0.1178608536 \pm 0.2520819223\, i$.
(The latter three points are the roots of the polynomial
$1 + 8w + 28w^2 + 64w^3$.)
The segment and the arc cross
at the multiple point at $w_{0Q} \approx -0.2251972448$.

%
%
\subsection{$L=3$} \label{sec.tri3P}

The connectivity basis is three-dimensional, and the transfer matrix
$\T$, and the vectors $\vv_{\rm id}$ and $\uu$ are given by
\begin{subeqnarray}
\T &=&
   \left( \begin{array}{ccc}
8w^3 & 6w^3(12w+1)  & 9w^4(6w+1)\\
4w^2 & 2w^2(19w+3)  & w^2(30w^2+10w+1)\\
6w+1 & 3(20w^2+8w+1)& 50w^3+33w^2+9w+1\\
          \end{array} \right) \\
\uu^{\rm T}          &=&  \left( 1, 3(2w+1), 3w^2+3w+1 \right) \\
\vv_{\rm id}^{\rm T} &=&  \left( 0, 0,       1    \right)
\end{subeqnarray}
We display in Figure~\ref{figure_tri_1}(b) the zeros of the polynomials
$F_{3_{\rm P} \times n_{\rm F}}(w)$ with $n=15,30$, along with
the corresponding limiting curve ${\cal B}$.
The curve ${\cal B}$ has three connected components.
One of them runs between the complex-conjugate endpoints
$w\approx -0.1763241163 \pm 0.1013423678\, i$
and intersects the real $w$-axis at $w_{0Q} \approx -0.1921127637$.
The other two run from
$w\approx -0.1662139535 \pm 0.0960130540\, i$
to $w_B \approx -0.0100833905 \pm 0.2849353892\, i$.

%
%
\subsection{$L=4$} \label{sec.tri4P}

The connectivity basis is six-dimensional; the transfer matrix
$\T$, and the vectors $\vv_{\rm id}$ and $\uu$ are given by
\begin{subeqnarray}
\T &=&
   \left( \! \begin{array}{cccccc}
16w^4& 16w^4 C_{12} &96w^5     &48w^5 C_9     &2w^4 C_{12}^2 &36w^6 C_6\\
8w^3 & 12w^3 C_8  &2w^3C_{24}&w^3 T_{2,4}   &2w^3C_6 C_{12}&3w^4D_{36,11}\\
0    & 8w^4       &8w^4      &8w^4C_6       &3w^4C_8       &w^4C_6^2\\
4w^2 & 2w^2T_{4,2}&4w^2C_8   &2w^2T_{4,4}   &w^2T_{4,5}    &T_{4,6}\\
4w^2 & 8w^2C_6    &2w^2C_{12}&2w^2D_{52,14} &2w^2C_6^2     &2w^3D_{24,9}\\
C_8  & 4D_{28,10} &2D_{32,10}&T_{6,4}       &2C_4D_{24,8}  &T_{6,6}\\
\end{array} \! \right) \qquad \\
\uu^{\rm T}           &=&  \left(1, 4C_2, 2C_2, 4D_{3,3}, 2C_2^2, C_2 D_{2,2}
                         \right) \\
\vv_{\rm id}^{\rm T}  &=&  \left(0,0,0,0,0,1 \right)
\end{subeqnarray}
where we have used the shorthand notations \reff{def_Ck}/\reff{def_Dk} and
\begin{subeqnarray}
T_{2,4}(w) &=& (8w+1)(27w+2)\\
T_{4,2}(w) &=& 28w+5 \\
T_{4,4}(w) &=& 80w^2+28w+3 \\
T_{4,5}(w) &=& (6w+1) (16w+3) \\
T_{4,6}(w) &=& w^2(96w^3+54w^2+12w+1) \\
T_{6,4}(w) &=& 4(80w^3+47w^2+11w+1)   \\
T_{6,6}(w) &=& 192 w^4 + 164w^3 + 62w^2 + 12w + 1
\end{subeqnarray}

The zeros of the polynomials $F_{4_{\rm P} \times n_{\rm F}}(w)$
with $n=20,40$ are displayed in Figure~\ref{figure_tri_1}(c), along with
the limiting curve ${\cal B}$.
The curve ${\cal B}$ has three connected components
separated by two very small gaps.
One of the components is the union of a horizontal segment
running from $w_{0-} \approx -0.1846154722$ to $w_{0+} = -0.1833753245$
and an arc running between the complex-conjugate endpoints
$w\approx -0.1343195918 \pm 0.1422271304\, i$.
The segment and the arc cross
at the multiple point at $w_{0Q} \approx -0.1839945026$.
The other two components are complex-conjugate arcs running from
$w \approx -0.1325113192 \pm 0.1408071373\, i$
to $w_B \approx 0.0588726934 \pm 0.3024953798\, i$.

%
%
\subsection{$L=5,6,7,8,9,10$} \label{sec.tri5P}

The transfer matrices for $L\geq 5$ are too lengthy to be quoted here.
Those for $L \le 9$ can be found in the {\sc Mathematica} file
{\tt forests\_tri\_2-9P.m} that is available with the electronic version
of this paper in the cond-mat archive at {\tt arXiv.org}.
The file for $L=10$, which is 31.9 MB long, can be obtained on request
from the authors.

\begin{figure}
\centering
\begin{tabular}{cc}
   \includegraphics[width=200pt]{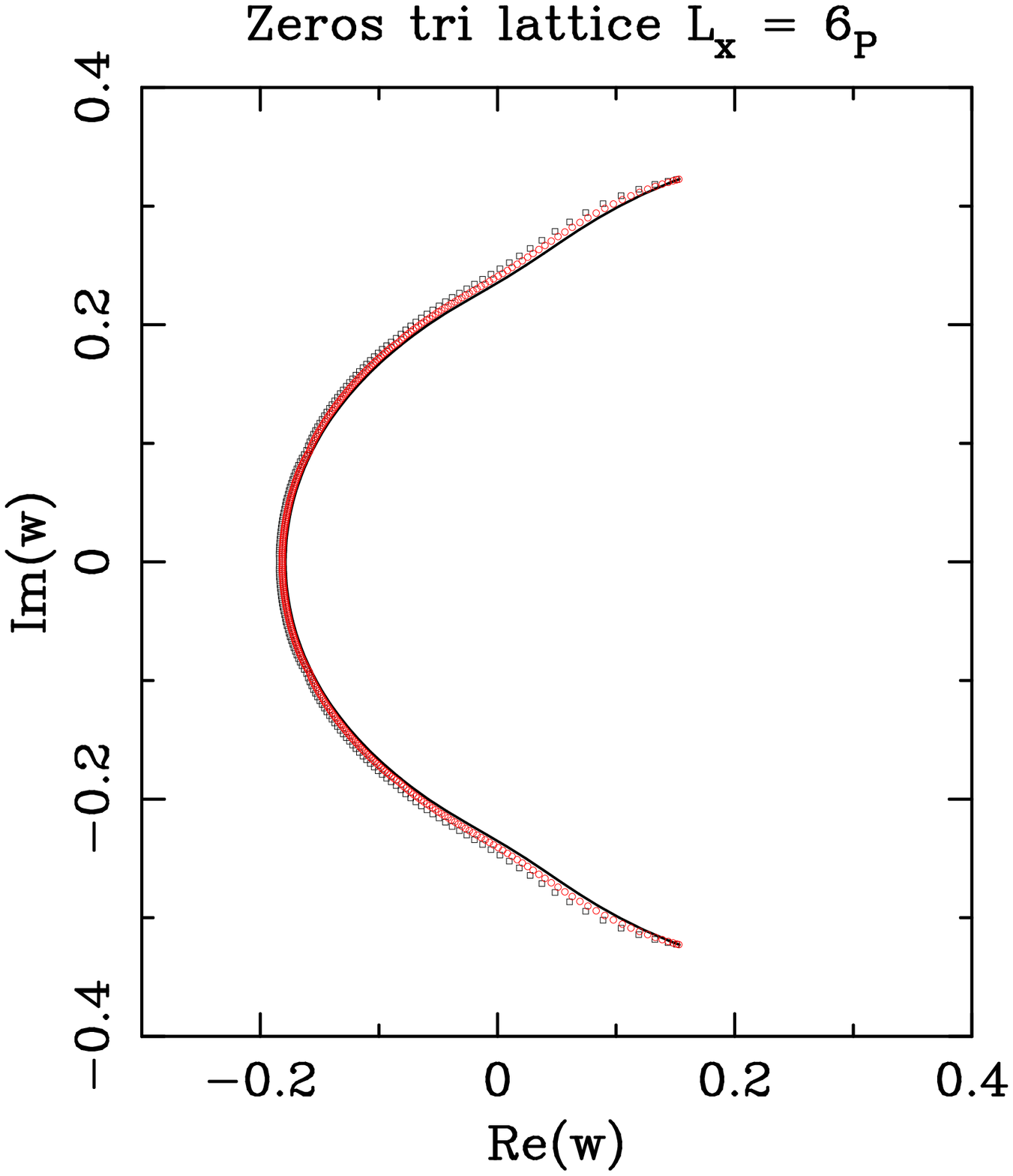} &
   \includegraphics[width=200pt]{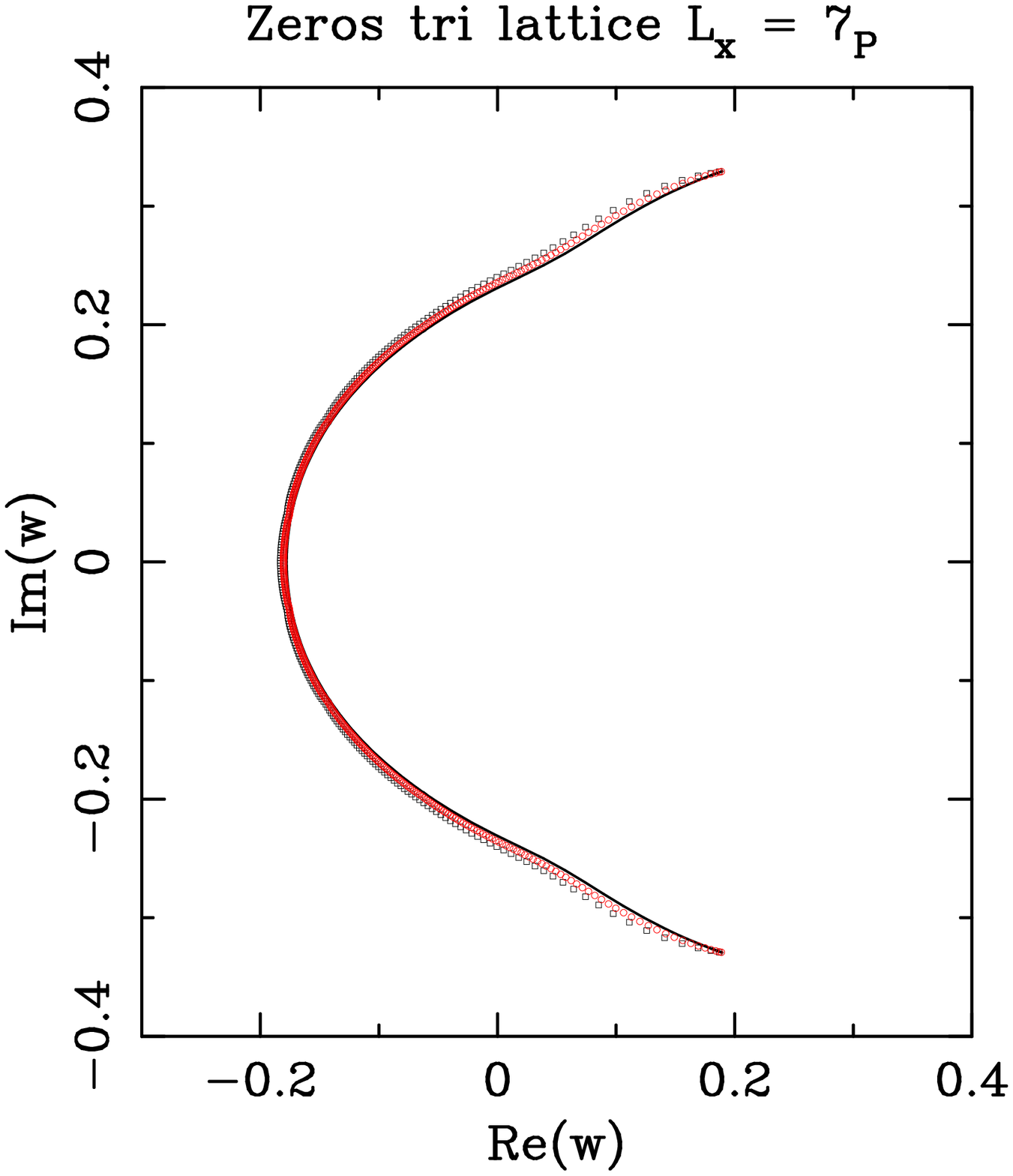}
   \\[1mm]
   \phantom{(((a)}(a)    & \phantom{(((a)}(b) \\[5mm]
   \includegraphics[width=200pt]{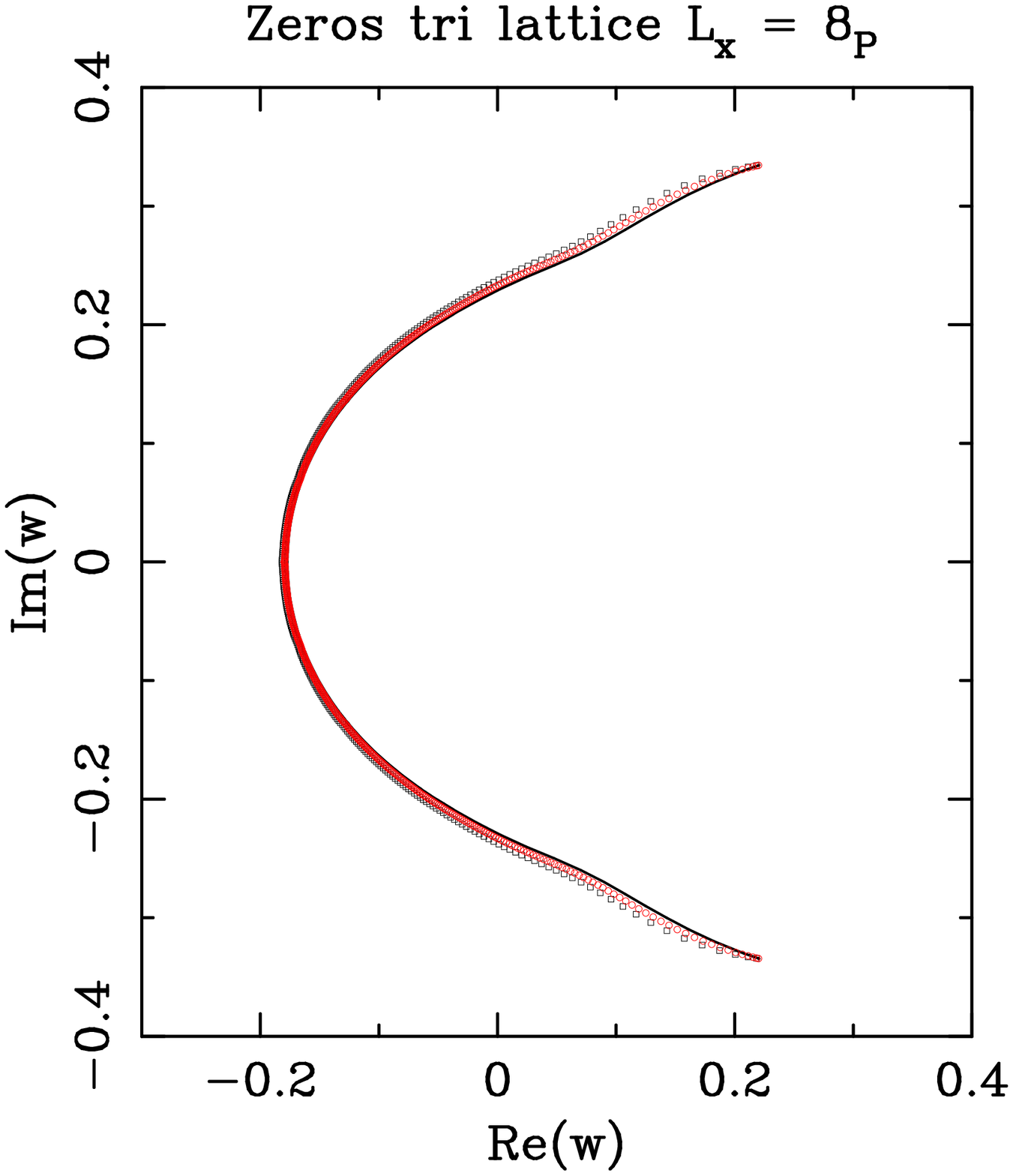} &
   \includegraphics[width=200pt]{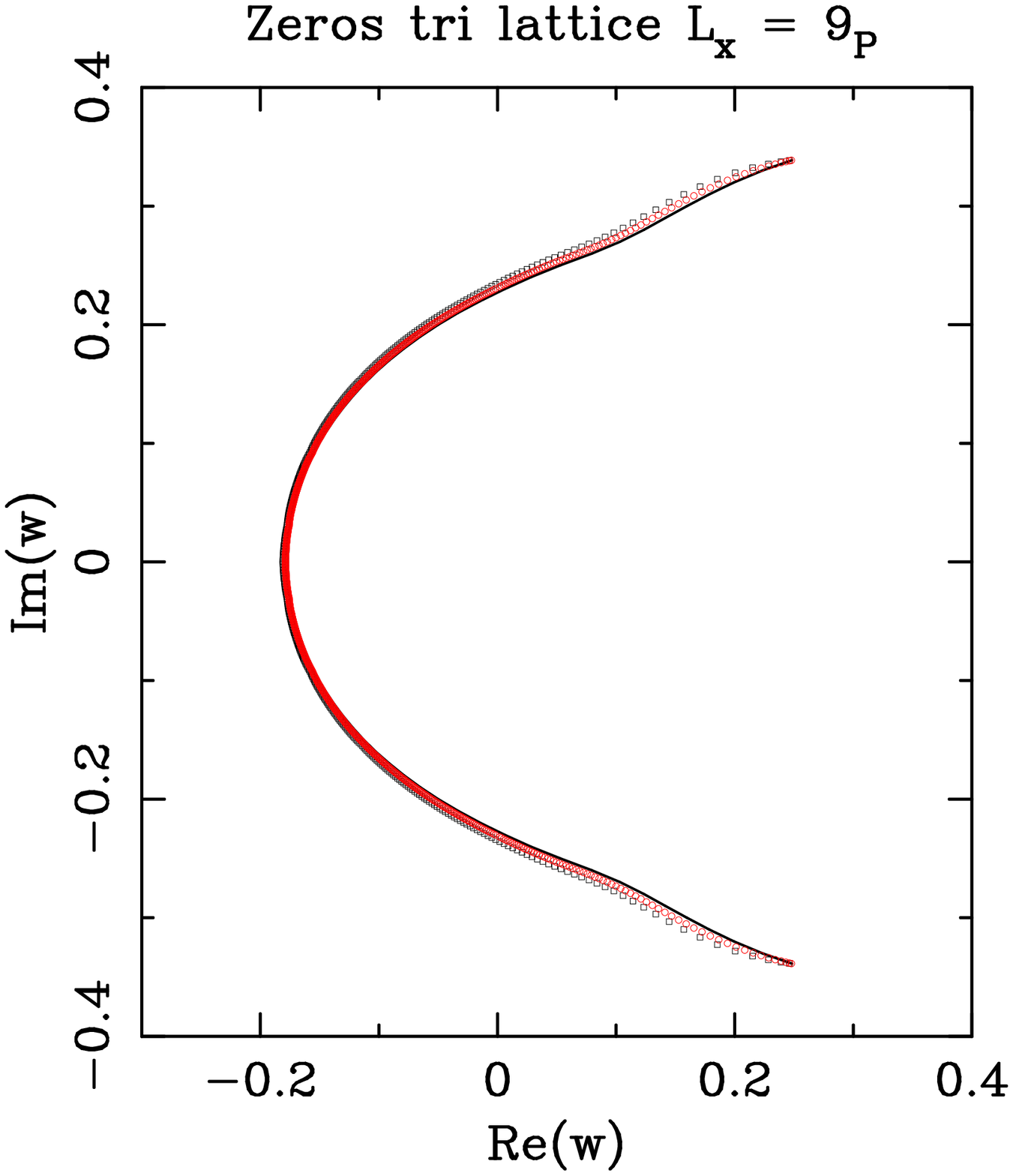}
   \\[1mm]
   \phantom{(((a)}(c) & \phantom{(((a)}(d)  \\
\end{tabular}
\caption{\label{figure_tri_2}
Limiting curves for triangular-lattice strips of width (a) $L=6$, (b) $L=7$,
(c) $L=8$, and (d) $L=9$ with cylindrical boundary conditions. We also show the
zeros for the strips $L_{\rm P} \times (5L)_{\rm F}$ (black $\Box$) and
$L_{\rm P} \times (10L)_{\rm F}$ (red $\circ$) for the same values of $L$.
}
\end{figure}

We have plotted for each $L$ ($=5,6,7,8$) the zeros
of $F_{L_{\rm P} \times (\rho L)_{\rm F}}(w)$ for aspect ratios
$\rho=5,10$ as well as the limiting curves ${\cal B}$ ($\rho=\infty)$.
See Figure~\ref{figure_tri_1}(d) for $L=5$,
Figure~\ref{figure_tri_2}(a) for $L=6$,
Figure~\ref{figure_tri_2}(b) for $L=7$,
and Figure~\ref{figure_tri_2}(c) for $L=8$.

The principal features of the limiting curves ${\cal B}$ are summarized in
Table~\ref{table_tri}. For $L=5$, there are ten endpoints located at
$w\approx -0.1732373428 \pm  0.0559213543\, i$,
$ -0.1725608663 \pm  0.0557343311\, i$,
$ -0.1046922573 \pm  0.1666741778\, i$,
$ -0.1043151240 \pm  0.1662893120\, i$, and  \linebreak
$  0.1110085784 \pm  0.3142261926\, i$
(the latter is $w_B$).
The limiting curve crosses the real axis at $w_{0Q} \approx --0.1805863920$.

For $L=6$ there is a multiple point at $w_{0Q} \approx -0.1788458357$.
This point belongs to a very short horizontal segment running from
$w_{0-} \approx -0.1788596197$ to $w_{0+} \approx -0.1788320527$.
There are ten more endpoints at
$w\approx -0.1572400315 \pm  0.0919540390\, i$,
$-0.1570191889 \pm  0.0918929205\, i$,
$-0.0812379488 \pm  0.1829264297\, i$,
$-0.0813151207 \pm  0.1830202393\, i$, and  \linebreak
$ 0.1534513521 \pm  0.3227719569\, i$
(the latter is $w_B$).

For $L=7$ there are fourteen endpoints located at
$-0.173950 \pm  0.039380\, i$,
$-0.173885 \pm  0.039369\, i$,
$-0.141195 \pm  0.117208\, i$,
$-0.141154 \pm  0.117209\, i$,
$-0.061954 \pm  0.194837\, i$,
$-0.061949 \pm  0.194832\, i$, and
$ 0.1894308555 \pm 0.3293185497\, i$
(the latter is $w_B$).
The limiting curve crosses the real axis at $w_{0Q} \approx -0.1778368691$.

For $L=8$ there is a multiple point at $w_{0Q} \approx -0.1772011941$.
This point belongs to a very short horizontal segment running from
$w_{0-} \approx -0.1772035681$ to $w_{0+} \approx -0.1771988202$.
There are fourteen additional endpoints at
$-0.165087 \pm 0.068518\, i$,
$-0.165105 \pm 0.068521\, i$,
$-0.126193 \pm 0.135891\, i$,
$-0.126182 \pm 0.135887\, i$,
$-0.045467 \pm 0.203914\, i$,
$-0.045470 \pm 0.203918\, i$,
$ 0.2207308275 \pm 0.3345156821\,i$
(the latter is $w_B$).

For $L=9$ we were unable to compute the full limiting curve ${\cal B}$,
but as in the square-lattice case, we were able to obtain a rough estimate by
computing the points corresponding at fixed values of $|\imag w|=0,0.01,0.02$,
etc. In particular, the point where ${\cal B}$ crosses the real axis is
$w_{0Q} \approx -0.1767753501$. We have also computed the rightmost
endpoints $w_B \approx 0.2484623416 \pm 0.3387563998\, i$.
Using the same method as in the square-lattice case, we have concluded that
${\cal B}$ is formed by at least 9 connected components and contains at least
18 endpoints (see Table~\ref{table_tri}).

For $L=10$, we have computed the points of the limiting curve corresponding to
$|\imag w| = 0,0.01,0.2$, etc by using the {\sc arpack} subroutines. We find
a multiple point at $w_{0Q}\approx -0.1764764073$. This point belongs
to the horizontal segment running from $w_{0-}\approx -0.1764769384$ to
$w_{0+}\approx -0.1764758762$. Using the same method as for $L=9$, we
conclude that the limiting curve is formed by at least 9 connected components
and contains at least 20 endpoints. Finally, we have computed the
rightmost endpoints  $w_B \approx 0.2733731381 \pm 0.3422936097\, i$.

%
%
\section{Extrapolation to infinite width} \label{sec.res}

In this section we analyze the finite-strip data
from Sections~\ref{sec.sq} and \ref{sec.tri}
and study the limiting behavior as the strip width $L$ tends to infinity.
Our goal is to determine the nature of the phase transition at $w=w_0$
and to extract numerical estimates of the critical point $w_0$,
the free energy $f(w)$, the central charge $c(w)$,
and the thermal scaling dimensions $x_{T,i}(w)$
for each of the two lattices.

\subsection{Procedure I: Estimates of critical points}  \label{sec.res.proc1}

One goal of this paper is to obtain the limiting value as $L \to\infty$
of the quantities $w_{0+}(L)$, $w_{0-}(L)$ and $w_{0Q}(L)$.
We expect that all three quantities converge to the same limit $w_0$:
\be
w_0 \;=\; \lim\limits_{L\to\infty} w_{0j}(L)
  \qquad\hbox{for } j=+,-,Q   \;.
\label{def_w0}
\ee
This limit is also expected to be independent of the choice of
boundary conditions (e.g.\ free, cylindrical, cyclic, toroidal, etc.).
We have estimated $w_0$ by fitting our finite-width data to the Ansatz
\be
w_{0j}(L) \;=\; w_0 + A_j L^{-\Delta} \;.
\label{def_ansatz_w0}
\ee
At least for the square lattice, this Ansatz is theoretically justified
due to the vicinity of the critical theory at $w_0 = 1/4$,
and we should have $\Delta = 1/\nu$.
Thus, for each value $L_{\rm min}$ we fit the data for the triplet
$L=L_{\rm min},L_{\rm min}+2,L_{\rm min}+4$
to the Ansatz \reff{def_ansatz_w0} and extract the estimates
of $w_0$, $A_j$ and $\Delta$.
(For $j=Q$ we can also use the triplets
$L=L_{\rm min},L_{\rm min}+1,L_{\rm min}+2$.)
The observed variations in the estimates as a function of $L_{\rm min}$
arise from higher-order correction-to-scaling terms
that are neglected in \reff{def_ansatz_w0}.

It is also of interest to estimate the exponent $\omega$
associated to the width of the interval $[w_{0-}, w_{0+}]$
for even $L$:
\be
w_{0+}(L) - w_{0-}(L) \;\approx\; A L^{-\omega} \;.
\label{def_ansatz_omega}
\ee
We fit to the Ansatz \reff{def_ansatz_omega}
using pairs $L=L_{\rm min},L_{\rm min}+2$.

By definition, the endpoints $w_{0\pm}(L)$ correspond,
at any given finite $L$, to the collision (in modulus)
of the two dominant eigenvalues.
Now, for {\em even}\/ widths $L$ on the {\em square}\/ lattice,
we have found that $w_{0+}(L) = 1/4$ {\em exactly}\/.
Therefore, at $w=w_0 = -1/4$ we have $\xi_1^{-1}(L) = 0$
{\em exactly}\/ for all even $L$
(and not merely in the limit $L\to\infty$).
{}From \reff{def_xi_critical} we can conclude that
the leading thermal scaling dimension $x_{T,1}$ is {\em zero}\/,
and hence that $\nu=1/2$.
We expect this latter conclusion to hold (by universality)
also for the triangular lattice.
In view of this prediction,
we can expect to obtain more accurate estimates for $w_0$
by fixing $\Delta$ in the Ansatz \reff{def_ansatz_w0} to its predicted
value $\Delta=1/\nu=2$.

More generally, we have also tried to extrapolate the whole limiting
curve $\scrb$ to $L\to\infty$:
for fixed $\imag w = 0.01,0.02, \ldots$,
we extrapolated the values of $\real w$ using an Ansatz
of the form \reff{def_ansatz_w0}.
We performed several different extrapolations,
e.g.\ including all data points or including only even or odd widths.
We obtained consistent estimates for
$|\imag w| \ltapprox 0.33$ on the square lattice and
$|\imag w| \ltapprox 0.23$ on the triangular lattice
(see Table~\ref{table_limiting_curves}).
The results are given by the black dots of Figures~\ref{figure_sq_all}
and \ref{figure_tri_all}.

By the same method we can obtain the limit as $L \to\infty$
of the endpoints with largest real part, which we have denoted $w_B(L)$.
As a first guess, we used the Ansatz
\be
w_B(L) \;=\; w_B + A L^{-\Delta}
\label{def_ansatz_wB}
\ee
separately for the real and imaginary parts of $w_B(L)$.
Then, after examining the results, we conjectured better Ans\"atze
(see Sections~\ref{sec.res.sq1} and \ref{sec.res.tri1} below).

\subsection{Procedure II: Estimates of the free energy}  \label{sec.res.proc2}

\begin{figure}
\centering
\includegraphics[width=400pt]{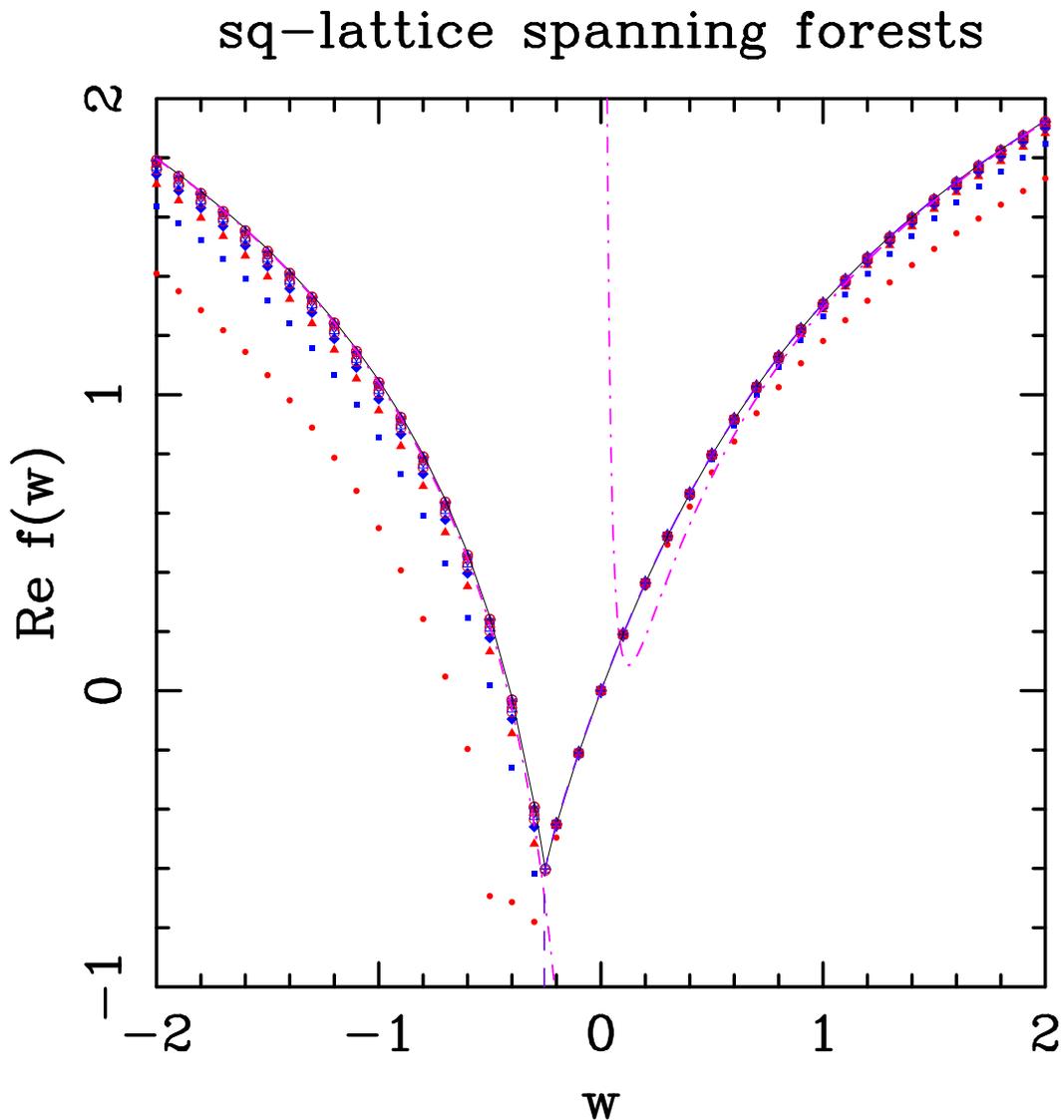}
\caption{\label{figure_f_sq}
   Real part of the free energy for square-lattice spanning forests
   as a function of $w$, for strips of width
   $L=2$ ($\bullet$),
   $3$ ($\blacksquare$), $4$ ($\blacktriangle$), $5$ ($\blacklozenge$),
   $6$ ($\circ$),        $7$ ($\Box$),           $8$ ($\triangle$),
   $9$ ($\diamondsuit$), $10$ ($\times$),        $11$ ($+$),
   $12$ ($*$), $13$ ($\oplus$) and               $14$ ($\odot$).
   To make clearer any even-odd effect we have displayed in red 
   (resp.\  blue) the points corresponding to even (resp.\ odd) $L$. 
   The black solid curve is obtained by
   extrapolating the finite-width data to $L\to\infty$
   and then joining the points.
   The violet dashed line is the Pad\'e $[20,20]$
   approximant to our longest small-$w$ series. Finally, the pink dot-dashed
   curve corresponds to the large-$w$ expansion
   \protect\reff{def_large_w_series} through order $w^{-1}$.
}
\end{figure}

The second goal of this paper is to estimate the bulk free energy
\be
f(w)  \;=\; \lim\limits_{L\to\infty} f_L(w)
\label{def_f}
\ee
and its derivatives as a function of $w$.
For each of the two lattices (square and triangular),
we have computed the five largest eigenvalues (in modulus)
of the transfer matrix ${\sf T}(w)$
for selected real values of $w$ and for all $L\leq 14$.
The $w$ values are taken between $w=-2$ and $w=2$ in steps of $0.1$.
{}From the dominant eigenvalue $\lambda_\star(w)$
we can extract the strip free energy
\be
   f_L(w) \;=\; {1 \over L} \log \lambda_\star(w)
\ee
[cf.\ \reff{def_fm}].
In Figures~\ref{figure_f_sq} and \ref{figure_f_tri}
we have plotted the real part of the strip free energy,
\be
   \real f_L(w) \;=\; {1 \over L} \log |\lambda_\star(w)|
   \;,
\ee
for the square and triangular lattices, respectively.
The imaginary part of the free energy, not shown in the plot,
arises from the following qualitative properties of the eigenvalues:
\begin{itemize}
   \item[(a)]  For odd $L$:
      \begin{itemize}
         \item For $w > w_{0Q}$, there is a unique dominant eigenvalue
             $\lambda_\star(w)$, which is real and positive.
         \item At $w = w_{0Q}$, the dominant eigenvalue $\lambda_\star(w) > 0$
             becomes equimodular with an eigenvalue $-\lambda_\star(w) < 0$.
         \item For $w < w_{0Q}$, there is a unique dominant eigenvalue
             $\lambda_\star(w)$, which is real and negative.
      \end{itemize}
   \item[(b)]  For even $L$:
      \begin{itemize}
         \item For $w > w_{0+}$, there is a unique dominant eigenvalue
             $\lambda_\star(w)$, which is real and positive.
         \item At $w = w_{0+}$,  the two leading eigenvalues collide.
             The common value $\lambda_\star(w)$ is still real and positive.
         \item For $w_{0-} < w < w_{0+}$, there is a complex-conjugate pair
             of dominant eigenvalues.
         \item At $w = w_{0-}$,  the two leading eigenvalues again collide.
             The common value $\lambda_\star(w)$ is real and positive.
         \item For $w < w_{0-}$, there is a unique dominant eigenvalue
             $\lambda_\star(w)$, which is real and positive.
      \end{itemize}
\end{itemize}
The only exception to the above pattern (at least for $L \le 8$)
is the square-lattice strip with $L=4$ at $w=w_{0-}$
(see Figure~\ref{figure_sq_4P_zoom}).
In this case, the point
$w_{0-}$ happens to be a T point, {\em not}\/ a regular endpoint.
At $w_{0-}$, {\em three}\/ eigenvalues attain the same modulus:
one is real and positive, while the other two form a complex-conjugate pair.

The limit $L\to\infty$ of the finite-width free energy $f_L(w)$
can be extracted
by using the Ansatz \reff{def_fm_critical} or \reff{def_fm_non_critical},
depending on whether the model is critical or not.  As we do not know
{\em a priori} to which regime a given value of $w$ belongs (this is
especially true for the triangular-lattice model, for which we do not know
the exact value of $w_0$), we can try first the preliminary Ansatz
\be
f_L(w)  \;=\; f(w) \,+\, A(w) L^{-\Delta(w)}   \;.
\label{def_ansatz0_fm}
\ee
We fit to triplets $L=L_{\rm min},L_{\rm min}+2,L_{\rm min}+4$
in order to minimize the effects of even-odd oscillations.
The variations of the estimates for $f(w)$, $A(w)$ and $\Delta(w)$
as a function of $L_{\rm min}$ will give us an idea about the critical or
non-critical nature of the model.
If the model is critical, we expect that
the estimates for $\Delta(w)$ will converge to 2 as $L_{\rm min} \to\infty$;
if the model is non-critical, we expect that the estimates
for $\Delta(w)$ will grow without bound.
Once we determine the regime to which a given value of $w$ belongs,
we can use the more appropriate Ansatz
\reff{def_fm_critical} or \reff{def_fm_non_critical} to obtain
more accurate estimates of the physical parameters,
notably the central charge $c(w)$ for the critical models.

Finally, by studying the $L$-dependence of the gap between the
dominant eigenvalue $\lambda_\star(w)$
and the subdominant eigenvalues $\lambda_i(w)$,
we can estimate the thermal scaling dimensions $x_{T,i}$
using the Ansatz \reff{def_xi_critical}.

%
%
\subsection{Square lattice: Estimates of critical points and
   phase boundary} \label{sec.res.sq1}

\begin{figure}
\centering
\includegraphics[width=380pt]{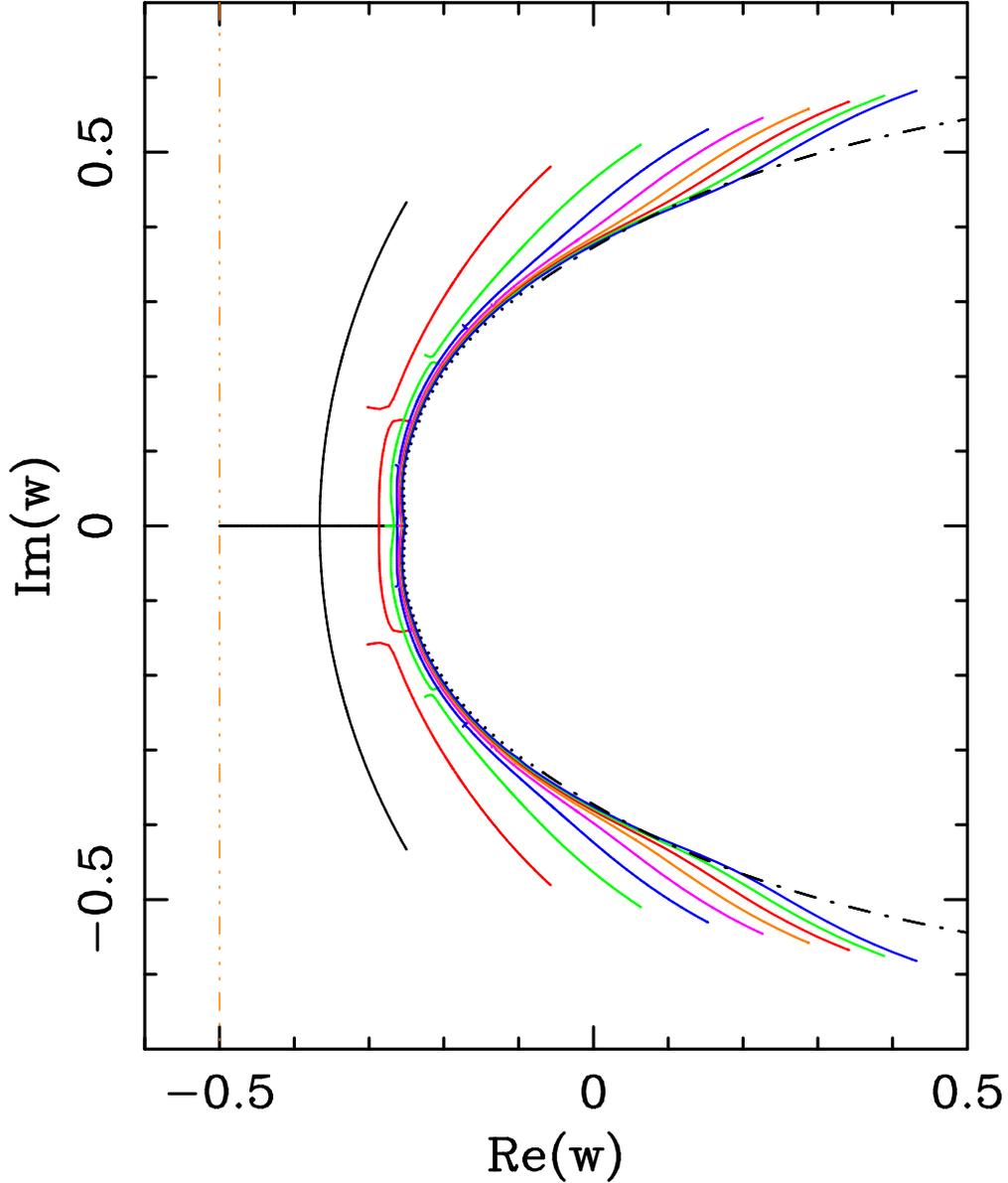}
\caption{\label{figure_sq_all}
Limiting curves for square-lattice strips of widths $L$: 2 (black),
3 (red), 4 (green), 5 (blue), 6 (pink), 7 (brown), 8 (red), 9 (green), and
10 (blue).
The dotted-dashed vertical (brown) line marks the line  $\real (w) = -1/2$.
As $L$ increases, the limiting curve moves towards the right.
We have also depicted our estimate for the $L \to\infty$ limiting curve
in the interval $-0.33 \ltapprox \imag w \ltapprox 0.33$ (black dots)
and our very rough estimate beyond this (black dotted-dashed curve).
}
\end{figure}

The square-lattice data that we wish to extrapolate
are collected in Table~\ref{table_sq}
and depicted graphically in Figure~\ref{figure_sq_all}.
The quantities $w_{0-}(L)$ and $w_{0+}(L)$ are defined only for even $L$.
The quantity $w_{0Q}(L)$ is defined for all $L$,
but it exhibits such strong even-odd oscillations
that it makes sense to fit it separately for odd and even $L$.
We therefore perform the fits to the Ansatz \reff{def_ansatz_w0}
using $L=L_{\rm min},L_{\rm min}+2,L_{\rm min}+4$.
The results are displayed in Table~\ref{table_fits_sq}.
The estimates of $w_0$ are well converged,
and from $L_{\rm min}=5,6$ we can estimate
\be
w_0({\rm sq}) \;=\; -0.2501 \pm 0.0002   \;.
\ee
This result agrees very well with the theoretical prediction \reff{def_w0_sq}:
\be
w_0({\rm sq}) \;=\; -1/4  \;.
\ee
It is also noteworthy that for all even values of $L\leq 10$,
we have found that $w_{0+}(L) =-1/4$ {\em exactly}\/.
Therefore, the exponent $\omega$ defined by \reff{def_ansatz_omega}
is identical to the exponent $\Delta$ extracted from $w_{0-}(L)$.

The exponent $\Delta$ in Table~\ref{table_fits_sq}
seems to be approaching $\Delta = 1/\nu \approx 2$ (at least roughly).
This behavior is consistent with the interpretation that at $w=w_0$
there is a first-order phase transition
(see Section~\ref{sec.res.sq2a} for overwhelming evidence that the
 first derivative of the free energy is in fact discontinuous at $w=w_0$):
such a phase transition can be described in the renormalization-group framework
by a discontinuity fixed point characterized by $\nu = 1/d$, where $d$
is the dimensionality of the lattice ($d=2$ in our case)
\cite{Nienhuis_75,Fisher_82}.
The behavior $\Delta=2$ is also consistent with an ordinary critical point
having $x_T=0$
(see Section~\ref{sec.res.sq3} for an analysis of the correlation
length in this model).
Note, finally, that the only possible values of $\Delta$
consistent with conventional finite-size-scaling theory
are $\Delta \le d$ (corresponding to a critical point or
a first-order transition)
and $\Delta=\infty$ (corresponding to a non-critical point).
Estimates of $\Delta$ that are slightly larger than 2
but {\em decreasing}\/ with $L$
(or, at any rate, not strongly increasing with $L$)
therefore suggest that $\Delta=2$.

Numerical investigations of first-order phase transitions
\cite[and references therein]{Billoire_92} suggest that the higher-order
corrections to the $L^{-d}$ behavior are simply integer powers of $L^{-d}$.
This observation motivates the Ansatz
\be
w_{0j}(L)  \;=\; w_0 + A_2 L^{-2} + A_4 L^{-4}  \;.
\label{def_ansatz_w0_bis}
\ee
The results of fitting the data to this Ansatz are shown in
Table~\ref{table_fits_sq_bis}. In this table we observe that the
estimates for $w_0({\rm sq})$ converge slightly faster to values close to the
exact result than do the estimates in Table~\ref{table_fits_sq}
(at least for small $L$).

Let us next extrapolate the endpoints $w_B(L)$,
handling separately their real and imaginary parts.
The real part of $w_B(L)$ seems to diverge as $L\to\infty$:
indeed, as we increase $L_{\rm min}$, the estimate for $\real w_B$
using the Ansatz \reff{def_ansatz_wB} appears to grow without bound
(e.g., $w_B \approx 1.19$ for $L_{\rm min}=2$,
$w_B \approx 2.75$ for $L_{\rm min}=3$,
and $w_B \approx 9.18$ for $L_{\rm min}=4$).
We therefore tried to guess the behavior of $\real w_B(L)$
as a function of $L$, and found that it fits very closely to the Ansatz
\be
   \real w_B(L)  \;=\;  c_1 \log L \,+\, c_2   \;.
 \label{def_ansatz_Re_wB}
\ee
The estimates for $c_1$ and $c_2$ as a function of $L_{\rm min}$
are shown in Table~\ref{table_fits_wB_sq}(a).
For $L_{\rm min} \ge 5$, the estimate of $c_1$
is clearly rising with $L_{\rm min}$ and is not yet slowing down.
This suggests that the true value of $c_1$ is quite a bit larger
than the value $\approx 0.405$ observed at our largest $L_{\rm min}$;
how much larger is difficult to say, especially since
it is very difficult to fit a slowly-varying function (such as $\log L$)
with data points in the (extremely narrow) interval $2\leq L \leq 10$.
One expects, in any case, that the true $c_1$ may exceed the value
$\approx 0.405$ by an amount considerably larger than the
observed variations in Table~\ref{table_fits_wB_sq}(a).
If forced to guess, we might estimate
\begin{subeqnarray}
   c_1  & = &   0.41 \pm 0.03 \\
   c_2  & = &  -0.52 \pm 0.06
\end{subeqnarray}
but these estimates should be taken with a grain of salt!

A more precise Ansatz can be tried if we recall the
recently-discovered relationship \cite{Caracciolo_PRL}
of this model with the $N$-vector model
with $N=-1$ and a sign change in the coupling constant.
This relation implies that the spanning-forest model is
perturbatively asymptotically free, and yields the theoretical prediction
\be
   \real w_B(L)  \;=\;  c_1 \log L \,+\, c_3 \log\log L \,+\, c_2
                 \,+\,  O\!\left( {\log\log L \over \log L} \right)
 \label{def_ansatz_Re_wB_bis}
\ee
with $c_1 = 3/(2\pi) \approx 0.477465$
and $c_3 = -1/(2\pi) \approx -0.159155$ \cite{Caracciolo_PRL}.\footnote{
   Indeed, it was our numerical observation of the behavior
   \reff{def_ansatz_Re_wB} that led us to conjecture that
   the spanning-forest model is asymptotically free in two dimensions
   --- a conjecture that played an important role
   in catalyzing the work reported in \cite{Caracciolo_PRL}.
}
The estimates obtained by performing a one-parameter fit
to \reff{def_ansatz_Re_wB_bis}
with $c_1$ and $c_3$ fixed to their theoretical values
(and the correction term $O(\log\log L/\log L)$ neglected)
are displayed in the last column of Table~\ref{table_fits_wB_sq}(a).
{}From these values we can estimate that
\be
c_2  \;=\;  -0.535 \pm 0.005
 \label{def_ansatz_Re_wB_c2}
\ee
Please note, however, that the correction terms omitted in
\reff{def_ansatz_Re_wB_bis}
--- $\log\log L/\log L$, $1/\log L$ and so forth ---
are extremely slowly varying functions of $L$.
Hence, as we have access only to a very narrow range of $L$ values, we cannot 
rule out the possibility that the actual value of $c_2$ 
may differ from the above estimate by many times our estimated error.

The imaginary part of $w_B(L)$, by contrast,
seems to converge to a finite limit:
the estimates using the Ansatz \reff{def_ansatz_wB}
are displayed in Table~\ref{table_fits_wB_sq}(b).
We conclude that $\imag w_B = 0.69 \pm 0.03$
and $\Delta \approx 1/2$.
Guessing that $\Delta = 1/2$ exactly,
we can refine the estimate for $\imag w_B$ by fitting the data to the
Ansatz
\be
   \imag w_B(L)  \;=\;  \imag w_B \,+\, A  L^{-1/2}
 \label{def_ansatz_Im_wB}
\ee
The results are displayed in Table~\ref{table_fits_wB_sq}(b).
The estimate for $\imag w_B ({\rm sq})$ is still decreasing with
$L_{\rm min}$, and our best estimate is
\be
\imag w_B ({\rm sq}) \;=\; 0.70 \pm 0.02  \;.
 \label{eq.wB.sq.v2}
\ee

Finally, let us try to extrapolate the limiting curve $\scrb$
to $L\to\infty$;
the result is a curve $\scrb_\infty$ that can be interpreted as the
phase boundary in the complex $w$-plane.
We began by extrapolating $\real w$
at the selected values $\imag w = 0.01,0.02, \ldots$,
using the Ansatz \reff{def_ansatz_wB} for $\real w$.
We performed several different extrapolations,
e.g.\ including all data points or including only even or odd widths.
We obtained consistent estimates for $|\imag w| \ltapprox 0.33$
(see Table~\ref{table_limiting_curves}).
The results are given by the black dots of Figure~\ref{figure_sq_all}.

Next we attempted to complete this curve at larger $\real w$.
The behaviors \reff{def_ansatz_Re_wB}--\reff{def_ansatz_Re_wB_c2}
and \reff{def_ansatz_Im_wB}--\reff{eq.wB.sq.v2}
suggest that the curve $\scrb_\infty$ continues all the way
to $\real w = +\infty$,
asymptotically approaching $\imag w \approx \pm 0.70$
as $\real w \to +\infty$.
This behavior is in agreement with the theoretical prediction arising from
renormalization-group computations in the four-fermion model
describing spanning forests at small $g \equiv 1/w$ \cite{Caracciolo_PRL}.
These computations show \cite{Caracciolo_PRL}
that the fixed point at $g=0$ (a free fermion theory)
is marginally repulsive for $g > 0$ (i.e., perturbatively asymptotically free)
and marginally attractive for $g < 0$,
with a renormalization-group flow given by
\be
   {dg \over dl} \;=\;  b_2 g^2 \,+\, b_3 g^3 \,+\, b_4 g^4 \,+\, b_5 g^5
          \,+\, \ldots
 \label{eq.RGflow_g}
\ee
or equivalently
\be
   {dw \over dl} \;=\;  - b_2 \,-\, {b_3 \over w} \,-\, {b_4 \over w^2}
      \,-\, {b_3 \over w^3} \,-\, \ldots
 \label{eq.RGflow_w}
\ee
with $b_2 > 0$,
where $l$ is the logarithm of the length rescaling factor.
The phase boundary $\scrb_\infty$ is a special RG flow curve:
namely, it is a {\em separatrix}\/
that divides the complex $w$-plane into two ``phases''.
Initial conditions belonging to the ``phase'' containing $w=0$
are attracted to the ``high-temperature'' fixed point at $w=0$
and are therefore noncritical.
Initial conditions belonging to the ``phase'' containing $w=-\infty$
are attracted to the free-fermion fixed point at $\real w=-\infty$
and are thus critical.
{}From \reff{eq.RGflow_w} we see that all the RG flow curves
(not only the separatrix)
tend to a constant value $\imag w$ as $\real w \to +\infty$.
It is a nonperturbative question to determine which one of these
RG flow curves is the separatrix.
But our transfer-matrix calculations yield an approximate
answer to that question ---
namely, $\imag w \approx \pm 0.70$ ---
which we can combine with the perturbative calculations
to estimate the shape of the curve $\scrb_\infty$ at large $\real w$.
To do this, let us write $w = \alpha+i\beta$,
substitute into \reff{eq.RGflow_w}, separate real and imaginary parts,
and divide to obtain a differential equation
for the unparametrized flow curves:
\be
   {d\beta \over d\alpha}  \;=\;
   {   b_3 {\beta \over \alpha^2 + \beta^2} \,+\,
       b_4 {2\alpha\beta \over (\alpha^2 + \beta^2)^2} \,+\,
       b_5 {3\alpha^2 \beta - \beta^3 \over (\alpha^2 + \beta^2)^3}
       \,+\, \ldots
    \over
       -b_2 \,-\, b_3 {\alpha \over \alpha^2 + \beta^2}
            \,-\, b_4 {\alpha^2 - \beta^2 \over (\alpha^2 + \beta^2)^2}
            \,-\, b_5 {\alpha^3 - 3\alpha\beta^2 \over (\alpha^2 + \beta^2)^3}
            \,-\, \ldots
   }
 \label{eq.RGflow_unparam}
\ee
Plugging into \reff{eq.RGflow_unparam} the Ansatz
\be
   \beta  \;=\;
   \beta_0 \left[ 1 \,+\, {A_1 \over \alpha} \,+\, {A_2 \over \alpha^2}
                    \,+\, {A_3 \over \alpha^3} \,+\, \ldots
           \right]
   \;,
\ee
we obtain
\begin{subeqnarray}
   A_1  & = &  b_3/b_2   \\[1mm]
   A_2  & = &  b_4/b_2   \\[1mm]
   A_3  & = &  b_5/b_2 \,-\, \beta_0^2 b_3 / (3b_2)
\end{subeqnarray}
Finally, we need to determine the RG coefficients $b_2,b_3,\ldots\;$.
This can be done by applying the
recently-discovered mapping \cite{Caracciolo_PRL}
of the spanning-forest model onto the $N$-vector model
with $N=-1$ and a sign change in the coupling constant
(i.e., $w_{\rm forests} = -\beta_{N{\rm -vector}}$).
Using the known coefficients of the $N$-vector RG beta function
through four loops \cite{Caracciolo_95,Shin_99,Alles_99},
we obtain
\begin{subeqnarray}
   b_2  & = & \left. - w_0^{\hphantom{latt}} \right| _{N=-1}
              \;=\; 3/(2\pi)  \\[2mm]
   b_3  & = & \left. \hphantom{-} w_1^{\hphantom{latt}} \right| _{N=-1}
              \;=\; -3/(2\pi)^2  \\[2mm]
   b_4  & = & \left. - w_2^{latt} \right| _{N=-1}
              \;\approx\; 2.34278457/(2\pi)^3  \\[2mm]
   b_5  & = & \left. \hphantom{-} w_3^{latt} \right| _{N=-1}
              \;\approx\; 1.43677/(2\pi)^4
\end{subeqnarray}
and hence
\begin{subeqnarray}
   A_1  & =       &  -1/(2\pi)           \\[1mm]
   A_2  & \approx &  0.78092819/(2\pi)^2   \\[1mm]
   A_3  & \approx &  6.92706/(2\pi)^3 \quad\hbox{if $\beta_0=0.7$}
 \label{eq.A1A2A3}
\end{subeqnarray}

For numerical purposes, we found it convenient to use
the variant Ansatz
\be
   \beta  \;=\;
   \beta_0 \exp\!\left[ {B_1 \over \alpha-\alpha_0} \,+\,
                        {B_2 \over (\alpha-\alpha_0)^2} \,+\,
                        {B_3 \over (\alpha-\alpha_0)^3} \,+\,
                        {B_4 \over (\alpha-\alpha_0)^4} \,+\, \ldots
           \right]
   \;.
 \label{def.exp.Ansatz}
\ee
If we impose $B_3=B_4=0$ and fit to the
theoretical prediction \reff{eq.A1A2A3}
of the first three derivatives at $\alpha=\infty$,
we obtain $\alpha_0 = -0.479224$, $B_1 = -1/(2\pi)$, $B_2 = -0.069155$;
the resulting fit \reff{def.exp.Ansatz} evaluated at $\alpha=-0.0674$
has value 0.316354 and derivative 0.923329,
compared to the correct value 0.33 and derivative 0.6849
deduced from Table~\ref{table_limiting_curves}.
This is quite good agreement between extrapolated perturbation theory
and our nonperturbative estimates.
We therefore used the Ansatz \reff{def.exp.Ansatz},
imposing the first three derivatives at $\alpha=\infty$
along with the estimated value and derivative at $\alpha=-0.0674$;
we obtain
\begin{subeqnarray}
   \alpha_0  & = &  -1.06145   \\
   B_1       & = &  -1/(2\pi) \;\approx\; -0.159155  \\
   B_2       & = &  -0.161818  \\
   B_3       & = &  -0.134478  \\
   B_4       & = &  -0.284334
\end{subeqnarray}
The corresponding curve \reff{def.exp.Ansatz}
is depicted as a black dotted-dashed curve in Figure~\ref{figure_sq_all}.
This curve is presumably not quite right, because it intersects the $L=10$
curve, whereas the monotonicity of the finite-$L$ curves suggests that
the curve $\scrb_\infty$ lies to the right of all of them.
Therefore, the true curve $\scrb_\infty$ probably lies slightly to the
right of the dotted-dashed curve in Figure~\ref{figure_sq_all}.
Nevertheless, we suspect that this latter curve is a fairly good
approximation to the true curve $\scrb_\infty$,
especially at large positive $\real w$.

\subsection[Square lattice: Behavior of the free energy and its derivatives
   as $w \to w_0$]{Square lattice: Behavior of the free energy and
   its derivatives as \mbox{\protect\boldmath $w \to w_0$}}
 \label{sec.res.sq2a}

We have computed the real part of the strip free energy,
$\real f_L(w) = (1/L) \log |\lambda_\star(w)|$,
for widths $2 \le L \le 14$ and $w = -2,-1.9,\ldots,1.9,2$
as well as $w = -1/4$.
These results are plotted in Figure~\ref{figure_f_sq}.
We have extrapolated these values to $L=\infty$ (solid black line)
using the Ansatz
\be
   \real f_L(w) \;=\; \real f(w) \,+\, A(w) L^{-\Delta(w)}   \;.
 \label{eq.Ansatz.f.Delta}
\ee
We show also the [20,20] Pad\'e approximant
to the small-$w$ series for this quantity (dashed violet curve), and the
large-$w$ series \reff{def_large_w_series} through order $w^{-1}$
(dot-dashed pink curves).
It is interesting to note that the small-$w$ series agrees very well
with the free-energy data in the regime for $w>w_0({\rm sq})=-1/4$
(in fact, the two curves lie one over the other, at least out to $w=2$);
but it says nothing whatsoever about the behavior for $w<w_0$.
On the other hand, the large-$w$ series agrees well with the free energy
for $|w|\gtapprox 1$, for both signs of $w$;
indeed, it gives a fairly good description of the whole regime
$-\infty < w\ltapprox w_0$, except very close to the transition point.

\begin{figure}
\centering
\includegraphics[width=400pt]{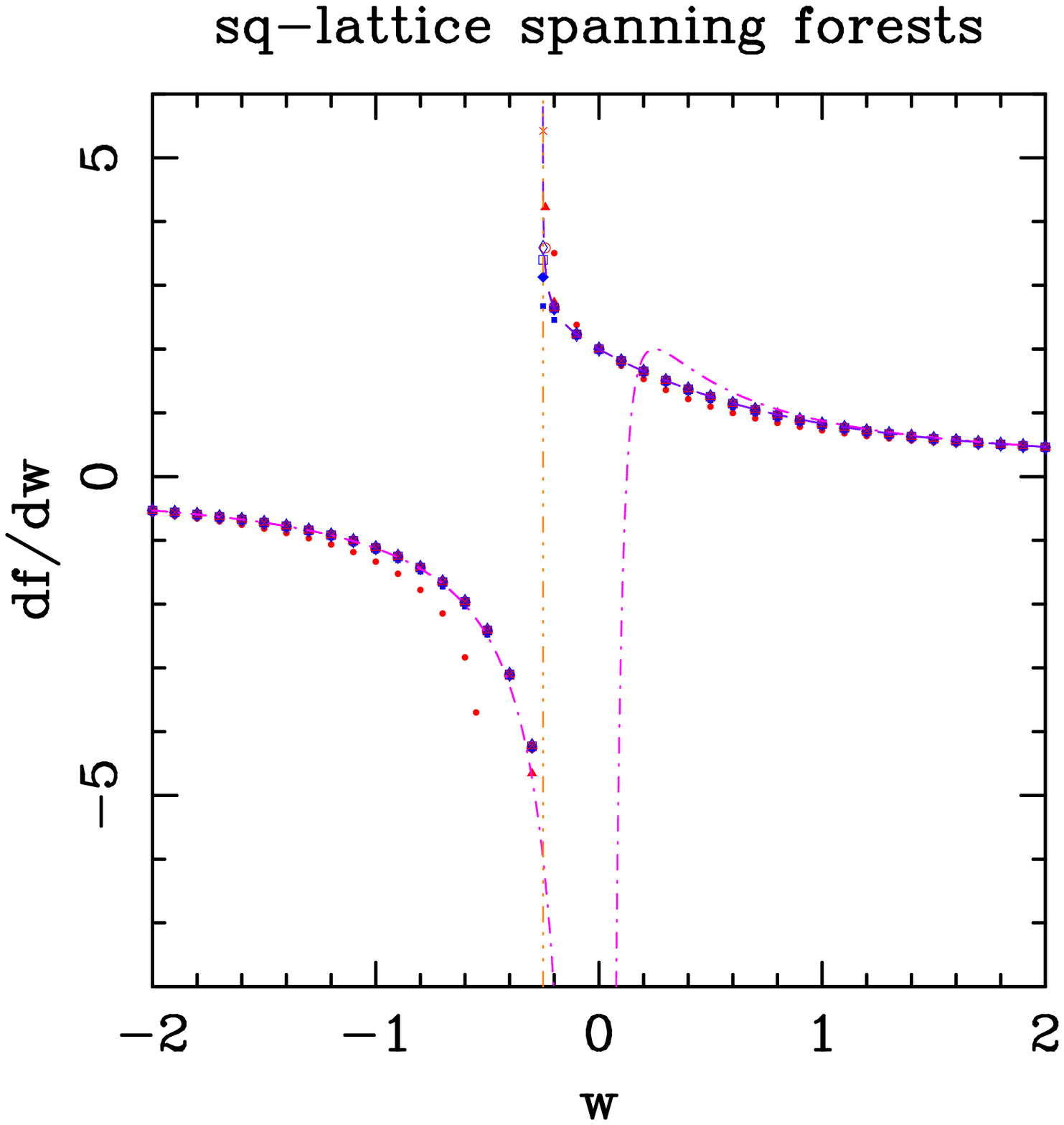}
\caption{\label{figure_E_sq}
   First derivative $f'_L(w)$ of the square-lattice free energy
   for strips of width $L=2$ ($\bullet$),
   $3$ ($\blacksquare$),
   $4$ ($\blacktriangle$), $5$ ($\blacklozenge$),
   $6$ ($\circ$),          $7$ ($\Box$),
   $8$ ($\triangle$),      $9$ ($\diamondsuit$) and
   $10$ ($\times$).
   Points with even (resp.\ odd) $L$ are shown in red (resp.\ blue).
   The violet dashed curve on the right corresponds to the
   Pad\'e approximant $[20,20]$ to our longest small-$w$ series.
   The pink dot-dashed curve corresponds to the derivative of the
   large-$w$ expansion \protect\reff{def_large_w_series},
   through order $w^{-2}$.
   The vertical brown dot-dot-dashed line marks the point $w_0=-1/4$.
}
\end{figure}

The real part of the free energy clearly has a minimum at $w=-1/4$.
{}From the plot it appears that the free energy is continuous at $w=-1/4$,
while its derivative with respect to $w$ 
(the ``internal energy'') has a jump discontinuity there.
To verify this latter point, we computed the derivative of the
free energy with respect to $w$ at the same list of values of $w$,
for widths $L \le 10$, by numerical differentiation.\footnote{
   For $L\leq 9$ we have done this computation
   using {\sc Mathematica} with high-precision arithmetic (200 digits).
   For $L=10$ we have used the {\sc arpack} package \protect\cite{arpack}
   with double precision. In the former case we have used the
   three-point interpolation formulas for central differentiation
   \protect\cite{Dvornikov} with mesh width $\Delta w = 0.5 \times 10^{-20}$
   (resp.\  $\Delta w = 10^{-20}$) for the first (resp.\   second)
   derivative of the free energy. In the latter case, we have used
   seven-point interpolation formulas with mesh widths
   $\Delta w = 0.5 \times 10^{-3}$ and $\Delta w = 10^{-3}$,
   respectively. This is enough to get results with at least nine-digit
   precision.
}
The results are shown in Figure~\ref{figure_E_sq}. In this
figure we have also shown the $[20,20]$ Pad\'e approximant
to the small-$w$ series for the internal energy $f'(w)$ [dashed violet curve]
and the first derivative of the large-$w$ series \reff{def_large_w_series},
namely $f'(w) = 1/w - 1/(8w^2)$ [dot-dashed pink curves].

\begin{figure}
\centering
\begin{tabular}{c}
\includegraphics[width=400pt]{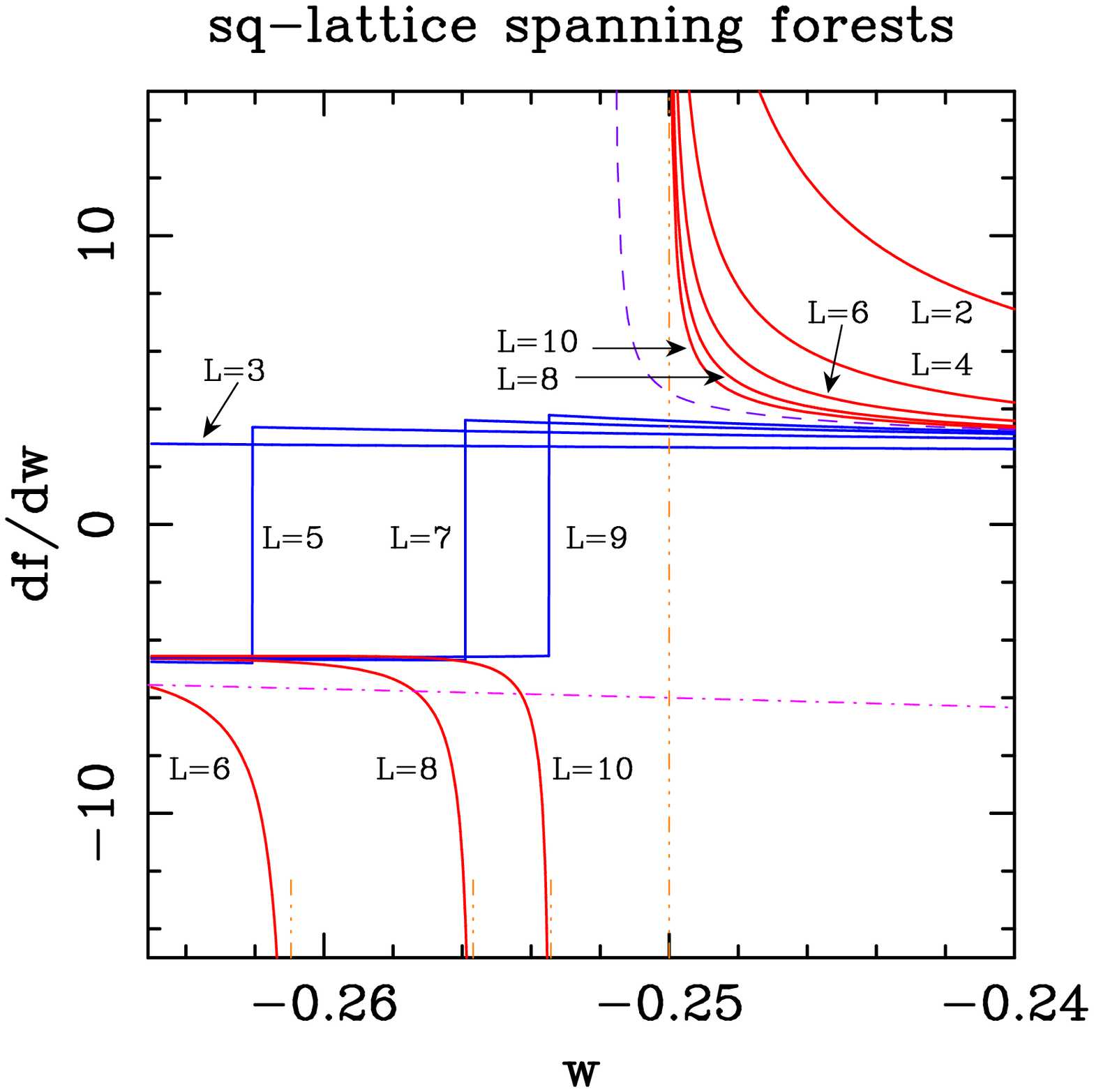} \\[2mm]
\phantom{(aa)}\large{(a)}
\end{tabular}
\caption{\label{figure_Ec_sq}
   First derivative $f'_L(w)$ of the square-lattice free energy
   for strips of widths $2\leq L \leq 10$ close to $w_0({\rm sq})=-1/4$.  
   (a) Curves for even (resp.\ odd) $L$ are shown in red (resp.\ blue).
   The violet dashed curve on the right corresponds to the
   Pad\'e approximant $[20,20]$ to our longest small-$w$ series.
   The pink dot-dashed curve corresponds to the derivative of the
   large-$w$ expansion \protect\reff{def_large_w_series},
   through order $w^{-2}$.
   The vertical brown dot-dot-dashed line marks the point $w_0=-1/4$.
   The vertical brown dot-dot-dashed lines near the bottom of the figure
   mark the position of the points $w_{0-}(L)$ for $L=6,8,10$.
   (b) Curves for strips of widths $L=2,4,6,8,10$
   are plotted versus $w-w_{0-}(L)$.
}
\end{figure}

It is clear that $f'(w)$ has a jump discontinuity at $w=w_0 = -1/4$:
when $w \uparrow w_0$ it decreases towards a negative limit,
and when $w \downarrow w_0$ it increases towards a positive limit.
What is less clear from Figure~\ref{figure_E_sq}
is whether these limiting values are finite or infinite.
In order to address this question, we have plotted in 
Figure~\ref{figure_Ec_sq}(a)
a blow-up of the region very near $w=w_0=-1/4$, showing $f'_L(w)$ together 
with the same small-$w$ and large-$w$ curves
as in Figure~\ref{figure_E_sq}.
The curves $f'_L(w)$ show a clear parity effect due to the existence
for even $L$ of the interval $[w_{0-},w_{0+}]$
where the two dominant eigenvalues form a complex-conjugate pair,
and due to the existence for odd $L$
of the point $w_{0Q}$ where the two dominant eigenvalues cross in modulus.
More precisely, for even $L$ the free energy has square-root branch points
at $w = w_{0\pm}(L)$,
so that $f'_L(w)$ diverges to $\pm \infty$ at those two points,
at a rate $\sim |w-w_{0\pm}(L)|^{-1/2}$.
These divergences are seen clearly in Figure~\ref{figure_Ec_sq}(a)
(note that $w_{0+}(L) = -1/4$ for all $L$,
while $w_{0-}(L)$ for $L=6,8,10$ is marked with a brown  
dot-dot-dashed vertical line).
For odd $L$, by contrast, the free energy switches at $w = w_{0Q}(L)$
between two distinct eigenvalues,
so that $f'_L(w)$ has a jump discontinuity there.
This discontinuity is also seen clearly in Figure~\ref{figure_Ec_sq}(a).

We now wish to infer from Figure~\ref{figure_Ec_sq}(a)
the infinite-volume behavior of of the internal energy $f'(w)$
in the two regimes $w > w_0$ and $w < w_0$.
To do this, we must disregard the behavior too near the points
$w = w_{0\pm}(L)$, where $f'_L(w)$ is dominated by the finite-$L$ divergence,
and focus instead on extracting the limit $L \to\infty$ at fixed $w$.
Fortunately, a fairly clear picture emerges:

1) For $w > w_0$, $f'_L(w)$ is decreasing (resp.\ increasing) in $L$
for even (resp.\ odd) $L$;
the two data sets seem to be converging to a common limiting curve
that lies very near the [20,20] Pad\'e approximant to the small-$w$ series.
Furthermore, this limiting curve very likely remains {\em finite}\/
as $w \downarrow w_0$.
Indeed, for odd $L$ there is not the slightest indication that $f'_L(w)$
could be diverging;  the data seem, rather, to indicate a limiting
value around 4, and certainly no more than 5.
For even $L$ the data are less clear,
but at least for $w \gtapprox -0.248$
we see that $f'_L(w)$ is rapidly decreasing with $L$
and apparently approaching a limiting value
close to the Pad\'e-approximant curve;
furthermore, this convergence is rapid when $w$ is well above $w_0$
and less rapid when $w$ is nearer to $w_0$.
It is therefore reasonable to expect that a similar convergence occurs
for {\em all}\/ $w > w_0$, but which for $w$ very near $w_0$
requires very large $L$ to be seen definitively.
We would guess a limiting value
$f'_+(w_0) \equiv \lim_{w \downarrow w_0} f'(w)$ in the range 4--5.
However, it is at least conceivable that the infinite-volume
internal energy $f'(w)$ is weakly divergent as $w \downarrow w_0$,
e.g.\ like $f'(w) \sim \log (w-w_0)$ or $f'(w) \sim \log \log (w-w_0)$.

2) For $w < w_0$ the behavior is even clearer:
both the odd-$L$ and even-$L$ curves stay close to each other
(at a value $\approx -4.7$)
until they peel away due to the discontinuity at $w=w_{0Q}(L)$ for odd $L$
or due to the proximity of the divergence at $w=w_{0-}(L)$ for even $L$;
furthermore, these ``peel away'' points get closer to $w_0$ as $L$ grows.
This strongly suggests that a similar convergence occurs
for {\em all}\/ $w < w_0$.
Finally, the apparent limiting value $\approx -4.7$
is not far from the curve corresponding to the large-$w$ expansion
(namely, about 20\% above it).
Once again, there is no evidence that $f'(w)$ could be diverging
as $w \uparrow w_0$.
Rather, the data for both odd and even $L$ suggest a limiting
value $f'_-(w_0) \equiv \lim_{w \uparrow w_0} f'(w) \approx -4.7$.

Since $w_{0-}(L)$ varies with $L$, it is useful to plot
$f'_L(w)$ versus $w - w_{0-}(L)$ in order to compare
the behavior at ``comparable'' values of $w$.
Such a plot is shown in Figure~\ref{figure_Ec_sq_bis}(b).
We see clearly that the region of divergence grows narrower as $L$ grows.
In particular, at any {\em fixed}\/ value of $w - w_{0-}(L) < 0$,
$f'_L(w)$ appears to tend to a limiting value around $-4.7$:
this is seen clearly for $w - w_{0-}(L) \ltapprox -0.003$,
and would presumably be seen also for smaller $|w - w_{0-}(L)|$
if we were to go to larger $L$.
This plot confirms, from a slightly different point of view,
our conclusion that
$f'_-(w_0) \equiv \lim_{w \uparrow w_0} f'(w) \approx -4.7$.

In summary, there is strong evidence that the infinite-volume
internal energy $f'(w)$ tends to a finite value as $w \uparrow w_0$.
The evidence is less clear concerning the behavior as $w \downarrow w_0$,
but the most likely scenario is also a finite limit,
with weak divergences also possible.
In any case, the existence of a discontinuity ---
whether finite or infinite --- in the first derivative of the
free energy is a clear signal of a first-order phase transition at $w_0$.

We can also compute the second derivative of the free energy with respect to
$w$ (i.e., the analogue of the specific heat in the usual Potts model),
again by numerical differentiation.
The results are shown in Figure~\ref{figure_CH_sq}, along with
the $[20,20]$ Pad\'e approximant for the corresponding small-$w$ series
for $f''(w)$ [dashed violet curve]
and the second derivative of the large-$w$ series
\reff{def_large_w_series}, namely $f''(w) = -1/w^2 + 1/(4w^3)$
[dot-dashed pink curves].
This derivative is clearly getting large and negative as we approach $w_0$
from both sides, and it may well be diverging to $-\infty$.
To check this point more carefully,
we made a blow-up plot of $f''(w)$ very near $w=w_0$,
using a {\em logarithmic}\/ vertical scale
[see Figure~\ref{figure_CHc_sq}(a)].
Just as for $f'$, we find a different behavior for odd and even $L$.
The curves $f''_L(w)$ for odd $L$ have a jump discontinuity at 
$w=w_{0Q}(L)$. Their value at $w=w_0 = -1/4$ grows with $L$
and it is not clear how these values behave in the limit $L \to\infty$.
For fixed even $L$, we find that $f''_L(w)$ grows rapidly
as $w \uparrow w_{0-}(L)$ and as $w \downarrow w_{0+}(L) = -1/4$;
this rapid growth is once again explainable
as arising from the square-root branch points in $f_L(w)$,
causing a divergence $f''_L(w) \sim |w-w_{0\pm}(L)|^{-3/2}$.
However, the magnitude of $f''_L(w)$ at fixed $w$ is decreasing as $L$ grows,
and it is unclear how $f''(w) = \lim_{L\to\infty} f''_L(w)$ behaves.
Once again, let us consider this question separately on the two sides
of $w_0$:

1) For $w > w_0$, it is plausible that the finite-$L$ curves
are converging from opposite sides (i.e., from above when $L$ is even
and from below when $L$ is odd) to a limiting curve $f'(w)$
that is close to the [20,20] Pad\'e approximant to the small-$w$ series
(i.e., the dashed violet curve in Figure~\ref{figure_CH_sq}).
Analysis of this Pad\'e approximant (see below) then suggests
that $f''(w)$ diverges as $w \downarrow w_0$,
with an exponent near $\alpha \approx 1$.

2) For $w < w_0$, let us plot $f''_L(w)$ versus $w - w_{0-}(L)$
in order to compare the behavior at ``comparable'' values of $w$.
Such a plot is shown in Figure~\ref{figure_CHc_sq_bis}(b).
Careful inspection shows that the ratios of $f''(L)$ for different $L$
--- i.e., the vertical distances between curves on this logarithmic plot ---
are {\em decreasing}\/ as $w \downarrow w_0$.
(The strange behavior for $L=4$ very near $w_{0-}(L)$
 arises from the fact that this point is {\em not}\/ an endpoint,
 but rather a T point,
 as shown in Figure~\ref{figure_sq_4P_zoom}.)
This suggests that the infinite-volume specific heat $f''(w)$ may diverge
as $w \uparrow w_0$ in a manner somewhat similar to that shown
by the $L=10$ curve.
But the evidence is admittedly weak, and it is fair to say
that we cannot really be sure how $f''(w)$ behaves as $w \uparrow w_0$.

Finally, further information about the behavior of
$f(w)$, $f'(w)$ and $f''(w)$ as $w \downarrow w_0$ can be
obtained by analyzing the small-$w$ series for these quantities. This is
done in detail in Appendix~\ref{sec_series_analysis}; here we simply
summarize the results. We find that the behavior of these three quantities 
is consistent with
\begin{subeqnarray}
f  (w)&\sim& \left(w + \smfrac{1}{4} \right)^{2-\alpha} \quad
             \hbox{with $\alpha = 0.94 \pm 0.10$} \\
f' (w)&\sim& \left(w + \smfrac{1}{4} \right)^{1-\alpha} \quad
             \hbox{with $\alpha = 0.90 \pm 0.04$} \\
f''(w)&\sim& \left(w + \smfrac{1}{4} \right)^{-\alpha\phantom{1}} \quad
             \hbox{with $\alpha = 0.91 \pm 0.02$}
  \slabel{eq.series.CH}
\end{subeqnarray}
We thus have strong evidence that $f''(w)$ diverges as $w\downarrow w_0$,
but it is not clear whether $f'(w)$ diverges there or not.
Taking our estimates literally,
the corresponding critical exponent is $1-\alpha=0.10\pm 0.04 > 0$,
so that $f'$ is finite as $w\downarrow w_0$;  but we cannot
make a firm conclusion as the error bar is very large!
Taking into account the uncertainties of series analysis,
it seems fair to say that \reff{eq.series.CH} is consistent with
the theoretically predicted value $\alpha = 1$,
especially if we recall the possibility of multiplicative logarithmic
corrections.
Note that, depending on the form of the latter corrections,
$f''$ could be either integrable or nonintegrable at $w=w_0$,
i.e., $f'$ could be either finite or infinite there.

\subsection{Square lattice: Estimates of the free energy and
   central charge} \label{sec.res.sq2b}

Baxter's exact solution \cite{Baxter_82} for the Potts-model free energy
on the curve \reff{curve_sq_2} implies the exact value
\be
 f({\rm sq},w=-1/4) \;=\;
 2 \log \left( {\Gamma(1/4) \over 4 \Gamma(3/4)} \right)
                    \;\approx\; -0.6031055757   \;.
 \label{def_f_sq_w=-0.25}
\ee
In order to make a high-precision test of this prediction,
we computed the strip free energies $f_L(w=-1/4)$ for all widths $L$ up to 15.
For $2 \le L \le 9$ we used our symbolic transfer matrices,
while for $10 \le L \le 15$ we used a numerical transfer-matrix algorithm.
These data are reported in Table~\ref{table_f_sq}.
The free energies $f_L(w=-1/4)$ show strong even-odd oscillations;
moreover, they appear to be monotonically increasing in $L$ for even $L$,
and monotonically decreasing for odd $L$.
If this monotonicity persists for larger $L$, then from $L=14,15$
we can deduce the bounds
\be
   -0.6067349394  \;\le\; f({\rm sq},w=-1/4) \;\le\; -0.6019335900
    \;.
\ee
Separate extrapolations of the even and odd subsequences
using the Ansatz \reff{eq.Ansatz.f.Delta} yield the results
reported in Table~\ref{table_f_fits_sq}.
Putting them together, we conclude that
\be
   f({\rm sq},w=-1/4)  \;=\;  -0.60310 \pm 0.00010
   \;,
\ee
in excellent agreement with the exact result \reff{def_f_sq_w=-0.25}.
We also find a rate of convergence compatible with $\Delta=2$
(see below for a more detailed analysis).

We have similarly extended up to width $L=16$
the data for the points $w = \pm 1$ of interest to mathematicians:
see Table~\ref{table_f_sq}.
In both cases, $\real f(w)$ appears to be monotonically increasing in $L$.
If this monotonicity really holds for all $L$, it would imply that
\begin{subeqnarray}
   \real f({\rm sq},w=1) & \ge &  \real f_{L=16}({\rm sq},w=1)
      \phantom{-}  \;\approx\; 1.3079471010
\slabel{conjecture_N_sq_w=1} \\
   \real f({\rm sq},w=-1) & \ge &  \real f_{L=16}({\rm sq},w=-1)
                 \;\approx\; 1.0432380373
\slabel{conjecture_N_sq_w=-1}
\label{conjecture_N_sq}
\end{subeqnarray}
The lower bound \reff{conjecture_N_sq_w=1} lies between
the rigorous lower and upper bounds \reff{Noy_bounds_improved}.
Extrapolation of these sequences using the Ansatz \reff{eq.Ansatz.f.Delta}
for triplets $L=L_{\rm min},L_{\rm min}+1,L_{\rm min}+2$
yields the results reported in Table~\ref{table_f_fits_sq}.
The convergence for $w=-1$ is slower ($\Delta\approx 2$)
than for $w=1$ ($\Delta\approx 3.6$).
Our best estimates $f({\rm sq},w=\pm 1)$ are
\begin{subeqnarray}
f({\rm sq},w=1)    &=&  1.30819 \pm 0.00010 \\
f({\rm sq},w=-1)   &=&  1.04870 \pm 0.00010  \slabel{eq.fsq.w=-1.first}
\end{subeqnarray}
These values are of course larger than the conjectured lower bounds
\reff{conjecture_N_sq};
and the estimate for $f({\rm sq},w=1)$ lies
between the rigorous lower and upper bounds \reff{Noy_bounds_improved}.

The different values found for the correction exponent $\Delta$
can be explained in the following way.
For $w > -1/4$, we expect that the system
is governed by the high-temperature fixed point, which is non-critical.
As a consequence, we expect an exponentially rapid convergence
of the strip free energy to its infinite-volume limit.
This explains why for $w=1$ the estimates of $\Delta$
grow with $L_{\rm min}$, apparently without bound.
For $w<-1/4$, by contrast,
we expect that the system will be governed by the fixed point
located inside the Berker--Kadanoff phase \cite{Saleur_90,Saleur_91},
which is critical.
Conformal field theory (CFT) then predicts that
\be
f_L(w) \;=\;  f(w) +  {\pi \over 6} c(w) L^{-2} + \ldots
\label{def_ansatz_f}
\ee
where $c(w)$ is the central charge (see Section~\ref{sec.setup.3});
we therefore expect $\Delta=2$.
Moreover, the central charge associated to the critical point
of the $q=0$ Potts model in the Berker--Kadanoff phase
is predicted from \reff{def_c_BK} to be $c=-2$.
Finally, at $w=-1/4$ there is a phase transition
that we expect also to be a critical point;
we therefore again expect $\Delta=2$.
Here, however, the central charge is predicted from \reff{def_c_antiferro}
to be $c=-1$.

Let us therefore use our finite-size data to estimate the central charge
$c(w)$. As a first approach, we have fit the data to the Ansatz
\reff{def_ansatz_f} with no higher-order corrections. The results are
displayed in Figure~\ref{figure_c_sq}(a). We observe that there are two clear
regimes separated by $w=w_0=-1/4$, in qualitative agreement with the above
discussion.

For $w>w_0$, the estimates for the central charge are very close to
$c=0$ when $w_0 < w \ltapprox 0.4$. This agrees well with the
prediction that the system is non-critical (the convergence of
$f_L(w)$ to the bulk free energy $f(w)$ is exponentially fast, so that $c=0$).
For $w\gtapprox 0.4$ we observe sizable downward deviations from $c=0$,
even for large values of $L_{\rm min}$;
but the estimates of $c$ increase monotonically and are at least consistent
with convergence to the value $c=0$.
These deviations from $c=0$ can be explained as due to
crossover from the {\em marginally repulsive}\/
(i.e., asymptotically free \cite{Caracciolo_PRL}) ferromagnetic fixed point
at $w=+\infty$, which by \reff{def_c_ferro} is a $c=-2$ theory.
In this regime, we expect large finite-size corrections that
are not adequately handled by our Ansatz \reff{def_ansatz_f}.

For $w<w_0$, we find that the estimates for the central charge are
quite far away from the predicted value $c=-2$
(at least up to $L_{\rm min} = 13$).
However, these estimates are increasing monotonically with $L_{\rm min}$,
and it is at least plausible that they are approaching $c=-2$
as $L_{\rm min} \to \infty$.
Furthermore, the estimates are closer to $c=-2$
for larger negative values of $w$
(i.e., for points deeper inside the Berker--Kadanoff phase).
Indeed, for large negative values of $w$ we are closer to the
renormalization-group fixed point (\ref{curve_sq_1}$_-$), which lies at
$w=-\infty$;
therefore, the corrections to scaling are expected to be weaker for larger
negative values of $w$.
All these findings reflect the {\em marginally attractive}\/ nature
of the RG fixed point at $w=-\infty$ \cite{Caracciolo_PRL}.
Indeed, the theory {\em predicts}\/ very slow convergence
to the infinite-volume limit,
as a consequence of the $\log\log L/\log L$ and $1/\log L$
corrections to scaling arising from the marginally irrelevant operator.
It is quite plausible that the behavior exhibited in 
Figure~\ref{figure_c_sq}(a) is a manifestation of
such logarithmic corrections to scaling.

It is instructive to plot the estimates of $c(w)$ versus $1/w$,
as shown in Figure~\ref{figure_c_sq}(b).
The smooth behavior of the plot near $1/w = 0$
suggests that the fixed points lying at $w=\pm\infty$
are in fact identical and have central charge $c=-2$,
in agreement with the prediction \cite{Caracciolo_PRL}
that this is a theory of a pair of free scalar fermions.
The $L$-dependence of the estimates of $c(w)$ ---
tending towards $-2$ as $L \to\infty$ when $1/w < 0$,
and towards $0$ as $L \to\infty$ when $1/w > 0$ ---
is likewise in agreement with the prediction \cite{Caracciolo_PRL}
that this fixed point is marginally repulsive for $g \equiv 1/w > 0$
and marginally attractive for $g < 0$.

Somewhat stabler estimates for the central charge can be obtained if we
fit the numerical data to the Ansatz
\be
f_L(w) \;=\;  f(w) +  {\pi \over 6} c(w) L^{-2} + A L^{-4}   \;.
\label{def_ansatz_f_bis}
\ee
As explained in Section~\ref{sec.setup.3},
the term $L^{-4}$ is predicted by conformal-field-theory arguments;
but we cannot rule out the existence of additional correction terms,
lying between $L^{-2}$ and $L^{-4}$,
arising from irrelevant operators with critical exponents
between $y=-2$ and $y=-4$.
In practice, inclusion of the $L^{-4}$ term usually tends to accelerate
the convergence of the estimates for the central charge
\cite{Bloete_82,Cardy_98}. The resulting plot of $c(w)$ versus $w$ is shown
in Figure~\ref{figure_c_sq}(c).
We observe that the estimates for $c(w)$ converge more rapidly
than with Ansatz \reff{def_ansatz_f},
although we are still far away from
the predicted value $c=-2$ in the regime $w<-1/4$.
For example, at $w=-2$ we obtain $c \approx -2.487$ at $L_{\rm min}=12$
with the Ansatz \reff{def_ansatz_f_bis},
compared to $c \approx -2.543$ at $L_{\rm min}=13$
with the Ansatz \reff{def_ansatz_f}.
The corresponding  plot of $c(w)$ versus $1/w$ is shown
in Figure~\ref{figure_c_sq}(d).

The mapping of the spanning-forest model to the $N$-vector model at $N=-1$
\cite{Caracciolo_PRL} suggests that a more correct Ansatz for the free energy
in the critical phase would be
\be
f_L(w) \;=\;  f(w) +  {\pi \over 6} c(w) L^{-2} + 
      A L^{-2} \frac{\log\log L}{\log L} + B \frac{L^{-2}}{\log L}  
   \;.
\label{def_ansatz_f_tris}
\ee
The plot of $c(w)$ obtained from a fit to \reff{def_ansatz_f_tris},
restricted to the regime $w < -1/4$, is shown in 
Figure~\ref{figure_c_sq2}.\footnote{
  For $w=-0.3$ we found strong even--odd oscillations.
  Therefore, for this particular value we performed the fits using the data
  with $L=L_{\rm min},L_{\rm min}+2,L_{\rm min}+4,L_{\rm min}+6$. 
}
We obtain values for the central charge closer to the expected result
$c=-2$ than for the simpler Ans\"atze 
\reff{def_ansatz_f}/\reff{def_ansatz_f_bis}. For instance, for $w=-2$ we
get $c=-2.161$ at $L_{\rm min}=11$. 

For $w=-1$ we have extended the above analysis to include finite-size data
up to $L=16$ (see Table~\ref{table_f_fits_c_sq_w=-1}).
The estimates for $c(w=-1)$ continue to increase monotonically,
but we are still far away from the predicted value $c=-2$. We obtain
values slightly closer to $c=-2$ using the Ansatz \reff{def_ansatz_f_bis}
than with the original Ansatz \reff{def_ansatz_f},
but we are still far away.
As a side benefit, we obtain an estimate for $f({\rm sq},w=-1)$
that is slightly more stable than \reff{eq.fsq.w=-1.first}:
\be
   f({\rm sq},w=-1)  \;=\; 1.04883 \pm 0.00008   \;.
\ee
If we use the improved Ansatz \reff{def_ansatz_f_tris},
the estimates for $c(w=-1)$ still increase monotonically
for $L_{\rm min} \ge 5$
(see Table~\ref{table_f_fits_c_sq_w=-1_bis}),
but these estimates are now much closer to the predicted value $c=-2$
(e.g., for $L_{\rm min}=13$, we get $c \approx -2.046$). 
Finally, this Ansatz provides a more stable and precise estimate
for $f({\rm sq},w=-1)$: 
\be
   f({\rm sq},w=-1)  \;=\; 1.048755 \pm 0.000008   \;.
\ee

For $w=-1/4$ we already know the exact value of the bulk free energy $f(w)$,
namely \reff{def_f_sq_w=-0.25}.  We therefore used this value and performed
a one-parameter fit to \reff{def_ansatz_f} to estimate $c(w=-1/4)$:
see Table~\ref{table_f_fits_c_sq_w=-0.25}.
For even $L$, the estimates of $c$ increase monotonically from
$c \approx -1.79$ at $L_{\rm min}=2$ to $c \approx -1.36$
at $L_{\rm min}=14$. This suggests that the central charge for
even widths would eventually converge to the predicted value $c=-1$
in the limit $L_{\rm min}\to\infty$.
For odd $L$, by contrast, the estimates of $c$ decrease monotonically from
$c \approx 0.556$ at $L_{\rm min}=3$ to $c\approx 0.5036$
at $L_{\rm min}=15$.
This latter value strongly suggests an effective central charge
\be
c_{\rm effective}({\rm sq},w=-1/4) \;=\;  1/2  \;.
\label{conjecture_c_sq_w=-0.25}
\ee
Similar results are obtained by the two-parameter fit to the Ansatz
\reff{def_ansatz_f_bis} with the bulk free energy fixed to its exact value
\reff{def_f_sq_w=-0.25}. For even $L$, we obtain values
of $c(w=-1/4)$ slightly closer to the predicted value $c=-1$
(the estimates increase monotonically from
$c\approx -1.57$ at $L_{\rm min}=2$ to $c \approx -1.31$ at $L_{\rm min}=12$).
For odd $L$ the estimates are stable around $c\approx 0.501$.

The free energies for even $L$ are related to the ground state
of the corresponding continuum theory, since frustration effects
are absent. Similarly, it is natural to assume that the free
energies for odd $L$
are related to the ground state of a modified continuum theory
in which the fields are subjected to a twist in the spatial
direction (antiperiodic boundary conditions). If this is indeed the case,
conformal field theory \cite{Itzykson_collection,DiFrancesco_97,Henkel_99}
predicts that for $L$ odd,
\be
 c_{\rm effective} \;=\; c - 24 h_{\rm twist} \;.
 \label{h_twist0}
\ee
Assuming the conjecture (\ref{conjecture_c_sq_w=-0.25}) to be correct,
the conformal weight of the twist operator must then be
\be
 h_{\rm twist} \;=\; -1/16 \;.
 \label{h_twist}
\ee
The twist operator was not discussed in the original paper of
Saleur \cite{Saleur_91} on the CFT analysis of antiferromagnetic Potts models.
However, recent work \cite{Saleur_04}
on the antiferromagnetic critical line \reff{curve_sq_2}
has identified the twist operator with
the fundamental disorder operator \cite{Zam_86}
of $Z_\delta$ parafermions, where $\delta$ is defined by \reff{def_delta}.
In the limit $q \to 0$ this confirms the conjecture \reff{h_twist}.

\subsection{Square lattice: Estimates of the thermal scaling dimensions}
   \label{sec.res.sq3}

The first step in obtaining the thermal exponent $\nu$ is to obtain the
inverse correlation lengths for a square-lattice strip of width $L$
[cf.\ \reff{def_inverse_xi}]:
\be
\xi^{-1}_i(w;L) = \log \left| {\lambda_\star(w) \over \lambda_i(w)}\right|
\label{def_inverse_xi_2}
\ee
where $\lambda_\star(w)$ is the largest eigenvalue (in modulus) of
the transfer matrix associated to a square-lattice strip of width $L$
and cylindrical boundary conditions,
and $\lambda_i(w)$ is the $i$-th subdominant eigenvalue.
In Figures~\ref{figure_xi1_sq}(a) and~(b) we have plotted the
first two inverse correlation lengths $\xi^{-1}_i(w;L)$
for all widths $L\leq 10$ in the range $-2 \le w \le 2$.

The second step is to extract from this finite-width data the thermal
scaling dimensions $x_{T,i}$.
If $w$ is a noncritical point, we expect as usual
an exponentially rapid convergence:
\be
\xi_i^{-1}(w;L)  \;=\;  \xi_i^{-1}(w) \,+\,  O(e^{-A_i L})
\label{def_xi_noncritical}
\ee
where $0 < A_i(w)<\infty$.
If $w$ is a critical point, then the behavior of the inverse correlation
length can be described by the CFT prediction \reff{def_xi_critical}.
In order to fit the data in an unified fashion, we began by using the Ansatz
\be
\xi_i^{-1}(w;L)  \;=\;  \xi_i^{-1}(w) + {2 \pi x_{T,i}(w) \over L}   \;.
\label{def_xi_critical_sq}
\ee
For a critical system, we expect $\xi^{-1}_i=0$,
and then $x_{T,i}$ is the true scaling dimension for that system,
modulo higher-order finite-size-scaling corrections that are neglected
in \reff{def_xi_critical_sq}.
For a noncritical system, we expect that $\xi_i^{-1} > 0$.
In this case, the $L^{-1}$ term in \reff{def_xi_critical_sq}
tries to mimic the exponentially small corrections,
so that the estimates of $x_{T,i}$ should tend rapidly to zero
as $L$ grows.

We began by fitting our finite-width data to the Ansatz
\reff{def_xi_critical_sq}. We found, as expected, three different regimes,
according as $w < -1/4$, $w=-1/4$ or $w > -1/4$.

Let us start by discussing the first correlation length $\xi_1$:
see Figure~\ref{figure_xi1_sq}(a) for the finite-width data
as well as the extrapolated infinite-width values (black solid circles).
Throughout the regime $w<-1/4$,
the estimated value of $\xi_1^{-1}$ is small in magnitude,
in agreement with the prediction that this phase is critical.
Indeed, for $w \ltapprox -0.5$ we find that
$\xi_1^{-1}(w)\ltapprox 0.095$ and the estimates
decrease steadily with $L_{\rm min}$.
For $-0.5\ltapprox w\ltapprox -1/4$,
the estimates of $\xi_1^{-1}$ are larger,
but they again decrease as $L_{\rm min}$ grows.
We interpret these deviations of $\xi_1^{-1}$ from zero
as arising from corrections to scaling,
which not surprisingly become stronger as the transition point
$w_0 = -1/4$ is approached.
We take all this as an indication that the true value
is $\xi_1^{-1}(w)=0$ throughout the region $w < -1/4$.
We therefore repeated the fit, imposing $\xi_1^{-1} = 0$.
The resulting estimates for the scaling dimension $x_{T,1}$ are displayed in
Figure~\ref{figure_x1_sq}(a).
We see that the scaling dimension is close to the value $x_{T,1}(w) = 2$
predicted by Saleur \cite{Saleur_91} for the Berker--Kadanoff phase $w < -1/4$
[cf.\ \reff{def_xT_BK}].
Furthermore, as we proceed deeper into the Berker--Kadanoff phase
(i.e., towards $w$ more negative), the estimates also become closer to
$x_{T,1}=2$.

For $w> -1/4$, the estimated value of $\xi_1^{-1}$ is strictly positive
(corresponding to a non-critical theory),
at least up to $w \approx 0.8$.
Around $w=0$ the inverse correlation length becomes very large, as expected;
as $w$ increases beyond this point, $\xi_1^{-1}$ decreases.
For $w\gtapprox 0.8$ our estimates of $\xi_1^{-1}$ become small
($\ltapprox 0.18$);
but we do not take this fact as an indication of a critical system.
Rather, it simply reveals the fact that the correlation length $\xi_1$
has become comparable to the strip widths we are considering,
and that the Ansatz \reff{def_xi_critical_sq} is not adequate
to describe accurately the corrections to scaling in this regime
(indeed, there is no evidence of a phase transition around $w\approx 0.8$).
We expect, in fact, that the correlation length $\xi_1$ is finite for all
$w > 0$, tending to infinity as $w \to +\infty$.
Our estimates for $\xi_1^{-1}$ are depicted in Figure~\ref{figure_xi1_sq}(a) as
black solid dots.

The case $w=w_0=-1/4$ is rather special.
We have already seen that for even widths,
this point is an {\em exact}\/ endpoint of the limiting curve.
Therefore, $\xi_1^{-1}(w=-1/4;L)=0$ {\em exactly}\/ for all even $L$.
In particular, from even $L$ we conclude that $x_{T,1} = 0$.
The values of $\xi_1^{-1}(w=-1/4;L)$ for odd $L$
are displayed in Table~\ref{table_inverse_xi1_sq},
where we also show (in columns 3--5) the results of fitting the
data with odd $L$ to the Ansatz \reff{def_xi_critical_sq}.
We see that the estimate for $\xi_1^{-1}$ is a small negative number,
which decreases in modulus as $L_{\rm min}$ is increased.
This suggests that $\xi^{-1} = 0$ exactly,
so we repeated the fit with this value fixed.
The results are shown in the last two columns of
Table~\ref{table_inverse_xi1_sq}.
As we increase $L_{\rm min}$, the estimate for $x_{T,1}$ decreases;
but it is not clear whether it is tending to zero
or to a nonzero value around $x_{T,1}\approx 0.3$.
What is clear is that Saleur's conjecture \reff{def_xT_antiferro}
for the thermal exponent along the critical antiferromagnetic curve,
which reduces to $x_{T,1}=2$ at $q=0$, is {\em incorrect}\/,
as our estimates are clearly smaller than this value.
If it is indeed the case that $x_{T,1}=0$ (as we suspect),
then the correlation-length exponent is
\be
  \nu  \;=\; {1\over 2-x_{T,1}}  \;=\; {1\over 2} \;,
\ee
in agreement with our findings from the finite-size shift of the
critical point (Section~\ref{sec.res.sq1} and Table~\ref{table_fits_sq})
and with the existence of a first-order phase transition at $w=w_0$.

The second inverse correlation length $\xi^{-1}_2$ is displayed in
Figure~\ref{figure_xi2_sq}(b). If we try to estimate the thermodynamic limit
of $\xi^{-1}_2$ using the Ansatz \reff{def_xi_critical_sq},
we find that $0.56 \ltapprox \xi^{-1}_2 \ltapprox 0.74$ in the regime
$w<w_0$. At first sight this could imply that this correlation length is
finite when $L\to\infty$. However, because the correlation length $\xi_1$
actually diverges in this regime, we expect that $\xi_2$ will do so as well.
On the other hand, we expect that as the index $i$ grows,
the finite-size corrections to $\xi_i(L)$ will become larger.
We therefore repeated the fit, fixing $\xi_2^{-1}=0$ for $w<w_0$
and attempting to extract $x_{T,2}$ from the coefficient of the
linear term in $1/L$. The results are displayed in Figure~\ref{figure_x2_sq}(b).
We see that in the Berker--Kadanoff phase the value of this scaling dimension
is close to $x_{T,2}=4$.

For $w>w_0$, we find a pattern similar to what was observed for $\xi_1$:
namely, the inverse correlation length grows as we approach $w=0$,
and then it decreases, getting stable around $\xi^{-1}_2 \approx 2.5$.
As we concluded that $\xi_1$ remains finite in this regime,
$\xi_2$ should stay finite too;
but the finite-size corrections might be larger than for $\xi_1$ (as
happens in the $w<w_0$ regime).

The values of $\xi^{-1}_2$ at $w=w_0 = -1/4$ are displayed
in Table~\ref{table_inverse_xi2_sq}. We observe two different behaviors
depending on the parity of the width $L$. For even $L$ we find 
that the estimates for $x_{T,2}$ decrease as $L_{\rm min}$ increases.
Furthermore those estimates are not far away from those obtained for 
$x_{T,1}$ with odd $L$ (see Table~\ref{table_inverse_xi1_sq}).
For a fixed value of $L_{\rm min}$,
the estimates for $x_{T,2}$ (even $L$)
are systematically larger than
the corresponding ones for $x_{T,1}$ (odd $L$);
but it is unclear whether the limiting values coincide or not.\footnote{
  Please notice that if we merge the $\xi_1^{-1}$ data for odd $L$ (displayed 
  in Table~\protect\ref{table_inverse_xi1_sq}) with the $\xi_2^{-1}$ data 
  for even $L$ (displayed in Table~\protect\ref{table_inverse_xi2_sq}), we
  obtain a data set with notable even-odd effects that prevent us from 
  doing any joint fit. 
}
On the other hand, the estimates for $x_{T,2}$ for odd $L$ grows as
$L_{\rm min}$ increases, and we can conclude that $x_{T,2}\gtapprox 1.25$.  

As the finite-size effects in the quantities $\xi^{-1}_i(L)$ seem to grow
with the index $i$, and we have seen that these corrections are
important already for $i=2$ (at least for the widths considered here),
we did not try to analyze the correlation lengths $i \ge 3$.
However, it is worth noting the possibility that some of the
operators corresponding to $i \ge 3$ are
(like the ones corresponding to $i=1,2$)
more relevant than the operator with $x_T = 2$
that was conjectured (incorrectly, we now know)
by Saleur \cite{Saleur_90,Saleur_91}
to be the leading thermal operator [cf. \reff{def_xT_antiferro}].

%
%
\subsection{Triangular lattice: Estimates of critical points and
   phase boundary}
 \label{sec.res.tri1}

\begin{figure}
\centering
\includegraphics[width=400pt]{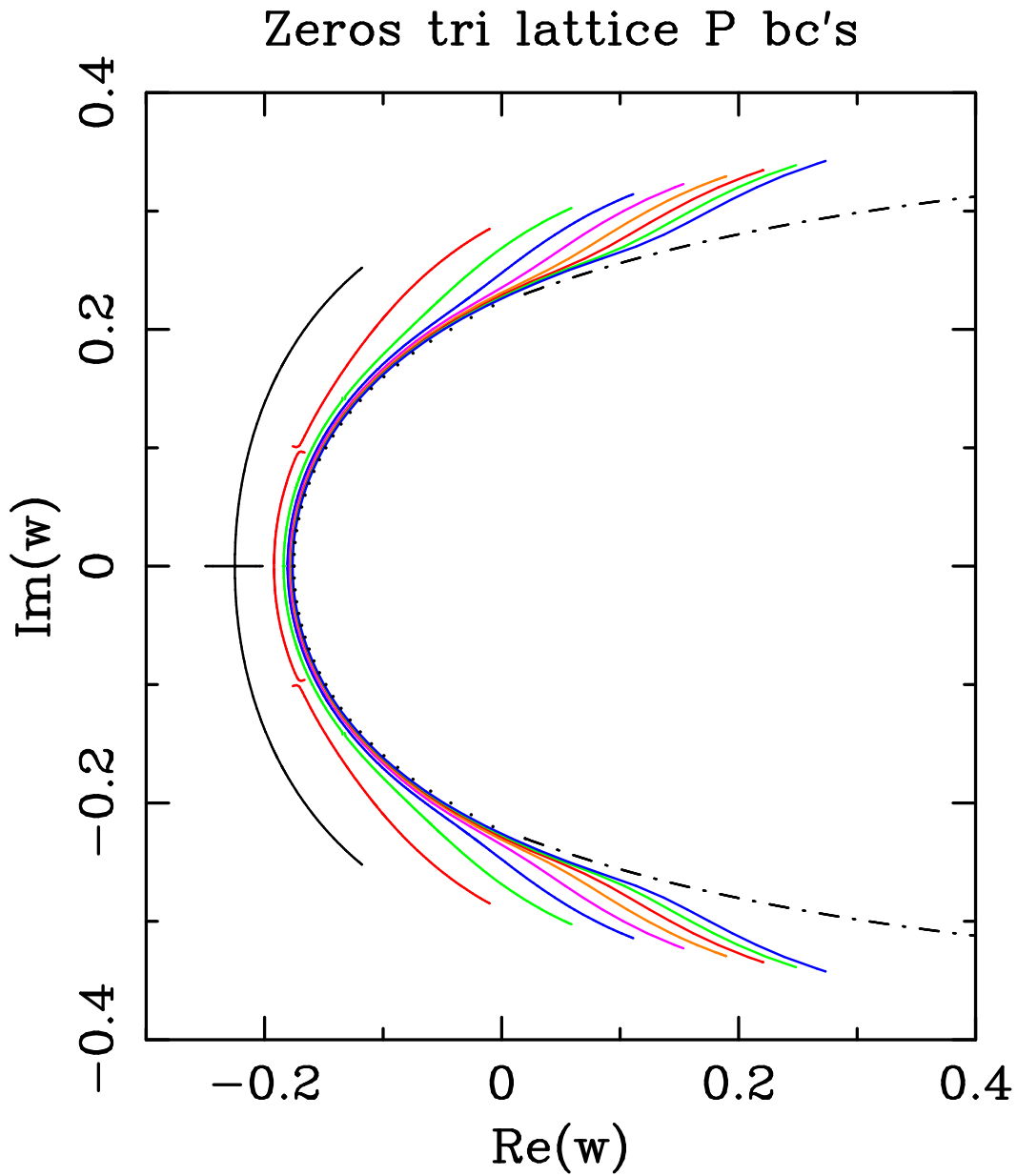}
\caption{\label{figure_tri_all}
Limiting curves for triangular-lattice strips of widths $L$: 2 (black),
3 (red), 4 (green), 5 (blue), 6 (pink), 7 (brown), 8 (red), and 9 (green).
As $L$ increases, the limiting curve moves towards the right.
We have also depicted our estimate for the $L \to\infty$ limiting curve
in the interval $-0.23 \ltapprox \imag w \ltapprox 0.23$ (black dots)
and our very rough estimate beyond this (black dotted-dashed curve).
}
\end{figure}

The triangular-lattice data that we wish to extrapolate
are collected in Table~\ref{table_tri}
and depicted graphically in Figure~\ref{figure_tri_all}.
The quantities $w_{0-}(L)$ and $w_{0+}(L)$ are defined only for even $L$,
so we must perform fits to the Ansatz \reff{def_ansatz_w0}
using $L=L_{\rm min},L_{\rm min}+2,L_{\rm min}+4$.
The quantity $w_{0Q}(L)$ is defined for all $L$;
and for the triangular lattice (contrary to what happened for the
square lattice) there do {\em not}\/ seem to be significant
even-odd oscillations.
We can therefore perform fits to \reff{def_ansatz_w0}
using $L=L_{\rm min},L_{\rm min}+1,L_{\rm min}+2$.

The results of these fits are displayed in Table~\ref{table_fits_tri}.
The estimates of $w_0$ are very well behaved
(particularly those based on $w_{0Q}$),
and from $L_{\rm min}=6,7,8$ we can estimate
\be
   w_0({\rm tri}) \;=\; -0.1753 \pm 0.0002  \;.
\ee
This value is far away from $-1/6 \approx -0.1667$,
and so rules out the naive conjecture that $w_0({\cal L}) = -1/r$
whenever ${\cal L}$ is a regular two-dimensional lattice
of coordination number $r$.
As in the square-lattice case, the exponent $\Delta$ seems to approach
$\Delta=1/\nu=2$ as $L_{\rm min}$ increases.
This fact agrees with the interpretation that there is a first-order phase
transition at $w=w_0$. In Table~\ref{table_fits_tri_bis} we present
the results of fitting the data to the modified Ansatz
\reff{def_ansatz_w0_bis}: the estimates converge slightly faster,
and we again find
\be
   w_0({\rm tri}) \;=\; -0.1753 \pm 0.0002  \;.
\ee

If we fit the width of the interval $[w_{0-}, w_{0+}]$ for even $L$
to the Ansatz $A L^{-\omega}$,
we find 
\be
   \omega_{\rm tri}  \;\approx\; 6.0   \;,
 \label{eq.omega.tri}
\ee
which is three times larger than $\omega_{\rm sq} = \Delta \approx 2$.
In other words, on the triangular lattice the interval $[w_{0-}, w_{0+}]$
is extremely narrow, and gets rapidly narrower as $L$ grows.
We do not know the reason for this different behavior on the two lattices.

Next let us extrapolate the endpoints $w_B(L)$,
handling separately their real and imaginary parts.
Once again, the real part of $w_B(L)$ seems to diverge as $L\to\infty$
like $c_1 \log L + c_2$. We have used the Ansatz \reff{def_ansatz_Re_wB}, and
the estimates for $c_1$ and $c_2$ as a function of $L_{\rm min}$
are shown in columns 2--3 of Table~\ref{table_fits_wB_tri}(a).
As for the square lattice, the estimates of $c_1$
are rising with $L_{\rm min}$ and not yet slowing down.
This suggests that the true value of $c_1$ is quite a bit larger
than the value $\approx 0.236$ observed at our largest $L_{\rm min}$.
If forced to guess, we might estimate
\begin{subeqnarray}
c_1 &=&  0.24 \pm 0.02 \\
c_2 &=& -0.27 \pm 0.04
 \label{eq.tri.c1c2}
\end{subeqnarray}

The relation with the $N=-1$ vector model \cite{Caracciolo_PRL} predicts that
\be
   \real w_B(L)  \;=\;  c_1 \log L \,+\, c_3 \log\log L \,+\, c_2
                 \,+\,  O\!\left( {\log\log L \over \log L} \right)
\label{def_ansatz_Re_wB_bis_tri}
\ee
with $c_1 = \sqrt{3}/(2\pi) \approx 0.275664$
and $c_3 = -1/(2\sqrt{3}\pi) \approx -0.091888$ \cite{Caracciolo_PRL}.
Indeed, the ratio 
\be
{c_1({\rm sq}) \over c_1({\rm tri}) } \;=\;
{c_3({\rm sq}) \over c_3({\rm tri}) } \;=\;
 \sqrt{3} \approx 1.732051
\label{prediction_ratio}
\ee
is characteristic of asymptotically free theories,
and simply comes from comparing the lattice free propagators
with their continuum limits
(see e.g.\ \cite[eq.~(5.7)]{Caracciolo_83}).

In the last column of Table~\ref{table_fits_wB_tri}(a),
we display the estimates of the parameter
$c_2$ obtained by performing the one-parameter fit to the Ansatz
\reff{def_ansatz_Re_wB_bis_tri}
with $c_1$ and $c_3$ fixed to their theoretical values
(and the correction terms $O(\log\log L/\log L)$ neglected).
Our estimate is
\be
c_2  \;=\;  -0.285 \pm 0.005
 \label{eq.tri.c2}
\ee
However, the very-slowly-varying nature of the expected correction terms
and the narrow range of $L$ values available here
make plausible that the true value for 
$c_2$ differs from the above estimate by many times our estimated error.

The imaginary part of $w_B(L)$, by contrast, seems
to converge to a finite value as $L\to\infty$:
the estimates using the Ansatz \reff{def_ansatz_wB}
are displayed in Table~\ref{table_fits_wB_tri}(b).
We find 
\be
   \imag w_B  \;=\;  0.394 \pm 0.004
 \label{eq.tri.ImwB}
\ee
with $\Delta = 0.63 \pm 0.05$.

Finally, we tried to extrapolate the whole limiting curve $\scrb$
to the infinite-volume limit, by the same methods used
for the square lattice (Section~\ref{sec.res.sq1}).
We began by extrapolating $\real w$
at the selected values $\imag w = 0.01,0.02, \ldots$,
using the Ansatz \reff{def_ansatz_wB} for $\real w$.
We obtained consistent estimates of $\real w$ for
$|\imag w| \ltapprox 0.23$ (see Table~\ref{table_limiting_curves}).
The results are shown by black dots in Figure~\ref{figure_tri_all}.
Next we attempted to complete this curve at larger $\real w$,
using once again the theoretical predictions from the
perturbative renormalization group at small $g = 1/w$.
Unfortunately, for the triangular lattice the three-loop and four-loop
computations have not yet been performed;
therefore, all we have available are the universal values
for the first two beta-function coefficients.
Using the relation $\beta_{\rm continuum} = \sqrt{3} \, \beta_{\rm tri}$
\cite[eq.~(5.7)]{Caracciolo_83}
to match normalizations, we obtain
\begin{subeqnarray}
   b_2  & = & \left. - w_0/\sqrt{3} \right| _{N=-1}
              \;=\; \sqrt{3}/(2\pi)  \\[2mm]
   b_3  & = & \left. \hphantom{-} w_1/3 \right| _{N=-1}
              \;=\; -1/(2\pi)^2
\end{subeqnarray}
and hence
\be
   A_1  \;=\;  b_3/b_2 \;=\; -1/(2\sqrt{3}\pi)
   \;.
\ee
Using the Ansatz \reff{def.exp.Ansatz} with $\beta_0 = 0.394$
and $B_3=B_4=0$
and imposing the first derivative at $\alpha=\infty$
along with the estimated value and derivative at $\alpha=0.0198$,
we obtain
\begin{subeqnarray}
   \alpha_0  & = &  -0.550842   \\
   B_1       & = &  -1/(2\sqrt{3}\pi) \;\approx\; -0.091888  \\
   B_2       & = &  -0.122843
\end{subeqnarray}
The corresponding curve \reff{def.exp.Ansatz}
is depicted as a black dotted-dashed curve in Figure~\ref{figure_tri_all}.

\subsection[Triangular lattice: Behavior of the free energy and its derivatives
   as $w \to w_0$]{Triangular lattice: Behavior of the free energy and
   its derivatives as \mbox{\protect\boldmath $w \to w_0$}}
 \label{sec.res.tri2a}

We have computed the real part of the strip free energy $\real f_L(w)$
for widths $2\leq L \leq 13$ and $w=-2,-1.9,\ldots,2$, as well as
$w=-0.175$. These results are displayed in Figure~\ref{figure_f_tri}.
The solid black line represents the extrapolation of the free energy to
$L=\infty$ using the Ansatz~\reff{eq.Ansatz.f.Delta}. We also show the
$[10,10]$ Pad\'e approximant to the small-$w$ series for this quantity
(dashed violet line), and the large-$w$ series \reff{def_large_w_series} 
through order $w^{-1}$ (dot-dashed pink curves).
Once again, the small-$w$ series expansion agrees very well with
the free-energy data in the regime $w>w_0$.
On the other hand, the large-$w$ expansion gives very accurately the
free energy in the whole regime $w\ltapprox w_0$.
The free energy has clearly a minimum at a value
close to the estimate $w_0({\rm tri})\approx -0.1753$.

The first derivative of the free energy clearly has a jump discontinuity
at $w_0$, as is seen on Figure~\ref{figure_E_tri}.
Again, we have made a blow-up picture (Figure~\ref{figure_Ec_tri})
in an effort to determine whether this discontinuity is finite or infinite.
As for the square-lattice case we find a clear parity effect: for odd 
$L$, $f''_L(w)$ has a jump discontinuity at $w=w_{0Q}$, while for
even $L$ we find divergences $f''_L(w) \sim |w - w_{0\pm}|^{-1/2}$ 
due to the square-root branch points $w=w_{0\pm}(L)$. 
Note that for the triangular lattice the intervals $[w_{0-},w_{0+}]$
for even $L$ are extremely narrow [cf.\ \reff{eq.omega.tri}];
this makes it much easier, compared to the square lattice,
to extract the limiting behavior as $L\to\infty$ with $w$ fixed,
as it suffices to exclude from consideration a very narrow region
surrounding the interval $[w_{0-},w_{0+}]$.
In fact, Figure~\ref{figure_Ec_tri} strongly suggests that the discontinuity
in $f'(w)$ is finite at $w=w_0$. For $w\downarrow w_0$, we observe that $f'$
is very well approximated by the $[10,10]$ Pad\'e approximant to the small-$w$
series for $f'$.
Thus, we can conclude that
$f'_+(w_0) \equiv \lim_{w \downarrow w_0} f'(w) \approx 3.389$
(this value comes from evaluating the Pad\'e approximant
 at $w=w_0 \approx -0.1753$).
On the other side, the curves $f'_L(w)$ for $w < w_{0-}$ (even $L$) or 
$w < w_{0Q}$ (odd $L$) stay rather close to (and slightly above) the first
derivative of the large-$w$ series \reff{def_large_w_series},
i.e.\ $f'(w) = 1/w - 1/(12w^2)$.  We conclude that
$f'_-(w_0) \equiv \lim_{w \uparrow w_0} f'(w) \approx -8.14$.\footnote{
   We have empirically found that for large values of $L$
   and for $w\uparrow w_0$, 
   the quantity $\delta f'(w)\equiv f'_L(w)-[1/w -1/(12w^2)]$
   behaves approximately linearly in $w$.
   In fact, a rough fit of our data for $L=9,10$ gives
   $\delta f'(w) \approx 0.727 + 2.580 w$.
   The value $f'_-(w_0)\approx -8.14$ comes from evaluating
   both the large-$w$ series and the fit for $\delta f'(w)$
   at $w=w_0\approx -0.1753$.
}
Thus, the jump in the internal energy is apparently finite and takes the
value $\Delta f'(w_0) \approx 11.53$.
The existence of a discontinuity in the first derivative of the free energy
implies that there is a first-order phase transition at $w=w_0$ also for the
triangular-lattice model.

Since $w_{0-}(L)$ and $w_{0+}(L)$ vary with $L$,
it is useful to plot $f'_L(w)$ versus
$w - w_{0-}(L)$ or $w - w_{0+}(L)$ in order to compare
the behavior at ``comparable'' values of $w$.
Such plots are shown in Figure~\ref{figure_Ec_tri_bisa}(a,b).
For both $w < w_0$ and $w > w_0$,
we see that the region of divergence grows rapidly narrower as $L$ grows,
so that at any {\em fixed}\/ value of $w - w_{0-}(L) < 0$
[resp.\ $w - w_{0+}(L) > 0$],
$f'_L(w)$ clearly tends to a limiting value around $-8.0$ (resp.\ $3.4$).
This plot confirms, in a most striking way,
our conclusions in the preceding paragraph.
The evidence for a finite limit of $f'_L(w)$
as $w \to w_0$ from either side is, in fact,
vastly more compelling for the triangular lattice
than for the square lattice.

We have also computed the second derivative of the free energy:
see Figure~\ref{figure_CH_tri} for an overview plot,
and Figure~\ref{figure_CHc_tri} for a blow-up plot near $w_0$
using a logarithmic vertical scale.
Once again,
for odd $L$, $f''_L(w)$ has a jump discontinuity at $w=w_{0Q}(L)$,
while for even $L$ we find divergences
$f''_L(w) \sim | w - w_{0\pm}(L)|^{-3/2}$
due to the square-root branch points at $w=w_{0\pm}(L)$.
However, the divergences of $f''_L(w)$ for $w \approx w_{0\pm}$
are extremely narrow, which allows us to extract with fair confidence
the limit $L\to\infty$ at $w$ fixed $\neq w_0$, as follows:

1) For $w$ slightly above $w_0$,
the specific heat $f''_L(w)$ stays very small
(and almost independent of $L$) until $w$ goes fairly close to $w_{0+}(L)$.
Moreover, the width of this ``divergence'' region
becomes smaller as $L$ grows.
This suggests that the infinite-volume specific heat $f''(w)$
remains bounded (and in fact) small as $w \downarrow w_0$,
in sharp contrast to what was observed for the square lattice.
Indeed, the behavior for $w\gtapprox w_0$ is well described for $L\geq 5$
by the $[10,10]$ Pad\'e approximant to the small-$w$ series,
from which we can conclude that
$f''_+(w_0) \equiv \lim_{w \downarrow w_0} f''(w) \approx -1.187$
(based on evaluating the Pad\'e approximant at $w=w_0\approx -0.1753$).

2) For $w$ slightly below $w_0$, by contrast,
we find large negative values for $f''(w)$,
although these large values do seem to remain finite as $w \uparrow w_0$.
Indeed, the data points for $w < w_{0-}$ (even $L\geq 6$) or 
$w < w_{0Q}$ (odd $L \geq 7$) remain very close to the second derivative 
of the large-$w$ expansion \reff{def_large_w_series},
namely $f''(w) = -1/w^2 + 1/(6w^3)$.
This suggests that $f''(w)$ does {\em not}\/ tend to $-\infty$
as $w\uparrow w_0$, but rather to the finite value
$f''_-(w_0) \equiv \lim_{w \uparrow w_0} f''(w) \approx -63.5$
(based on evaluating the large-$w$ series at $w=w_0\approx -0.1753$).

In Figure~\ref{figure_CHc_tri_bisa}(a,b) we plot $f''_L(w)$ versus
$w - w_{0-}(L)$ or $w - w_{0+}(L)$ in order to compare
the behavior at ``comparable'' values of $w$.
These plots confirm our conclusion that $f''_L(w)$ stays
almost independent of $L$ except in a narrow region around $w_{0\pm}(L)$,
which furthermore becomes narrower as $L$ grows,
so that the infinite-volume specific heat $f''(w)$
remains uniformly bounded as $w \to w_0$ from either side.

These findings on the behavior of the free-energy derivatives merit some
discussion. We have seen that, on the triangular lattice,
both $f'$ and $f''$ apparently remain finite as $w \to w_0$
from either side;
at $w_0$ these quantities simply jump from one finite value to another.
This behavior is very different from what was found for the square lattice
(Section~\ref{sec.res.sq2a}),
where $f''$ very likely diverges at $w_0$
(at least when approached from above),
and $f'$ may conceivably diverge as well.
At least for {\em fixed}\/ $w<w_0$ (i.e., inside the Berker--Kadanoff phase)
one would expect some universal behavior (as indeed happens
for the central charge and the thermal scaling dimensions $x_{T,i}$). 
So the discrepancy between the square and triangular lattices
is puzzling;  see Section~\ref{sec.discussion.free_energy} for
further discussion.

\subsection{Triangular lattice: Estimates of the free energy and
   central charge}  \label{sec.res.tri2b}

The real part of the triangular-lattice free energy at $w=\pm 1$
is monotonic in $L$, at least as far as we have been able to compute it
(see Table~\ref{table_f_tri}).
If this property continues to hold for larger $L$, we can conclude that
\begin{subeqnarray}
\real f({\rm tri},w=1) &\geq& \real f_{L=15}({\rm tri},w=1)\approx 1.7010224200
\slabel{conjecture_N_tri_w=1} \\
\real f({\rm tri},w=-1)&\geq& \real f_{L=13}({\rm tri},w=-1)\approx 1.5273746626
\slabel{conjecture_N_tri_w=-1}
\label{conjecture_N_tri}
\end{subeqnarray}
The lower bound \reff{conjecture_N_tri_w=1} falls between the rigorous
lower and upper bounds \reff{Noy_bounds_tri}.

If we fit the finite-$L$ data at $w = \pm 1$
to the Ansatz~\reff{def_ansatz0_fm},
we get the results displayed in Table~\ref{table_f_fits_tri}.
At $w=-1$, the estimates of the exponent $\Delta$ are very close
to $\Delta=2$ and apparently tending to it,
exactly as expected for the Berker--Kadanoff phase.
At $w=1$, the estimated exponent $\Delta$
is slightly larger than 2, and slowly rising;
this is compatible with the idea that $w=1$ is a non-critical point
which has, however, a reasonably large correlation length,
so that the system is strongly affected by crossover from
the critical point at $w=+\infty$.\footnote{
  In Figure~\protect\ref{figure_xi1_tri}(a) we observe that the extrapolated
  value for the inverse correlation length at $w=1$ is very close to zero.
  In fact, our best estimate is a small negative number,
  $\xi^{-1}_1(w=1)\approx -0.01$.
}
Our preferred estimates for $\real f({\rm tri},w)$ are:
\begin{subeqnarray}
\real f({\rm tri},w=1)    &=&  1.70255 \pm 0.00010 \\
\real f({\rm tri},w=-1)   &=&  1.53413 \pm 0.00007
\end{subeqnarray}
These values are indeed larger than the conjectured lower bounds
\reff{conjecture_N_tri}; and the estimate for $\real f({\rm tri},w=1)$
is smaller than the rigorous upper bound \reff{Noy_bounds_tri}.

Finally, we can estimate the central charge of the model as a function of
$w$, using the same method as for the square lattice.
In this case, the Ansatz~\reff{def_ansatz_f} has to be modified
to include the geometrical factor \reff{def_G}:
\be
f(L_{\rm P} \times \infty_F,w)   \;=\;
   f(w) +   {\sqrt{3} \pi c(w) \over 12} L^{-2} + \ldots  \;.
\label{def_ansatz_f_tri}
\ee
We began by fitting the data to the Ansatz \reff{def_ansatz_f_tri}
with no higher-order correction terms. The results are displayed in
Figure~\ref{figure_c_tri}(a). There are two clear regimes separated by
$w_0({\rm tri})\approx-0.1753$, and the values of the central charge
agree qualitatively with those found for the square lattice.
In Figure~\ref{figure_c_tri}(b) we show the same plot versus $1/w$;
the behavior near $1/w=0$ is once again
in agreement with the prediction \cite{Caracciolo_PRL}
that the fixed point at $1/w = 0$ has central charge $c=-2$
and is marginally repulsive for $g \equiv 1/w > 0$
and marginally attractive for $g < 0$.

In Figure~\ref{figure_c_tri}(c) we show the estimates for
the central charge $c(w)$ coming from the
modified Ansatz in which an $L^{-4}$ term is included.
The estimates seem to converge faster in the region $w<w_0$,
but they are still far from the expected value of $c=-2$.
In Figure~\ref{figure_c_tri}(d) we show the corresponding plot
versus $1/w$. Finally, we show in Figure~\ref{figure_c_tri2} the 
estimates for $c(w)$ for $w<w_0$ using the improved Ansatz 
\reff{def_ansatz_f_tris}. As for the square-lattice case, these results
agree better with the expected result $c=-2$.

For $w=-1$ we have extended the above analysis to include finite-size data
up to $L=13$ (see Tables~\ref{table_f_fits_c_tri_w=-1} 
and~\ref{table_f_fits_c_tri_w=-1_bis}). As for the square-lattice case,
we find a good agreement with the predicted value $c=-2$ only for the
improved Ansatz \reff{def_ansatz_f_tris}: for $L_{\rm min}=10$ we 
find $c(w=-1)\approx -2.077$.  
Finally, we obtain a more stable and precise estimate for $f({\rm tri},w=-1)$:
\be
   f({\rm tri},w=-1)  \;=\; 1.534166 \pm 0.000008   \;.
\ee

\subsection{Triangular lattice: Estimates of the thermal scaling dimensions}
   \label{sec.res.tri3}

As for the square lattice, we began by obtaining the
inverse correlation lengths $\xi_i^{-1}(w;L)$
for finite-width strips using the definition \reff{def_inverse_xi_2}.
The first and second inverse correlation lengths for widths $L\leq 10$ are
plotted in Figures~\ref{figure_xi1_tri}(a) and \ref{figure_xi2_tri}(b),
respectively.

The scaling exponent $x_{T,i}$ is obtained by fitting the finite-width data
to the Ansatz
\be
\xi^{-1}_i(w;L)  \;=\;  \xi_i^{-1}(w) + {\sqrt{3} \pi x_{T,i}(w) \over L}  \;,
\label{def_xi_critical_tri}
\ee
which includes the geometrical factor \reff{def_G}.
The applicability of this Ansatz is subject to the same comments made
in Section~\ref{sec.res.sq3}.
The estimates of the infinite-volume correlation length $\xi_i^{-1}$
are depicted in Figures~\ref{figure_xi1_tri}(a) and~\ref{figure_xi2_tri}(b)
by black solid dots.

Let us begin by discussing the first correlation length $\xi_1$.
As expected, we found clearly different behavior in the regimes
$w < w_0$ and $w>w_0$.

For $w < w_0$, the estimated infinite-volume inverse correlation length
is compatible with the value $\xi_1^{-1}=0$.
Indeed, we find estimates that are small negative numbers,
whose absolute value decreases as we increase $L_{\rm min}$;
we therefore expect that they will converge to zero as $L\to\infty$.
We therefore repeated the fit, fixing $\xi_1^{-1}=0$.
The results for $x_{T,1}$ are displayed in Figure~\ref{figure_x1_tri}(a).
Inside the Berker--Kadanoff phase $w < w_0$, we again find that
the scaling dimension is close to Saleur's prediction $x_{T,1}=2$.
The scaling dimension at the transition point $w=w_0$ is difficult to obtain,
as we do not know the exact value of $w_0$
and the correction-to-scaling effects are very large near this point.

For $w>w_0$, by contrast, we find a clear non-zero estimate for $\xi_1^{-1}$,
at least up to $w \approx 0.5$.
For $w\gtapprox 0.6$, our estimates for $\xi_1^{-1}$
are rather small ($\xi_1^{-1}\ltapprox 0.07$); but as before, we believe this
indicates only that $\xi_1$ has become comparable to the strip widths
and that we are unable to handle the strong corrections to scaling
in this regime.
(Indeed, we have not found any evidence of a phase transition near
$w\approx 0.5$.)

For $\xi_2$ the overall picture is very similar, although the
finite-size corrections are larger (as happened in the square-lattice case).
For $w\ltapprox -0.5$, the data are compatible with $\xi^{-1}_2=0$:
indeed, our estimates for $\xi^{-1}_2$ are small negative numbers,
whose absolute value decreases as we increase $L_{\rm min}$.
In the interval $-0.5 \ltapprox w < w_0$,
we find strong correction-to-scaling effects;
but again we expect that the correlation length will diverge,
as it does for $\xi_1$.
We therefore repeated the fit, fixing $\xi_2^{-1}=0$.
The values obtained for $x_{T,2}$ are displayed in
Figure~\ref{figure_x2_tri}(b). In this regime the estimates are close
to $x_{T,2}=4$, as for the square lattice.
In the regime $w>w_0$, the behavior is similar to that of $\xi_1$:
$\xi^{-1}_2(L)$ tends to a finite value, which goes to zero as $w\to\infty$.

\section{Discussion}  \label{sec.discussion}

\subsection[Behavior of dominant-eigenvalue-crossing curves
   ${\cal B}$]{Behavior of dominant-eigenvalue-crossing curves
            {\boldmath ${\cal B}$}}

We have computed the symbolic transfer matrices for square- and
triangular-lattice strips of widths $2\leq L \leq 10$ with
cylindrical boundary conditions. From these matrices one can
compute (a) the generating polynomial of spanning forests $F_G(w)$ for a
cylindrical strip of width $L$ and arbitrary length $n$,
along with its complex zeros,
and (b) the accumulation points of such zeros in the limit $n\to\infty$.
According to the Beraha--Kahane--Weiss theorem (Section~\ref{sec.setup.4}),
these zeros accumulate at certain points
(the isolated limiting points, which seem not to exist in the models
 studied here)
and along certain curves (the limiting curves ${\cal B}$).
By studying the behavior of these limiting curves as
a function of the strip width $L\to\infty$, we hope to shed light on the
thermodynamic limit of the model, in which $L,n\to\infty$.

For strips of widths $L \leq 6$,
we were able to compute the limiting curves ${\cal B}$
using the resultant method \cite{transfer1};
this method allows a complete determination of ${\cal B}$,
including all its endpoints.
For widths $7\leq L \leq 10$, we used the direct-search method,
which gives only lower bounds on the numbers of endpoints,
as it could miss some.
For $L=7,8$ we were able to obtain a fairly good description of ${\cal B}$,
but for $L\geq 9$ we were only able to obtain the points
at the discrete set $|\imag w| = 0, 0.01,0.02$, etc.
These limiting curves ${\cal B}$ are shown
in Figures~\ref{figure_sq_all} and \ref{figure_tri_all}
for the square and triangular lattices, respectively,
and their characteristics are summarized in
Tables~\ref{table_sq} and~\ref{table_tri}.
In all cases we can verify the identity
\begin{eqnarray}
\hbox{\rm endpoints} &=&
 \hbox{\rm (2 $\times$ components) + (2 $\times$ double points) + (T points)}
 \nonumber \\
 & & \qquad -\, \hbox{\rm (2 $\times$ enclosed regions)}   \;,
\end{eqnarray}
which can be derived by simple topological/graph-theoretic arguments.
Furthermore, for all but one of the strips considered in this paper,
the limiting curves ${\cal B}$ do not contain any enclosed region or T point
(the only exception is the $L=4$ square-lattice strip,
 which contains two enclosed regions and a single T point,
 as shown in Figure~\ref{figure_sq_4P_zoom}).
Finally, the number of double points $\#Q$ is given simply by
\be
\hbox{\rm \#Q} \;=\; \mod(L-1,2)  \;=\;
    \cases{1  & if $L$ is even  \cr
           0  & if $L$ is odd   \cr
          }
\label{conjectures_B2}
\ee
Therefore, the topological structure of the sets ${\cal B}$
can be summarized very simply:
for odd $L$, the set ${\cal B}$ consists of
a collection of arcs separated by extremely small gaps
(these gaps are in fact nearly invisible
 in Figures~\ref{figure_sq_all} and \ref{figure_tri_all}
 for $L \gtapprox 5$);
for even $L$, it is the same,
except that the component touching the real axis
consists of a complex arc and a small segment of the real axis
intersecting at a double point at $w=w_{0Q}$.

Furthermore, the number of endpoints (\#E) and
the number of components (\#C) seem to grow with the strip width
in a very regular way.
In fact, for all the strips considered in this paper
except the $L=4$ square lattice, we find
\begin{subeqnarray}
\hbox{\rm \#E} &=& 2 L \\
\hbox{\rm \#C} &=& L - \mod(L-1,2)
\label{conjectures_B1}
\end{subeqnarray}
We conjecture that \reff{conjectures_B2} and \reff{conjectures_B1}
hold for larger $L$ as well.
If this conjecture is indeed true, it suggests that the endpoints
(whose number grows linearly with $L$) will become dense in ${\cal B}$.
This may imply in turn that the
infinite-volume limiting curve ${\cal B}_\infty$ will be
a {\em natural boundary}\/ for the infinite-volume free energy $f(w)$,
i.e., a boundary through which analytic continuation of $f(w)$ is impossible.

Finally, we have found that the endpoint $w_B(L)$ behaves with $L$
in a very regular way:  namely, its real part tends logarithmically
to $+\infty$
[cf.\ \reff{def_ansatz_Re_wB}--\reff{def_ansatz_Re_wB_c2}
 and \reff{eq.tri.c1c2}--\reff{eq.tri.c2}]
while its imaginary part tends to a finite value
[cf.\ \reff{def_ansatz_Im_wB}--\reff{eq.wB.sq.v2} and \reff{eq.tri.ImwB}].
This suggests that the infinite-volume limiting curve $\scrb_\infty$
will extend to infinity.
This behavior is consistent with the idea of an asymptotically free
theory defined around the critical point at $w=+\infty$ \cite{Caracciolo_PRL}.

\subsection[Conformal field theory of the $q\to 0$ limit]
    {Conformal field theory of the \mbox{\protect\boldmath $q \to 0$} limit}

In this paper we have obtained several new pieces of information
concerning the phase diagram of the $q\to0$ limit of the Potts model
on the square and triangular lattices,
and on the interpretation of this model in terms of
conformal field theory (CFT).
This information provides evidence concerning the behavior
of the Potts model also at small positive $q$,
confirming some conjectures of Saleur \cite{Saleur_90,Saleur_91}
and refuting others.

We have found that the phase diagram for the two lattices is similar:
for $w>w_0$, the model is non-critical with exponential decay of correlations;
for $w<w_0$, the model is critical and can be interpreted as a
Berker--Kadanoff phase.  Finally, at $w=w_0$ the model displays a 
{\em first-order critical point}\/ \cite{Fisher_82},
that is, a first-order transition point for
which the correlation length diverges as $w\downarrow w_0$.
The transition at $w=w_0$ corresponds to the $q\to0$ limit of the 
antiferromagnetic transition curve of the Potts model.

It is well-known that the Potts model in the
antiferromagnetic ($-1 \le v < 0$) and unphysical ($v < -1$) regimes
may display non-universal (i.e.\ lattice-dependent) behavior;
this is true in particular for the zero-temperature antiferromagnetic
(chromatic polynomial) limit $v=-1$ \cite{transfer1,transfer2,transfer3}.
Since both the Berker--Kadanoff phase and the $w=w_0$ transition point
lie in the antiferromagnetic regime
$w < 0$ of the spanning-forest model,
it is therefore interesting to assess the extent (if any)
to which they display universality.

For the Berker--Kadanoff phase we have indeed found a number of universal
features.  For both the square- and triangular-lattice models,
the central charge is found to be consistent with the $q\to0$ limit
of Saleur's prediction \reff{def_c_BK} for the Berker--Kadanoff phase:
\be
c(\mbox{Berker--Kadanoff})  \;=\;  -2 \;.
 \label{eq.BK.c}
\ee
Furthermore, the first two thermal critical exponents appear to be given by
\begin{subeqnarray}
 x_{T,1} &=& 2  \slabel{subeq.xT1} \\
 x_{T,2} &=& 4  \slabel{subeq.xT2}
\end{subeqnarray}
for both lattices.
The first exponent agrees with the $q\to0$ limit of Saleur's prediction
\reff{def_xT_BK}.
The second exponent can presumably be derived from
a more detailed analysis of the free fermion theory
that is thought to govern the $w\to-\infty$ limit.

It is interesting that the results \reff{eq.BK.c} and \reff{subeq.xT1}
are identical to what is predicted \cite{Caracciolo_PRL}
for the ferromagnetic fixed point at $w = +\infty$.
Indeed, all our results are consistent with the idea \cite{Caracciolo_PRL}
that the fixed points lying at $w=\pm\infty$ are {\em identical}\/
and are given by a theory of a pair of free scalar fermions,
with central charge $c=-2$
and marginal first thermal exponent $x_{T,1} = 2$.
Moreover, renormalization-group computations in the four-fermion model
describing spanning forests at $g \equiv 1/w \neq 0$ \cite{Caracciolo_PRL}
show that the $g=0$ fixed point is marginally repulsive for $g > 0$
and marginally attractive for $g < 0$.
Therefore, the Berker--Kadanoff phase at $-\infty < w < w_0$
renormalizes onto this free fermion theory and hence has $c=-2$,
while the phase at $w_0 < w < +\infty$
is noncritical (hence $c=0$) and obeys asymptotic freedom as $w \to +\infty$.
More generally, there is a separatrix in the complex $g$-plane
separating these two regimes:  it is given by the curve ${\cal B}_\infty$.
In Figures~\ref{figure_sq_all_1overw} and \ref{figure_tri_all_1overw}
we have depicted the relevant phase diagrams,
by mapping Figures~\ref{figure_sq_all} and \ref{figure_tri_all}
from the $w$-plane to the $g$-plane.
Our best estimate for the separatrix ${\cal B}_\infty$ is given
by the black dots and, more roughly, by the black dotted-dashed curve,
which were obtained by the methods described in
Sections~\ref{sec.res.sq1} and \ref{sec.res.tri1}.
This phase diagram in the complex $g$-plane
(where $g$ is the {\em bare}\/ coupling)
is typical of what is expected for an asymptotically free theory,
whose renormalization-group flow satisfies
\be
   {dg \over dl} \;=\;  b_2 g^2 \,+\, b_3 g^3 \,+\, \ldots
 \label{eq.RGflow}
\ee
with $b_2 > 0$.
Points lying inside the cardioid ${\cal B}_\infty$
are attracted as $l \to +\infty$ to the fixed point at $g=0$,
and hence belong to the complex Berker--Kadanoff phase.
Points lying outside the cardioid ${\cal B}_\infty$
flow away from the fixed point at $g=0$,
and hence belong to a different phase (i.e.\ the noncritical phase).

Let us now turn to the $w=w_0$ transition point, which corresponds to
the $q\to0$ limit of the antiferromagnetic transition curve
and gives information about the probable behavior of this curve
at small positive $q$.
For the square lattice, we have strong evidence that the
correlation length $\xi(w)$ diverges as $w \downarrow w_0$,
so that this transition point is indeed a critical point.
For the triangular lattice, however, the analysis of this point
is difficult, since the exact value of $w_0$ is unknown
and we know fewer terms in the small-$w$ expansion.
The following pieces of information are therefore based on
the square-lattice case only:

1) For even $L$ we find a central charge that is consistent with
the $q\to0$ limit of Saleur's prediction \reff{def_c_antiferro}
for the antiferromagnetic critical curve:
\be
c \;=\; -1 \;.
\ee
For odd $L$, we find [cf.\ \reff{conjecture_c_sq_w=-0.25}]
that the effective central charge is consistent with 
\be 
c_{\rm eff}  \;=\;  1/2  \;,
\ee
which we explain by conjecturing the existence of a twist operator
with conformal weight
\be
h_{\rm twist} \;=\;  -1/16
\ee
[cf.~\reff{h_twist0}--\reff{h_twist}].
The role of boundary-condition-changing operators in the type of CFT proposed
by Saleur \cite{Saleur_90,Saleur_91} 
is currently under study \cite{Saleur_04}.

2) Our data at $w > w_0$ and even $L$ are consistent with the following
behavior as $w \downarrow w_0 = -1/4$ of the
infinite-volume correlation length $\xi(w)$
and the infinite-volume free energy $f(w)$:
\begin{subeqnarray}
   \xi(w)  & \sim &  (w-w_0)^{-1/2}   \\[3mm]
   f''(w)  & \sim &  (w-w_0)^{-1}     \\
   f'(w)   & \sim &  \log(w-w_0)      \\
   f(w)    & \sim &  (w-w_0) \log(w-w_0) \;.
\end{subeqnarray}

3) We find that the dominant thermal eigenvalue is consistent with
\be
x_{T,1} \;=\; 0 
\ee
and hence $\nu=1/2$.
This result is in clear {\em disagreement}\/
with Saleur's prediction $x_{T,1}=2$
arising from the $q\to 0$ limit of his conjecture
\reff{def_xT_antiferro}/\reff{def_nu_antiferro};
it indicates that the thermal operators along
the antiferromagnetic critical curve need to be restudied \cite{JSS_RGflow}.
Our data for the second thermal eigenvalue indicates that
\be
x_{T,2} \;\gtapprox\; 1.25  \;.
\ee

An important open question is to obtain the same quantities for the 
triangular-lattice model and compare the to the above ones. This will
provide a direct check of universality for these two models.

\subsection{Behavior of the free energy and its derivatives} 
  \label{sec.discussion.free_energy}

We have studied the behavior of the free energy and its derivatives
near the transition point $w=w_0$ using two complementary methods:
extrapolation of small-$w$ series expansions
(Appendix~\ref{sec_series_analysis})
and finite-size analysis of the transfer-matrix data
(Sections~\ref{sec.res.sq2a} and \ref{sec.res.tri2a}).
 
Series expansions give valuable information concerning the behavior
as $w\downarrow w_0$, directly in the infinite-volume limit.
Our results show that for the square lattice $f''$ diverges
as $w\downarrow w_0$ with an exponent $\alpha \approx 1$,
while $f'$ is most likely finite
(but we cannot rule out a very weak divergence).
For the triangular lattice, 
both $f''$ and $f'$ appear to take finite values when $w\downarrow w_0$. 
The finite-size data are consistent with these conclusions.

For the regime $w\uparrow w_0$, by contrast,
only the finite-size data give useful information. 
For the square lattice we have \emph{weak} evidence for a diverging $f''$. 
For the triangular lattice, we have clear evidence that
$f''$ and $f'$ stay finite as $w$ approaches $w_0$ from below. 

These puzzling differences of behavior between the square and
triangular lattices have implications for universality;
and conversely, the sundry (and more or less plausible)
universality conjectures have implications for the interpretation
of our numerical results.
On the one hand, the analytical results of \cite{Caracciolo_PRL}
show that, at least for the spanning-forest model
(i.e., the limit $q,v\to 0$ with $w=v/q$ fixed of the Potts model),
the behavior inside the Berker--Kadanoff phase is universal:
more precisely, at each {\em fixed}\/ $w < w_0$,
the spanning-forest model is simply the $c=-2$ theory
of a pair of free scalar fermions, perturbed by a four-fermion operator
that is (in this phase) marginally irrelevant.
On the other hand, as noted in the Introduction,
Saleur \cite[p.~669]{Saleur_90} expects universality not only for
the Berker--Kadanoff phase, but also for the critical theories forming its
upper and lower boundaries.
The validity of this latter conjecture is, however, an open question:
in particular, we do not yet understand very well
the new phase-transition curve that we have found \cite{JSS_RGflow}
in the triangular-lattice Potts model, so we are unable to say for sure
whether it belongs to the same universality class
as the square-lattice antiferromagnetic critical curve,
in the sense of having the same central charge $c(q)$ and
the same critical exponents $x_{T1}(q),x_{T2}(q)$, etc.

Now, the behavior as $w \to w_0$ is controlled by the theory {\em at}\/ $w_0$.
Therefore, if universality does in fact hold (at least in the limit $q \to 0$)
for the upper boundary of the Berker--Kadanoff phase,
then (barring subtleties) the square and triangular lattices
should exhibit identical behavior as $w \to w_0$.
In particular, our clear results on the triangular lattice
($f''$ and $f'$ staying finite as $w\uparrow w_0$)
would suggest that the same behavior ought to occur for the square lattice;
if so, our contrary indications for the square lattice
would be an artifact of too-small strip widths $L$.
Likewise, our clear results on the square lattice
($f''$ diverging as $w \downarrow w_0$)
would suggest that the same behavior ought to occur for the triangular lattice;
if so, our contrary indications for the triangular lattice
would be an artifact of too-short small-$w$ series.

On the other hand, if universality does not hold for the theory at $w_0$,
then we are unable at present to draw any firm conclusion
about the behavior of $f''$ as $w\uparrow w_0$ on the square lattice,
or as $w \downarrow w_0$ on the triangular lattice.

\appendix

%
%
\section{Series analysis} \label{sec_series_analysis}

In this appendix we shall analyze the small-$w$ series expansions
displayed in Table~\ref{table_series}. For simplicity, we shall denote
the series expansions for the free energy and its derivatives as follows:
\begin{subeqnarray}
f(w)   &=& \sum\limits_{k=1}^\infty f_k w^k 
           \sim (w-w_0)^{2-\alpha} \slabel{F_series} \\
f'(w)  &=& \sum\limits_{k=0}^\infty e_k \, w^k  =
           \sum\limits_{k=0}^\infty (k+1) f_{k+1} w^k 
           \sim (w-w_0)^{1-\alpha} \slabel{E_series}\\
f''(w) &=& \sum\limits_{k=0}^\infty c_k w^k =
           \sum\limits_{k=0}^\infty (k+1)(k+2) f_{k+2} w^k
           \sim (w-w_0)^{-\alpha} \slabel{CH_series}
           \label{all_series}
\end{subeqnarray}
The rightmost side of each equation in \reff{all_series} shows the expected 
leading asymptotic behavior for $w$ near $w_0$
(assuming, for simplicity, that no multiplicative logarithms occur). 
We have used the standard notation $\alpha$ for the
critical exponent associated to the ``specific heat'' $f''(w)$.

Let us start with the square-lattice case, which is the one that leads
to the most interesting results, both because the series is twice as long
as for the triangular lattice
and because the putatively exact critical point $w_0 = -1/4$ is known.
The sign of the coefficients of the
free-energy expansion has a clear alternating pattern (at least up to $k=47$):
$f_k \sim (-1)^{k+1}$.
This behavior is a clear sign that the singularity nearest to the origin
lies on the negative $w$-axis.

A first (and very rough) approach can be obtained using the ratio method
\cite[and references therein]{Guttmann_book}. Let us suppose that
$F(w)$ is a function for which we can obtain the series expansion
$F(w) = \sum_{k=1}^\infty f_k w^k$, and whose behavior close to $w=w_0$ is
of the form
\be
F(w) \; \sim \; A \left( 1 - {w \over w_0} \right)^\lambda + B \; ,
\ee
Estimates of the critical point $w_0$ and the critical exponent $\lambda$
can be obtained from the series coefficients $\{f_k\}$ by computing the
ratios
\be
r_k \;=\; { f_k \over f_{k-1} }  \;.
\label{def_ratios_rn}
\ee
In absence of any competing singularity, a plot of $r_k$ versus $1/k$
is expected to be a straight line taking the value $1/w_0$ at $1/k=0$
and having slope $-(\lambda+1)/w_0$.
In Figure~\ref{figure_ratio_sq}(a) we show the ratio $r_k$ versus $1/k$
for the free energy and its first and second derivatives. In addition to
the expected linear behavior in $1/k$, we find a strongly oscillatory behavior 
that makes the analysis rather difficult.\footnote{
  It is not clear from Figure~\protect\ref{figure_ratio_sq}(a) what is the 
  period of these oscillations (if indeed they are periodic at all),
  but it is clear that the period is not 2.
} 
It is apparent from Figure~\ref{figure_ratio_sq}(a) that the limit of the
three sequences as $1/k\to 0$ is close to the expected value 
$w_0({\rm sq})=-1/4$. As a matter of fact, the value of $w_0$ can be estimated
from the sequence \cite{Guttmann_book} 
\be
(k+\epsilon) r_k \,-\, (k+\epsilon-1) r_{k-1}  \;=\;  {1\over w_0} \left[ 
  1 \,+\, O(1/k^2) \right]
\ee
where $\epsilon$ is any small $k$-shift. The oscillations are smaller for these
new sequences, and we conclude that $w_0= -0.250(1)$. We can therefore assume
that $w_0=-1/4$ exactly, and try to estimate the exponent $\lambda$
using the sequence \cite{Guttmann_book}
\be
\lambda_k  \;\equiv\;  k - 1 - k w_0 r_k  \;=\;
    \lambda \left[ 1 \,+\, O(1/k) \right]
\ee
with $w_0$ fixed to its theoretical value.
The estimates $\lambda_k$ as a function of $1/k$ for $f$, $f'$ and $f''$ 
are displayed in Figure~\ref{figure_ratio_sq}(b). Again, we observe 
oscillations in addition to the expected linear behavior in $1/k$.  
Rough estimates of $\lambda$ can be obtained from the corresponding sequences
$\{\lambda_k\}$: 
\begin{subeqnarray}
  2-\alpha &=&  1.10(8) \\ 
  1-\alpha &=&  0.10(3) \\
    \alpha &=&  0.90(3) \slabel{eqA.6c}
\end{subeqnarray}
These results are close to the expected value $\alpha=1$
for a first-order critical point \cite{Fisher_82},
although the error bars cannot be taken at face value.  

A more quantitative study can be performed using differential approximants
\cite[and references therein]{Guttmann_book}. The $K$-th order differential
approximant to a function $F(w)$ is built as follows: we choose polynomials
$Q_0,Q_1,\ldots,Q_K$ and $P$ of degrees $N_0,N_1,\ldots,N_K$ and $L$,
respectively, so that the solution $\widetilde{F}$
of the inhomogeneous linear differential equation
\be
\sum\limits_{j=0}^K Q_j(w) \, \left( w \, {d \over dw} \right) ^{\! j}
                           \widetilde{F}(w) \; = \; P(w)
\ee
agrees with the first coefficients of the series
$F(w)=\sum_{k=0}^N f_k w^k$.
The resulting equations can be solved if the number of unknown
coefficients in the polynomials is smaller than the order $N$
of the available series expansion for $F$.
The singularities of the function $\widetilde{F}(w)$
are located at the zeros $\{w_\ell\}$ of the polynomial $Q_K$ (along with
$w=0$ and $w=\pm\infty$). The critical exponent associated to a {\em simple}\/
zero $w_\ell$ of $Q_K$ is given by 
\be
\lambda_\ell \;=\;  K - 1 - {Q_{K-1}(w_\ell) \over w_\ell \, Q'_K(w_\ell) }
   \;.
\ee
When $w_\ell$ is a multiple zero of $Q_K$, this formula should 
be modified;  details can be found in reference \cite{Guttmann_book}.

In practice, we have used a modified version of the program {\tt newgrqd.f}
described in reference~\cite{Guttmann_book} to obtain the differential
approximants. Our program uses {\sc Mathematica} to obtain the polynomials
$Q_j$ {\em exactly}\/ (i.e., with exact rational coefficients),
and then MPSolve \cite{Bini_package,Bini-Fiorentino}
to compute the $N_K$ zeros of $Q_K$ to arbitrarily high precision
(100 digits in our case).
We have computed all the differential approximants of first and second order
(i.e., $K=1,2$) satisfying $|N_i - N_j| \leq 1$ and using at least 36 
coefficients of the corresponding series.
For each zero $w_\ell$ of $Q_K$, we have computed the corresponding critical 
exponents $\lambda_\ell$. Our procedure has the advantage,
over the double-precision Fortran program {\tt newgrqd.f},
that roundoff errors are under control.
We have checked the accuracy of our results 
with the help of an independent program by Y.~Chan, A.J.~Guttmann and 
A.~Rechnitzer written in {\sc C++}. 

Once all the approximants have been computed,
we need to discard the {\em defective}\/ ones.
We consider an approximant to be non-defective if
there is a real zero of $Q_K$ sufficiently near to
the expected value $w_0({\rm sq}) = -1/4$
and this zero is sufficiently well separated from all other zeros of $Q_K$.
In practice, we asked that the zero satisfy
$-0.251 \leq w_\ell \leq -0.249$ for $f''$, 
$-0.256 \leq w_\ell \leq -0.248$ for $f'$, and 
$-0.257 \leq w_\ell \leq -0.247$ for $f$,
and that no other singularity should appear in the region 
$-0.3 \le \real w \le 0$, $-0.05 \le \imag w \le 0.05$.
This is essentially what the program {\tt tabul.f}
(described in reference~\cite{Guttmann_book}) does. 

The non-defective approximants for the square-lattice spanning-forest 
free energy and its first and second derivatives are displayed in 
Figures~\ref{figure_DA_sq}(a)--(c), respectively. It is interesting to note 
that the estimates are not scattered uniformly over the corresponding plots; 
rather, they tend to accumulate along certain curves.  Unbiased estimates for
$w_0$ and $\lambda$ are obtained by averaging over the data points; 
crude estimates of their precision can be obtained by defining
the error bars to be one standard deviation of the corresponding
data distribution.  (But we emphasize that these are {\em not}\/ true
statistical errors as in a Monte Carlo simulation,
and the error bars should not be taken too seriously.)

As we know the exact value of $w_0({\rm sq})=-1/4$, we can also obtain biased 
estimates of the critical exponents $\lambda$. The idea is to construct
differential approximants for which $w_0$ is an {\em exact} simple zero of 
the polynomial $Q_K$. We have analyzed these approximants following the same
criteria as for the non-biased ones. 

Let us begin with the second derivative of the free energy, $f''(w)$.
The data points in Figure~\ref{figure_DA_sq}(c) fall over a 
narrow interval in both axes, and we see no significant differences
between the first- and second-order approximants.
Our preferred unbiased estimates come from merging both types of approximants:
\begin{subeqnarray}
w_0    &=&  -0.25011 \pm 0.00016 \\ 
\alpha &=&   0.933   \pm 0.030
\end{subeqnarray}
The estimate for the critical point agrees very well with the theoretical 
prediction $w_0({\rm sq})=-1/4$. The estimate for the critical exponent 
is close to the expected result $\alpha=1$; but it is barely compatible with 
it within errors (if one takes the error bars literally).
The biased estimates for $\alpha$ are also very similar for $K=1$ and $K=2$,
so that once again our preferred estimate comes from merging both data sets:
\be
\alpha  \;=\; 0.913   \pm 0.021 
\ee
This estimate is four standard deviations away from the expected value
$\alpha=1$.

The analysis of the first derivative $f'(w)$ is less clear: 
the spread of the data points in Figure~\ref{figure_DA_sq}(b) is 
much larger than it was for the second derivative.
We observe that most of the first-order approximants
give a singularity with $w_\ell \ltapprox -1/4$, 
while most of the second-order ones lie at somewhat higher values of $w_0$
(the latter approximants also have smaller dispersion).
The estimates obtained from the $K=2$ approximants are 
\begin{subeqnarray}
w_0      &=&  -0.25004 \pm 0.00063  \\ 
1-\alpha &=&   0.10    \pm 0.22
\end{subeqnarray}
while those coming from merging both data sets are 
\begin{subeqnarray}
w_0      &=&  -0.2506  \pm 0.0011  \\ 
1-\alpha &=&  -0.08    \pm 0.31 
\end{subeqnarray}
We observe that the error bars are 4--7 times larger than for
the second-derivative series. The estimates of $w_0$ are still compatible 
within errors with the expected result $w_0 = -1/4$. The estimates of
$\alpha$ are compatible with the expected value $\alpha=1$, but the 
error bars are so large that we are unable to tell whether
the first derivative diverges or not at $w=w_0$
(i.e., whether $1-\alpha$ is negative or positive).
If we look at the biased estimates of $1-\alpha$,
we see that the dispersion of the second-order approximants
is smaller than for the first-order ones.
Thus, we take as our preferred estimate the value coming from the
$K=2$ approximants:
\be
1-\alpha  \;=\; 0.104   \pm 0.040  
\ee
This value suggests that the first derivative $f'$ does not diverge at 
$w=w_0$; but this conclusion cannot be taken too seriously as the signal
is only 2.5 times larger than the alleged error bar.

We conclude this analysis with the free energy [Figure~\ref{figure_DA_sq}(a)].
The dispersion of the data is even larger than for the first derivative,
so we should expect even larger error bars.
The $K=1$ approximants yield estimates of $w_0$ quite a bit below $-1/4$,
and estimates of $2-\alpha$ that are rather near to zero than to 1.
Therefore, our preferred estimate comes from the $K=2$ data set: 
\begin{subeqnarray}
w_0      &=&  -0.2501  \pm 0.0011 \\ 
2-\alpha &=&   1.07    \pm 0.39 
\end{subeqnarray}
The estimates are compatible with the expected values,
but the error bars are huge.
The analysis of the biased estimates for $2-\alpha$ is similar
and our preferred estimate comes from the $K=2$ approximants:
\be
2-\alpha  \;=\; 1.063   \pm 0.095   \;,
\ee
which is again compatible within errors with the expected value $\alpha=1$.

Let us now summarize the results of this analysis. 
We have found that the the unbiased estimates of $w_0$ are in all cases 
compatible within errors with the expected value $w_0=-1/4$:
\be
w_0 \;=\; \left\{ \begin{array}{ll}
              -0.25011 \pm 0.00016 &\qquad \hbox{\rm for $f''$} \\ 
              -0.25004 \pm 0.00063 &\qquad \hbox{\rm for $f'$} \\ 
              -0.2501  \pm 0.0011  &\qquad \hbox{\rm for $f$} 
              \end{array}
      \right.
\ee
This result is an independent confirmation that $w_0({\rm sq})=-1/4$.
The unbiased estimates of the critical exponent $\alpha$ are given by
\be
\alpha \;=\; \left\{ \begin{array}{ll}
               0.933 \pm 0.030 &\qquad \hbox{\rm for $f''$} \\
               0.90  \pm 0.22  &\qquad \hbox{\rm for $f'$} \\
               0.93  \pm 0.39  &\qquad \hbox{\rm for $f$}
              \end{array}
      \right.
\ee
while the biased estimates are
\be
\alpha \;=\; \left\{ \begin{array}{ll}
               0.913 \pm 0.022 &\qquad \hbox{\rm for $f''$} \\
               0.896 \pm 0.040 &\qquad \hbox{\rm for $f'$} \\
               0.937 \pm 0.095 &\qquad \hbox{\rm for $f$}
              \end{array}
      \right.
\ee
The foregoing estimates are compatible among themselves within the quoted
errors. However, the estimates coming from $f''$ are at least a factor
of 2 more precise than those coming from $f'$ or $f$. It is 
intriguing that all the estimates of $\alpha$ are consistently smaller 
than the expected value $\alpha=1$. Furthermore, the difference between
the above estimates and the theoretical prediction is significant,
if one takes the error bars seriously.
Perhaps this indicates a multiplicative logarithmic correction
to the leading behavior $\alpha=1$.

The story for the triangular lattice is rather short. From
Table~\ref{table_series} we observe that the coefficients $f_k$
have a complicated sign pattern. This is an indication that the
dominant singularity (or singularities) is/are complex. This fact makes
the ratio method useless.
Furthermore, the differential-approximant method
does not give any sensible estimate of $w_0$
(or of the corresponding critical exponents).
We computed the differential approximants to the second derivative $f''$ 
(which is {\em a priori} the most favorable observable) using at least  
20 series coefficients and with the same constraints on the $N_i$ as for the
square lattice.
We found that the zero closest to $w_0({\rm tri}) \approx -0.1753$
varies over a range vastly wider than for the square lattice
(by two orders of magnitude).
We were therefore obliged to take a very lenient view
of what constitutes a non-defective approximant:
we considered an  approximant to be non-defective
if there is a real zero in the interval $-0.3 \leq w_\ell \leq -0.1$
and there is no other zero 
in the region $-0.5 \leq \real w \leq 0$ and $-0.05 \leq \imag w \leq 0.05$. 
The non-defective approximants are displayed in Figure~\ref{figure_DA_sq}(d).
We find that, even with this lenient definition,
there are many fewer non-defective appoximants 
than for the square-lattice case (45 for $K=1$ and 5 for $K=2$, compared 
to 91 and 164, respectively). The non-defective approximants seem to 
accumulate along a curve; but the low density of zeros prevents us
from drawing any reliable conclusion from these data.

\section*{Acknowledgments}

We wish to thank Dario Bini for supplying us the MPSolve 2.1.1 package
\cite{Bini_package,Bini-Fiorentino} for computing roots of polynomials,
and for many discussions about its use;
Ian Enting, Tony Guttmann and Iwan Jensen for correspondence concerning the
finite-lattice method; Yao-ban Chan and Andrew Rechnitzer for providing us 
their {\sc C++} code to compute differential approximants,
and for assistance in its use;
and Hubert Saleur and Robert Shrock for many helpful conversations
throughout the course of this work.
We also wish to thank two referees for thoughtful comments
on the first version of this paper.

The authors' research was supported in part
by U.S.\ National Science Foundation grants PHY--0099393, PHY--0116590
and PHY--0424082.



\clearpage

%
%
%
%
\begin{table}
\centering
\begin{tabular}{rrrrr}
\hline\hline
\multicolumn{1}{c}{$L$}           &
\multicolumn{1}{c}{SqCyl($L$)}    &
\multicolumn{1}{c}{TriCyl$'$($L$)}&
\multicolumn{1}{c}{SqFree($L$)}   &
\multicolumn{1}{c}{TriFree($L$)} \\
\hline
1  &         1  &        1 &          1 &          1 \\
2  &         2  &        2 &          2 &          2 \\
3  &         3  &        3 &          4 &          5 \\
4  &         6  &        6 &         10 &         14 \\
5  &        10  &       10 &         26 &         42 \\
6  &        24  &       28 &         76 &        132 \\
7  &        49  &       63 &        232 &        429 \\
8  &       130  &      190 &        750 &  {\bf 1430}\\
9  &       336  &      546 &  {\bf 2494}&       4862 \\
10 &  {\bf 980} &{\bf 1708}&       8524 &      16796 \\
11 &      2904  &     5346 &      29624 &      58786 \\
12 &      9176  &    17428 &     104468 &     208012 \\
13 &     29432  &    57148 &     372308 &     742900 \\
14 &     97356  &   191280 &    1338936 &    2674440 \\
15 &    326399  &   646363 &    4850640 &    9694845 \\
16 &   1111770  &  2210670 &   17685270 &   35357670 \\
17 &   3825238  &  7626166 &   64834550 &  129644790 \\
18 &  13293456  & 26538292 &  238843660 &  477638700 \\
19 &  46553116  & 93013854 &  883677784 & 1767263190 \\
20 & 164200028  &328215300 & 3282152588 & 6564120420 \\
\hline\hline
\end{tabular}
\caption{\label{table_dimensions}
Dimensions of the transfer matrices for square- and triangular-lattice strips
of width $L$ with cylindrical and free boundary conditions.
TriFree($L$) gives the total number of non-crossing partitions of the set
$\{1,2,\ldots,L\}$. The other three columns give the number of
equivalence classes of non-crossing partitions of $\{1,2,\ldots,L\}$
modulo some symmetries: SqFree($L$) [reflection],
TriCyl$'$($L$) [translation], and SqCyl($L$) [translation and reflection].
We show in boldface the dimensions of the largest transfer matrices that
we have used {\em in symbolic form}\/
in the computations reported in this paper.
}
\end{table}

%
%
\begin{table}
\centering
\scriptsize
\begin{tabular}{rrr}
\hline\hline
\multicolumn{1}{c}{$k$}     &
\multicolumn{1}{c}{$k f_k({\rm sq})$} &
\multicolumn{1}{c}{$k f_k({\rm tri})$} \\
\hline
1 &                           $2$&              $3$\\
2 &                          $-2$&             $-3$\\
3 &                           $2$&             $-3$\\
4 &                          $-6$&              $9$\\
5 &                          $22$&              $3$\\
6 &                         $-74$&            $-33$\\
7 &                         $254$&              $3$\\
8 &                        $-918$&            $393$\\
9 &                        $3422$&          $-2325$\\
10&                      $-12862$&           $9327$\\
11&                      $ 48138$&         $-24483$\\
12&                     $-178530$&           $1815$\\
13&                     $ 655826$&         $458253$\\
14&                    $-2391370$&       $-3497133$\\
15&                     $8688262$&       $17900097$\\
16&                   $-31600918$&      $-70905543$\\
17&                   $115606190$&      $209837565$\\
18&                  $-426864494$&     $-314875455$\\
19&                  $1593065490$&    $-1242857637$\\
20&                 $-6004037966$&    $13767148419$\\
21&                 $22795625582$&   $-73591894407$\\
22&                $-86925982926$&   $264618912819$\\
23&                $332053760646$&  $-455046987303$\\
24&              $-1268578680714$& $-2402029948737$\\
25&               $4844247521322$& $31594277221653$\\
26&             $-18494593884938$&                 \\
27&              $70644561464090$&                 \\
28&            $-270190727926594$&                 \\
29&            $1035346222307838$&                 \\
30&           $-3976144389096514$&                 \\
31&           $15304261265448246$&                 \\
32&          $-59027790560689238$&                 \\
33&          $228068574553887894$&                 \\
34&         $-882464779625379526$&                 \\
35&         $3418403165602360314$&                 \\
36&       $-13253811969559767270$&                 \\
37&        $51425605359158653378$&                 \\
38&      $-199663639129405278414$&                 \\
39&       $775682274057446798018$&                 \\
40&     $-3015326560156376960998$&                 \\
41&     $11728909932236608346846$&                 \\
42&    $-45652065541079598767758$&                 \\
43&    $177805114097058031764786$&                 \\
44&   $-692953582585445674377902$&                 \\
45&   $2702276554574318555870842$&                 \\
46& $-10544061987158176469650990$&                 \\
47&  $41164706135505931628292550$&                 \\
\hline\hline
\end{tabular}
\caption{\label{table_series}
Small-$w$ (high-temperature) expansions for the bulk free energy
$f({\cal L},w) = \sum_{k=1}^\infty f_k({\cal L}) w^k$ for the square and
triangular lattices.
For simplicity, we present here, instead of the coefficient $f_k({\cal L})$,
the product $k f_k({\cal L})$, which is always an integer.
}
\end{table}

%
%
\begin{table}
\centering
\begin{tabular}{crrrrrr}
\hline\hline
\multicolumn{1}{c}{Type} &
\multicolumn{1}{c}{$w_{\rm min}$} &
\multicolumn{1}{c}{$f_1$} &
\multicolumn{1}{c}{$f_2$} &
\multicolumn{1}{c}{$f_3$} &
\multicolumn{1}{c}{$f_4$} &
\multicolumn{1}{c}{$f_5$} \\
\hline
Non-Biased
  &2    &  0.12499 &  0.01005 &  0.00349&   0.00405 &  0.00020 \\
  &3    &  0.12499 &  0.01002 &  0.00366&   0.00357 &  0.00070 \\
  &4    &  0.12499 &  0.01004 &  0.00354&   0.00400 &  0.00013 \\
  &5    &  0.12499 &  0.01006 &  0.00330&   0.00500 &$-0.00144$\\
  &6    &  0.12499 &  0.01009 &  0.00297&   0.00662 &$-0.00441$\\
\hline
Biased
  & 2   & $1/8$    &  0.00995 &  0.00397&   0.00300  & 0.00101 \\
  & 3   & $1/8$    &  0.00992 &  0.00429&   0.00173  & 0.00263 \\
  & 4   & $1/8$    &  0.00990 &  0.00462&   0.00015  & 0.00521 \\
  & 5   & $1/8$    &  0.00988 &  0.00505& $-0.00244$ & 0.01029 \\
  & 6   & $1/8$    &  0.00985 &  0.00564& $-0.00652$ & 0.01966 \\
  & 7   & $1/8$    &  0.00982 &  0.00640& $-0.01260$ & 0.03571 \\
\hline\hline
Non-Biased
  &$-2$ &  0.12500 &  0.00998 &  0.00398  & 0.00250  & 0.00096 \\
  &$-3$ &  0.12500 &  0.01001 &  0.00415  & 0.00300  & 0.00149 \\
  &$-4$ &  0.12501 &  0.01003 &  0.00436  & 0.00373  & 0.00245 \\
  &$-5$ &  0.12501 &  0.01006 &  0.00462  & 0.00484  & 0.00420 \\
  &$-6$ &  0.12501 &  0.01009 &  0.00496  & 0.00652  & 0.00728 \\
\hline
Biased
  &$-2$ & $1/8$    &  0.00993 &  0.00374&   0.00198  &  0.00056 \\
  &$-3$ & $1/8$    &  0.00992 &  0.00358&   0.00135  &$-0.00024$\\
  &$-4$ & $1/8$    &  0.00990 &  0.00329& $-0.00007$ &$-0.00256$\\
  &$-5$ & $1/8$    &  0.00988 &  0.00286& $-0.00262$ &$-0.00756$\\
  &$-6$ & $1/8$    &  0.00985 &  0.00227& $-0.00671$ &$-0.01697$\\
  &$-7$ & $1/8$    &  0.00982 &  0.00150& $-0.01284$ &$-0.03314$\\
\hline\hline
\end{tabular}
\caption{\label{table_fit_sq_large_w}
Fits of the large-$w$ series for the square lattice.
We show the estimates of the fits of the (infinite-volume)
free-energy data (for large values of $|w| \geq 2$) to the Ansatz
\protect\reff{Ansatz_large_w} with $k_{\rm max}=5$.
The column ``Type'' shows whether $f_1$ has been left free in the fit
(``Non-Biased'') or it has been fixed to its conjectured value $1/8$
(``Biased''). Non-biased fits are based on data points with
$|w|=w_{\rm min},\ldots,w_{\rm min}+4$; biased fits are based on data points
with $|w|=w_{\rm min},\ldots,w_{\rm min}+3$.
}
\end{table}

%
%
\begin{table}
\centering
\begin{tabular}{crrrrrr}
\hline\hline
\multicolumn{1}{c}{Type} &
\multicolumn{1}{c}{$w_{\rm min}$} &
\multicolumn{1}{c}{$f_1$} &
\multicolumn{1}{c}{$f_2$} &
\multicolumn{1}{c}{$f_3$} &
\multicolumn{1}{c}{$f_4$} &
\multicolumn{1}{c}{$f_5$} \\
\hline
Non-Biased
 & 2    &  0.08333 &  0.00285 &  0.00080 &  0.00039 &$-0.00006$\\
 & 3    &  0.08333 &  0.00286 &  0.00076 &  0.00049 &$-0.00005$\\
 & 4    &  0.08333 &  0.00286 &  0.00070 &  0.00069 &$-0.00031$\\
 & 5    &  0.08333 &  0.00287 &  0.00062 &  0.00103 &$-0.00085$\\
 & 6    &  0.08333 &  0.00288 &  0.00052 &  0.00156 &$-0.00181$\\
\hline
Biased
 & 2    &  $1/12$  &  0.00283 &  0.00089 &  0.00019 &  0.00022\\
 & 3    &  $1/12$  &  0.00283 &  0.00095 &$-0.00005$&  0.00053\\
 & 4    &  $1/12$  &  0.00282 &  0.00105 &$-0.00052$&  0.00129\\
 & 5    &  $1/12$  &  0.00281 &  0.00118 &$-0.00134$&  0.00290\\
 & 6    &  $1/12$  &  0.00280 &  0.00137 &$-0.00264$&  0.00589\\
 & 7    &  $1/12$  &  0.00279 &  0.00161 &$-0.00459$&  0.01103\\
\hline\hline
Non-Biased
 &$-2$  &  0.08333 &  0.00285 &  0.00087 &  0.00036 &  0.00012 \\
 &$-3$  &  0.08333 &  0.00286 &  0.00091 &  0.00048 &  0.00025 \\
 &$-4$  &  0.08334 &  0.00286 &  0.00097 &  0.00069 &  0.00052 \\
 &$-5$  &  0.08334 &  0.00287 &  0.00105 &  0.00103 &  0.00106 \\
 &$-6$  &  0.08334 &  0.00288 &  0.00115 &  0.00156 &  0.00202 \\
\hline
Biased
 &$-2$  &  $1/12$  &  0.00283 &  0.00078 &  0.00017 &$-0.00003$\\
 &$-3$  &  $1/12$  &  0.00283 &  0.00072 &$-0.00006$&$-0.00033$\\
 &$-4$  &  $1/12$  &  0.00282 &  0.00062 &$-0.00054$&$-0.00109$\\
 &$-5$  &  $1/12$  &  0.00281 &  0.00048 &$-0.00136$&$-0.00271$\\
 &$-6$  &  $1/12$  &  0.00280 &  0.00030 &$-0.00267$&$-0.00572$\\
 &$-7$  &  $1/12$  &  0.00279 &  0.00005 &$-0.00463$&$-0.01089$\\
\hline\hline
\end{tabular}
\caption{\label{table_fit_tri_large_w}
Fits of the large-$w$ series for the triangular lattice.
We show the estimates of the fits of the (infinite-volume)
free-energy data (for large values of $|w| \geq 2$) to the Ansatz
\protect\reff{Ansatz_large_w} with $k_{\rm max}=5$.
The column ``Type'' shows whether $f_1$ has been left free in the fit
(``Non-Biased'') or it has been fixed to its conjectured value $1/12$
(``Biased''). Non-biased fits are based on data points with
$|w|=w_{\rm min},\ldots,w_{\rm min}+4$; biased fits are based on data points
with $|w|=w_{\rm min},\ldots,w_{\rm min}+3$.
}
\end{table}

\clearpage

\def\kk{\phantom{$\mbox{}^\dagger$}}
\def\dd{\dagger}
%
%
\begin{table}
\small
\hspace*{-1.1cm}
\begin{tabular}{rrrrrrrr}
\hline\hline
\multicolumn{1}{c}{$L$}    &
\multicolumn{1}{c}{\#C}    &
\multicolumn{1}{c}{\#E}    &
\multicolumn{1}{c}{\#Q}    &
\multicolumn{1}{c}{$w_{0-}(L)$}  &
\multicolumn{1}{c}{$w_{0Q}(L)$}  &
\multicolumn{1}{c}{$w_{0+}(L)$}  &
\multicolumn{1}{c}{$w_{B}(L)$}  \\
\hline
2 & 1\kk  &  4\kk  & 1\kk  &$-0.5000000000$ &$-0.3660254038$&$-0.2500000000$ &
    $-0.2500000000 \pm 0.4330127019\, i$ \\
3 & 3\kk  &  6\kk  & 0\kk  &                &$-0.2868497019$&                &
    $-0.0574443351 \pm 0.4806253161\, i$ \\
4 & 3\kk  &  7\kk  & 2\kk  &$-0.2776248333$ &$-0.2670015604$&$-0.2500000000$ &
    $0.0632994010  \pm 0.5099839130\, i$\\
5 & 5\kk  & 10\kk  & 0\kk  &                &$-0.2620754678$&                &
    $0.1533657968  \pm 0.5306112949\, i$ \\
6 & 5\kk  & 12\kk  & 1\kk  &$-0.2609570768$ &$-0.2563792782$&$-0.2500000000$ &
    $0.2262944917  \pm 0.5460127254\, i$ \\
7 &$7^\dd$&$14^\dd$&$0^\dd$&                &$-0.2559077691$&                &
    $0.2879810252  \pm 0.5579895735\, i$ \\
8 &$7^\dd$&$16^\dd$&$1^\dd$&$-0.2556827597$ &$-0.2532620078$&$-0.2500000000$ &
    $0.3415981731  \pm 0.5675966263\, i$ \\
9 &$9^\dd$&$18^\dd$&$0^\dd$&                &$-0.2534832041$&                &
    $0.3890914478  \pm 0.5754959494\, i$ \\
10&$9^\dd$&$20^\dd$&$1^\dd$&$-0.2534268314$ &$-0.2519570283$&$-0.2500000000$ &
    $0.4317571213  \pm 0.5821234989\, i$ \\
\hline\hline
\end{tabular}
\caption{\label{table_sq}
Characteristics of the limiting curves ${\cal B}$ for square-lattice strips
with cylindrical boundary conditions.
For each width $L$, we show the number of
connected components (\#C), the number of endpoints (\#E), the number of
multiple points (\#Q), and the $w_0$ and $w_B$ values defined in the text.
The numbers marked with a $\mbox{}^\dagger$ are only lower bounds
on the exact values.
}
\end{table}

%
%
\begin{table}
\small
\hspace*{-1.1cm}
\begin{tabular}{rrrrrrrr}
\hline\hline
\multicolumn{1}{c}{$L$}    &
\multicolumn{1}{c}{\#C}    &
\multicolumn{1}{c}{\#E}    &
\multicolumn{1}{c}{\#Q}    &
\multicolumn{1}{c}{$w_{0-}(L)$}  &
\multicolumn{1}{c}{$w_{0Q}(L)$}  &
\multicolumn{1}{c}{$w_{0+}(L)$}  &
\multicolumn{1}{c}{$w_{B}(L)$}  \\
\hline
2 & 1\kk  &  4\kk  &  1\kk &$-0.2500000000$ &$-0.2251972448$ &$-0.2017782928$ &
    $-0.1178608536 \pm 0.2520819223\, i$ \\
3 & 3\kk  &  6\kk  &  0\kk &                &$-0.1921127637$ &                &
    $-0.0100833905 \pm 0.2849353892\, i$ \\
4 & 3\kk  &  8\kk  &  1\kk &$-0.1846154722$ &$-0.1839945026$ &$-0.1833753245$ &
    $0.0588726934 \pm 0.3024953798\, i$ \\
5 & 5\kk  & 10\kk  &  0\kk &                &$-0.1805863920$ &                &
    $0.1110085784 \pm 0.3142261926\, i$ \\
6 & 5\kk  & 12\kk  &  1\kk &$-0.1788596197$ &$-0.1788458357$ &$-0.1788320527$ &
    $0.1534513521 \pm 0.3227719569\, i$ \\
7 &$7^\dd$&$14^\dd$&$0^\dd$&                &$-0.1778368691$ &                &
    $0.1894308555 \pm 0.3293185497\, i$ \\
8 &$7^\dd$&$16^\dd$&$1^\dd$&$-0.1772035681$ &$-0.1772011941$ &$-0.1771988202$ &
    $0.2207308275\pm 0.3345156821\, i$ \\
9 &$9^\dd$&$18^\dd$&$0^\dd$&                &$-0.1767753501$ &                &
    $0.2484623416 \pm 0.3387563998\, i$ \\
10&$9^\dd$&$20^\dd$&$1^\dd$&$-0.1764769384$ &$-0.1764764073$ &$-0.1764758762$ &
    $0.2733731381 \pm 0.3422936097\, i$ \\
\hline\hline
\end{tabular}
\caption{\label{table_tri}
Characteristics of the limiting curves for triangular-lattice strips with
cylindrical boundary conditions. For each width $L$, we show the number of
connected parts (\#C), the number of endpoints (\#E), the number of
multiple points (\#Q), and the $w_0$ values defined in the text.
The numbers marked with a $\mbox{}^\dagger$ are only lower bounds
on the exact values.
}
\end{table}

\clearpage

%
%
\begin{table}
\centering
\begin{tabular}{lll}
\hline\hline
\multicolumn{1}{c}{\hbox{}}    &
\multicolumn{1}{c}{Square lattice}&
\multicolumn{1}{c}{Triangular lattice} \\
\hline
\multicolumn{1}{c}{$\imag w$} &
\multicolumn{1}{c}{$\real w$} &
\multicolumn{1}{c}{$\real w$} \\
\hline
 0    & $-0.2501 \pm 0.0001$ & $ -0.1753 \pm 0.0004$ \\
 0.01 & $-0.2512 \pm 0.0012$ & $ -0.1751 \pm 0.0001$ \\
 0.02 & $-0.2531 \pm 0.0006$ & $ -0.1743 \pm 0.0001$ \\
 0.03 & $-0.2531 \pm 0.0010$ & $ -0.1730 \pm 0.0001$ \\
 0.04 & $-0.2537 \pm 0.0005$ & $ -0.1712 \pm 0.0001$ \\
 0.05 & $-0.2539 \pm 0.0005$ & $ -0.1688 \pm 0.0001$ \\
 0.06 & $-0.2535 \pm 0.0005$ & $ -0.1659 \pm 0.0001$ \\
 0.07 & $-0.2526 \pm 0.0006$ & $ -0.1624 \pm 0.0001$ \\
 0.08 & $-0.2512 \pm 0.0002$ & $ -0.1584 \pm 0.0001$ \\
 0.09 & $-0.2493 \pm 0.0011$ & $ -0.1537 \pm 0.0001$ \\
 0.10 & $-0.2469 \pm 0.0009$ & $ -0.1483 \pm 0.0001$ \\
 0.11 & $-0.2441 \pm 0.0007$ & $ -0.1423 \pm 0.0001$ \\
 0.12 & $-0.2409 \pm 0.0010$ & $ -0.1355 \pm 0.0001$ \\
 0.13 & $-0.2374 \pm 0.0012$ & $ -0.1279 \pm 0.0001$ \\
 0.14 & $-0.2331 \pm 0.0025$ & $ -0.1193 \pm 0.0001$ \\
 0.15 & $-0.2305 \pm 0.0018$ & $ -0.1099 \pm 0.0001$ \\
 0.16 & $-0.2258 \pm 0.0015$ & $ -0.0994 \pm 0.0001$ \\
 0.17 & $-0.2207 \pm 0.0010$ & $ -0.0877 \pm 0.0002$ \\
 0.18 & $-0.2153 \pm 0.0009$ & $ -0.0748 \pm 0.0006$ \\
 0.19 & $-0.2095 \pm 0.0006$ & $ -0.0599 \pm 0.0004$ \\
 0.20 & $-0.2033 \pm 0.0011$ & $ -0.0429 \pm 0.0021$ \\
 0.21 & $-0.1968 \pm 0.0006$ & $ -0.0257 \pm 0.0025$ \\
 0.22 & $-0.1895 \pm 0.0005$ & $ -0.0073 \pm 0.0030$ \\
 0.23 & $-0.1819 \pm 0.0003$ & $ \phantom{-}
                                  0.0198 \pm 0.0060$ \\
 0.24 & $-0.1736 \pm 0.0001$ &  \\
 0.25 & $-0.1646 \pm 0.0001$ &  \\
 0.26 & $-0.1550 \pm 0.0002$ &  \\
 0.27 & $-0.1447 \pm 0.0002$ &  \\
 0.28 & $-0.1337 \pm 0.0003$ &  \\
 0.29 & $-0.1220 \pm 0.0002$ &  \\
 0.30 & $-0.1094 \pm 0.0005$ &  \\
 0.31 & $-0.0969 \pm 0.0015$ &  \\
 0.32 & $-0.0820 \pm 0.0017$ &  \\
 0.33 & $-0.0674 \pm 0.0010$ &  \\
\hline\hline
\end{tabular}
\caption{\label{table_limiting_curves}
Extrapolated limiting curve ${\cal B}$ for the square and triangular lattices.
For each lattice and each fixed value of $\imag w$,
we show the extrapolated value of $\real w$
together with its subjective error bar.
}
\end{table}


%
%
\begin{table}
\centering
\begin{tabular}{rcrrr}
\hline\hline
\multicolumn{1}{c}{$O$}              &
\multicolumn{1}{c}{$L_{\rm min}$}    &
\multicolumn{1}{c}{$w_0$}            &
\multicolumn{1}{c}{$A$}              &
\multicolumn{1}{c}{$\Delta$}         \\
\hline
$w_{0-}$ & 2 & $-0.255549$ & $-2.706839$ & $3.468990$ \\
         & 4 & $-0.249990$ & $-0.651261$ & $2.279345$ \\
         & 6 & $-0.250067$ & $-0.673708$ & $2.302190$ \\
\hline
$w_{0Q}$ & 2 & $-0.251626$ & $-0.851180$ & $2.895388$ \\
         & 3 & $-0.250375$ & $-0.420673$ & $2.225741$ \\
         & 4 & $-0.250280$ & $-0.525761$ & $2.487308$ \\
         & 5 & $-0.250102$ & $-0.381902$ & $2.151375$ \\
         & 6 & $-0.250099$ & $-0.450073$ & $2.384256$ \\
\hline
$w_{0+}$ & 2 & $-0.250000$ &             &            \\
         & 4 & $-0.250000$ &             &            \\
         & 6 & $-0.250000$ &             &            \\
\hline\hline
\end{tabular}
\caption{\label{table_fits_sq}
   Fits for $w_0$ on the square lattice,
   using the Ansatz $w_{0j}(L) = w_0 + A L^{-\Delta}$ ($j=+,-,Q$).
   Each fit is based on data points with
   $L=L_{\rm min}, L_{\rm min}+2, L_{\rm min}+4$.
}
\end{table}

%
%
\begin{table}
\centering
\begin{tabular}{rcrrr}
\hline\hline
\multicolumn{1}{c}{$O$}              &
\multicolumn{1}{c}{$L_{\rm min}$}    &
\multicolumn{1}{c}{$w_0$}            &
\multicolumn{1}{c}{$A_2$}            &
\multicolumn{1}{c}{$A_4$}           \\
\hline
$w_{0-}$ & 2 &$-0.253138$& $-0.193231$ &$-3.176863$\\
         & 4 &$-0.249328$& $-0.391380$ &$-0.981983$\\
         & 6 &$-0.249706$& $-0.353564$ &$-1.853270$\\
\hline
$w_{0Q}$ & 2 &$-0.249617$& $-0.215651$ &$-0.999924$\\
         & 3 &$-0.249785$& $-0.292447$ &$-0.370198$\\
         & 4 &$-0.249712$& $-0.210752$ &$-1.054185$\\
         & 5 &$-0.249899$& $-0.284036$ &$-0.509425$\\
         & 6 &$-0.249852$& $-0.196667$ &$-1.378714$\\
\hline
$w_{0+}$ & 2 &$-0.250000$&             &           \\
         & 4 &$-0.250000$&             &           \\
         & 6 &$-0.250000$&             &           \\
\hline\hline
\end{tabular}
\caption{\label{table_fits_sq_bis}
   Fits for $w_0$ on the square lattice,
   using the Ansatz $w_{0j}(L) = w_0 + A_2 L^{-2} + A_4 L^{-4}$ ($j=+,-,Q$).
   Each fit is based on data points with
   $L=L_{\rm min}, L_{\rm min}+2, L_{\rm min}+4$.
}
\end{table}

%
%
\def\kkk{\phantom{$-$}}
\begin{table}
\centering
\begin{tabular}{r|cc|c}
\hline\hline
                                                        &
\multicolumn{2}{|c|}{$\real w_B(L)= c_1 \log L + c_2$} &
\multicolumn{1}{|c} {$\real w_B(L)= c_1 \log L + c_3 \log\log L + c_2$}\\
\hline
\multicolumn{1}{c|}{$L_{\rm min}$}&
\multicolumn{1}{|c}{$c_1$}        &
\multicolumn{1}{c|}{$c_2$}        &
\multicolumn{1}{|c}{$c_2$}        \\ 
\hline
 2  &  0.474901 & $-0.579176$  &$-0.639286$ \\
 3  &  0.419712 & $-0.518546$  &$-0.567025$ \\
 4  &  0.403625 & $-0.496244$  &$-0.546622$ \\
 5  &  0.400000 & $-0.490410$  &$-0.539345$ \\
 6  &  0.400170 & $-0.490715$  &$-0.536389$ \\
 7  &  0.401532 & $-0.493364$  &$-0.535168$ \\
 8  &  0.403227 & $-0.496888$  &$-0.534745$ \\
 9  &  0.404949 & $-0.500673$  &$-0.534720$ \\
10  &           &              &$-0.534906$ \\ 
\hline\hline
\end{tabular}

\bigskip
(a)
\bigskip
\bigskip

\begin{tabular}{r|ccc|cc}
\hline\hline
                                                               &
\multicolumn{3}{|c}{$\imag w_B(L) = \imag w_B + A L^{-\Delta}$}&
\multicolumn{2}{|c}{$\imag w_B(L) = \imag w_B + A L^{-1/2}$}   \\
\hline
\multicolumn{1}{c|}{$L_{\rm min}$} &
\multicolumn{1}{|c}{$\imag w_B$}   &
\multicolumn{1}{c} {$A$}           &
\multicolumn{1}{c} {$\Delta$}      &
\multicolumn{1}{|c}{\kkk$\imag w_B$}   &
\multicolumn{1}{c} {$A$}           \\
\hline
2 &   0.748618 & $-0.417404$ &  0.403321 & \kkk 0.692477   & $-0.366938$\\
3 &   0.759610 & $-0.426574$ &  0.386512 & \kkk 0.699761   & $-0.379554$\\
4 &   0.728743 & $-0.404746$ &  0.443835 & \kkk 0.705369   & $-0.390771$\\
5 &   0.708533 & $-0.395674$ &  0.496601 & \kkk 0.707377   & $-0.395260$\\
6 &   0.697576 & $-0.394585$ &  0.534017 & \kkk 0.707470   & $-0.395487$\\
7 &   0.691678 & $-0.396335$ &  0.558477 & \kkk 0.706739   & $-0.393553$\\
8 &   0.688422 & $-0.398635$ &  0.574048 & \kkk 0.705719   & $-0.390668$\\
9 &            &             &           & \kkk 0.704646   & $-0.387450$\\
\hline\hline
\end{tabular}

\bigskip
(b)

\caption{\label{table_fits_wB_sq}
   Fits for $w_B$ on the square lattice.
   (a) For the real part, we use the Ans\"atze
   $\real w_B(L) = c_1\log L + c_2$ with $c_1$ and $c_2$ free,
   and $\real w_B(L) = c_1\log L + c_3 \log\log L + c_2$
   with fixed $c_1=3/(2\pi)$ and $c_3=-1/(2\pi)$
   [see \protect\reff{def_ansatz_Re_wB_bis} and surrounding text].
   (b) For the imaginary part, we first use
   $\imag w_B(L) = \imag w_B + A L^{-\Delta}$,
   and then we perform the fit with $\Delta=1/2$.
   Two-parameter fits are based on data points with
   $L=L_{\rm min}, L_{\rm min}+1$; three-parameter fits are based on
   data points with $L=L_{\rm min}, L_{\rm min}+1, L_{\rm min}+2$.
}
\end{table}

%
%
\begin{table}
\centering
\begin{tabular}{rrrr}
\hline\hline
\multicolumn{1}{c}{$L$}   &
\multicolumn{1}{c}{$\real f_L(w=-1)$} &
\multicolumn{1}{c}{$\real f_L(w=-1/4)$}   &
\multicolumn{1}{c}{$\real f_L(w=1)$} \\
\hline
2  & 0.5493061443 & $-$0.8369882168 & 1.1815331056 \\
3  & 0.8549831192 & $-$0.5707633924 & 1.2632940309 \\
4  & 0.9464522368 & $-$0.6562577390 & 1.2874485986 \\
5  & 0.9856542144 & $-$0.5922057762 & 1.2970413323 \\
6  & 1.0059956327 & $-$0.6249471934 & 1.3016030310 \\
7  & 1.0178936568 & $-$0.5976444627 & 1.3040407836 \\
8  & 1.0254486325 & $-$0.6148600247 & 1.3054521450 \\
9  & 1.0305421492 & $-$0.5998257544 & 1.3063180519 \\
10 & 1.0341369549 & $-$0.6104343745 & 1.3068731127 \\
11 & 1.0367673970 & $-$0.6009181846 & 1.3072413122 \\
12 & 1.0387494019 & $-$0.6081078906 & 1.3074923528 \\
13 & 1.0402794617 & $-$0.6015429512 & 1.3076673968 \\
14 & 1.0414849726 & $-$0.6067349394 & 1.3077917458 \\
15 & 1.0424514545 & $-$0.6019335900 & 1.3078814782 \\
16 & 1.0432380373 &               & 1.3079471010 \\
\hline
$\infty$ &        & $-$0.6031055757 &              \\
\hline\hline
\end{tabular}
\caption{\label{table_f_sq}
   Free energies $f_L(w)$ [cf.\ \protect\reff{def_fm}]
   for square-lattice strips of width $L$ and cylindrical boundary conditions,
   for $w=-1,-1/4,+1$. The row labelled
   $\infty$ shows the {\em exact} value \reff{def_f_sq_w=-0.25}
   in the thermodynamic limit $L=\infty$.
}
\end{table}

%
%
\begin{table}
\vspace*{-0.6cm}
\centering
\begin{tabular}{rrrr}
\hline\hline
\multicolumn{1}{c}{$L_{\rm min}$}             &
\multicolumn{1}{c}{$\real f({\rm sq},w=-1/4)$}      &
\multicolumn{1}{c}{$A(w=-1/4)$}        &
\multicolumn{1}{c}{$\Delta(w=-1/4)$}  \\
\hline
3  & $-$0.6026552580 &  0.3514454435 &  2.1843042660 \\
5  & $-$0.6029925573 &  0.3092573584 &  2.0851083253 \\
7  & $-$0.6030693569 &  0.2909915798 &  2.0464952763 \\
9  & $-$0.6030901419 &  0.2826853879 &  2.0304080199 \\
11 & $-$0.6030977473 &  0.2779994113 &  2.0219793544 \\
\hline
2  & $-$0.6018136184 & $-$1.0158505843 &  2.1108840738 \\
4  & $-$0.6036426651 & $-$1.1575753432 &  2.2297430523 \\
6  & $-$0.6034714802 & $-$1.1161060082 &  2.2049156257 \\
8  & $-$0.6032861810 & $-$1.0321405587 &  2.1595425257 \\
10 & $-$0.6031958219 & $-$0.9688529236 &  2.1266061029 \\
\hline
$\infty$&
     $-$0.6031055757 &               &               \\
\hline\hline
\multicolumn{4}{c}{\quad} \\[-1mm]
\hline\hline
\multicolumn{1}{c}{$L_{\rm min}$}          &
\multicolumn{1}{c}{$\real f({\rm sq},w=-1)$}     &
\multicolumn{1}{c}{$A(w=-1)$}       &
\multicolumn{1}{c}{$\Delta(w=-1)$}  \\
\hline
2  & 1.0366588370 & $-$2.6329852431 &  2.4336613152  \\
3  & 1.0442846989 & $-$2.3545416365 &  2.2944945872  \\
4  & 1.0467035384 & $-$2.1844090010 &  2.2227750669  \\
5  & 1.0477001325 & $-$2.0688028151 &  2.1789288156  \\
6  & 1.0481526051 & $-$1.9898682648 &  2.1511947863  \\
7  & 1.0483784334 & $-$1.9340682142 &  2.1327570133  \\
8  & 1.0485022480 & $-$1.8926046519 &  2.1197454120  \\
9  & 1.0485758832 & $-$1.8603185323 &  2.1100523544  \\
10 & 1.0486226265 & $-$1.8342219236 &  2.1025132464  \\
11 & 1.0486538611 & $-$1.8125199233 &  2.0964523136  \\
12 & 1.0486756030 & $-$1.7940748336 &  2.0914535840  \\
13 & 1.0486912486 & $-$1.7781266664 &  2.0872465655  \\
14 & 1.0487028225 & $-$1.7641447178 &  2.0836471061  \\
\hline\hline
\multicolumn{4}{c}{\quad} \\[-1mm]
\hline\hline
\multicolumn{1}{c}{$L_{\rm min}$}          &
\multicolumn{1}{c}{$\real f({\rm sq},w=1)$}      &
\multicolumn{1}{c}{$A(w=1)$}        &
\multicolumn{1}{c}{$\Delta(w=1)$}  \\
\hline
2  & 1.3107968081 & $-$0.7156482298 &  2.4689333981 \\
3  & 1.3093320071 & $-$0.7881463786 &  2.5852771137 \\
4  & 1.3089601334 & $-$0.8429753668 &  2.6461540521 \\
5  & 1.3087423951 & $-$0.9168496742 &  2.7098055866 \\
6  & 1.3085813800 & $-$1.0307347078 &  2.7878824576 \\
7  & 1.3084637131 & $-$1.1971341766 &  2.8782847595 \\
8  & 1.3083793230 & $-$1.4306376635 &  2.9776440535 \\
9  & 1.3083189900 & $-$1.7531876469 &  3.0836969829 \\
10 & 1.3082756267 & $-$2.1979960385 &  3.1951197122 \\
11 & 1.3082442022 & $-$2.8146824223 &  3.3111199451 \\
12 & 1.3082212320 & $-$3.6772081777 &  3.4311777392 \\
13 & 1.3082043053 & $-$4.8963374373 &  3.5549124381 \\
14 & 1.3081917409 & $-$6.6392455365 &  3.6820180432 \\
\hline\hline
\end{tabular}
\caption{\label{table_f_fits_sq}
  Fits of $\real f_L(w)$ on square-lattice strips
  to the Ansatz~\protect\reff{eq.Ansatz.f.Delta}.
  For each value of $w$ ($+1$, $-1$, $-1/4$),
  we show the estimates of $\real f(w)$, $A(w)$ and $\Delta(w)$
  obtained by fitting three consecutive data points
  with $L\geq L_{\rm min}$.
  This means $L_{\rm min},L_{\rm min}+1,L_{\rm min}+2$
  for $w = \pm 1$, but $L_{\rm min},L_{\rm min}+2,L_{\rm min}+4$ for $w=-1/4$.
  The row labelled $\infty$ shows the known exact value
  \protect\reff{def_f_sq_w=-0.25} in the infinite-volume limit.
}
\end{table}

%
%
\begin{table}
\centering
\begin{tabular}{r|rr|rrr}
\hline\hline
\multicolumn{1}{c|}{\hbox{}}                   &
\multicolumn{2}{c|}{$f_L=f + (c\pi/6)L^{-2}$}  &
\multicolumn{3}{c} {$f_L=f + (c\pi/6)L^{-2} + A L^{-4}$} \\
\cline{2-6}
\multicolumn{1}{c|}{$L_{\rm min}$}     &
\multicolumn{1}{c }{$f({\rm sq},w=-1)$}&
\multicolumn{1}{c|}{$c(w=-1)$}         &
\multicolumn{1}{c }{$f({\rm sq},w=-1)$}&
\multicolumn{1}{c }{$c(w=-1)$}         &
\multicolumn{1}{c} {$A(w=-1)$}        \\
\hline
2 &1.0995246990&$-$4.2033601324&1.0522322844&$-$3.0291759709 &$-$1.7025269242\\
3 &1.0640553881&$-$3.5936875870&1.0504479365&$-$2.9439796309 &$-$1.9594730321\\
4 &1.0553466191&$-$3.3275672082&1.0497297366&$-$2.8877415435 &$-$2.2467529705\\
5 &1.0522261288&$-$3.1785747743&1.0494002632&$-$2.8493574206 &$-$2.5432790654\\
6 &1.0508420314&$-$3.0834112435&1.0492104613&$-$2.8185453467 &$-$2.8780896721\\
7 &1.0501282194&$-$3.0166105063&1.0490891975&$-$2.7923749159 &$-$3.2583728412\\
8 &1.0497177416&$-$2.9664373949&1.0490078464&$-$2.7698463825 &$-$3.6800970526\\
9 &1.0494621793&$-$2.9269022644&1.0489513545&$-$2.7503180147 &$-$4.1376812436\\
10&1.0492933116&$-$2.8946509060&1.0489109249&$-$2.7332535305 &$-$4.6268794806\\
11&1.0491764712&$-$2.8676499122&1.0488811967&$-$2.7182077146 &$-$5.1448634916\\
12&1.0490926063&$-$2.8445853935&1.0488588065&$-$2.7048231509 &$-$5.6897520658\\
13&1.0490305778&$-$2.8245646824&1.0488415841&$-$2.6928174237 &$-$6.2602275770\\
14&1.0489835393&$-$2.8069566713&1.0488280902&$-$2.6819676896 &$-$6.8553060059\\
15&1.0489471059&$-$2.7913005767&            &                &               \\
\hline\hline
\end{tabular}
\caption{\label{table_f_fits_c_sq_w=-1}
Fits of $\real f_L(w=-1)$ on square-lattice strips
to the Ans\"atze \protect\reff{def_ansatz_f}/\protect\reff{def_ansatz_f_bis}.
We show the estimates of $f({\rm sq},w=-1)$, $c(w=-1)$, and $A(w=-1)$
obtained by fitting two or three consecutive data points
with $L\geq L_{\rm min}$.
}
\end{table}

%
%
\begin{table}
\centering
\begin{tabular}{r|rrrrr}
\hline\hline
\multicolumn{1}{c|}{$L_{\rm min}$}     &
\multicolumn{1}{c }{$f({\rm sq},w=-1)$}&
\multicolumn{1}{c }{$c(w=-1)$}         &
\multicolumn{1}{c }{$A(w=-1)$}         &
\multicolumn{1}{c }{$B(w=-1)$}         \\
\hline
 2 &  1.0481659817 & $-$1.8395026016 & $-$0.2962391773 & $-$0.8240960988\\
 3 &  1.0486170833 & $-$2.0667855660 & $-$0.1728865530 & $-$0.7094169716\\
 4 &  1.0487861877 & $-$2.2333779695 & $-$0.0629303046 & $-$0.6281602055\\
 5 &  1.0487985937 & $-$2.2540969965 & $-$0.0473045479 & $-$0.6186355145\\
 6 &  1.0487835568 & $-$2.2152906621 & $-$0.0796357366 & $-$0.6352167690\\
 7 &  1.0487715509 & $-$2.1704302257 & $-$0.1200678725 & $-$0.6528623630\\
 8 &  1.0487645572 & $-$2.1343787723 & $-$0.1547235661 & $-$0.6658127726\\
 9 &  1.0487606108 & $-$2.1073401566 & $-$0.1821652813 & $-$0.6746153901\\
10 &  1.0487582804 & $-$2.0867372527 & $-$0.2040744739 & $-$0.6806453142\\
11 &  1.0487568178 & $-$2.0704483013 & $-$0.2221173626 & $-$0.6848921285\\
12 &  1.0487558491 & $-$2.0571283328 & $-$0.2374141788 & $-$0.6879523513\\
13 &  1.0487551802 & $-$2.0459596860 & $-$0.2506617512 & $-$0.6901835157\\
\hline\hline                                                        
\end{tabular}                                                       
\caption{\label{table_f_fits_c_sq_w=-1_bis}
Fits of $\real f_L(w=-1)$ on square-lattice strips
to the Ansatz \protect\reff{def_ansatz_f_tris}.
We show the estimates of $f({\rm sq},w=-1)$, $c(w=-1)$, $A(w=-1)$, and
$B(w=-1)$ obtained by fitting four consecutive data points with 
$L\geq L_{\rm min}$.
}
\end{table}

%
%
\def\jj{\phantom{$-$}}
\begin{table}
\centering
\begin{tabular}{r|c|rr}
\hline\hline
\multicolumn{1}{c|}{\hbox{}}                   &
\multicolumn{1}{c|}{$f_L=f + (c\pi/6)L^{-2}$}  &
\multicolumn{2}{c} {$f_L=f + (c\pi/6)L^{-2} + A L^{-4}$} \\
\cline{2-4}
\multicolumn{1}{c|}{$L_{\rm min}$}     &
\multicolumn{1}{c|}{$c(w=-1/4)$}       &
\multicolumn{1}{c }{$c(w=-1/4)$}       &
\multicolumn{1}{c} {$A(w=-1/4)$}       \\
\hline
2  &$-$1.7867317647&$-$1.5700367024 & $-$0.4538450772\\
3  &\jj0.5559211806&   0.5004616666 &    0.2613468027\\
4  &$-$1.6242104680&$-$1.4037258487 & $-$1.8471276270\\
5  &\jj0.5204270917&   0.5013187856 &    0.2501271410\\
6  &$-$1.5017190128&$-$1.3532367216 & $-$2.7988252525\\
7  &\jj0.5110679214&   0.5017424517 &    0.2392574198\\
8  &$-$1.4367580104&$-$1.3338122127 & $-$3.4497467921\\
9  &\jj0.5073837852&   0.5016573071 &    0.2428685331\\
10 &$-$1.3996975232&$-$1.3212758647 & $-$4.1061484386\\
11 &\jj0.5054907346&   0.5015180462 &    0.2516914702\\
12 &$-$1.3757353498&$-$1.3111049796 & $-$4.8730151126\\
13 &\jj0.5043623971&   0.5013945821 &    0.2626165842\\
14 &$-$1.3585885169&                &                \\
15 &\jj0.5036237409&                &                \\
\hline\hline
\end{tabular}
\caption{\label{table_f_fits_c_sq_w=-0.25}
Fits of $\real f_L(w=-1/4)$ on square-lattice strips
to the Ans\"atze \protect\reff{def_ansatz_f}/\protect\reff{def_ansatz_f_bis},
using the exact value \protect\reff{def_f_sq_w=-0.25} of $f({\rm sq},w=-1/4)$
in both Ans\"atze.
We show the estimates of $c(w=-1)$ and $A(w=-1)$
obtained by fitting one or two consecutive data points of the same parity.
}
\end{table}

%
%
\begin{table}
\centering
\begin{tabular}{cc|c|cc|c}
\hline\hline
\multicolumn{3}{c|} {}                                 &
\multicolumn{2}{|c|}{$\xi_1^{-1} + (2\pi x_{T,1})/L$} &
\multicolumn{1}{|c} {$(2\pi x_{T,1})/L$}              \\
\cline{4-6}
\multicolumn{1}{c} {$L$}            &
\multicolumn{1}{c} {$\xi_1^{-1}(L)$}&
\multicolumn{1}{c|}{$L_{\rm min}$}  &
\multicolumn{1}{|c}{$\xi_1^{-1}$}   &
\multicolumn{1}{c|}{$x_{T,1}$}      &
\multicolumn{1}{|c}{$x_{T,1}$}      \\
\hline
 3 &  0.8447211436 & 3 &$-0.04367$& 0.42418 & 0.40332  \\
 5 &  0.4893646433 & 5 &$-0.03794$& 0.41962 & 0.38942  \\
 7 &  0.3387055144 & 7 &$-0.02746$& 0.40793 & 0.37735  \\
 9 &  0.2573363732 &   &          &         & 0.36861  \\
\hline
$2n$& 0            &   &          &         &          \\
\hline\hline
\end{tabular}
\caption{\label{table_inverse_xi1_sq}
  Inverse correlation length $\xi_1^{-1}(L)$ for the square lattice at $w=-1/4$.
  We also show the results of fitting the odd-width data to the
  Ansatz \protect\reff{def_xi_critical_sq}, either  
  with $\xi_1^{-1}$ variable (columns 4--5) or $\xi_1^{-1} = 0$ fixed
  (column 6).
  The last row shows that for all even widths $L$, we have $\xi_1^{-1}(L)=0$.
}
\end{table}

%
%
\begin{table}
\centering
\begin{tabular}{cc|c|cc|c}
\hline\hline
\multicolumn{3}{c|}{}                                 &
\multicolumn{2}{|c|}{$\xi_2^{-1} + (2\pi x_{T,2})/L$} &
\multicolumn{1}{|c} {$(2\pi x_{T,2})/L$}              \\
\cline{4-6}
\multicolumn{1}{c}  {$L$}            &
\multicolumn{1}{c}  {$\xi_2^{-1}(L)$}&
\multicolumn{1}{c|} {$L_{\rm min}$}  &
\multicolumn{1}{|c} {$\xi_2^{-1}$}   &
\multicolumn{1}{c|} {$x_{T,2}$}      &
\multicolumn{1}{|c} {$x_{T,2}$}      \\
\hline
 3 &  2.1006084199 &  3 &\kkk 0.51802  & 0.75563 & 1.00297 \\
 5 &  1.4675739282 &  5 &\kkk 0.19309  & 1.01421 & 1.16786 \\
 7 &  1.1034345588 &  7 &\kkk 0.06489  & 1.15703 & 1.22932 \\
 9 &  0.8726454559 &  9 &              &         & 1.24997 \\
\hline
 4 &  0.7229219112 &  4 &\kkk  0.09380 & 0.40051 & 0.46023 \\
 6 &  0.5132143892 &  6 &    $-0.01991$& 0.50910 & 0.49008 \\
 8 &  0.3799322185 &  8 &    $-0.04373$& 0.53942 & 0.48374 \\
10 &  0.2952007610 & 10 &              &         & 0.46983 \\
\hline\hline
\end{tabular}
\caption{\label{table_inverse_xi2_sq}
  Inverse correlation length $\xi^{-1}_2(L)$ for the square lattice at $w=-1/4$.
  We also show the results of fitting the odd-width data to the
  Ansatz \protect\reff{def_xi_critical_sq},
  either with $\xi^{-1}$ variable (columns 4--5) or $\xi^{-1} = 0$ fixed
  (column 6). The fits are done separately for $L$ odd and $L$ even.
}
\end{table}


%
%
\begin{table}
\centering
\begin{tabular}{rcrrr}
\hline\hline
\multicolumn{1}{c}{$O$}              &
\multicolumn{1}{c}{$L_{\rm min}$}    &
\multicolumn{1}{c}{$w_0$}            &
\multicolumn{1}{c}{$A$}              &
\multicolumn{1}{c}{$\Delta$}         \\
\hline
$w_{0-}$ & 2 & $-0.176711$ & $-0.679548$ &  3.212914  \\
         & 4 & $-0.175667$ & $-0.303430$ &  2.541774  \\
         & 6 & $-0.175334$ & $-0.183161$ &  2.204638  \\
\hline
$w_{0Q}$ & 2 & $-0.178005$ & $-0.371851$ &  2.978114  \\
         & 3 & $-0.175706$ & $-0.222577$ &  2.373516  \\
         & 4 & $-0.175505$ & $-0.205877$ &  2.299951  \\
         & 5 & $-0.175380$ & $-0.189033$ &  2.231829  \\
         & 6 & $-0.175345$ & $-0.182054$ &  2.205216  \\
         & 7 & $-0.175323$ & $-0.175936$ &  2.183217  \\
         & 8 & $-0.175311$ & $-0.171338$ &  2.167416  \\
\hline
$w_{0+}$ & 2 & $-0.173481$ & $-0.080930$ &  1.516018  \\
         & 4 & $-0.174992$ & $-0.120997$ &  1.925676  \\
         & 6 & $-0.175316$ & $-0.171962$ &  2.171007  \\
\hline\hline
\end{tabular}
\caption{\label{table_fits_tri}
   Fits for $w_0$ on the triangular lattice,
   using the Ansatz $w_{0j}(L) = w_0 + A L^{-\Delta}$ ($j=+,-,Q$).
   Each fit for $w_{0\pm}$ is based on data points with
   $L=L_{\rm min}, L_{\rm min}+2, L_{\rm min}+4$;
   fits for $w_{0Q}$ are based on data points with
   $L=L_{\rm min}, L_{\rm min}+1, L_{\rm min}+2$.
}
\end{table}

%
%
\begin{table}
\centering
\begin{tabular}{rcrrr}
\hline\hline
\multicolumn{1}{c}{$O$}              &
\multicolumn{1}{c}{$L_{\rm min}$}    &
\multicolumn{1}{c}{$w_0$}            &
\multicolumn{1}{c}{$A_2$}            &
\multicolumn{1}{c}{$A_4$}           \\
\hline
$w_{0-}$ & 2 & $-0.175684$  &  $-0.091446$  &  $-0.823270$\\
         & 4 & $-0.175347$  &  $-0.108955$  &  $-0.629315$\\
         & 6 & $-0.175247$  &  $-0.118958$  &  $-0.398856$\\
\hline
$w_{0Q}$ & 2 & $-0.176194$  &  $-0.101075$  &  $-0.379755$\\
         & 3 & $-0.175074$  &  $-0.129083$  &  $-0.218428$\\
         & 4 & $-0.175180$  &  $-0.124719$  &  $-0.260998$\\
         & 5 & $-0.175202$  &  $-0.123381$  &  $-0.280737$\\
         & 6 & $-0.175230$  &  $-0.121007$  &  $-0.330011$\\
         & 7 & $-0.175245$  &  $-0.119307$  &  $-0.377189$\\
         & 8 & $-0.175255$  &  $-0.117857$  &  $-0.429037$\\
\hline
$w_{0+}$ & 2 & $-0.174942$  &  $-0.144129$  &  $ 0.147137$\\
         & 4 & $-0.175066$  &  $-0.137675$  &  $ 0.075637$\\
         & 6 & $-0.175242$  &  $-0.120065$  &  $-0.330093$\\
\hline\hline
\end{tabular}
\caption{\label{table_fits_tri_bis}
   Fits for $w_0$ on the triangular lattice,
   using the Ansatz $w_{0j}(L) = w_0 + A_2 L^{-2} + A_4 L^{-4}$ ($j=+,-,Q$).
   Each fit for $w_{0\pm}$ is based on data points with
   $L=L_{\rm min}, L_{\rm min}+2, L_{\rm min}+4$; and those for $w_{0Q}$
   are based on data points with
   $L=L_{\rm min}, L_{\rm min}+1, L_{\rm min}+2$.
}
\end{table}

%
%
\begin{table}
\centering
\begin{tabular}{r|cc|c}
\hline\hline
                                                       &
\multicolumn{2}{|c|}{$\real w_B(L)= c_1\log L + c_2$} &
\multicolumn{1}{|c} {$\real w_B(L)= c_1\log L + c_3 \log\log L + c_2$}\\
\hline
\multicolumn{1}{c|}{$L_{\rm min}$}&
\multicolumn{1}{|c}{$c_1$}        &
\multicolumn{1}{c|}{$c_2$}        &
\multicolumn{1}{|c}{$c_2$}        \\
\hline
 2  &  0.265812 & $-0.302108$  &$-0.342615$ \\
 3  &  0.239695 & $-0.273416$  &$-0.304290$ \\
 4  &  0.233643 & $-0.265025$  &$-0.293266$ \\
 5  &  0.232791 & $-0.263654$  &$-0.288928$ \\
 6  &  0.233405 & $-0.264754$  &$-0.286884$ \\
 7  &  0.234402 & $-0.266694$  &$-0.285815$ \\
 8  &  0.235446 & $-0.268865$  &$-0.285226$ \\
 9  &  0.236434 & $-0.271036$  &$-0.284900$ \\
10  &           &              &$-0.284730$ \\
\hline\hline                     
\end{tabular}

\bigskip
(a)
\bigskip
\bigskip

\begin{tabular}{r|ccc}
\hline\hline
                                                               &
\multicolumn{3}{|c}{$\imag w_B(L) = \imag w_B + A L^{-\Delta}$}\\
\hline
\multicolumn{1}{c|}{$L_{\rm min}$} &
\multicolumn{1}{|c}{$\imag w_B$}   &
\multicolumn{1}{c} {$A$}           &
\multicolumn{1}{c} {$\Delta$}      \\
\hline
 2  &   0.369334   & $-0.205690$   &  0.810857  \\
 3  &   0.398666   & $-0.215786$   &  0.582965  \\
 4  &   0.401549   & $-0.216752$   &  0.564881  \\
 5  &   0.398620   & $-0.216574$   &  0.585571  \\
 6  &   0.396188   & $-0.217407$   &  0.605898  \\
 7  &   0.394800   & $-0.218515$   &  0.619295  \\
 8  &   0.394137   & $-0.219347$   &  0.626437  \\
\hline\hline
\end{tabular}

\bigskip
(b)

\caption{\label{table_fits_wB_tri}
   Fits for $w_B$ on the triangular lattice.
   (a) For the real part, we use the Ans\"atze
   $\real w_B(L) = c_1\log L + c_2$ with $c_1$ and $c_2$ free,
   and $\real w_B(L) = c_1\log L + c_3 \log\log L + c_2$
   with fixed $c_1=\sqrt{3}/(2\pi)$ and $c_3=-1/(2\sqrt{3}\pi)$
   [see \protect\reff{def_ansatz_Re_wB_bis_tri} and surrounding text].
   (b) For the imaginary part, we use
   $\imag w_B(L) = \imag w_B + A L^{-\Delta}$.
   Two-parameter fits are based on data points with
   $L=L_{\rm min}, L_{\rm min}+1$; three-parameter fits are based on
   data points with $L=L_{\rm min}, L_{\rm min}+1, L_{\rm min}+2$.
}
\end{table}

%
%
\begin{table}
\centering
\begin{tabular}{rrr}
\hline\hline
\multicolumn{1}{c}{$L$}   &
\multicolumn{1}{c}{$\real f_L(w=-1)$}    &
\multicolumn{1}{c}{$\real f_L(w=1)$}   \\
\hline
2  &  1.2070144097 & 1.5619770404  \\
3  &  1.3954889413 & 1.6442216134  \\
4  &  1.4579588323 & 1.6714852756  \\
5  &  1.4861038633 & 1.6835902358  \\
6  &  1.5011384591 & 1.6899375329  \\
7  &  1.5100964902 & 1.6936439914  \\
8  &  1.5158576372 & 1.6959785719  \\
9  &  1.5197784405 & 1.6975342156  \\
10 &  1.5225658002 & 1.6986169734  \\
11 &  1.5246173608 & 1.6993972288  \\
12 &  1.5261706619 & 1.6999756628  \\
13 &  1.5273746626 & 1.7004147101  \\
14 &               & 1.7007546649  \\
15 &               & 1.7010224200  \\
\hline\hline
\end{tabular}
\caption{\label{table_f_tri}
Values for the quantities $\real f_L(w)$ \protect\reff{def_fm} on
triangular-lattice strips of width $L$, infinitely length and cylindrical
boundary conditions for $w=\pm 1$.
}
\end{table}

%
%
\begin{table}
\centering
\begin{tabular}{rrrr}
\hline\hline
\multicolumn{1}{c}{$L_{\rm min}$}         &
\multicolumn{1}{c}{$\real f({\rm tri},w=-1)$}    &
\multicolumn{1}{c}{$A(w=-1)$}       &
\multicolumn{1}{c}{$\Delta(w=-1)$}  \\
\hline
2 &  1.5308739278 & $-$1.4384536794 &  2.1510787148 \\
3 &  1.5330939357 & $-$1.3873709566 &  2.1033613397 \\
4 &  1.5337115052 & $-$1.3570788682 &  2.0815319808 \\
5 &  1.5339268515 & $-$1.3384597021 &  2.0701440440 \\
6 &  1.5340208601 & $-$1.3256428931 &  2.0631760539 \\
7 &  1.5340693211 & $-$1.3159780873 &  2.0583757793 \\
8 &  1.5340972882 & $-$1.3082485775 &  2.0548049206 \\
9 &  1.5341147600 & $-$1.3018238947 &  2.0520093666 \\
10&  1.5341263362 & $-$1.2963375714 &  2.0497401388 \\
11&  1.5341343629 & $-$1.2915577847 &  2.0478477572 \\
\hline\hline
\multicolumn{4}{c}{\quad} \\[-1mm]
\hline\hline
\multicolumn{1}{c}{$L_{\rm min}$}          &
\multicolumn{1}{c}{$\real f({\rm tri},w=1)$}    &
\multicolumn{1}{c}{$A(w=1)$}       &
\multicolumn{1}{c}{$\Delta(w=1)$}  \\
\hline
2   & 1.7033149783 & $-$0.6276028679  & 2.1507031799 \\
3   & 1.7031440137 & $-$0.6317621745  & 2.1593529516 \\
4   & 1.7030024774 & $-$0.6399100527  & 2.1718288870 \\
5   & 1.7028803279 & $-$0.6534158534  & 2.1887282099 \\
6   & 1.7027905253 & $-$0.6704079915  & 2.2069423251 \\
7   & 1.7027253709 & $-$0.6899461332  & 2.2253789191 \\
8   & 1.7026770177 & $-$0.7116977010  & 2.2437647939 \\
9   & 1.7026401520 & $-$0.7355754454  & 2.2620579091 \\
10  & 1.7026113515 & $-$0.7615959157  & 2.2802754383 \\
11  & 1.7025883787 & $-$0.7898291690  & 2.2984470780 \\
12  & 1.7025697290 & $-$0.8203795382  & 2.3166023025 \\
13  & 1.7025543610 & $-$0.8533775965  & 2.3347671232 \\
\hline\hline
\end{tabular}
\caption{\label{table_f_fits_tri}
Fits of $\real f_L(w)$ on triangular-lattice strips
to the Ansatz~\protect\reff{eq.Ansatz.f.Delta}. For each value of $w=\pm 1$,
we show the estimators of $\real f({\rm tri},w)$, $A(w)$, and
$\Delta(w)$ obtained by fitting three consecutive data points
with $L\geq L_{\rm min}$.
}
\end{table}

%
%
\begin{table}
\centering
\begin{tabular}{c|rr|rrr}
\hline\hline
\multicolumn{1}{c|}{\hbox{}}                   &
\multicolumn{2}{c|}{$f_L=f + (c\pi\sqrt{3}/12)L^{-2}$}  &
\multicolumn{3}{c} {$f_L=f + (c\pi\sqrt{3}/12)L^{-2} + A L^{-4}$} \\
\cline{2-6}
\multicolumn{1}{c|}{$L_{\rm min}$}     &
\multicolumn{1}{c }{$f({\rm tri},w=-1)$}&
\multicolumn{1}{c|}{$c(w=-1)$}         &
\multicolumn{1}{c }{$f({\rm tri},w=-1)$}&
\multicolumn{1}{c }{$c(w=-1)$}         &
\multicolumn{1}{c} {$A(w=-1)$}         \\
\hline
2 &1.5462685665&$-$2.9926499123&1.5356134959&$-$2.6871786001&$-$0.3835825420\\
3 &1.5382772635&$-$2.8340398079&1.5349369676&$-$2.6498796414&$-$0.4810026139\\
4 &1.5361394741&$-$2.7586078037&1.5346428114&$-$2.6232826442&$-$0.5986650932\\
5 &1.5353079949&$-$2.7127659475&1.5344818361&$-$2.6016275649&$-$0.7435428479\\
6 &1.5349033457&$-$2.6806402993&1.5343868622&$-$2.5838245327&$-$0.9110768209\\
7 &1.5346773842&$-$2.6562227934&1.5343273821&$-$2.5690020579&$-$1.0976063455\\
8 &1.5345391118&$-$2.6367069990&1.5342881192&$-$2.5564469074&$-$1.3011455956\\
9 &1.5344487544&$-$2.6205664178&1.5342610311&$-$2.5456343907&$-$1.5205586626\\
10&1.5343866971&$-$2.6068808277&1.5342416477&$-$2.5361873825&$-$1.7550985119\\
11&1.5343423765&$-$2.5950541591&1.5342273504&$-$2.5278319025&$-$2.0042150985\\
12&1.5343097063&$-$2.5846792515&            &               &               \\
\hline\hline
\end{tabular}
\caption{\label{table_f_fits_c_tri_w=-1}
Fits of $\real f_L(w=-1)$ on triangular-lattice strips
to the Ans\"atze \protect\reff{def_ansatz_f}/\protect\reff{def_ansatz_f_bis}.
We show the estimates of $f({\rm sq},w=-1)$, $c(w=-1)$, and $A(w=-1)$
obtained by fitting two consecutive data points with $L\geq L_{\rm min}$.
}
\end{table}

%
%
\begin{table}
\centering
\begin{tabular}{r|rrrrr}
\hline\hline
\multicolumn{1}{c|}{$L_{\rm min}$}     &
\multicolumn{1}{c }{$f({\rm tri},w=-1)$}&
\multicolumn{1}{c }{$c(w=-1)$}         &
\multicolumn{1}{c }{$A(w=-1)$}         &
\multicolumn{1}{c }{$B(w=-1)$}         \\
\hline
 2 &  1.5342817466 & $-$2.2815122109 & $-$0.0987125628 & $-$0.2264608061\\
 3 &  1.5342534868 & $-$2.2650710611 & $-$0.1064401573 & $-$0.2336450376\\
 4 &  1.5342099124 & $-$2.2155031379 & $-$0.1347733865 & $-$0.2545830616\\
 5 &  1.5341878304 & $-$2.1729190238 & $-$0.1625865218 & $-$0.2715365779\\
 6 &  1.5341775460 & $-$2.1422717553 & $-$0.1846992056 & $-$0.2828772076\\
 7 &  1.5341723071 & $-$2.1196678362 & $-$0.2023424141 & $-$0.2905771445\\
 8 &  1.5341693702 & $-$2.1021864612 & $-$0.2168955988 & $-$0.2960154919\\
 9 &  1.5341675995 & $-$2.0881779553 & $-$0.2292081839 & $-$0.2999650620\\
10 &  1.5341664720 & $-$2.0766679965 & $-$0.2398080978 & $-$0.3028824071\\
\hline\hline                                                        
\end{tabular}                                                       
\caption{\label{table_f_fits_c_tri_w=-1_bis}
Fits of $\real f_L(w=-1)$ on triangular-lattice strips
to the Ansatz \protect\reff{def_ansatz_f_tris}.
We show the estimates of $f({\rm tri},w=-1)$, $c(w=-1)$, $A(w=-1)$, and
$B(w=-1)$ obtained by fitting four consecutive data points with 
$L\geq L_{\rm min}$.
}
\end{table}

\clearpage
%
%
%
%

%
%
\clearpage

%
%
\clearpage

%
%
\clearpage

%
%
\clearpage

%
%
\clearpage

%
%
\clearpage

%
%
\clearpage

%
%
\clearpage

%
%
\clearpage

%
%
\clearpage

%
%
\clearpage

%
%
\clearpage

%
%
\clearpage
\addtocounter{figure}{-1}

%
%
\clearpage
\begin{figure}
\centering
\begin{tabular}{c}
\includegraphics[width=400pt]{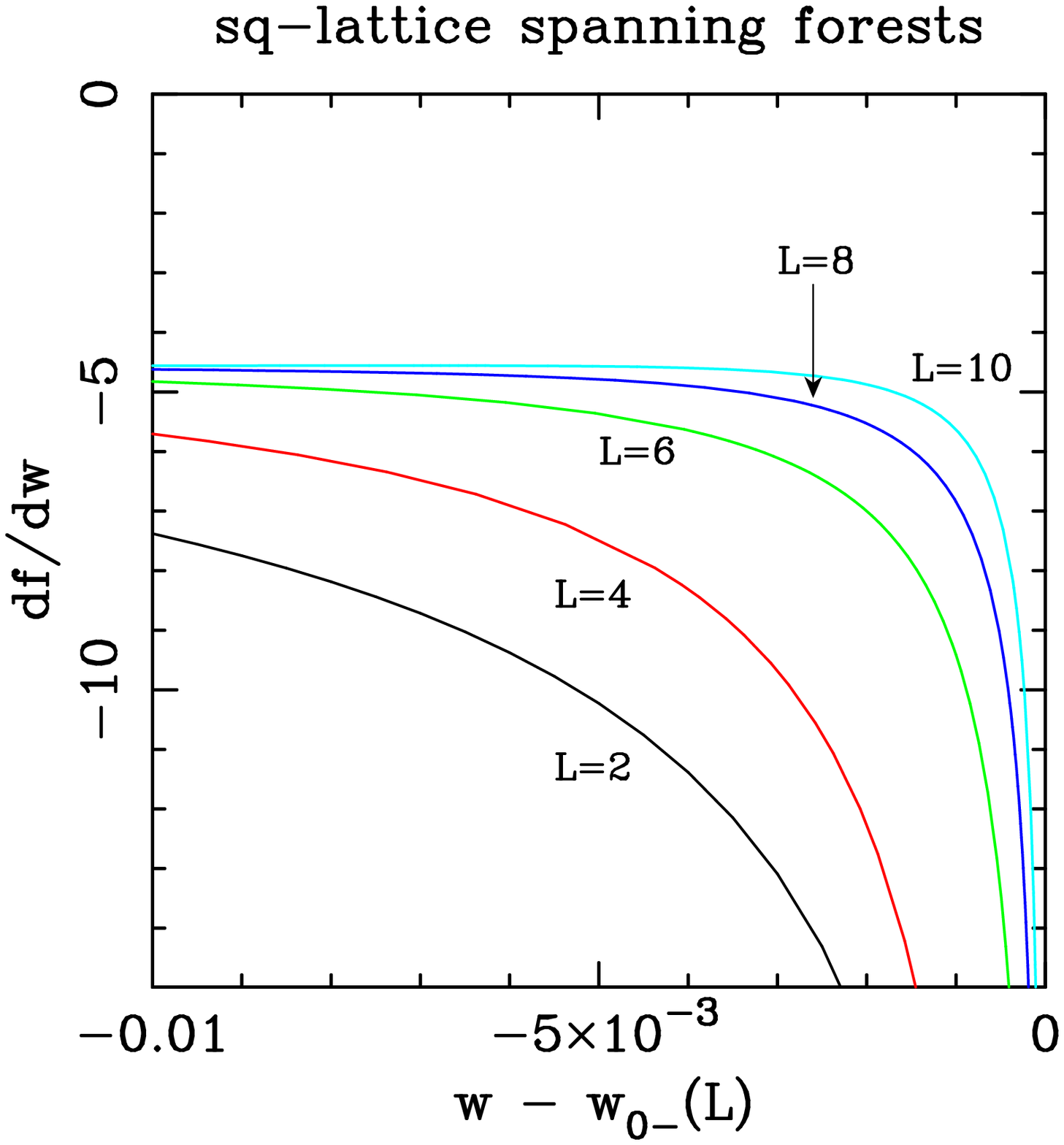} \\[2mm] 
\phantom{(aa)}\large{(b)}
\end{tabular}
\caption{\label{figure_Ec_sq_bis}
   First derivative $f'_L(w)$ of the square-lattice free energy
   for strips of widths $2\leq L \leq 10$ close to $w_0({\rm sq})=-1/4$.  
   (a) Curves for even (resp.\ odd) $L$ are shown in red (resp.\ blue).
   The violet dashed curve on the right corresponds to the
   Pad\'e approximant $[20,20]$ to our longest small-$w$ series.
   The pink dot-dashed curve corresponds to the derivative of the
   large-$w$ expansion \protect\reff{def_large_w_series},
   through order $w^{-2}$.
   The vertical brown dot-dot-dashed line marks the point $w_0=-1/4$.
   The vertical brown dot-dot-dashed lines near the bottom of the figure
   mark the position of the points $w_{0-}(L)$ for $L=6,8,10$.
   (b) Curves for strips of widths $L=2,4,6,8,10$
   are plotted versus $w-w_{0-}(L)$.
}
\end{figure}

%
%
\clearpage
\begin{figure}
\centering
\includegraphics[width=400pt]{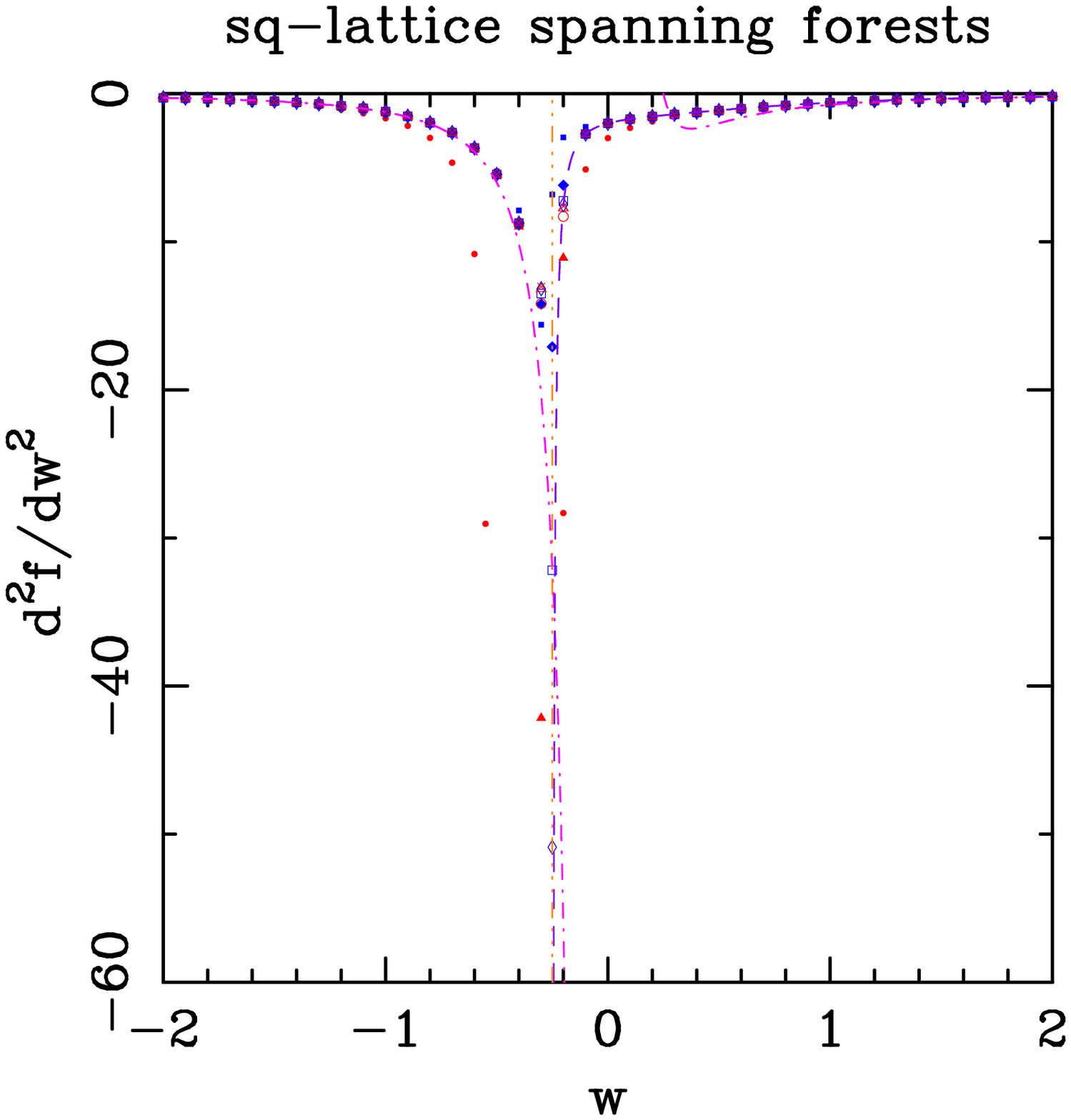}
\caption{\label{figure_CH_sq}
   Second derivative $f''_L(w)$ of the square-lattice free energy
   for strips of width
   $L=2$ ($\bullet$),
   $3$ ($\blacksquare$),
   $4$ ($\blacktriangle$), $L=5$ ($\blacklozenge$),
   $6$ ($\circ$),          $L=7$ ($\Box$),
   $8$ ($\triangle$),      $L=9$ ($\diamondsuit$), and
   $10$ ($\times$).
   Points with even (resp.\ odd) $L$ are shown in red (resp.\ blue).
   The violet dashed curve on the right corresponds to the
   Pad\'e approximant $[20,20]$ to our longest small-$w$ series.
   The pink dot-dashed curve corresponds to the second derivative of the
   large-$w$ expansion \protect\reff{def_large_w_series},
   through order $w^{-3}$.
   The vertical dot-dot-brown dashed line marks the point $w_0=-1/4$.
}
\end{figure}

%
%
\clearpage
\begin{figure}
\centering
\begin{tabular}{c}
\includegraphics[width=400pt]{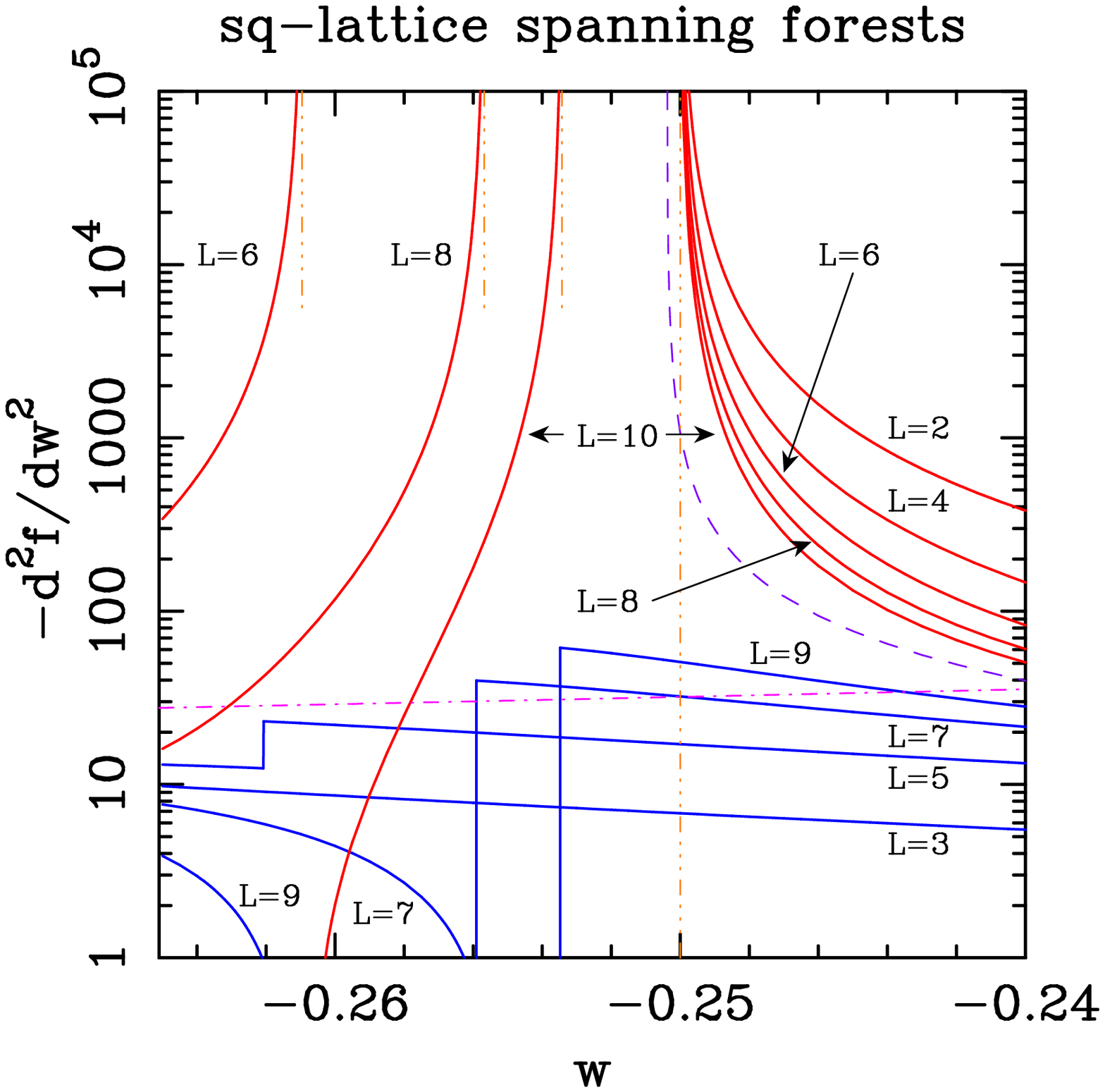} \\[2mm]
\phantom{(aa)}\large{(a)} 
\end{tabular}
\caption{\label{figure_CHc_sq}
   Second derivative $f''_L(w)$ of the square-lattice free energy
   for strips of widths $2\leq L \leq 10$ close to $w_0({\rm sq})=-1/4$.
   (a) Curves for even (resp.\ odd) $L$ are shown in red (resp.\ blue).
   The violet dashed curve on the right corresponds to the
   Pad\'e approximant $[20,20]$ to our longest small-$w$ series.
   The pink dot-dashed curve corresponds to the second derivative of the
   large-$w$ expansion \protect\reff{def_large_w_series},
   through order $w^{-3}$.
   The vertical brown dot-dot-dashed line marks the point $w_0=-1/4$.
   The vertical brown dot-dot-dashed lines near the top of the figure
   mark the position of the points $w_{0-}(L)$ for $L=6,8,10$.
   (b) Curves for strips of widths $L=2,4,6,8,10$
   are plotted versus $w-w_{0-}(L)$.
}
\end{figure}
\addtocounter{figure}{-1}

%
%
\clearpage
\begin{figure}
\centering
\begin{tabular}{c}
\includegraphics[width=400pt]{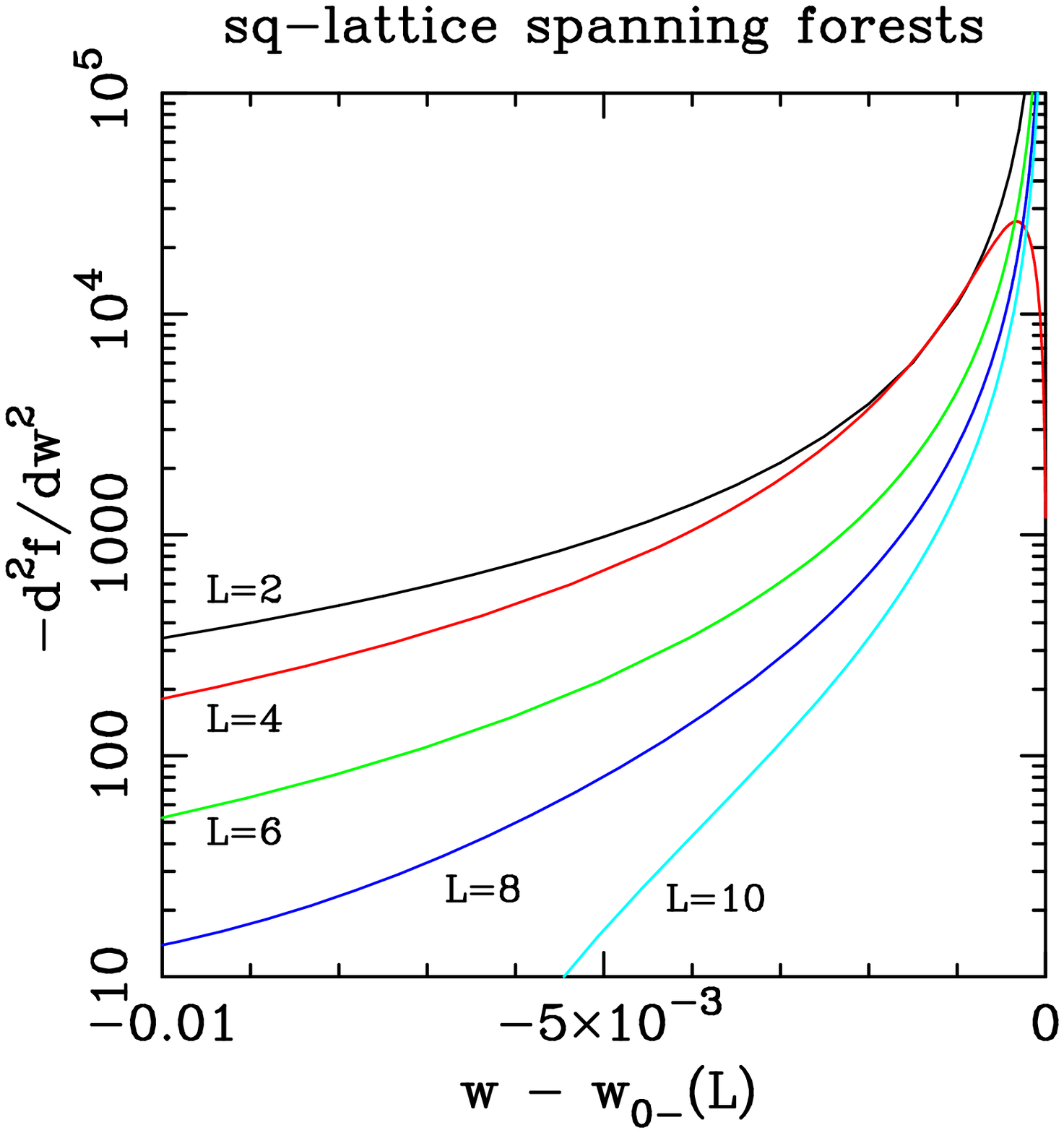}\\[2mm] 
\phantom{(aa)}\large{(b)} 
\end{tabular}
\caption{\label{figure_CHc_sq_bis}
   Second derivative $f''_L(w)$ of the square-lattice free energy
   for strips of widths $2\leq L \leq 10$ close to $w_0({\rm sq})=-1/4$.
   (a) Curves for even (resp.\ odd) $L$ are shown in red (resp.\ blue).
   The violet dashed curve on the right corresponds to the
   Pad\'e approximant $[20,20]$ to our longest small-$w$ series.
   The pink dot-dashed curve corresponds to the second derivative of the
   large-$w$ expansion \protect\reff{def_large_w_series},
   through order $w^{-3}$.
   The vertical brown dot-dot-dashed line marks the point $w_0=-1/4$.
   The vertical brown dot-dot-dashed lines near the top of the figure
   mark the position of the points $w_{0-}(L)$ for $L=6,8,10$.
   (b) Curves for strips of widths $L=2,4,6,8,10$
   are plotted versus $w-w_{0-}(L)$.
}
\end{figure}

%
%
\clearpage
\begin{figure}
\centering
\begin{tabular}{cc}
\includegraphics[width=200pt]{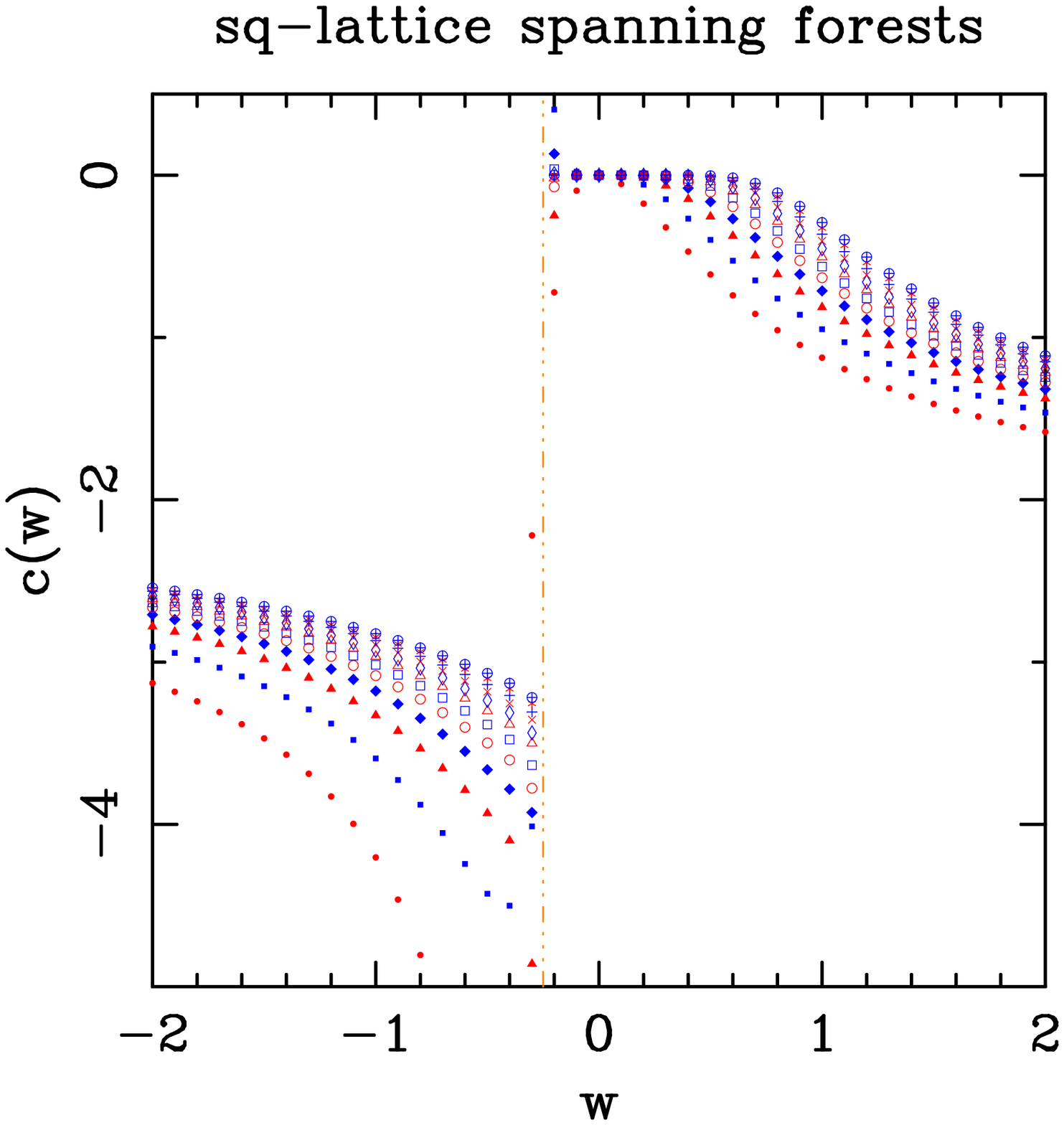} &
\includegraphics[width=200pt]{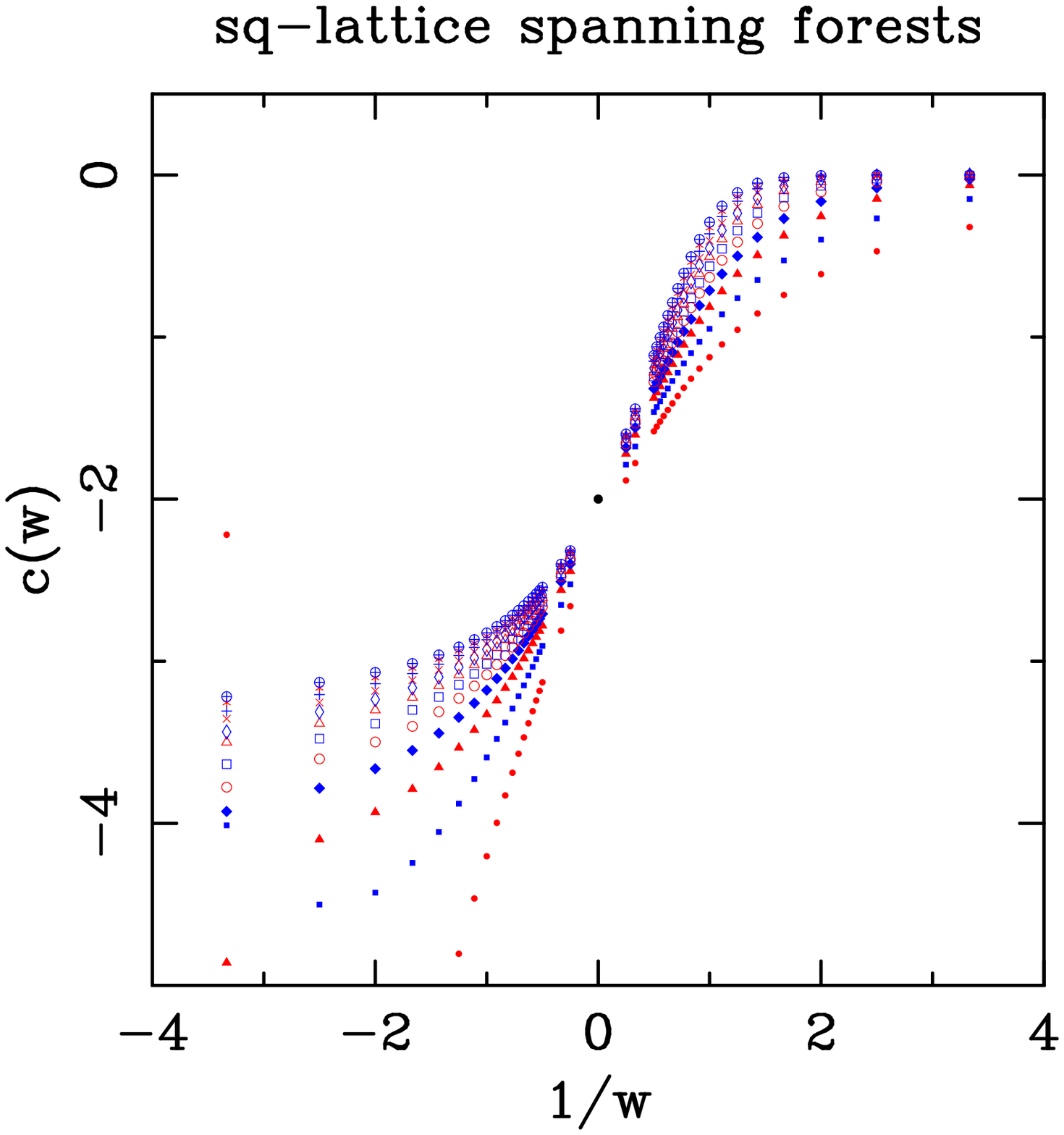}  \\[1mm]
   \phantom{(((a)}(a) & \phantom{(((a)}(b) \\[4mm]
\includegraphics[width=200pt]{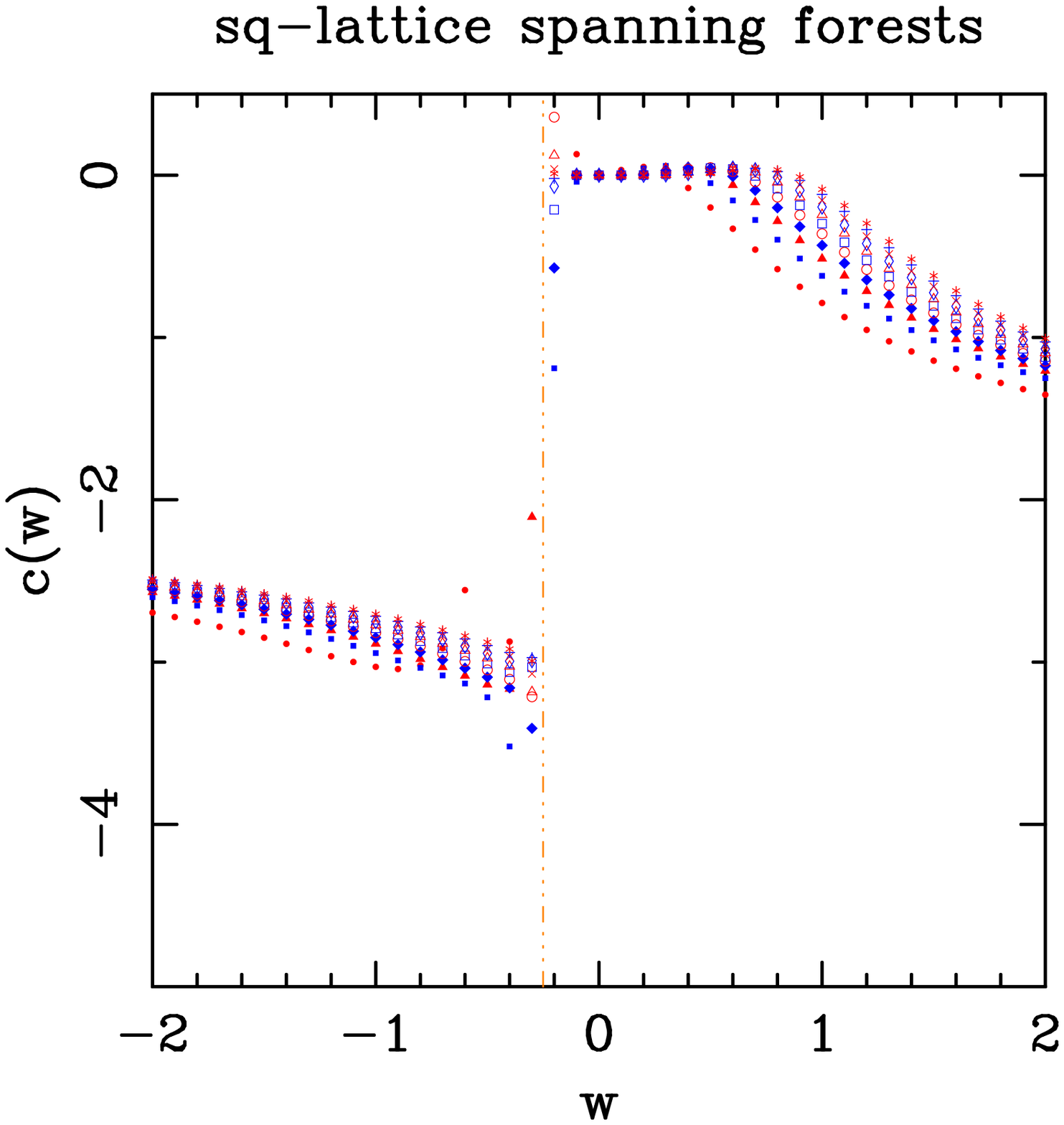} &
\includegraphics[width=200pt]{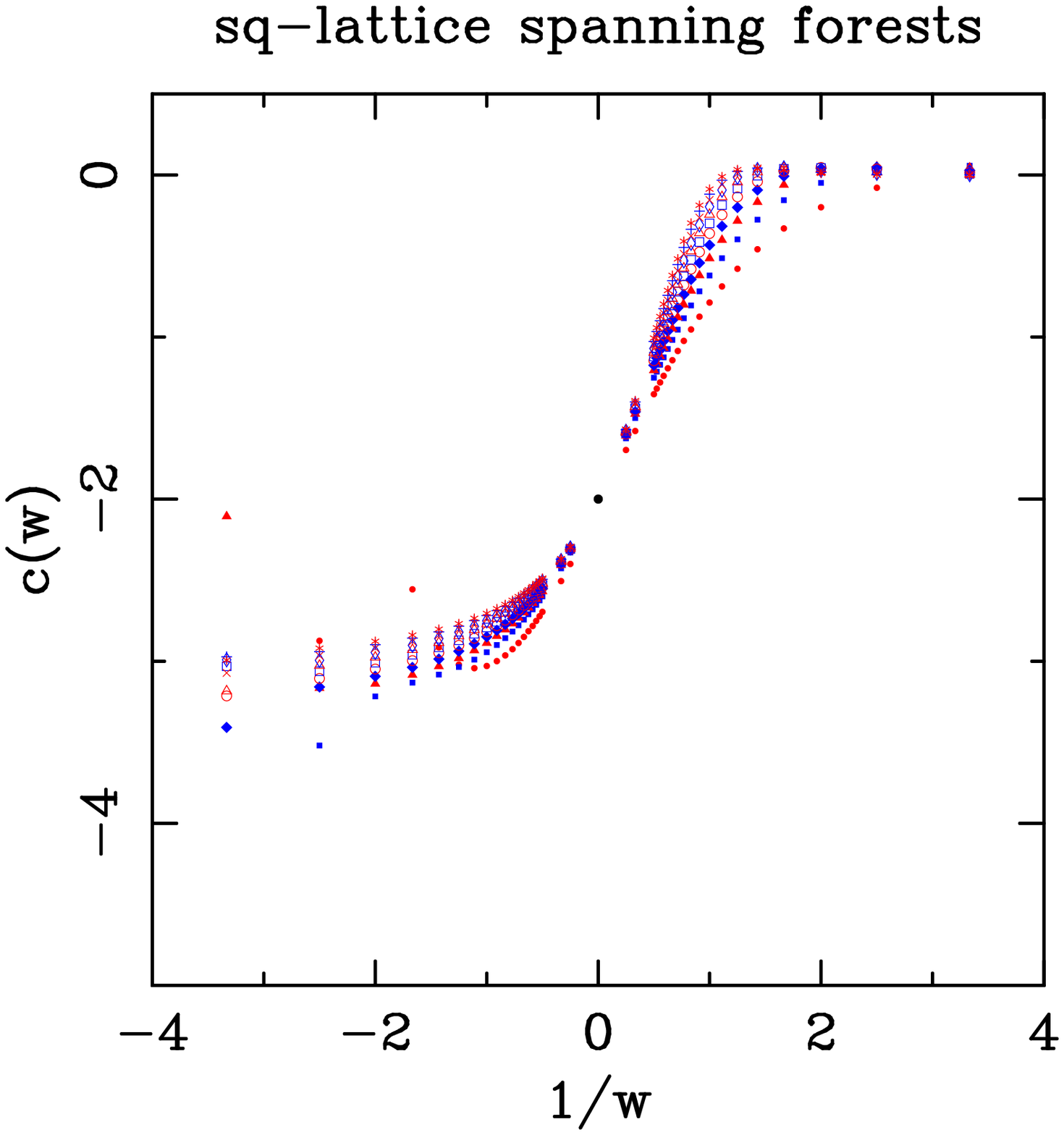}\\[1mm]
   \phantom{(((a)}(c) & \phantom{(((a)}(d) \\[4mm]
\end{tabular}
\caption{\label{figure_c_sq}
   Estimates for the square-lattice central charge $c(w)$
   obtained by fitting the free energy to the Ans\"atze (a,b)
   $\real f_L(w) = \real f(w) + [c(w)\pi/6] L^{-2}$, and
   (c,d) $\real f_L(w) = \real f(w) + [c(w)\pi/6] L^{-2} + AL^{-4}$.
   Fits are performed for
   $L_{\rm min}=2$ ($\bullet$), $3$ ($\blacksquare$),
   $4$ ($\blacktriangle$), $5$ ($\blacklozenge$),
   $6$ ($\circ$), $7$ ($\Box$),
   $8$ ($\triangle$), $9$ ($\diamondsuit$),
   $10$ ($\times$), $11$ ($+$),
   $12$ ($*$) and $13$ ($\oplus$).
   Points with even (resp.\ odd) $L$ are shown in red (resp.\ blue).
   The black dot at $1/w=0$, $c=-2$ marks the theoretical prediction.
   The vertical brown dot-dot-dashed line marks the point $w_0=-1/4$.
}
\end{figure}

%
%
\clearpage
\begin{figure}
\centering
\includegraphics[width=400pt]{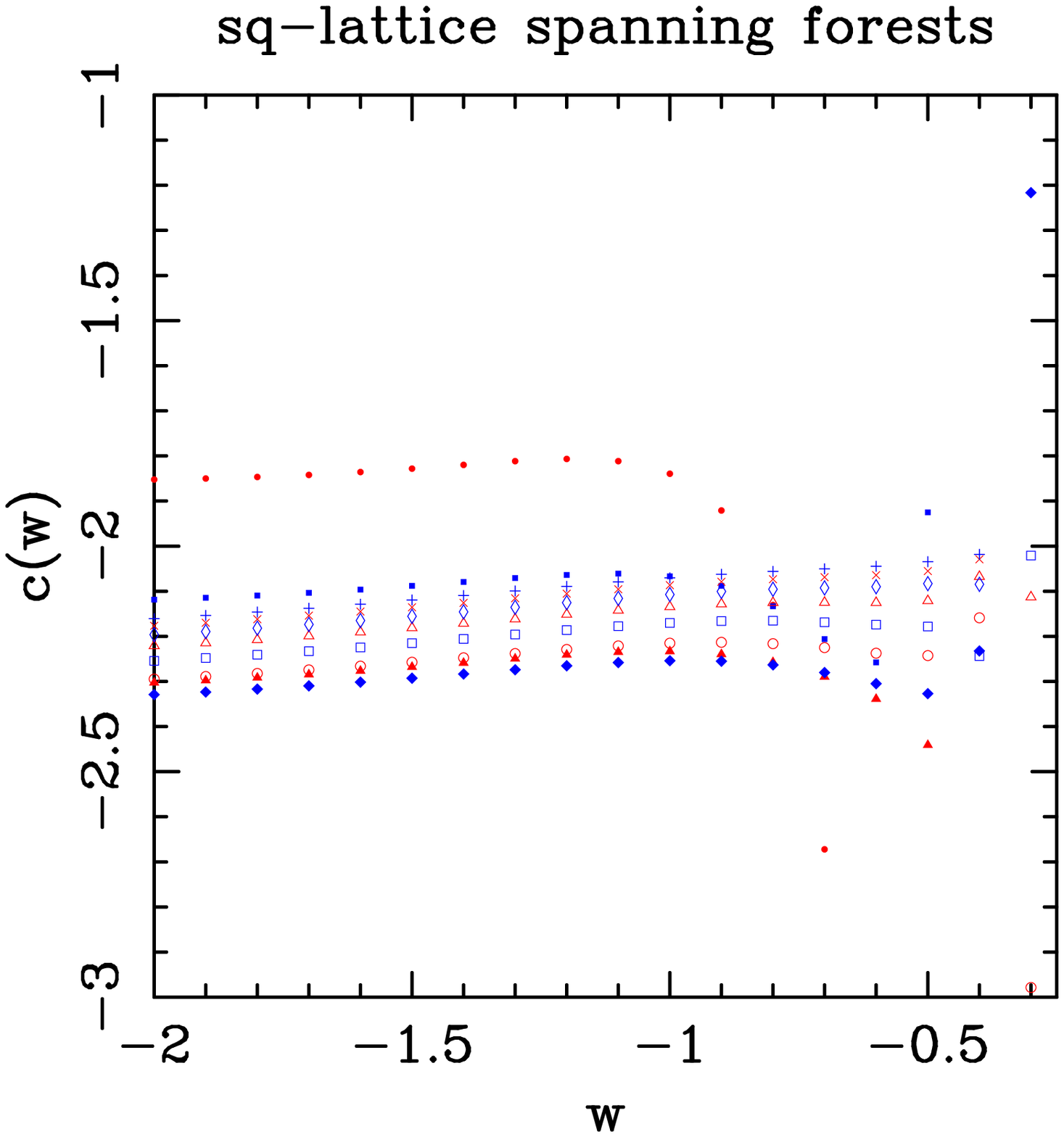} 
\caption{\label{figure_c_sq2}
   Estimates for the square-lattice central charge $c(w)$
   obtained by fitting the free energy to the Ansatz 
   $\real f_L(w) = \real f(w) + [c(w)\pi/6] L^{-2} + A \log\log L/(L^2 \log L)
    + B/(L^2 \log L)$. 
   Fits are performed for
   $L_{\rm min}=2$ ($\bullet$), $3$ ($\blacksquare$),
   $4$ ($\blacktriangle$), $5$ ($\blacklozenge$),
   $6$ ($\circ$), $7$ ($\Box$),
   $8$ ($\triangle$), $9$ ($\diamondsuit$),
   $10$ ($\times$), and $11$ ($+$).
   Points with even (resp.\ odd) $L$ are shown in red (resp.\ blue).
   For $w=-0.3$, to avoid parity effects, we have performed the fits 
   using data with $L=L_{\rm min},L_{\rm min}+2,L_{\rm min}+4,L_{\rm min}+6$.
}
\end{figure}

%
%
\clearpage
\begin{figure}
\centering
\begin{tabular}{c}
\includegraphics[width=400pt]{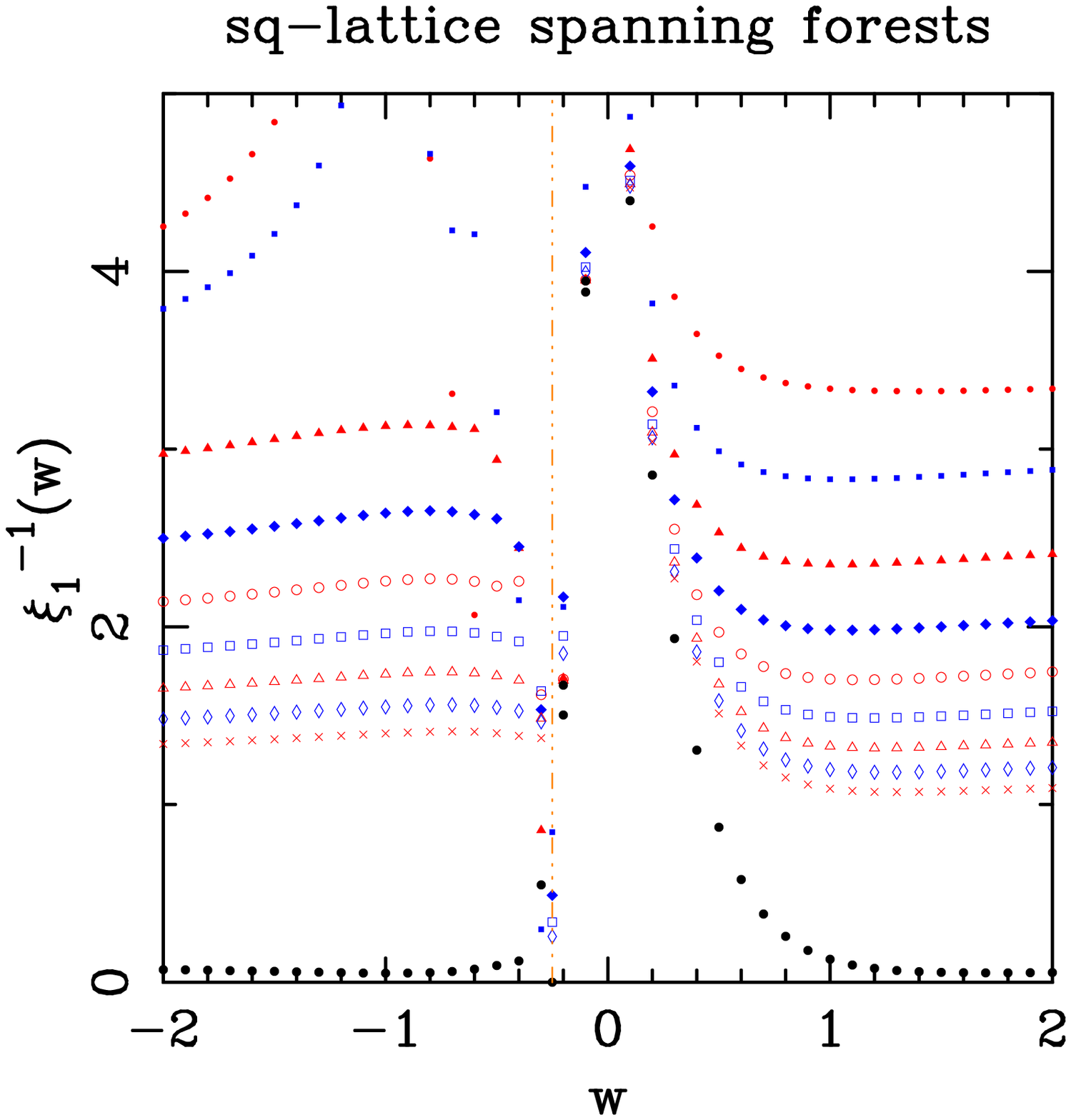}\\[2mm]
\phantom{(aaaaaaa)}\large{(a)}
\end{tabular}
\caption{\label{figure_xi1_sq}
   Values of the square-lattice inverse correlation length
   $\xi_j^{-1}(w) = \log |\lambda_\star/\lambda_j|$ for
   (a) $j=1$ and (b) $j=2$.
   Symbols indicate strip widths
   $L=2$ ($\bullet$),
   $3$ ($\blacksquare$),
   $4$ ($\blacktriangle$),
   $5$ ($\blacklozenge$),
   $6$ ($\circ$),
   $7$ ($\Box$),
   $8$ ($\triangle$),
   $9$ ($\diamondsuit$), and
   $10$ ($\times$).
   Points with even (resp.\ odd) $L$ are shown in red (resp.\ blue).
   The black solid circles ($\bullet$) correspond to the
   extrapolated infinite-volume limit of the finite-size data (see text).
   The vertical dot-dot-brown dashed line marks the point $w_0=-1/4$.
}
\end{figure}
\addtocounter{figure}{-1}

%
%
\clearpage
\begin{figure}
\centering
\begin{tabular}{c}
\includegraphics[width=400pt]{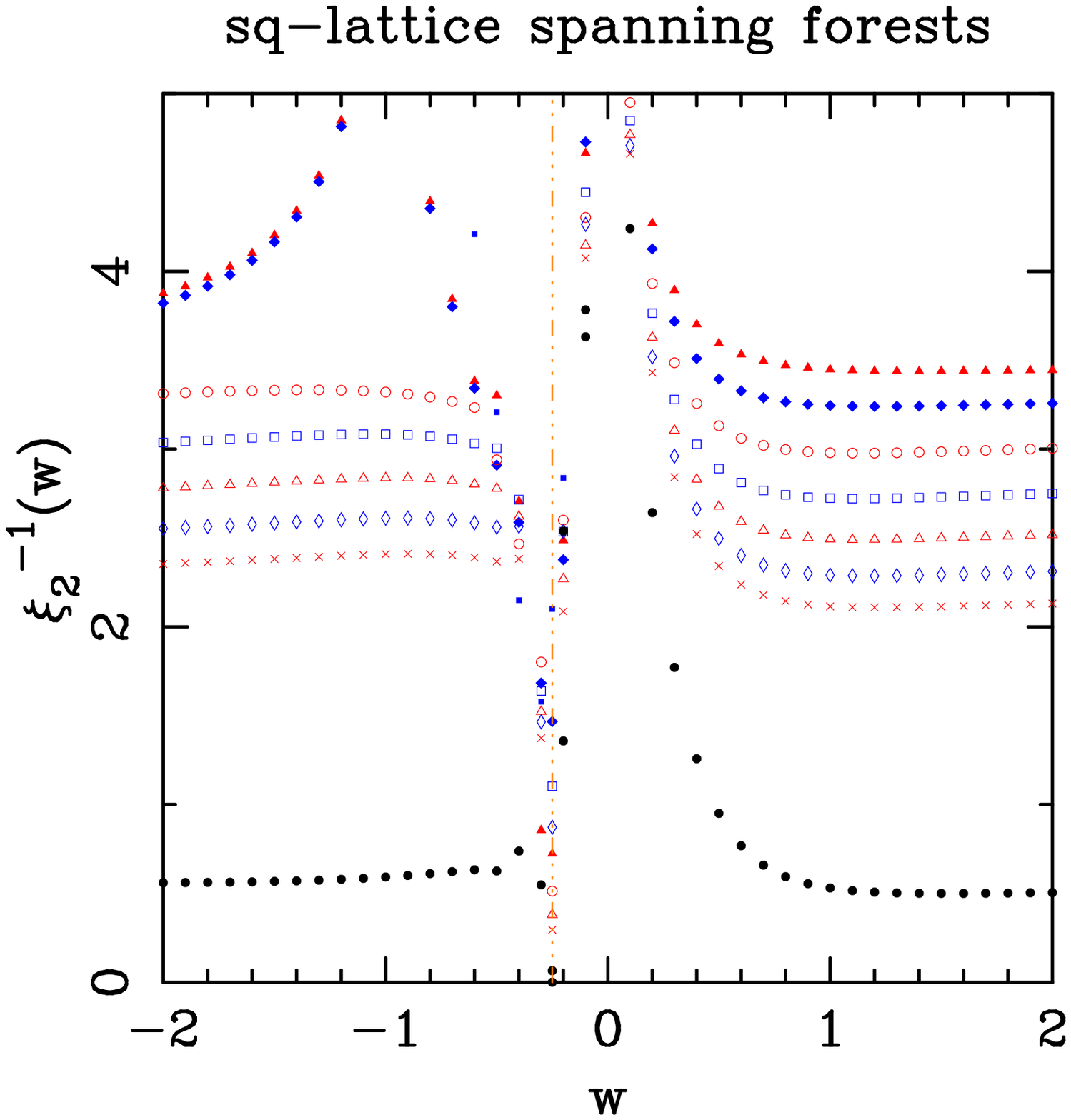}\\[2mm]
\phantom{(aaaaaaa)}\large{(b)}
\end{tabular}
\caption{\label{figure_xi2_sq}
   Values of the square-lattice inverse correlation length
   $\xi_j^{-1}(w) = \log |\lambda_\star/\lambda_j|$ for
   (a) $j=1$ and (b) $j=2$.
   Symbols indicate strip widths
   $L=2$ ($\bullet$),
   $3$ ($\blacksquare$),
   $4$ ($\blacktriangle$),
   $5$ ($\blacklozenge$),
   $6$ ($\circ$),
   $7$ ($\Box$),
   $8$ ($\triangle$),
   $9$ ($\diamondsuit$), and
   $10$ ($\times$).
   Points with even (resp.\ odd) $L$ are shown in red (resp.\ blue).
   The black solid circles ($\bullet$) correspond to the
   extrapolated infinite-volume limit of the finite-size data (see text).
   The vertical brown dot-dot-dashed line marks the point $w_0=-1/4$.
}
\end{figure}

%
%
\clearpage
\begin{figure}
\centering
\begin{tabular}{c}
\includegraphics[width=400pt]{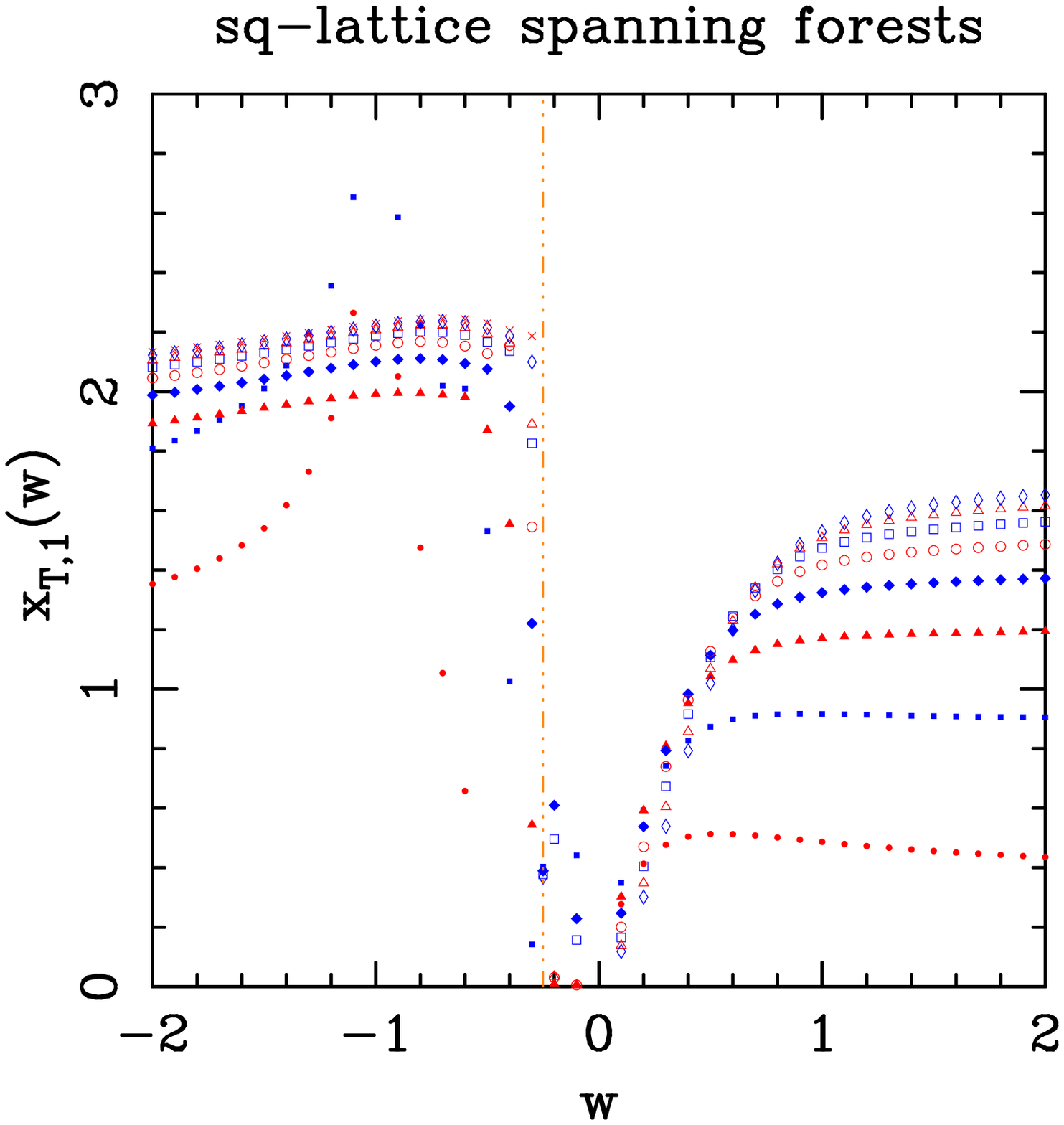}\\[2mm]
\phantom{(aaaaaaa)}\large{(a)}
\end{tabular}
\caption{\label{figure_x1_sq}
   Estimates for the square-lattice scaling dimension $x_{Tj}(w)$
   for (a) $j=1$ and (b) $j=2$,
   obtained by fitting the inverse correlation length to the Ansatz
   $\xi^{-1}_j(w) = \xi^{-1}_{j,\infty}(w)  + 2 \pi x_{Tj}(w) L^{-1}$.
   In the region $w \le -1/4$ we have fixed $\xi_{j,\infty}^{-1} = 0$;
   in the region $w > -1/4$ we have left it variable.
   Fits are performed for
   $L_{\rm min}=2$ ($\bullet$),
   $3$ ($\blacksquare$),
   $4$ ($\blacktriangle$),
   $5$ ($\blacklozenge$),
   $6$ ($\circ$),
   $7$ ($\Box$),
   $8$ ($\triangle$) and
   $9$ ($\diamondsuit$).
   Points with even (resp.\ odd) $L$ are shown in red (resp.\ blue).
   The vertical brown dot-dot-dashed line marks the point $w_0=-1/4$.
}
\end{figure}
\addtocounter{figure}{-1}

%
%
\clearpage
\begin{figure}
\centering
\begin{tabular}{c}
\includegraphics[width=400pt]{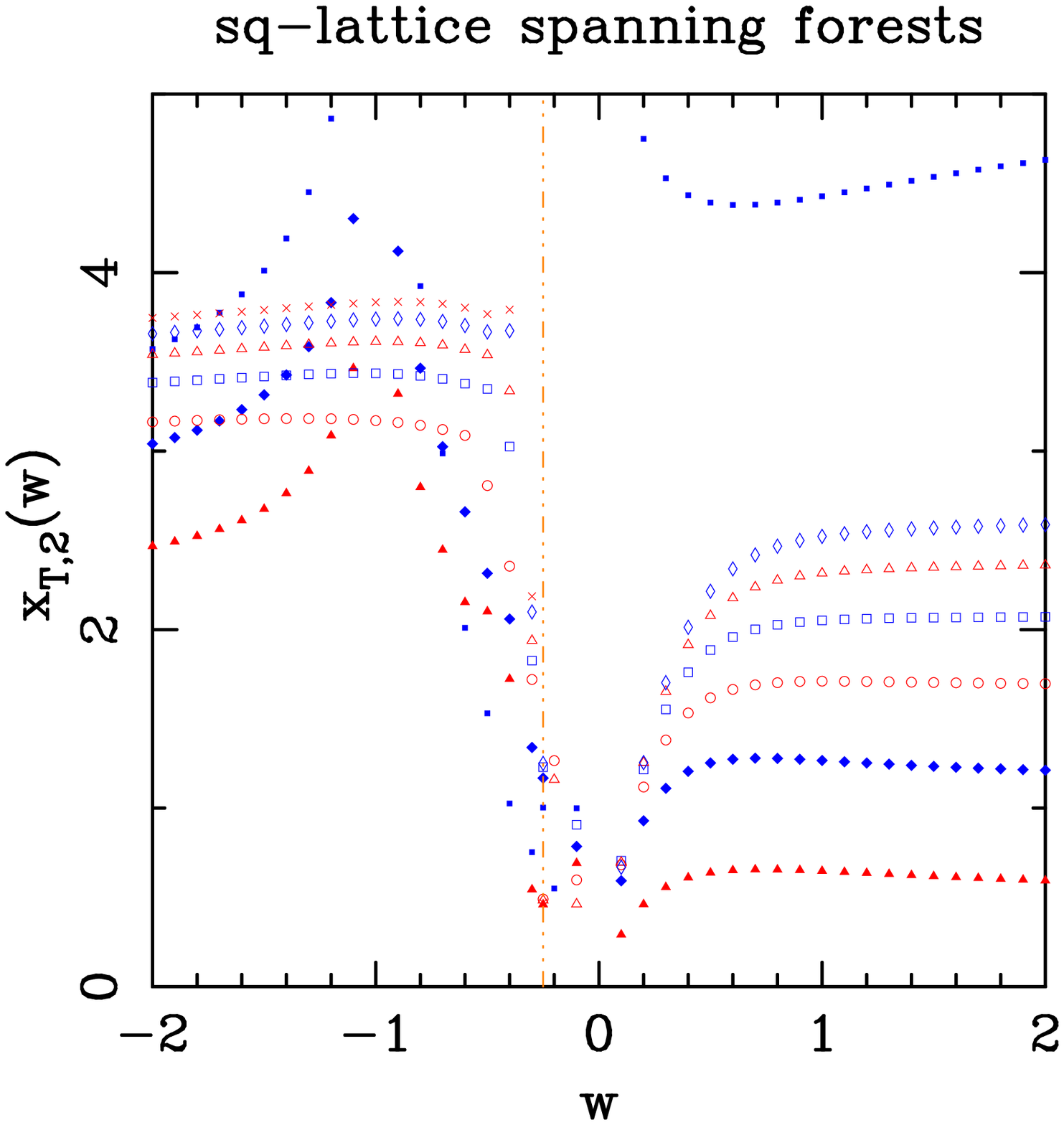}\\[2mm]
\phantom{(aaaaaaa)}\large{(b)}
\end{tabular}
\caption{\label{figure_x2_sq}
   Estimates for the square-lattice scaling dimension $x_{Tj}(w)$
   for (a) $j=1$ and (b) $j=2$,
   obtained by fitting the inverse correlation length to the Ansatz
   $\xi^{-1}_j(w) = \xi^{-1}_{j,\infty}(w)  + 2 \pi x_{Tj}(w) L^{-1}$.
   In the region $w \le -1/4$ we have fixed $\xi_{j,\infty}^{-1} = 0$;
   in the region $w > -1/4$ we have left it variable.
   Fits are performed for
   $L_{\rm min}=2$ ($\bullet$),
   $3$ ($\blacksquare$),
   $4$ ($\blacktriangle$),
   $5$ ($\blacklozenge$),
   $6$ ($\circ$),
   $7$ ($\Box$),
   $8$ ($\triangle$) and
   $9$ ($\diamondsuit$).
   Points with even (resp.\ odd) $L$ are shown in red (resp.\ blue).
   The vertical brown dot-dot-dashed line marks the point $w_0=-1/4$.
}
\end{figure}

%
%
\clearpage
\begin{figure}
\centering
\includegraphics[width=400pt]{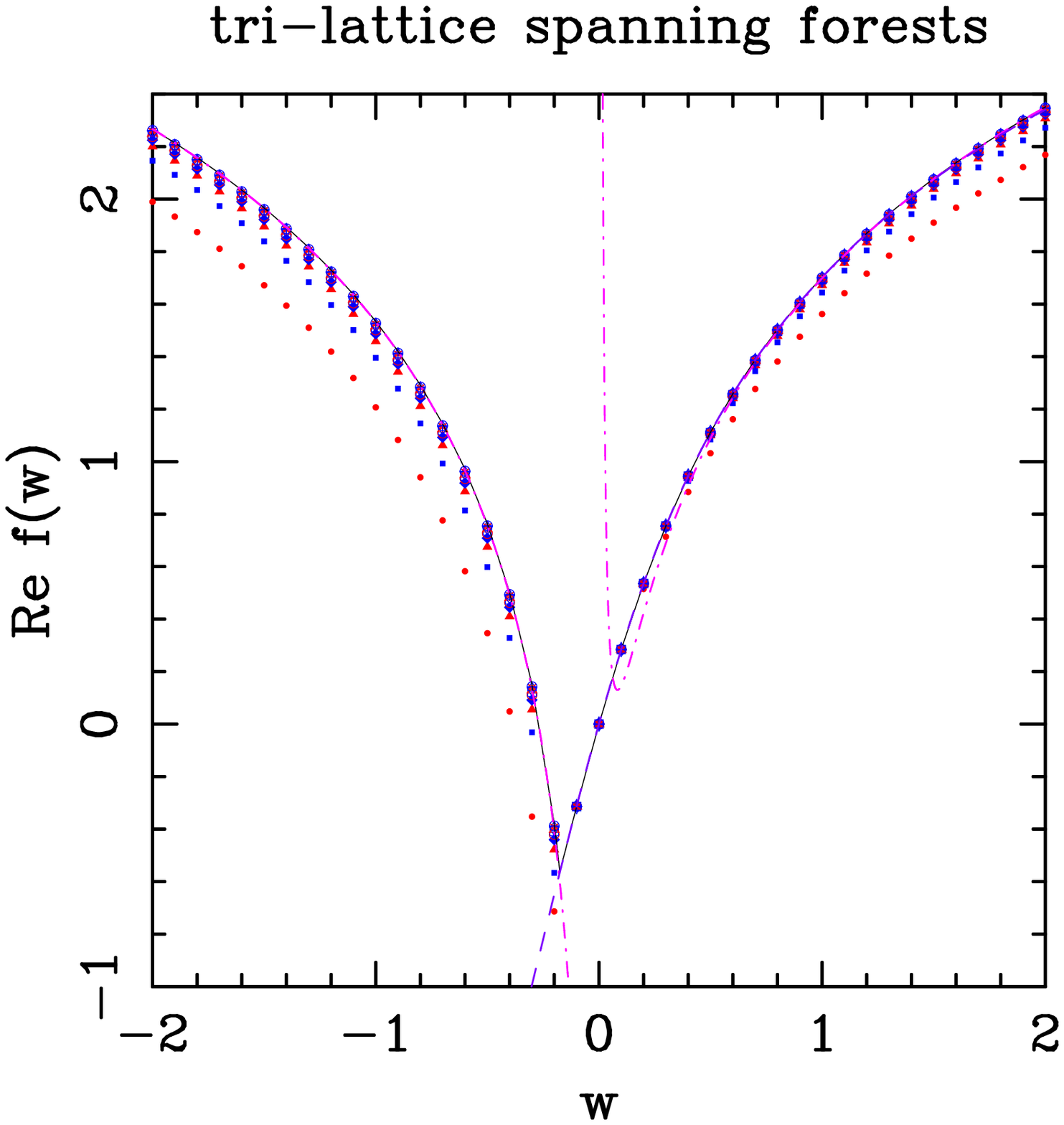}
\caption{\label{figure_f_tri}
   Real part of the free energy for triangular-lattice spanning forests
   as a function of $w$, for strips of width
   $L=2$ ($\bullet$),
   $3$ ($\blacksquare$), $L=4$ ($\blacktriangle$), $L=5$ ($\blacklozenge$),
   $6$ ($\circ$),        $L=7$ ($\Box$),           $L=8$ ($\triangle$),
   $9$ ($\diamondsuit$), $L=10$ ($\times$),        $L=11$ ($+$),
   $12$ ($*$), and       $L=13$ ($\oplus$).
   To make clearer any possible even-odd effect we have displayed in red 
   (resp.\ blue) the points corresponding to even (resp.\ odd) values of $L$.
   The black solid curve is obtained by
   extrapolating the finite-width data to $L\to\infty$
   and then joining the points.
   The violet dashed curve is the Pad\'e $[10,10]$
   approximant to our longest small-$w$ series.
   The pink dot-dashed curve corresponds to the large-$w$ expansion
   \protect\reff{def_large_w_series} through order $w^{-1}$.
}
\end{figure}

%
%
\clearpage
\begin{figure}
\centering
\includegraphics[width=400pt]{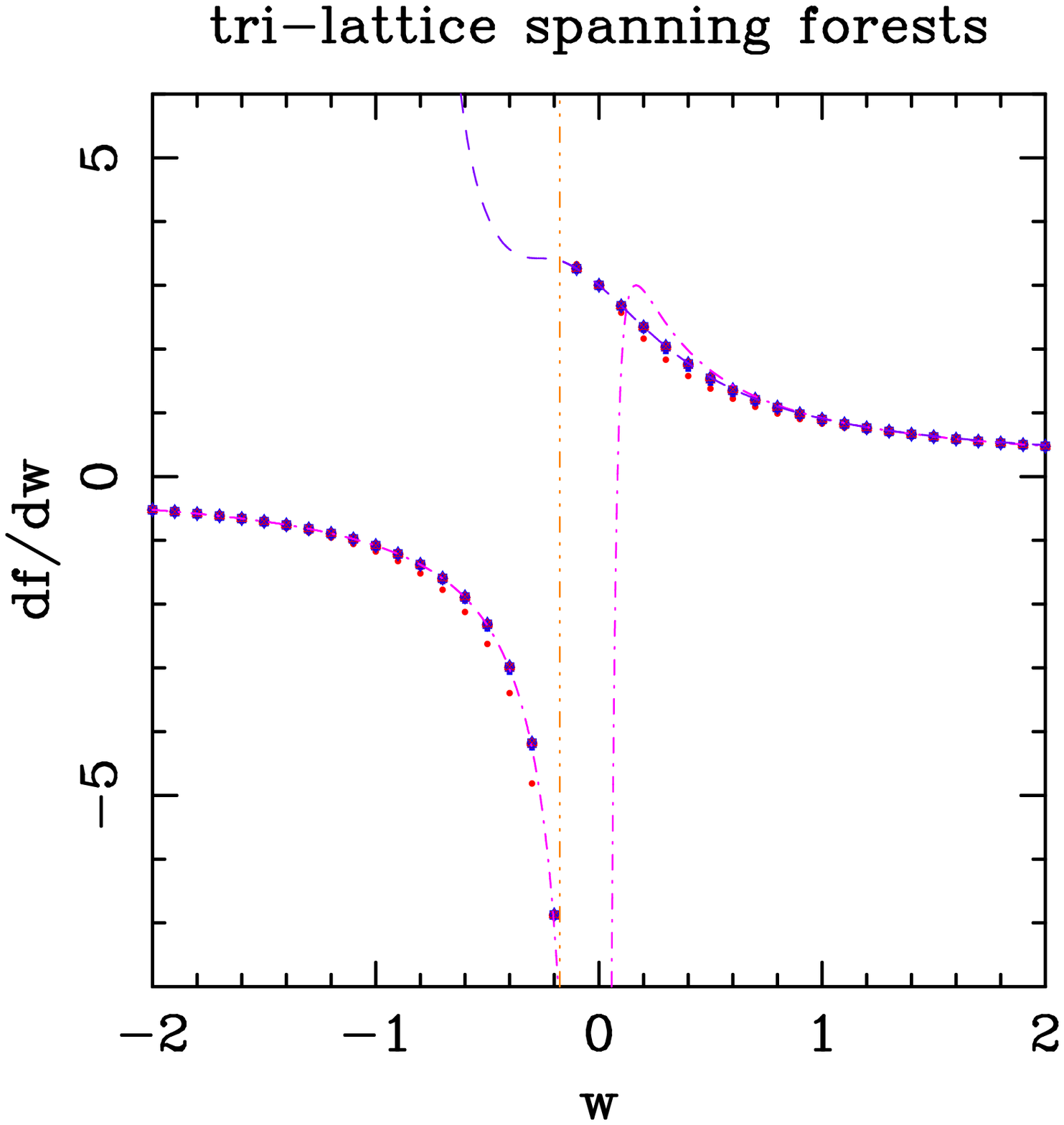}
\caption{\label{figure_E_tri}
   First derivative $f'_L(w)$ of the triangular-lattice free energy
   for strips of width
   $L=2$ ($\bullet$),
   $3$ ($\blacksquare$),
   $4$ ($\blacktriangle$), $5$ ($\blacklozenge$),
   $6$ ($\circ$),          $7$ ($\Box$),
   $8$ ($\triangle$)       $9$ ($\diamondsuit$), and $10$ ($\times$). 
   Points with even (resp.\ odd) $L$ are shown in red (resp.\ blue).
   The violet dashed curve on the right corresponds to the
   Pad\'e approximant $[10,10]$ to our longest small-$w$ series.
   The pink dot-dashed curve corresponds to the derivative of the
   large-$w$ expansion \protect\reff{def_large_w_series},
   through order $w^{-2}$.
   The vertical brown dot-dot-dashed line marks the point $w_0 \approx -0.1753$.
}
\end{figure}

%
%
\clearpage
\begin{figure}
\centering
\includegraphics[width=400pt]{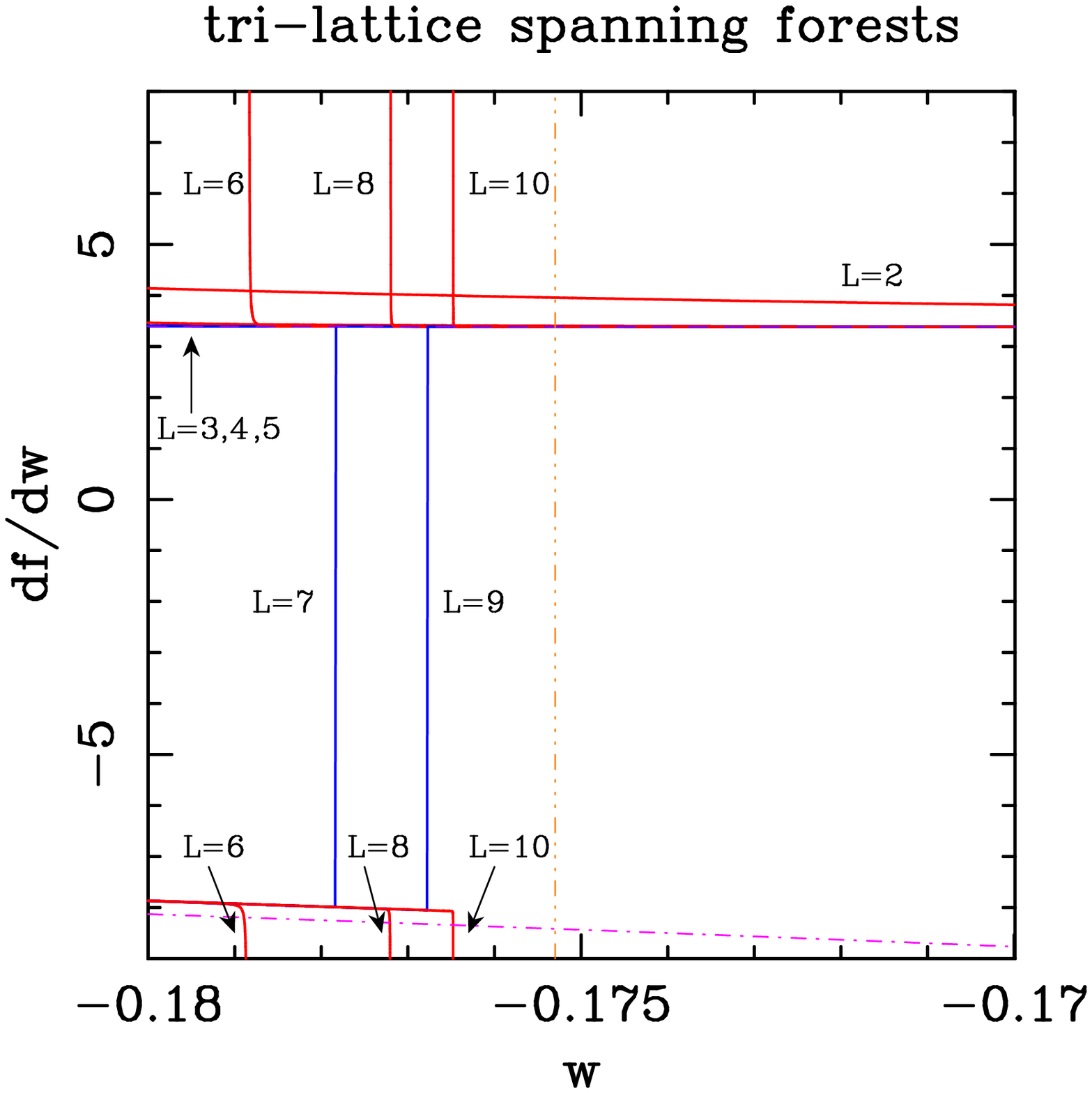}
\caption{\label{figure_Ec_tri}
   First derivative $f'_L(w)$ of the triangular-lattice free energy
   for strips of widths $2\leq L \leq 10$ close to 
   $w_0({\rm sq})\approx -0.1753$.
   Curves for even (resp.\ odd) $L$ are shown in red (resp.\ blue).
   The violet dashed solid curve on the right corresponds to the
   Pad\'e approximant $[10,10]$ to our longest small-$w$ series
   (this curve is barely visible, as it nearly coincides with those
   for $2\leq L \leq 5$).
   The pink dot-dashed curve corresponds to the derivative of the
   large-$w$ expansion \protect\reff{def_large_w_series},
   through order $w^{-2}$.
   The vertical brown dot-dot-dashed line marks the point $w_0\approx -0.1753$. 
}
\end{figure}

%
%
\clearpage
\begin{figure}
\centering
\begin{tabular}{c}
\includegraphics[width=400pt]{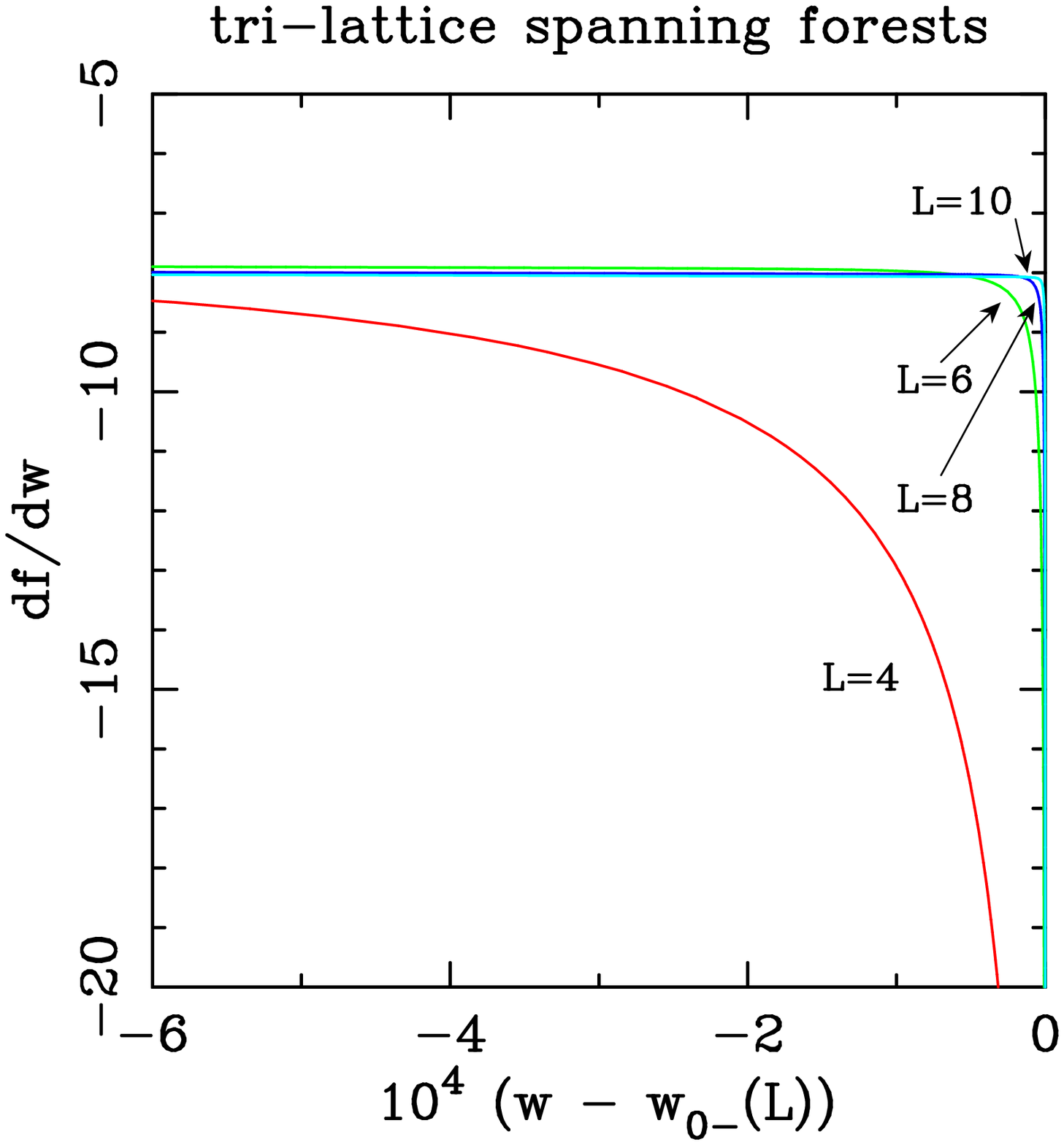} \\[1mm] 
\phantom{(aaaaaaa)}\large{(a)}
\end{tabular}
\caption{\label{figure_Ec_tri_bisa}
   First derivative $f'_L(w)$ of the triangular-lattice free energy
   for strips of widths $L=4,6,8,10$.
   (a) Regime $w < w_0$ plotted versus $w-w_{0-}(L)$.
   (b) Regime $w > w_0$ plotted versus $w-w_{0+}(L)$.
}
\end{figure}
\addtocounter{figure}{-1}

%
%
\clearpage
\begin{figure}
\centering
\begin{tabular}{c}
\includegraphics[width=400pt]{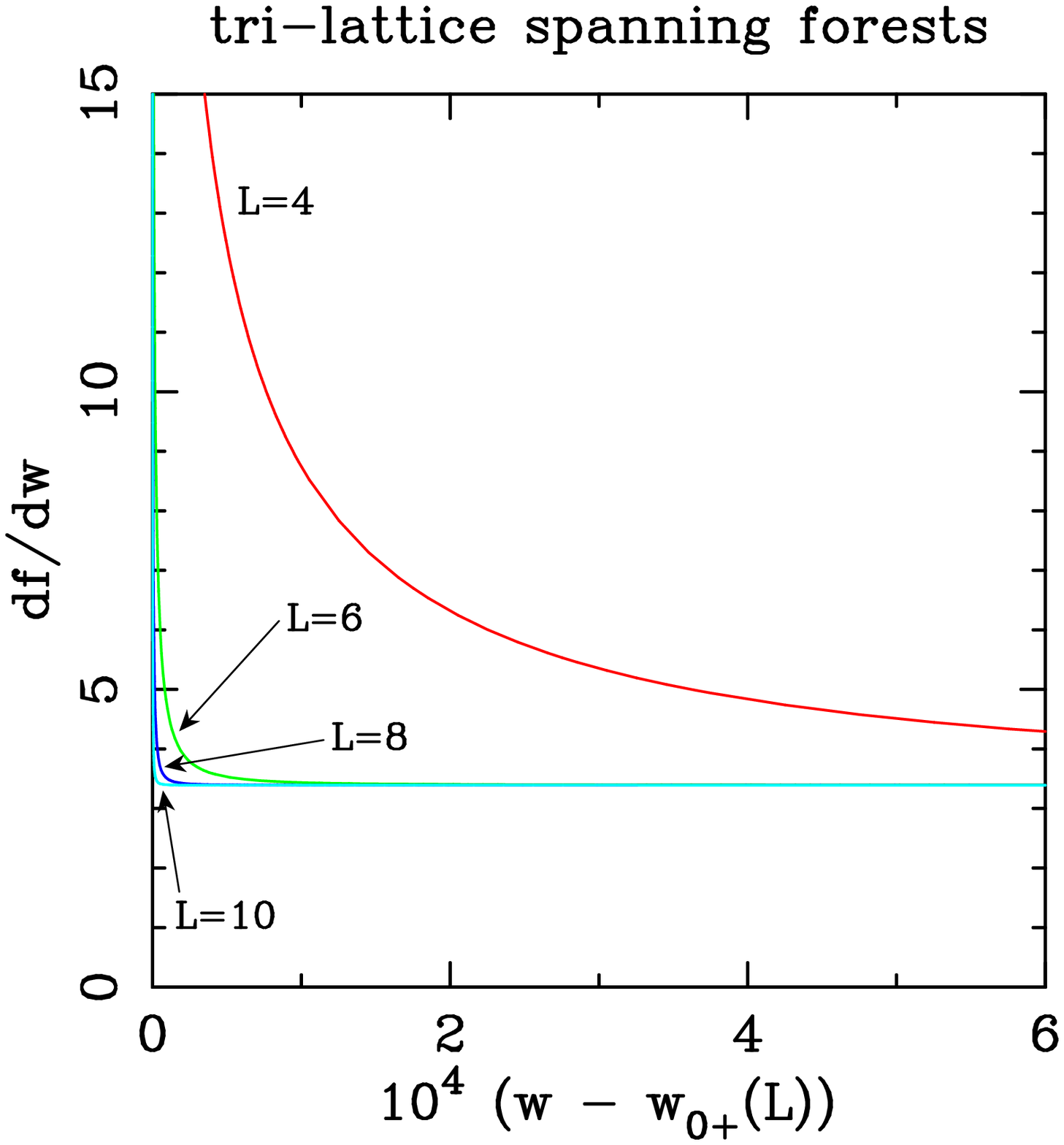} \\[1mm] 
\phantom{(aaaaaaa)}\large{(b)}
\end{tabular}
\caption{\label{figure_Ec_tri_bisb}
   First derivative $f'_L(w)$ of the triangular-lattice free energy
   for strips of widths $L=4,6,8,10$.
   (a) Regime $w < w_0$ plotted versus $w-w_{0-}(L)$.
   (b) Regime $w > w_0$ plotted versus $w-w_{0+}(L)$.
}
\end{figure}

%
%
\clearpage
\begin{figure}
\centering
\includegraphics[width=400pt]{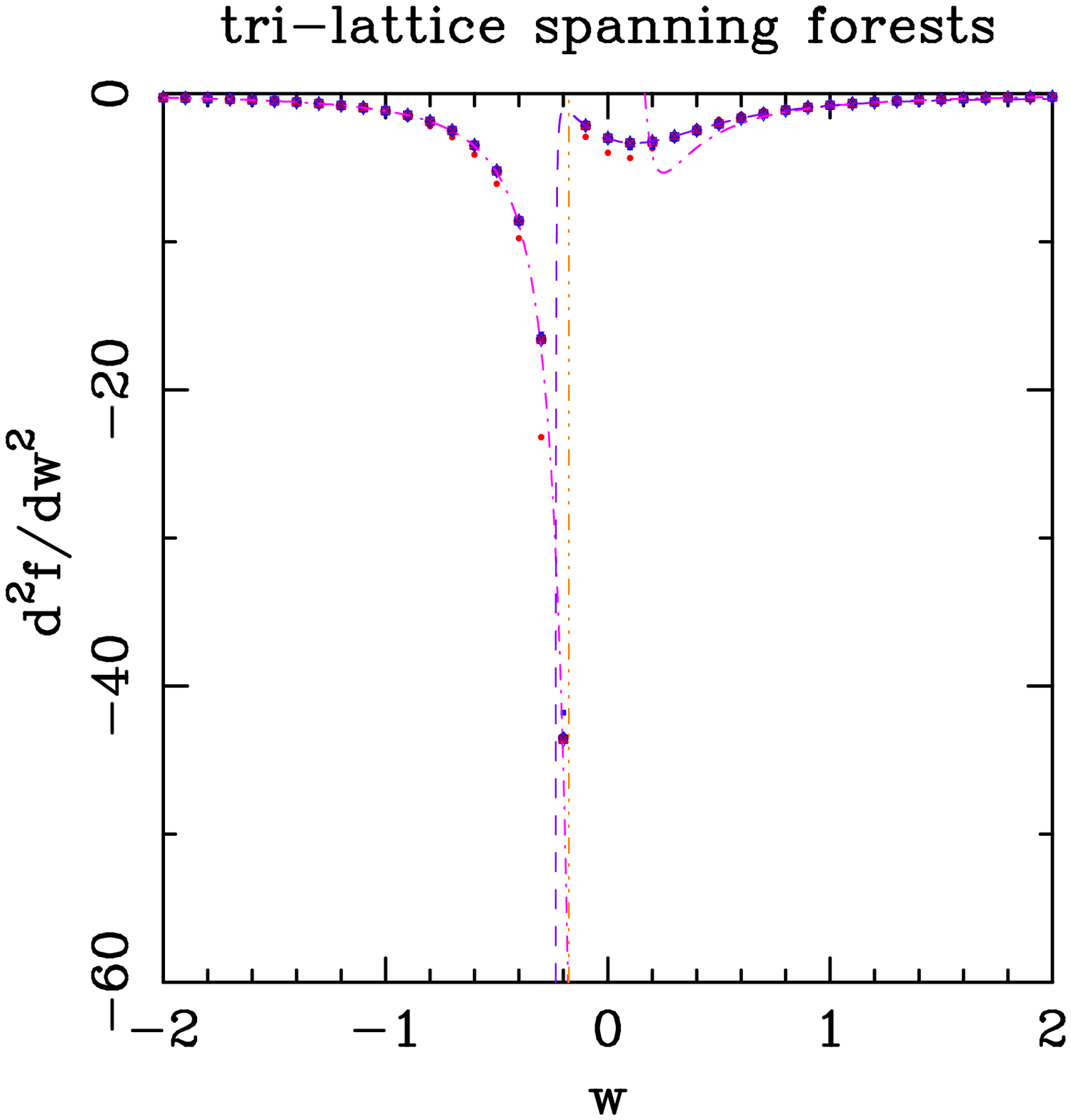}
\caption{\label{figure_CH_tri}
   Second derivative $f''_L(w)$ of the triangular-lattice free energy
   for strips of width
   $L=2$ ($\bullet$),
   $3$ ($\blacksquare$),
   $4$ ($\blacktriangle$), $5$ ($\blacklozenge$),
   $6$ ($\circ$),          $7$ ($\Box$),
   $8$ ($\triangle$),      $9$ ($\diamondsuit$), and $10$ ($\times$). 
   Points with even (resp.\ odd) $L$ are shown in red (resp.\ blue).
   The violet dashed curve on the right corresponds to the
   Pad\'e approximant $[10,10]$ to our longest small-$w$ series.
   The pink dot-dashed curve corresponds to the second derivative of the
   large-$w$ expansion \protect\reff{def_large_w_series},
   through order $w^{-3}$.
   The vertical brown dot-dot-dashed line marks the point $w_0 \approx -0.1753$.
}
\end{figure}

%
%
\clearpage
\begin{figure}
\centering
\includegraphics[width=400pt]{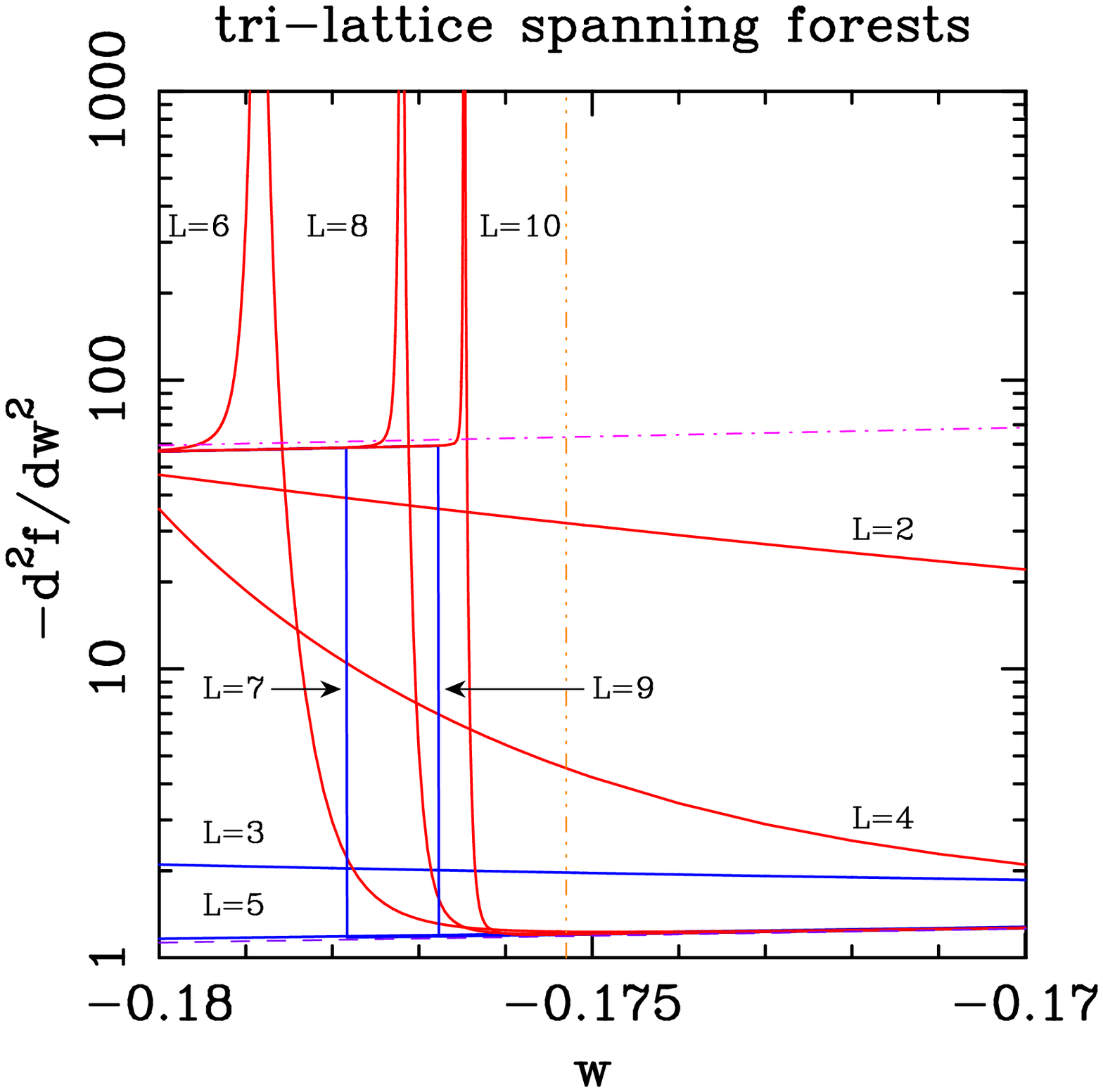}
\caption{\label{figure_CHc_tri}
   Second derivative $f''_L(w)$ of the triangular-lattice free energy
   for strips of widths $2\leq L \leq 10$ close to 
   $w_0({\rm sq})\approx -0.1753$.
   Curves for even (resp.\ odd) $L$ are shown in red (resp.\ blue).
   The violet dashed solid curve on the right corresponds to the
   Pad\'e approximant $[10,10]$ to our longest small-$w$ series.
   This curve is barely visible, as it very similar to that for $L=5$.
   The pink dot-dashed curve corresponds to the second derivative of the
   large-$w$ expansion \protect\reff{def_large_w_series},
   through order $w^{-3}$.
   The vertical brown dot-dot-dashed line marks the point $w_0\approx -0.1753$.
}
\end{figure}

%
%
\clearpage
\begin{figure}
\centering
\begin{tabular}{c}
\includegraphics[width=400pt]{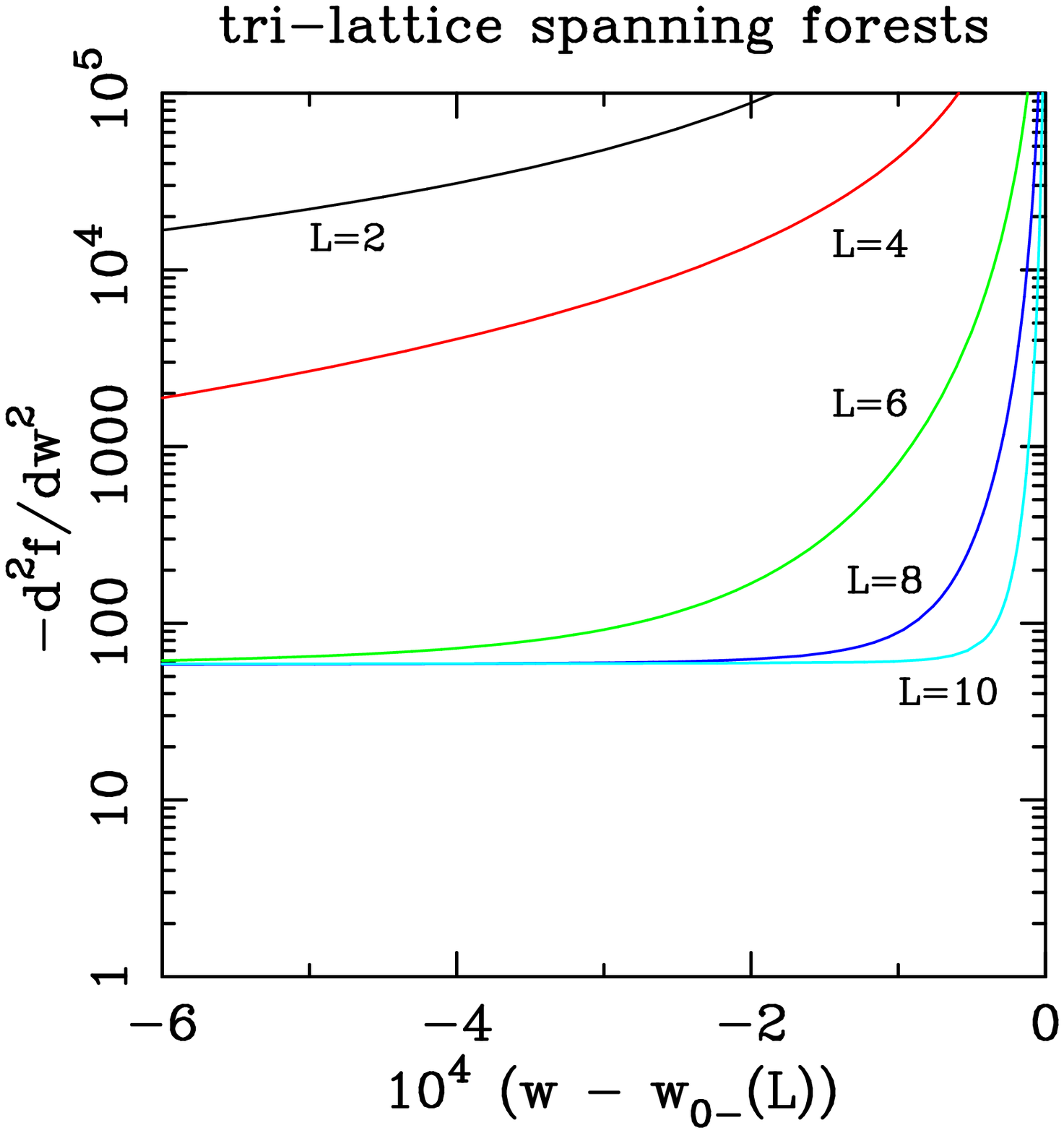} \\[1mm]
\phantom{(aaaaaaa)}\large{(a)}
\end{tabular}
\caption{\label{figure_CHc_tri_bisa}
   Second derivative $f''_L(w)$ of the triangular-lattice free energy
   for strips of widths $L=4,6,8,10$.
   (a) Regime $w < w_0$ plotted versus $w-w_{0-}(L)$.
   (b) Regime $w > w_0$ plotted versus $w-w_{0+}(L)$.
}
\end{figure}
\addtocounter{figure}{-1}

%
%
\clearpage
\begin{figure}
\centering
\begin{tabular}{c}
\includegraphics[width=400pt]{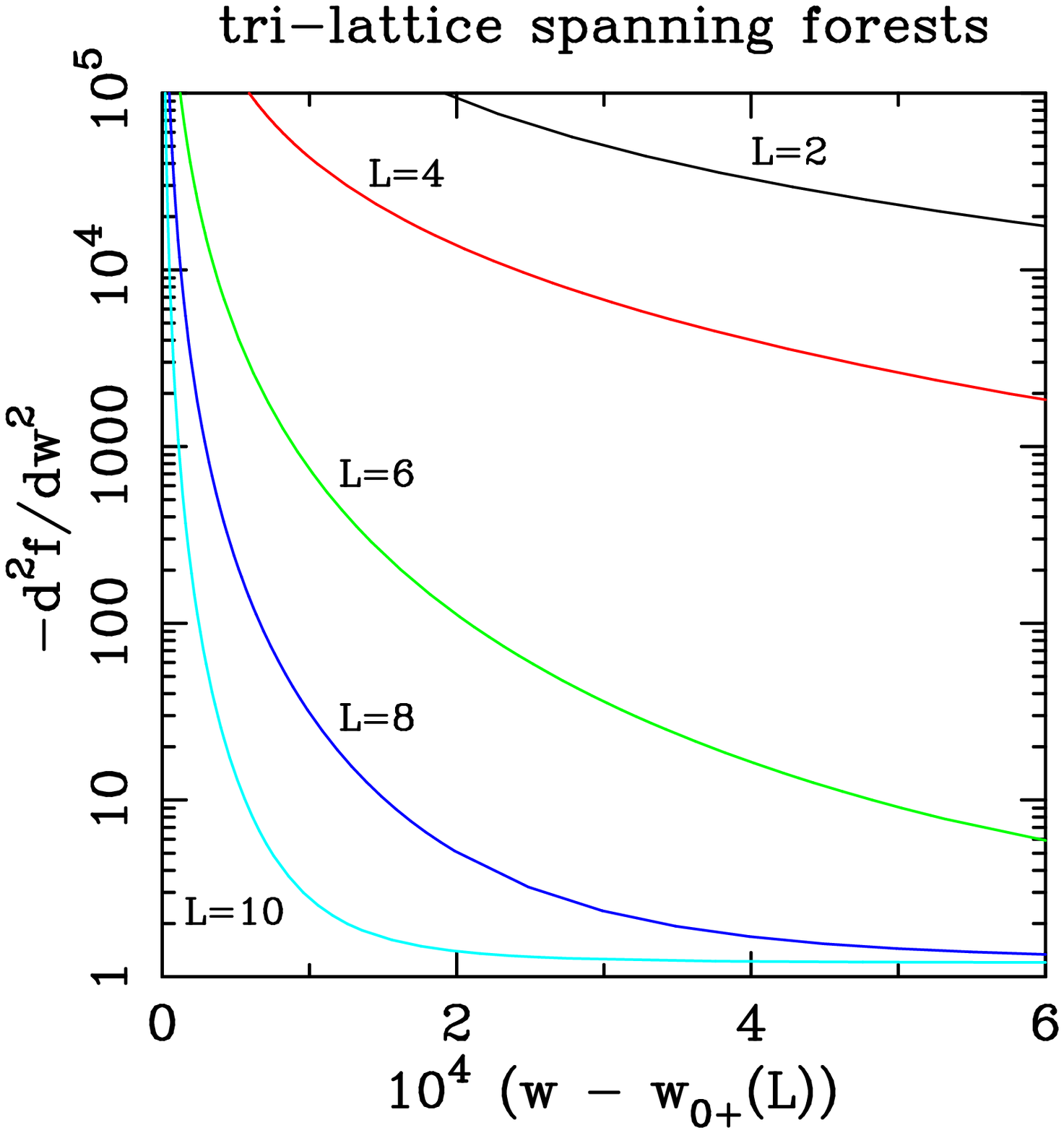} \\[1mm]
\phantom{(aaaaaaa)}\large{(b)}
\end{tabular}
\caption{\label{figure_CHc_tri_bisb}
   Second derivative $f''_L(w)$ of the triangular-lattice free energy
   for strips of widths $L=4,6,8,10$.
   (a) Regime $w < w_0$ plotted versus $w-w_{0-}(L)$.
   (b) Regime $w > w_0$ plotted versus $w-w_{0+}(L)$.
}
\end{figure}

%
%
\clearpage
\begin{figure}
\centering
\begin{tabular}{cc}
\includegraphics[width=200pt]{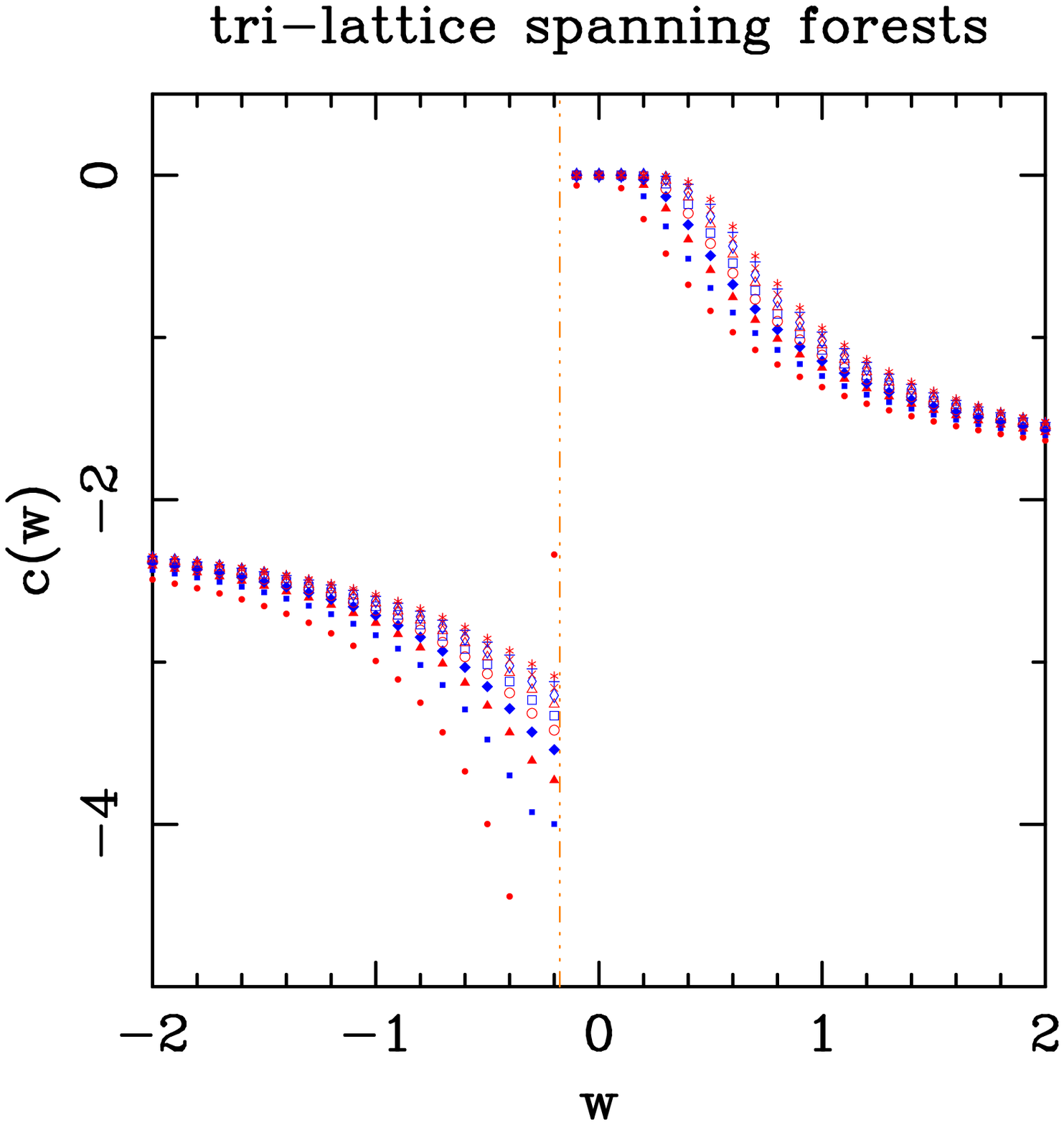} &
\includegraphics[width=200pt]{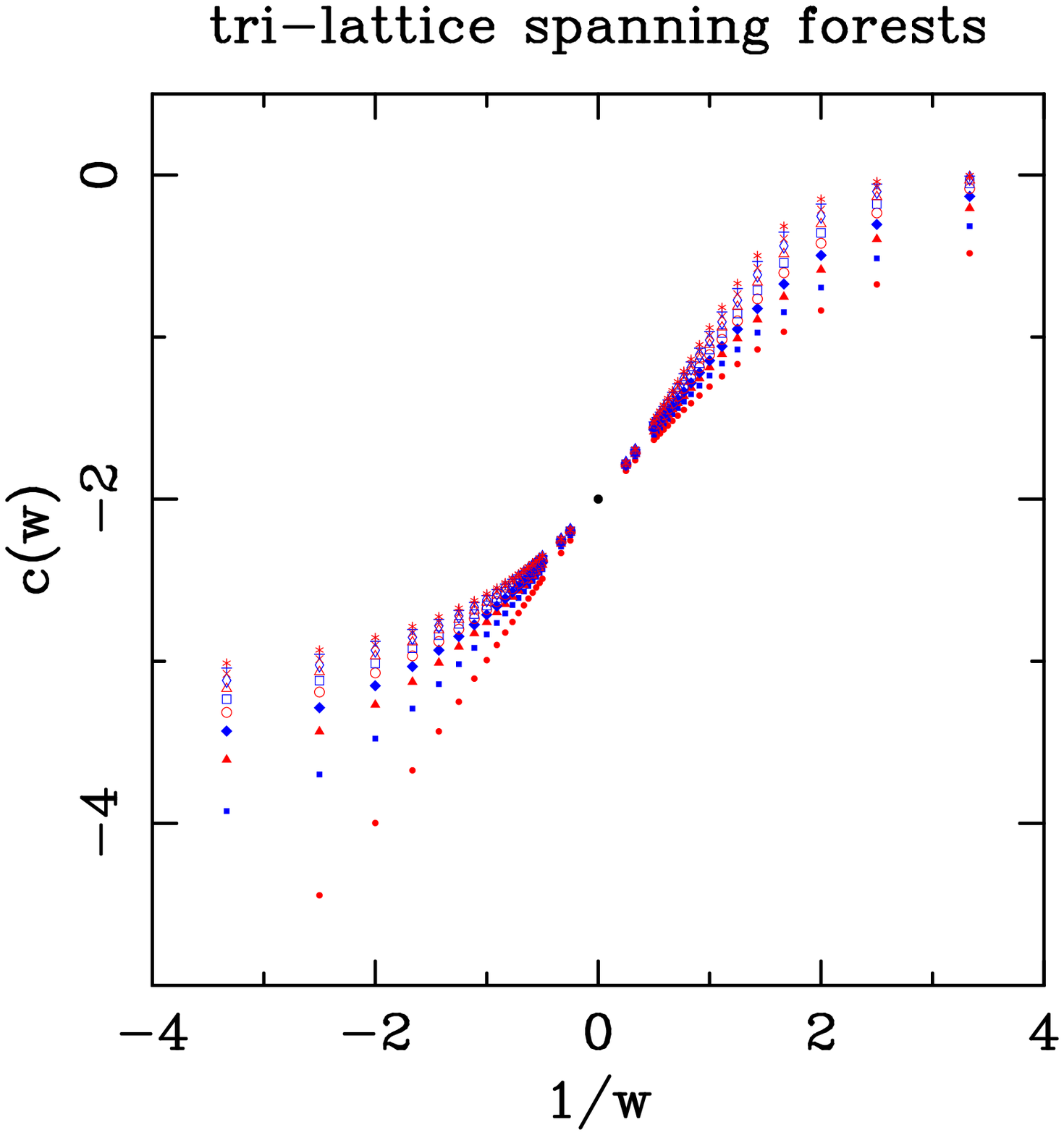} \\[1mm]
   \phantom{(((a)}(a)  & \phantom{(((a)}(b) \\[4mm]
\includegraphics[width=200pt]{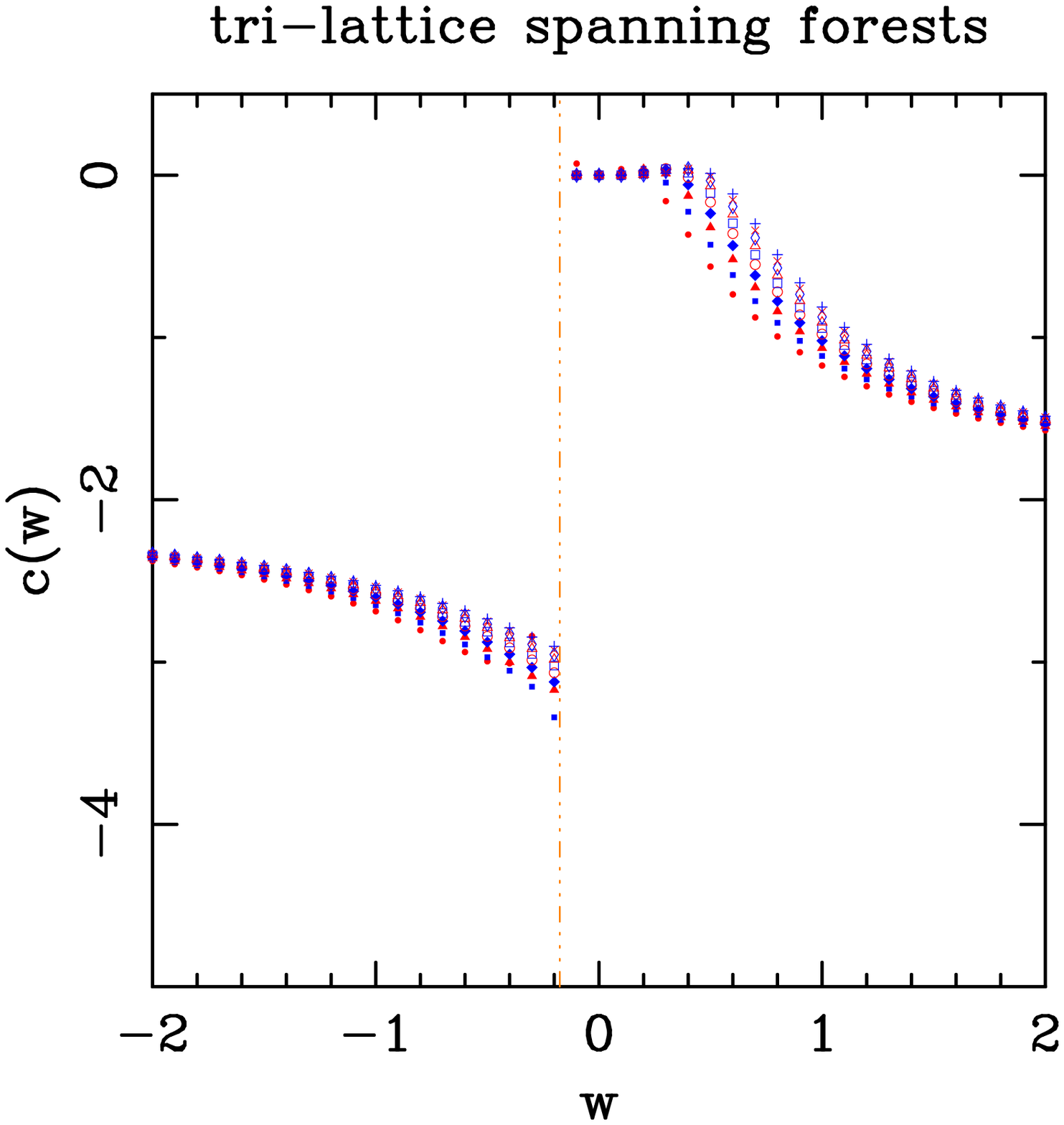} &
\includegraphics[width=200pt]{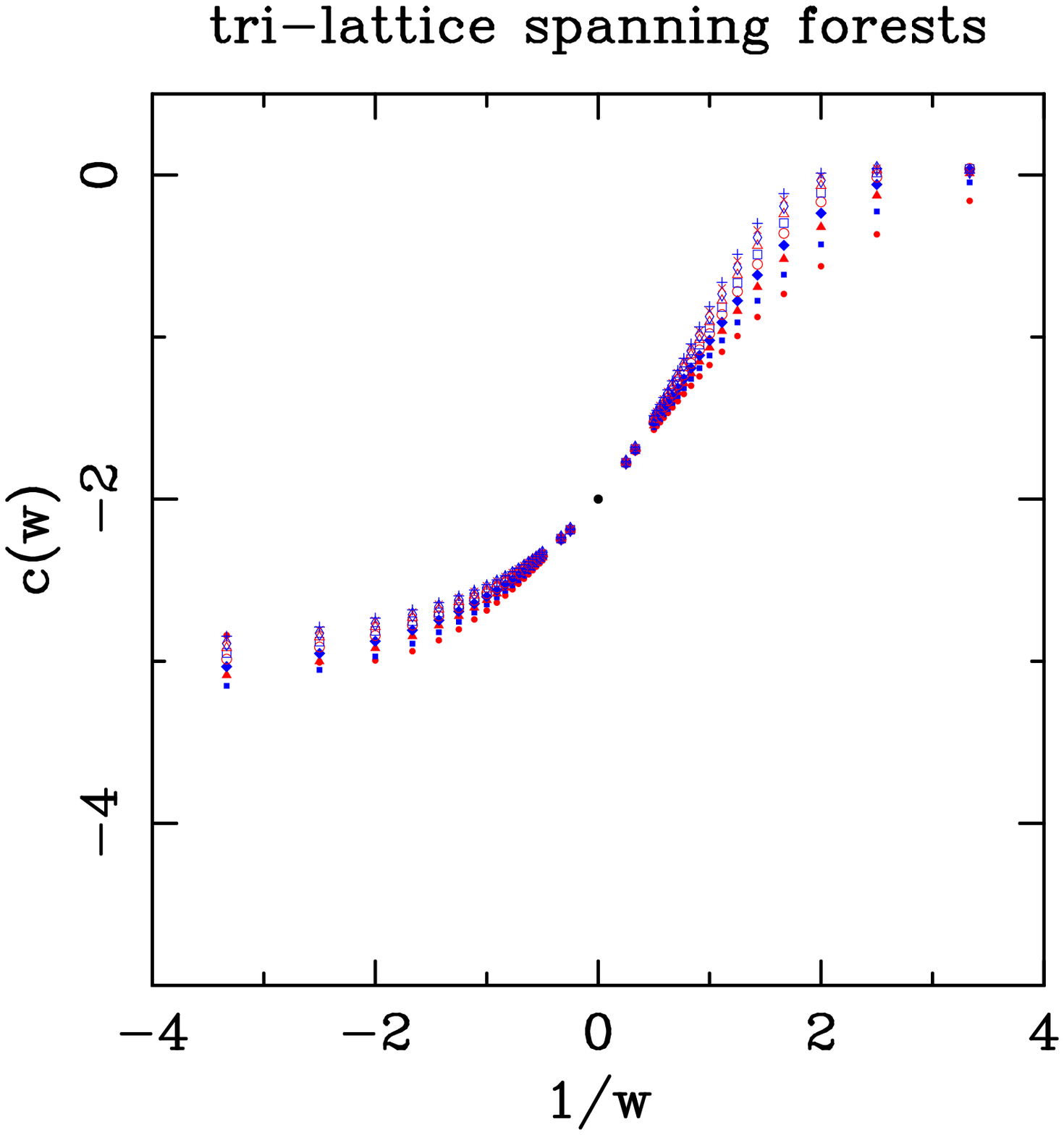}\\[1mm]
   \phantom{(((a)}(c)  & \phantom{(((a)}(d)
\end{tabular}
\caption{\label{figure_c_tri}
   Estimates for the triangular-lattice central charge $c(w)$
   obtained by fitting the free energy to the Ans\"atze (a,b)
   $\real f_L(w) = \real f(w) + [c(w)\sqrt{3}\pi/12] L^{-2}$, and
   (c,f) $\real f_L(w) = \real f(w) + [c(w)\sqrt{3}\pi/12] L^{-2} + A/L^4$.
   Fits are performed for
   $L_{\rm min}=2$ ($\bullet$),
   $3$ ($\blacksquare$),
   $4$ ($\blacktriangle$), $5$ ($\blacklozenge$),
   $6$ ($\circ$),          $7$ ($\Box$),
   $8$ ($\triangle$),      $9$ ($\diamondsuit$),
   $10$ ($\times$)        $11$ ($+$), and 
   $12$ ($*$).
   Points with even (resp.\ odd) $L$ are shown in red (resp.\ blue).
   The black dot at $1/w=0$, $c=-2$ marks the theoretical prediction.
   The vertical brown dot-dot-dashed line marks the point
   $w_0 \approx -0.1753$.
}
\end{figure}

%
%
\clearpage
\begin{figure}
\centering
\includegraphics[width=400pt]{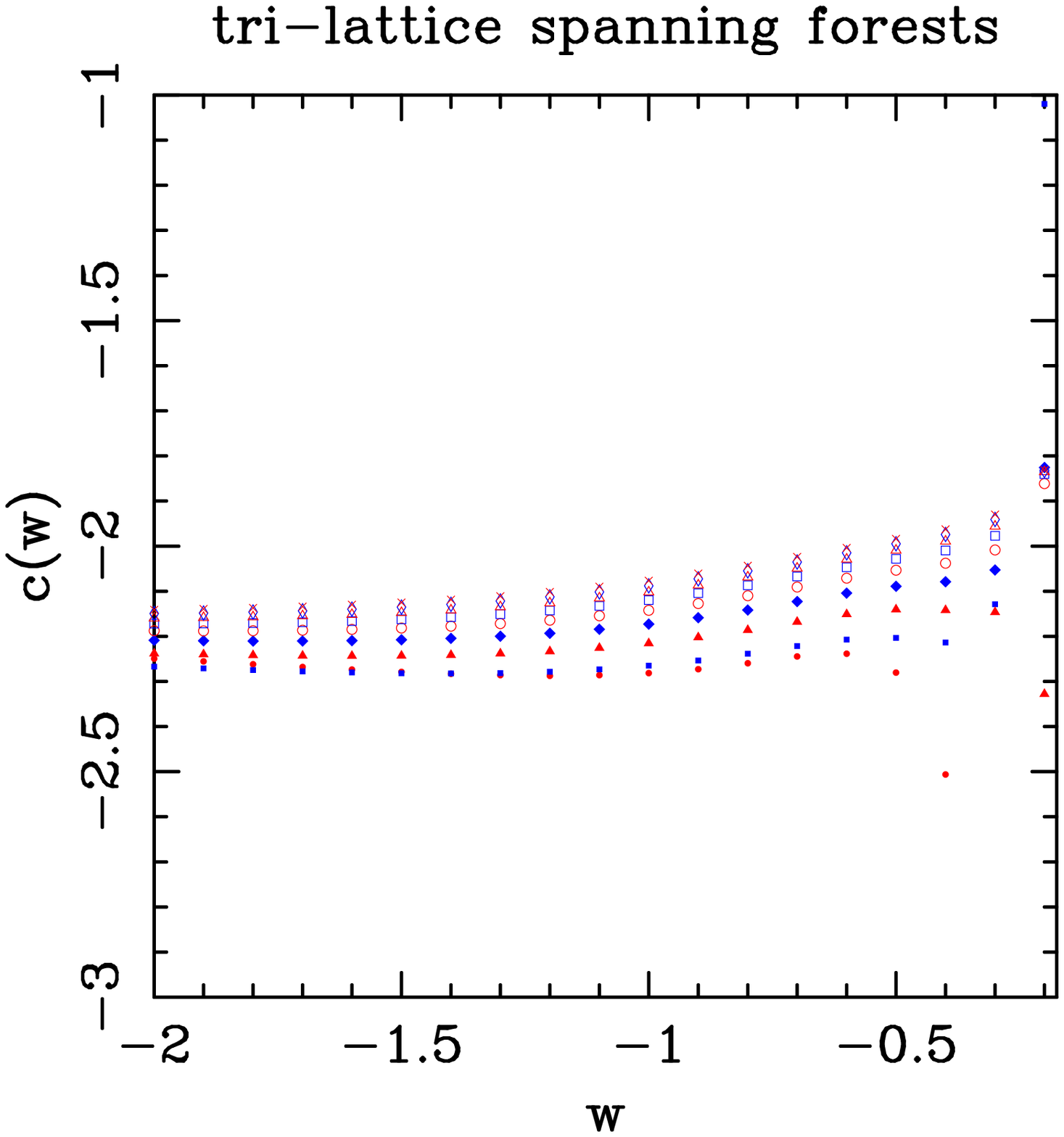} 
\caption{\label{figure_c_tri2}
   Estimates for the triangular-lattice central charge $c(w)$
   obtained by fitting the free energy to the Ansatz 
   $\real f_L(w) = \real f(w) + [c(w)\pi/6] L^{-2} + A \log\log L/(L^2 \log L) 
    + B/(L^2 \log L)$. 
   Fits are performed for
   $L_{\rm min}=2$ ($\bullet$), $3$ ($\blacksquare$),
   $4$ ($\blacktriangle$), $5$ ($\blacklozenge$),
   $6$ ($\circ$), $7$ ($\Box$),
   $8$ ($\triangle$), $9$ ($\diamondsuit$), and 
   $10$ ($\times$).
   Points with even (resp.\ odd) $L$ are shown in red (resp.\ blue).
}
\end{figure}

%
%
\clearpage
\begin{figure}
\centering
\begin{tabular}{c}
\includegraphics[width=400pt]{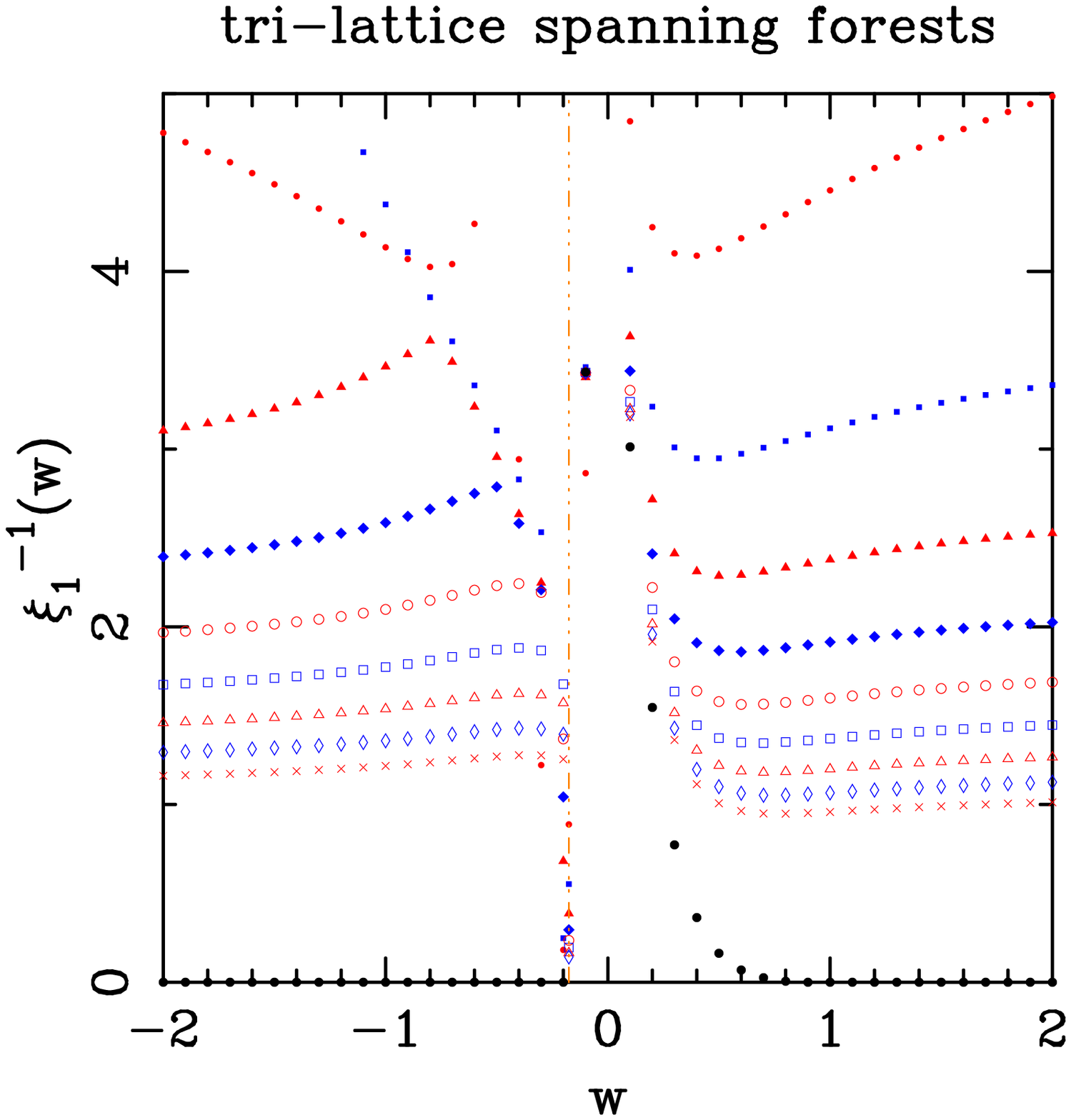}\\[2mm]
\phantom{(aaaaaaa)}\large{(a)}
\end{tabular}
\caption{\label{figure_xi1_tri}
   Values of the triangular-lattice inverse correlation length
   $\xi_j^{-1}(w) = \log |\lambda_\star/\lambda_j|$ for
   (a) $j=1$ and (b) $j=2$.
   Symbols indicate strip widths
   $L=2$ ($\bullet$),
   $3$ ($\blacksquare$),
   $4$ ($\blacktriangle$),
   $5$ ($\blacklozenge$),
   $6$ ($\circ$),
   $7$ ($\Box$),
   $8$ ($\triangle$),
   $9$ ($\diamondsuit$), and
   $10$ ($\times$).
   Points with even (resp.\ odd) $L$ are shown in red (resp.\ blue).
   The black solid circles ($\bullet$) correspond to the
   extrapolated infinite-volume limit of the finite-size data (see text).
   The vertical dot-dot-brown dashed line marks the point $w_0\approx -0.1753$. 
}
\end{figure}
\addtocounter{figure}{-1}

%
%
\clearpage
\begin{figure}
\centering
\begin{tabular}{c}
\includegraphics[width=400pt]{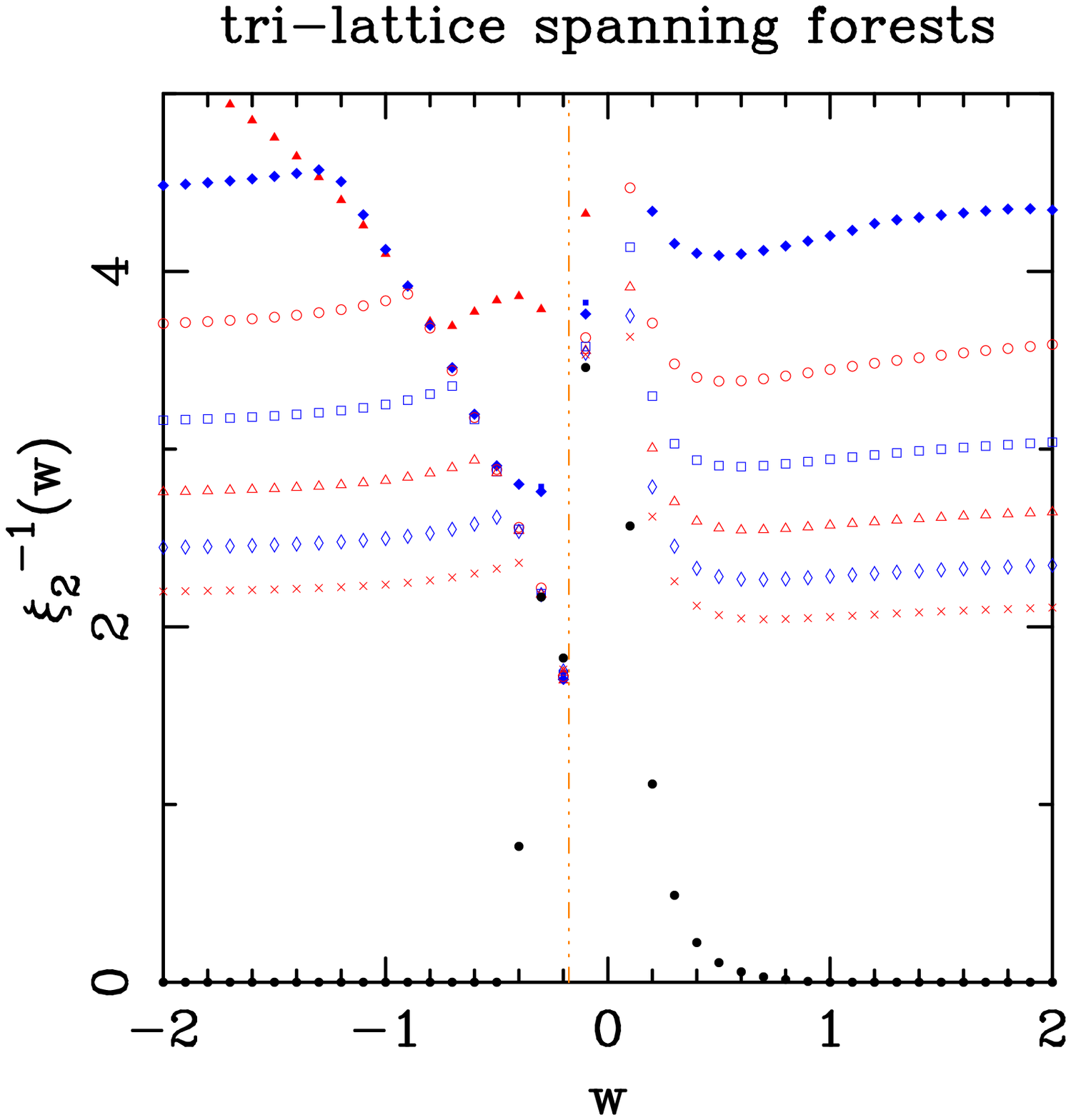}\\[2mm]
\phantom{(aaaaaaa)}\large{(b)}
\end{tabular}
\caption{\label{figure_xi2_tri}
   Values of the triangular-lattice inverse correlation length
   $\xi_j^{-1}(w) = \log |\lambda_\star/\lambda_j|$ for
   (a) $j=1$ and (b) $j=2$.
   Symbols indicate strip widths
   $L=2$ ($\bullet$),
   $3$ ($\blacksquare$),
   $4$ ($\blacktriangle$),
   $5$ ($\blacklozenge$),
   $6$ ($\circ$),
   $7$ ($\Box$),
   $8$ ($\triangle$),
   $9$ ($\diamondsuit$), and
   $10$ ($\times$).
   Points with even (resp.\ odd) $L$ are shown in red (resp.\ blue).
   The black solid circles ($\bullet$) correspond to the
   extrapolated infinite-volume limit of the finite-size data (see text).
   The vertical brown dot-dot-dashed line marks the point $w_0\approx -0.1753$. 
}
\end{figure}

%
%
\clearpage
\begin{figure}
\centering
\begin{tabular}{c}
\includegraphics[width=400pt]{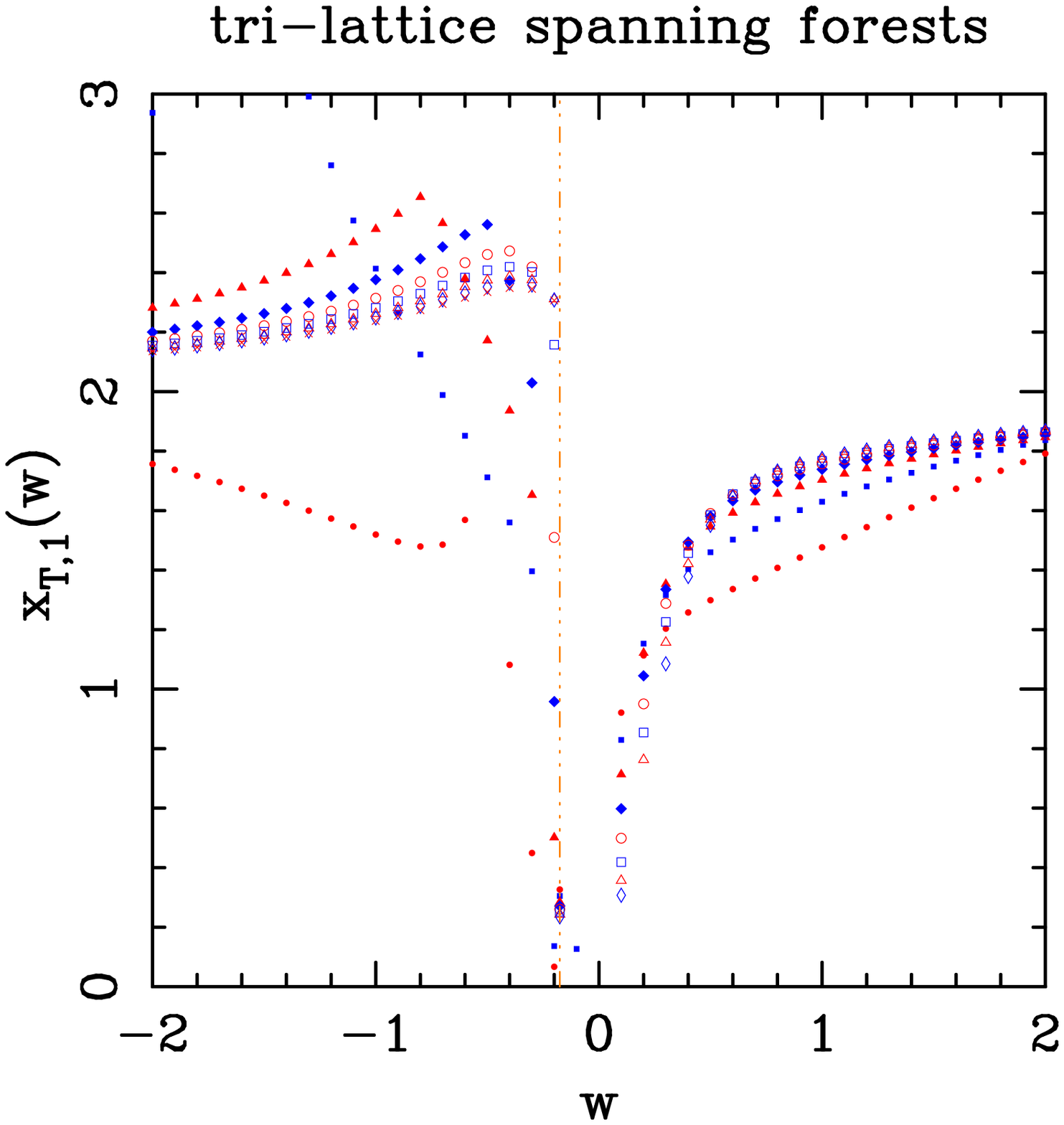}\\[2mm]
\phantom{(aaaaaaa)}\large{(a)}
\end{tabular}
\caption{\label{figure_x1_tri}
   Estimates for the triangular-lattice scaling dimension $x_{Tj}(w)$
   for (a) $j=1$ and (b) $j=2$,
   obtained by fitting the inverse correlation length to the Ansatz
   $\xi^{-1}_j(w) = \xi^{-1}_{j,\infty}(w)  + 2 \pi x_{Tj}(w) L^{-1}$.
   In the region $w \le -0.175$ we have fixed $\xi_{j,\infty}^{-1} = 0$;
   in the region $w > -0.175$ we have left it variable.
   Fits are performed for
   $L_{\rm min}=2$ ($\bullet$),
   $3$ ($\blacksquare$),
   $4$ ($\blacktriangle$),
   $5$ ($\blacklozenge$),
   $6$ ($\circ$),
   $7$ ($\Box$),
   $8$ ($\triangle$), and  
   $9$ ($\diamondsuit$). 
   Points with even (resp.\ odd) $L$ are shown in red (resp.\ blue).
   The vertical brown dot-dot-dashed line marks the point $w_0\approx-0.1753$.
}
\end{figure}
\addtocounter{figure}{-1}

%
%
\clearpage
\begin{figure}
\centering
\begin{tabular}{c}
\includegraphics[width=400pt]{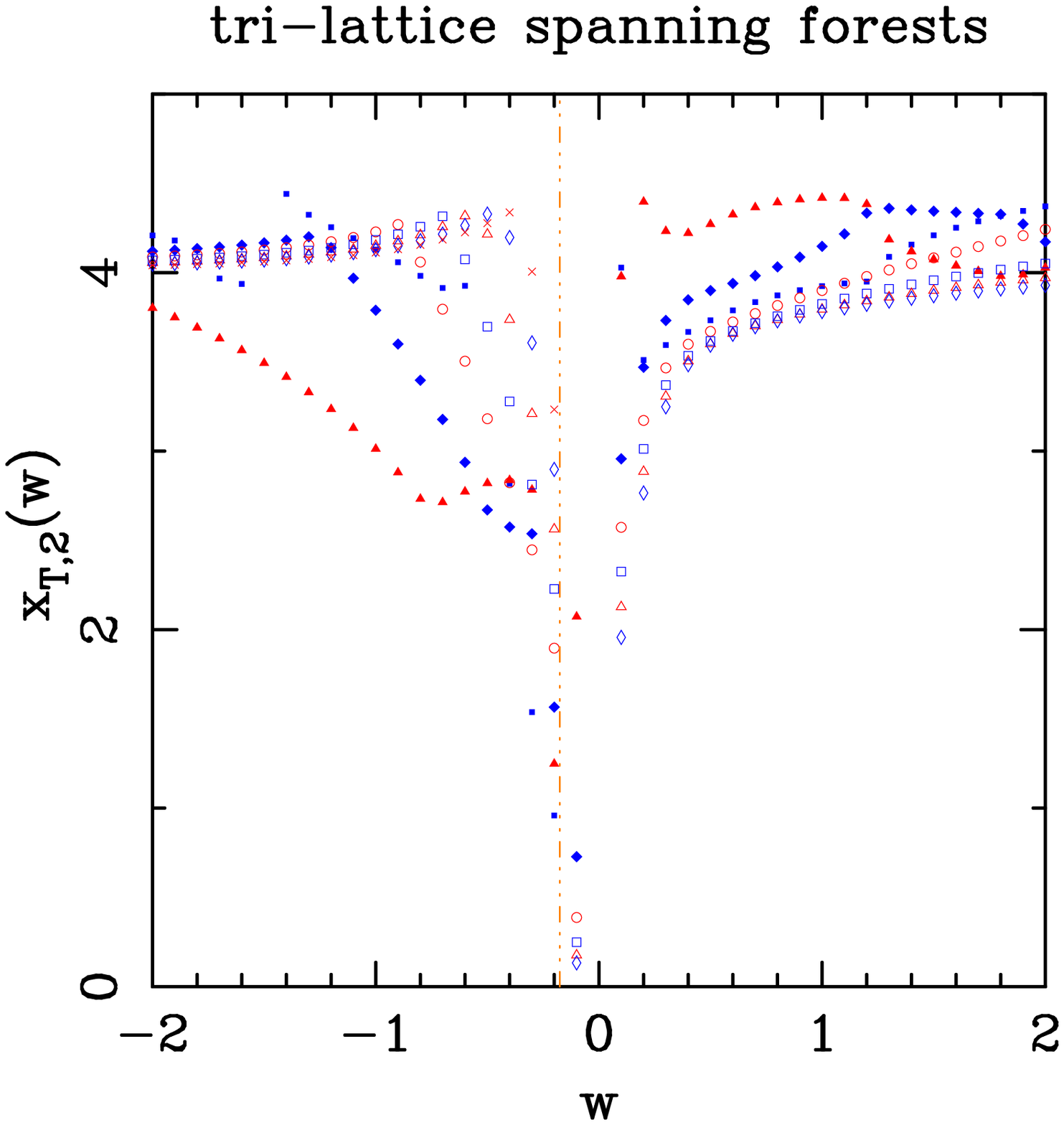}\\[2mm]
\phantom{(aaaaaaa)}\large{(b)}
\end{tabular}
\caption{\label{figure_x2_tri}
   Estimates for the triangular-lattice scaling dimension $x_{Tj}(w)$
   for (a) $j=1$ and (b) $j=2$,
   obtained by fitting the inverse correlation length to the Ansatz
   $\xi^{-1}_j(w) = \xi^{-1}_{j,\infty}(w)  + 2 \pi x_{Tj}(w) L^{-1}$.
   In the region $w \le -0.175$ we have fixed $\xi_{j,\infty}^{-1} = 0$;
   in the region $w > -0.175$ we have left it variable.
   Fits are performed for
   $L_{\rm min}=2$ ($\bullet$),
   $3$ ($\blacksquare$),
   $4$ ($\blacktriangle$),
   $5$ ($\blacklozenge$),
   $6$ ($\circ$),
   $7$ ($\Box$),
   $8$ ($\triangle$), and  
   $9$ ($\diamondsuit$). 
   Points with even (resp.\ odd) $L$ are shown in red (resp.\ blue).
   The vertical brown dot-dot-dashed line marks the point $w_0\approx -0.1753$.
}
\end{figure}

%
%
\clearpage
\begin{figure}
\centering
\includegraphics[width=380pt]{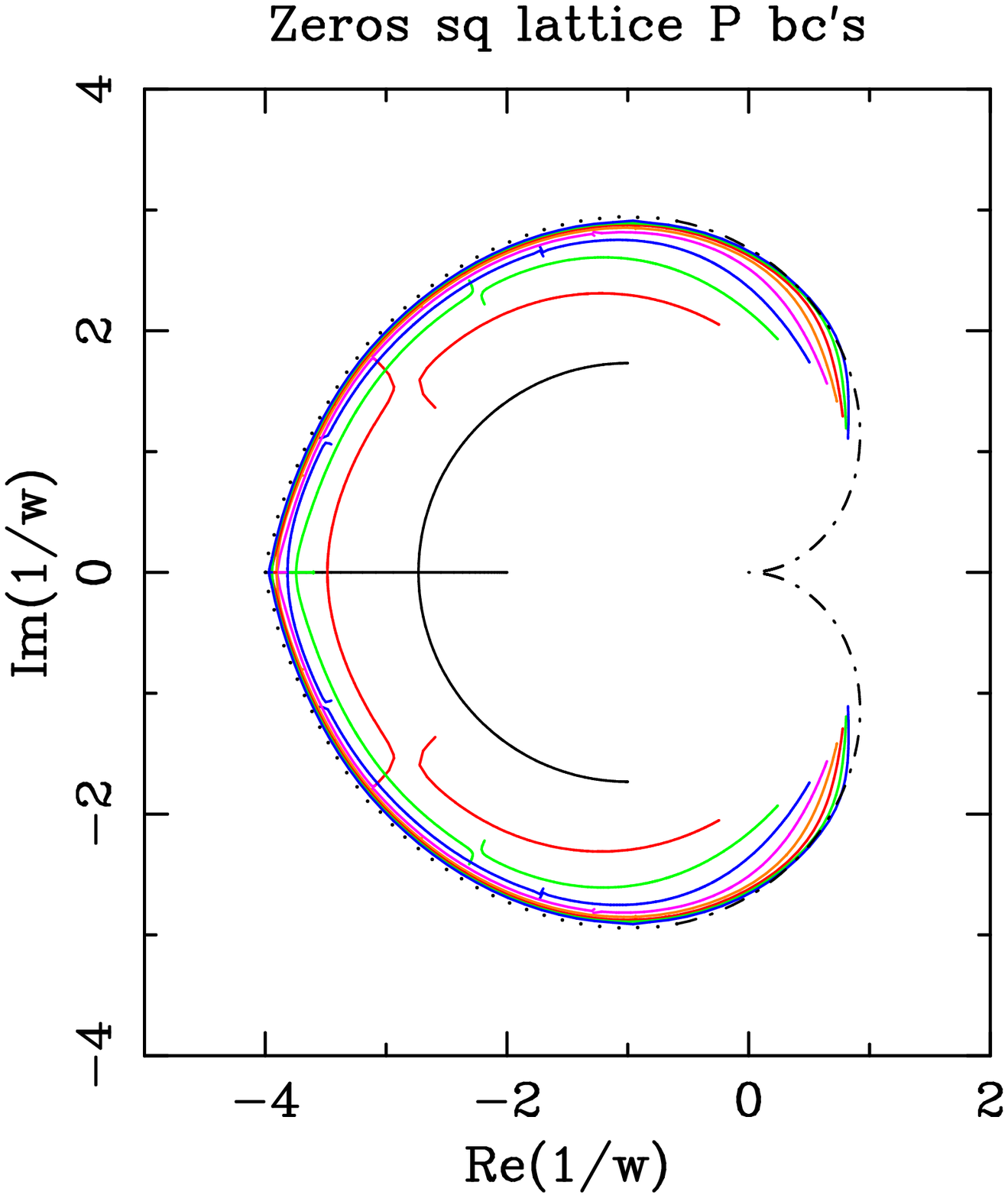}
\caption{\label{figure_sq_all_1overw}
Finite-$L$ limiting curves and phase diagram (Figure~\ref{figure_sq_all})
for the square lattice, mapped to the $1/w$ plane.
Our best estimate for the infinite-volume phase boundary ${\cal B}_\infty$
is depicted in black dots (for $-0.33 \ltapprox \imag w \ltapprox 0.33$)
and as a black dotted-dashed curve (very rough estimate).
}
\end{figure}

%
%
\clearpage
\begin{figure}
\centering
\includegraphics[width=380pt]{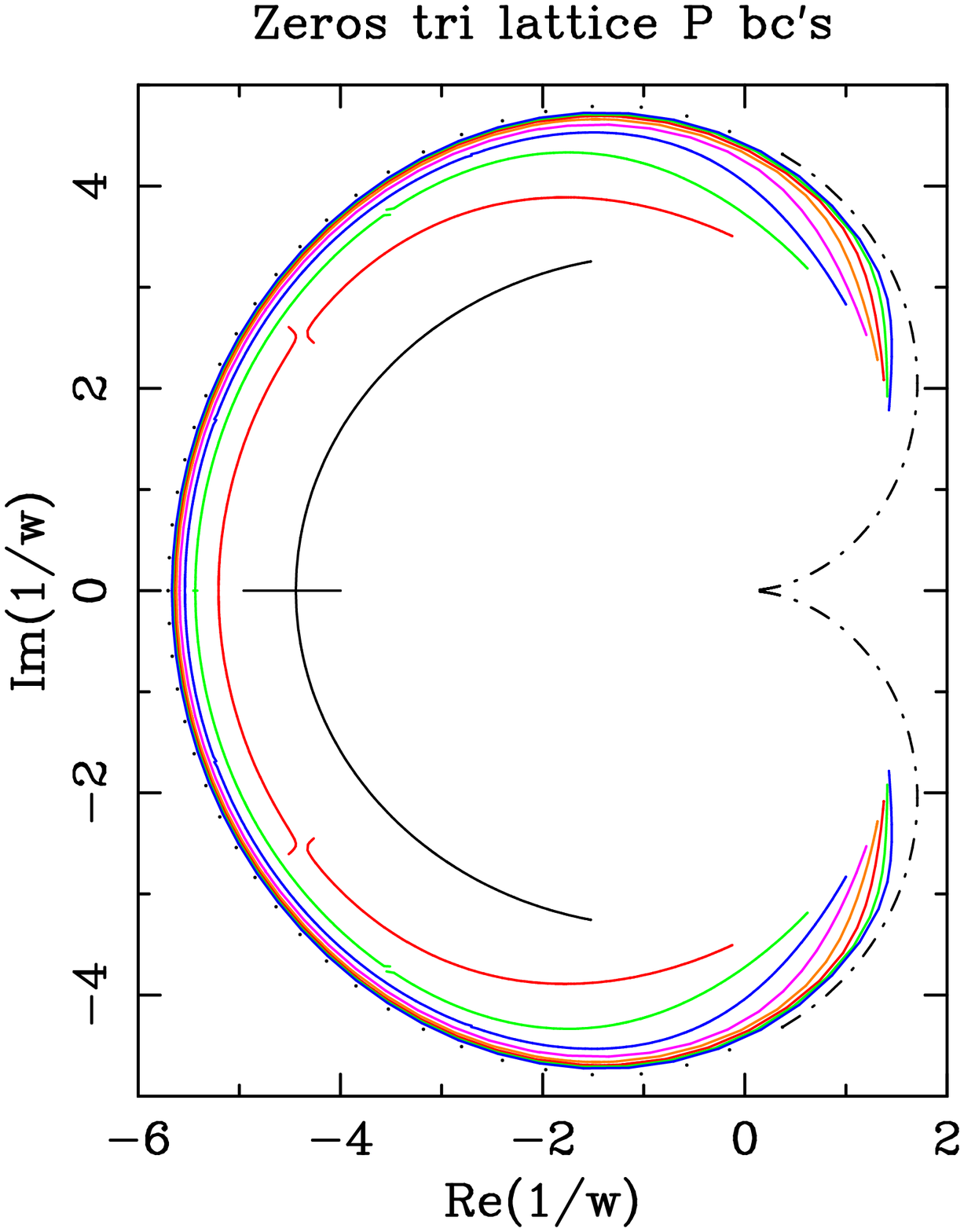}
\caption{\label{figure_tri_all_1overw}
Finite-$L$ limiting curves and phase diagram (Figure~\ref{figure_tri_all})
for the triangular lattice, mapped to the $1/w$ plane.
Our best estimate for the infinite-volume phase boundary ${\cal B}_\infty$
is depicted in black dots (for $-0.23 \ltapprox \imag w \ltapprox 0.23$)
and as a black dotted-dashed curve (very rough estimate).
}
\end{figure}

%
%
\clearpage
\begin{figure}
\centering
\begin{tabular}{cc}
\includegraphics[width=200pt]{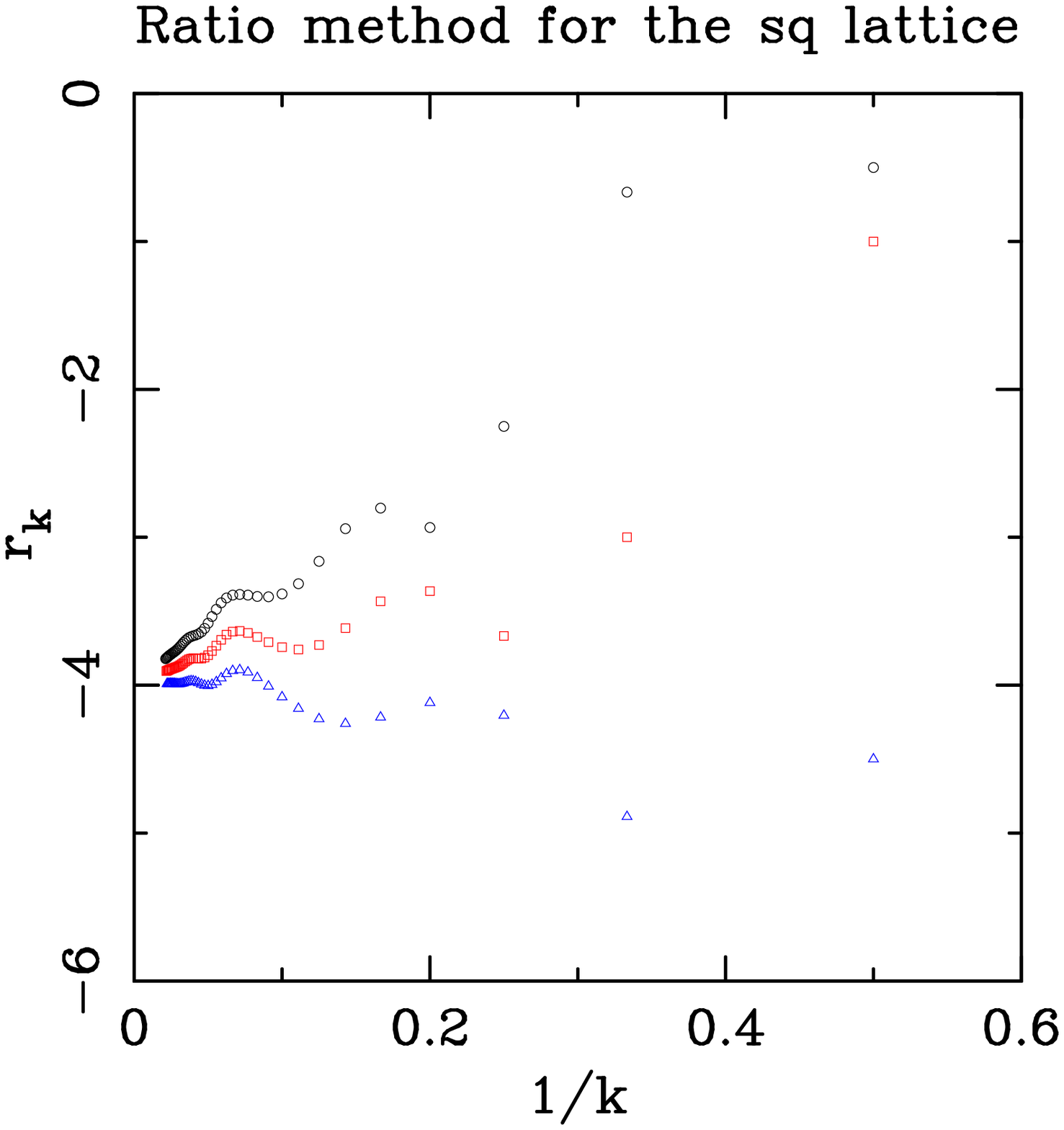}  &
\includegraphics[width=200pt]{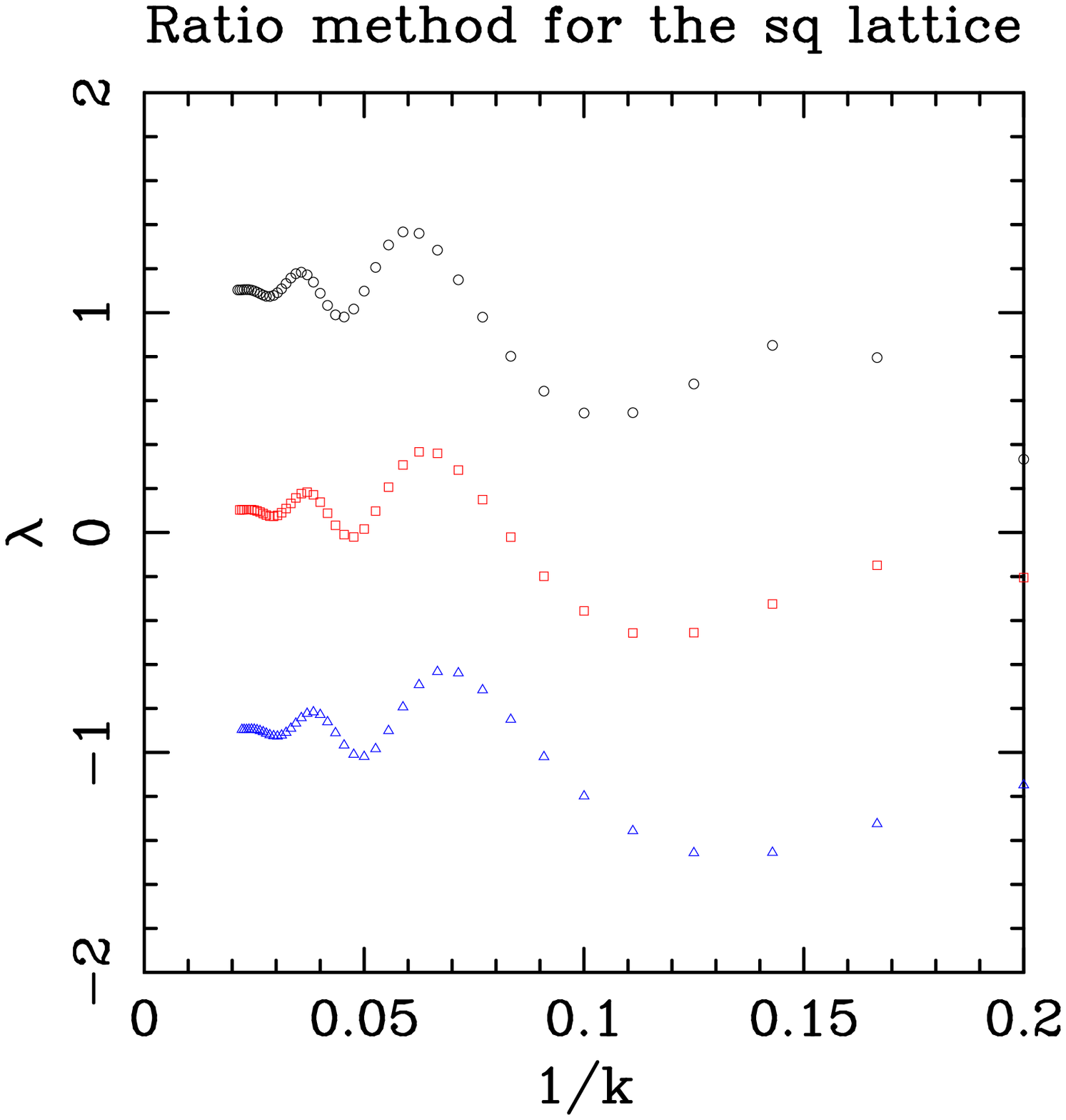} \\[1mm]
\phantom{(((a)}(a) & \phantom{(((a)}(b)
\end{tabular}
\caption{\label{figure_ratio_sq}
   Results of studying the small-$w$ series expansions for the square
   lattice using the ratio method. 
   (a) We show the ratio $r_k$ of two consecutive coefficients
   [cf.\ \protect\reff{def_ratios_rn}]
   as a function of $1/k$ for the 
   free energy $f(w)$ ($\circ$), its first 
   derivative $f'(w)$ ($\Box$), 
   and its second derivative $f''(w)$ ($\triangle$). 
   (b) We show the {\em biased}\/ estimate (based on $w_0 = -1/4$)
   for the critical parameter $\lambda$ 
   for the same three functions as above.
}
\end{figure}

%
%
\clearpage
\begin{figure}
\centering
\begin{tabular}{cc}
  \includegraphics[width=200pt]{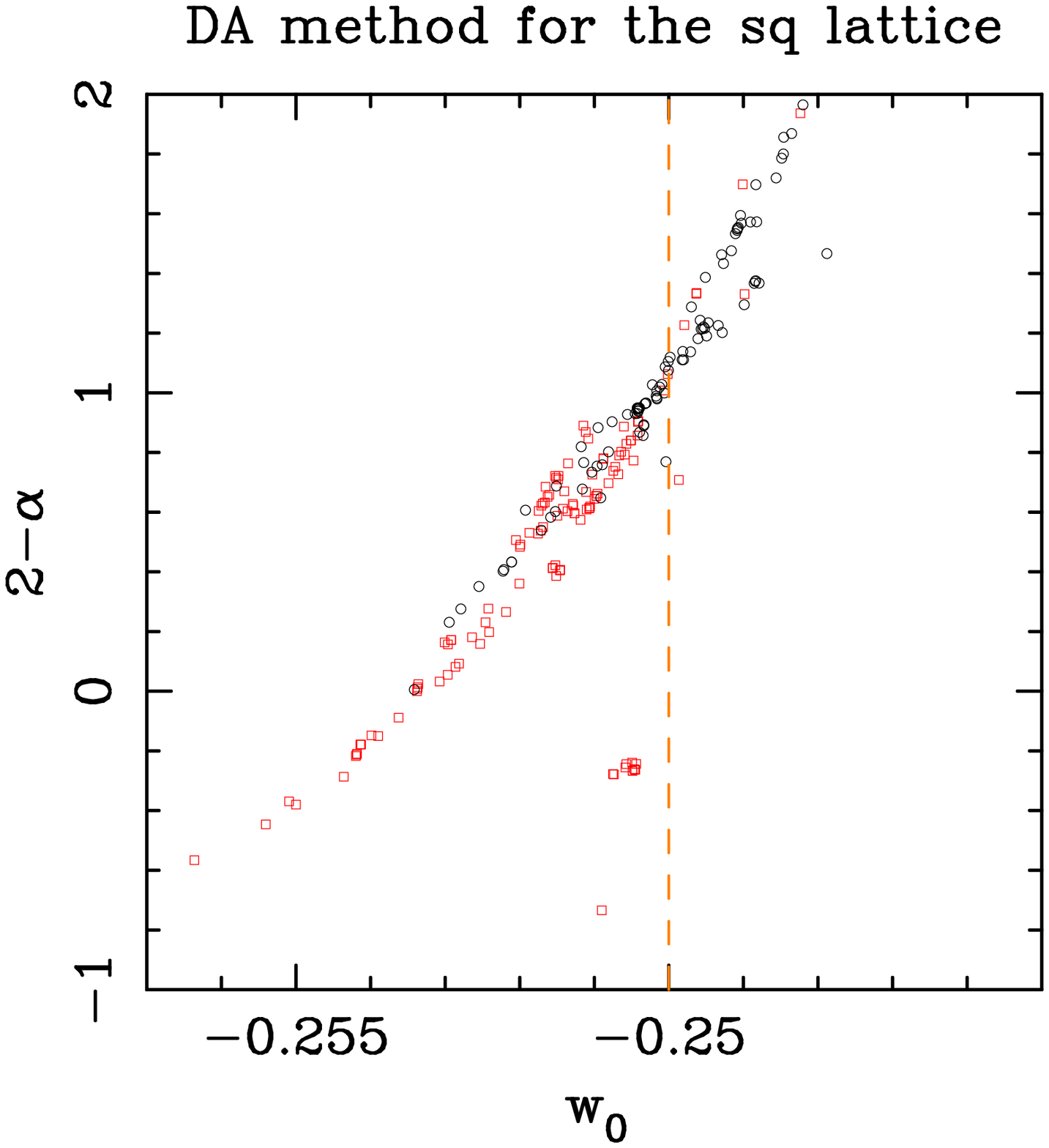} &  
  \includegraphics[width=212pt]{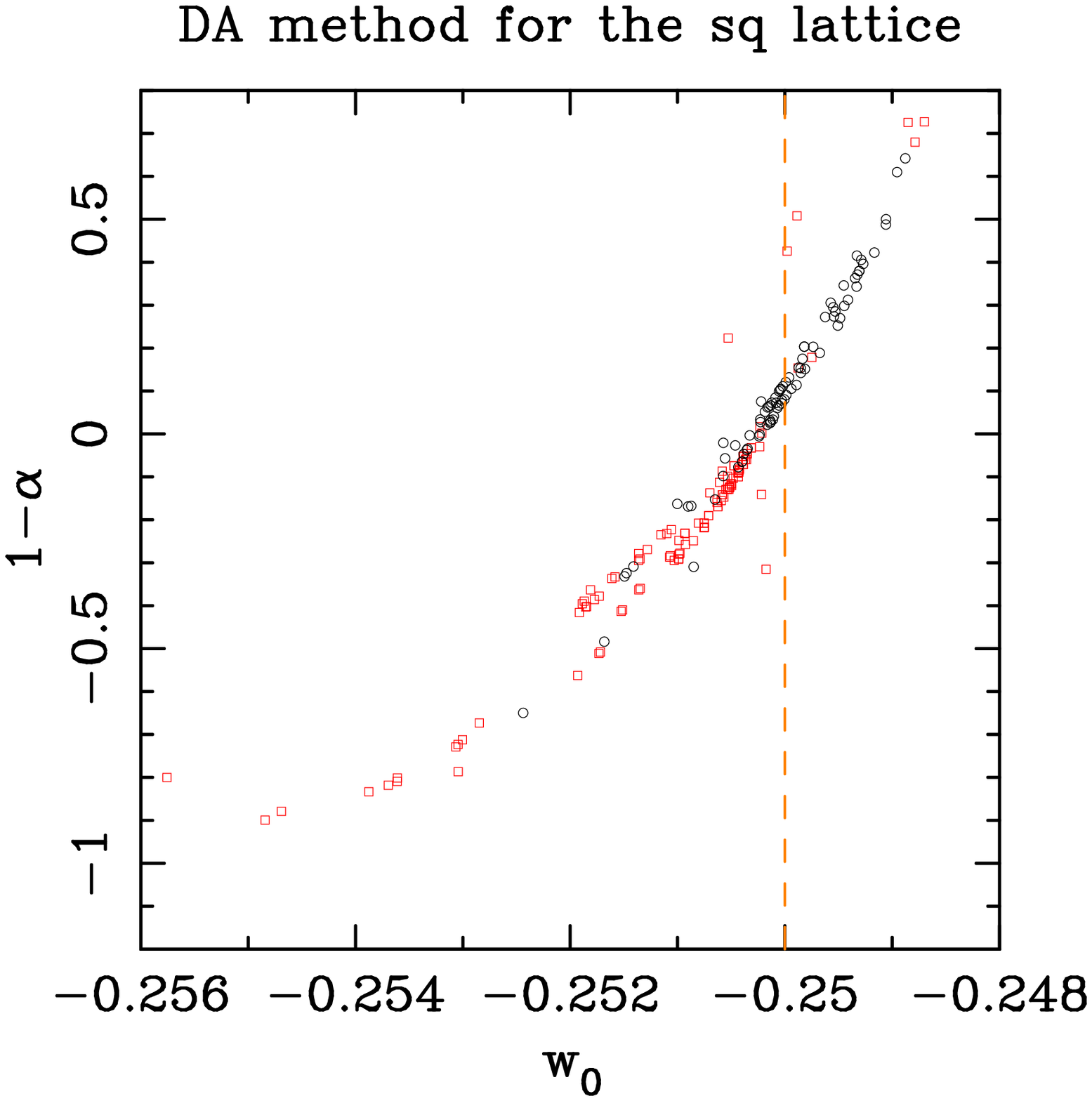} \\  
  \phantom{(((a)}(a) & \phantom{(((a)}(b)\\[1mm] 
  \includegraphics[width=200pt]{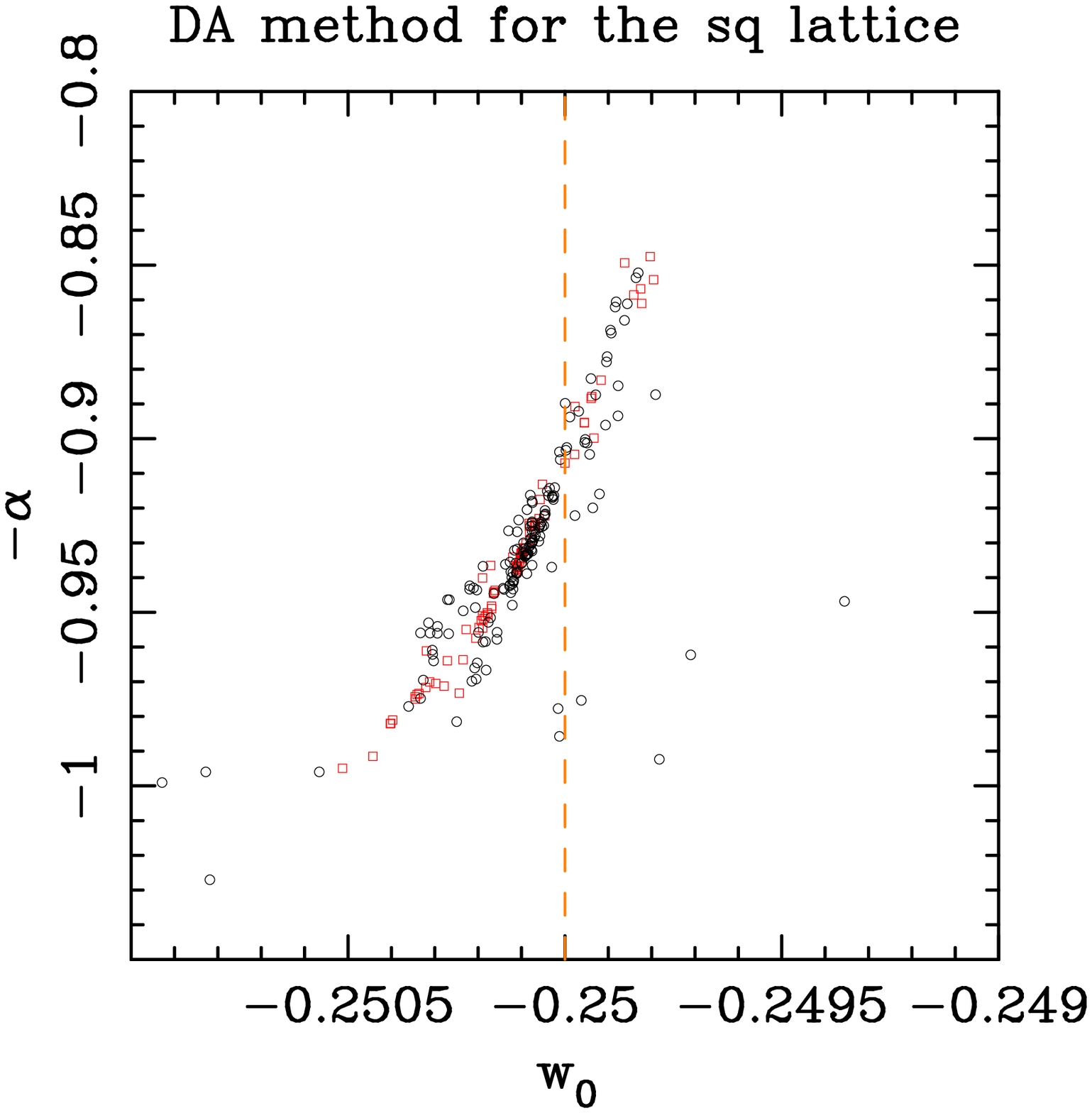} & 
  \includegraphics[width=200pt]{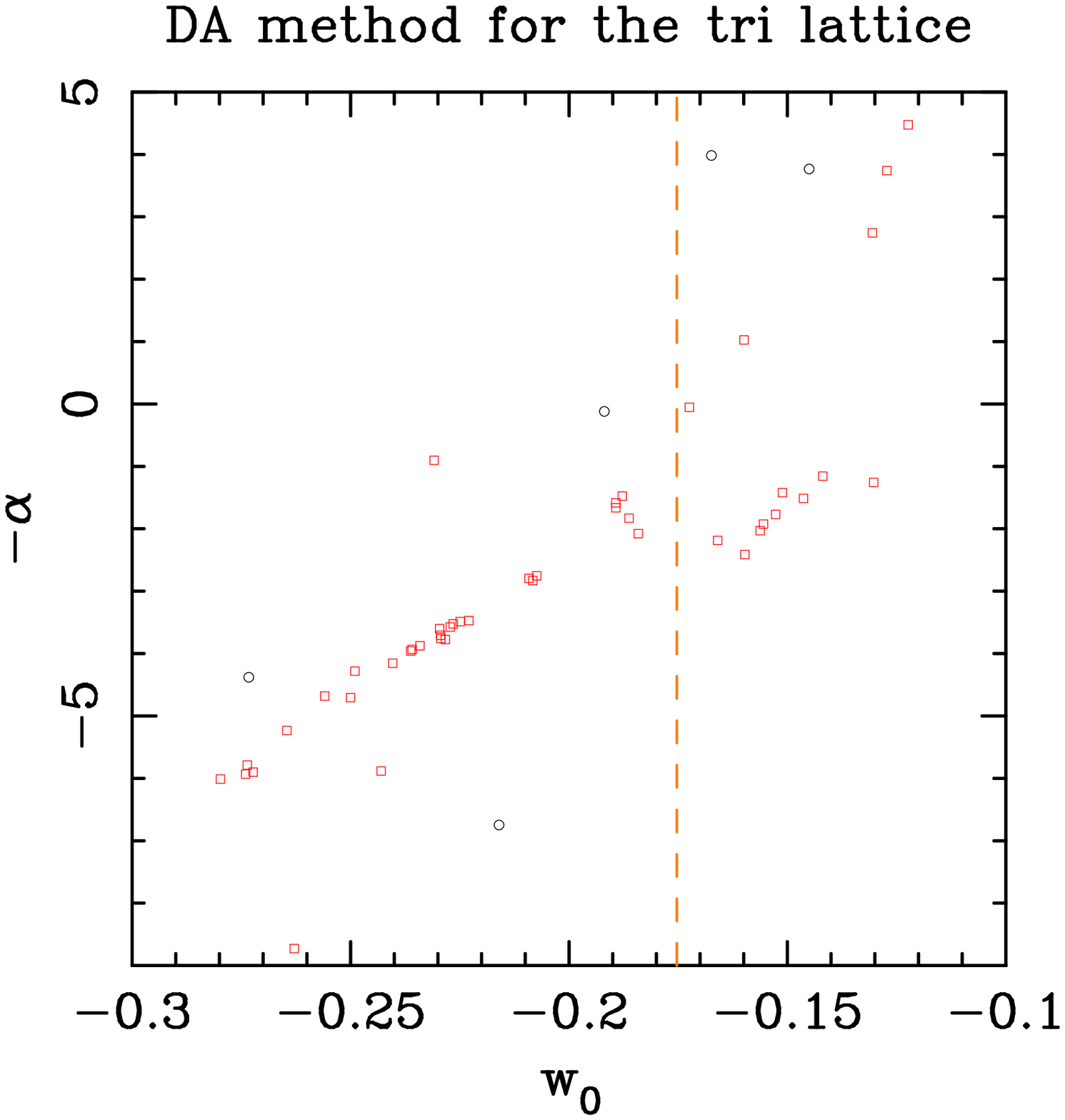} \\[1mm]
  \phantom{(((a)}(c) &  \phantom{(((a)}(d) 
\end{tabular}
\caption{\label{figure_DA_sq}
   Results of studying the small-$w$ series expansions for the square
   lattice using the differential-approximant method. 
   We show the estimate of the critical exponent $\lambda$ versus the
   location of the dominant singularity $w_0$
   for (a) the spanning-forest free energy $f(w)$,
   (b) its first derivative $f'(w)$, and
   (c) its second derivative $f''(w)$.
   The vertical dashed line marks the theoretical prediction
   $w_0({\rm sq}) = -1/4$. The results for first-order 
   (resp.\  second-order) approximants $K=1$ (resp.\  $K=2$) are denoted with 
   red squares $\Box$ (resp.\ black circles $\circ$). 
   In (d) we show the results for the second derivative $f''$ of the
   triangular-lattice free energy. In this case the vertical dashed line marks
   our best estimate for $w_0({\rm tri}) \approx -0.1753$.
}
\end{figure}


\end{document}